\documentclass[%
aps,
twocolumn,
10pt,
prd,
floatfix
]{revtex4-2}

\usepackage{graphicx}      
\usepackage{subcaption}    
\usepackage{amssymb}
\usepackage{amsmath}
\usepackage{dcolumn}       
\usepackage{bm}            
\usepackage{color}
\usepackage{comment}
\usepackage{enumerate}
\usepackage{booktabs}
\usepackage{multirow}
\usepackage{titlesec}

\titlespacing{\section}{5pt}{\baselineskip}{0.5\baselineskip}
\titlespacing{\subsection}{5pt}{\parskip}{-\parskip}

\usepackage{CJKutf8}
\usepackage{natbib}
\usepackage{verbatim}
\usepackage{lineno}
\usepackage{hyperref}
\hypersetup{
    hidelinks,
    colorlinks=true,
    allcolors=blue,
    pdfstartview=Fit,
    breaklinks=true
}

\DeclareUnicodeCharacter{0105}{\k{a}}
\newcommand{\ee}{e^{+}e^{-}}
\newcommand{\etap}{\eta^{\prime}}
\newcommand{\etac}{\eta_{c}}
\newcommand{\pip}{\pi^{+}}
\newcommand{\pim}{\pi^{-}}
\newcommand{\pippim}{\pi^{+}\pi^{-}}
\begin{document}

\preprint{APS/123-QED}
\makeatletter
\hypersetup{
    pdftitle={\@title},
}
\makeatother

\title{Search for $1^{-+}$ charmonium-like hybrid via \texorpdfstring{$e^{+}e^{-}\rightarrow \gamma \eta^{(\prime)} \eta_{c}$}{e+e- to gamma eta(eta') etac} at center-of-mass energies between 4.258 and 4.681 GeV } 

\author{
\begin{small}
  \begin{center}
M.~Ablikim$^{1}$, M.~N.~Achasov$^{4,c}$, P.~Adlarson$^{77}$, X.~C.~Ai$^{82}$, R.~Aliberti$^{36}$, A.~Amoroso$^{76A,76C}$, Q.~An$^{73,59,a}$, Y.~Bai$^{58}$, O.~Bakina$^{37}$, Y.~Ban$^{47,h}$
% ...
\\
\vspace{0.2cm}
(BESIII Collaboration)\\
\vspace{0.2cm} {\it
$^{1}$ Institute of High Energy Physics, Beijing 100049, People's Republic of China\\
% ...
$^{4}$ Budker Institute of Nuclear Physics SB RAS (BINP), Novosibirsk 630090, Russia\\
% ...
$^{36}$ Johannes Gutenberg University of Mainz, Johann-Joachim-Becher-Weg 45, D-55099 Mainz, Germany\\
$^{37}$ Joint Institute for Nuclear Research, 141980 Dubna, Moscow region, Russia\\
% ...
$^{47}$ Peking University, Beijing 100871, People's Republic of China\\
% ...
$^{58}$ Southeast University, Nanjing 211100, People's Republic of China\\
$^{59}$ State Key Laboratory of Particle Detection and Electronics, Beijing 100049, Hefei 230026, People's Republic of China\\
% ...
$^{73}$ University of Science and Technology of China, Hefei 230026, People's Republic of China\\
% ...
$^{76}$ University of Turin and INFN, (A)University of Turin, I-10125, Turin, Italy; (B)University of Eastern Piedmont, I-15121, Alessandria, Italy; (C)INFN, I-10125, Turin, Italy\\
$^{77}$ Uppsala University, Box 516, SE-75120 Uppsala, Sweden\\
% ...
$^{82}$ Zhengzhou University, Zhengzhou 450001, People's Republic of China\\

\vspace{0.2cm}
$^{a}$ Deceased\\
% ...
$^{c}$ Also at the Novosibirsk State University, Novosibirsk, 630090, Russia\\
% ...
$^{h}$ Also at State Key Laboratory of Nuclear Physics and Technology, Peking University, Beijing 100871, People's Republic of China\\
% ...

}

\end{center}
\end{small}
}

%
%
%


\begin{abstract}
Using $\ee$ collision data corresponding to an integrated luminosity of 10.6~fb$^{-1}$ collected at center-of-mass energies between 4.258 and 4.681 GeV with the BESIII detector at the BEPCII collider,
we search for the $1^{- +}$ charmonium-like hybrid via $e^{+}e^{-}\rightarrow\gamma\eta\eta_{c}$ and $e^{+}e^{-}\rightarrow\gamma\etap\eta_{c}$ decays for the first time. 
No significant signal is observed and the upper limits on the Born cross sections for both processes are set at the 90\% confidence level.

\end{abstract}

\maketitle

\section{INTRODUCTION} \label{sec:introduction}

\indent Quantum Chromodynamics (QCD) allows the existence of exotic hadron states beyond the quark model, such as glueballs, hybrids, multiquark states, and hadronic molecules. The search for these QCD exotic states is crucial for quantitatively testing the theory of strong interactions in non-perturbative regions and for understanding confinement. However, it is difficult to distinguish the exotic hadron states from the conventional mesons. Hadrons with exotic $J^{PC}$ quantum numbers, such as $0^{--}, even^{+-}, odd^{-+}$ cannot arise from ordinary $q\bar{q}$ states. Searching for such hadrons offers one of the most direct and unambiguous ways to establish the existence of exotic hadron states.\\
\indent Lattice QCD predicts the existence of a series of light hybrid multiplets ($J^{PC}=0^{-+},1^{--},1^{-+}$ and $2^{-+}$), with the exotic $1^{-+}$ nonet being the lightest~\cite{Dudek:2013yja}. Similar to the case of the hybrid multiplets composed of light quarks ($n\bar{n}g$), $c\bar{c}$ can also form analogous hybrid multiplets ($c\bar{c}g$)~\cite{HadronSpectrum:2012gic,Brambilla:2022hhi}.
In the charmonium sector, four confirmed states, $\psi(4230)$, $\psi(4360)$, $\psi(4390)$ and $\psi(4660)$, fall within the mass range expected for $1^{-+}$ hybrid charmonia~\cite{Brambilla:2019esw,ParticleDataGroup:2022pth,Olsen:2017bmm,Yuan:2021wpg}.\\
\indent Three light isovector states, the $\pi_{1}(1400)$, $\pi_{1}(1600)$, and $\pi_{1}(2015)$~\cite{Meyer:2010ku,Klempt:2007cp,JPAC:2018zyd,Woss:2020ayi}, have been experimentally identified as exotic hadrons exhibiting the quantum numbers $J^{PC}=1^{-+}$.
Recently, the BESIII collaboration performed a partial wave analysis of the $J/\psi \rightarrow \gamma \eta \eta^{\prime}$ decay process, and reported the first observation of a nonconventional state $\eta_{1}(1855)$ with the exotic quantum numbers $I^{G}J^{PC}=0^{+}1^{-+}$~\cite{BESIII:2022iwi,BESIII:2022riz}. However, no state with manifestly exotic $J^{PC}$ has been observed in the charmonium region. Analogous to the $\eta_1(1855)\rightarrow\eta\etap$ process, a possible $\eta_{c1}$ state could decay into $\eta^{(\prime)}\etac$ final states. Additionally, some theoretical predictions suggest that the $1^{- +}$ charmonium-like hybrid state might also decay to $\eta^{(\prime)}\eta_{c}$ final states~\cite{Dong:2019ofp,Shi:2023sdy}.\\
\indent In this paper, we report a study of $e^+e^-\rightarrow\gamma\eta^{(\prime)}\eta_c$ to search for the $1^{- +}$ charmonium-like hybrid state based on data samples collected with the BESIII detector at center-of-mass (c.m.) energies between 4.258 and 4.681 GeV, listed in Table~\ref{tab:summary}. The integrated luminosities of these data samples are determined by analyzing large-angle Bhabha scattering events with an uncertainty of $1.0\%$, and the c.m. energies are measured using the di-muon process~\cite{BESIII:2015qfd,BESIII:2022ulv}.\\
\section{BESIII DETECTOR AND MONTE CARLO SIMULATION} \label{sec:BESIII DETECTOR AND MONTE CARLO SIMULATION}
\indent The BESIII detector~\cite{BESIII} records symmetric $e^+e^-$ collisions provided by the BEPCII storage ring~\cite{BEPCII} in the c.m. energy range from 1.84 to 4.95 GeV,
with a peak luminosity of $1.1 \times 10^{33}\;\text{cm}^{-2}\text{s}^{-1}$ achieved at $\sqrt{s} = 3.773\;\text{GeV}$. 
BESIII has collected large data samples in this energy region~\cite{BESIII:2020nme,EventFilter}.
The cylindrical core of the BESIII detector covers 93\% of the full solid angle and consists of a helium-based multilayer drift chamber~(MDC), a time-of-flight
system~(TOF), and a CsI(Tl) electromagnetic calorimeter~(EMC), which are all enclosed in a superconducting solenoidal magnet providing a 1.0~T magnetic field.
The solenoid is supported by an octagonal flux-return yoke with resistive plate counter muon identification modules interleaved with steel. 
The charged-particle momentum resolution at $1~{\rm GeV}/c$ is
$0.5\%$, and the 
${\rm d}E/{\rm d}x$
resolution is $6\%$ for electrons
from Bhabha scattering. The EMC measures photon energies with a
resolution of $2.5\%$ ($5\%$) at $1$~GeV in the barrel (end cap)
region. The time resolution of the plastic scintillator TOF in the barrel region is 68~ps, while
that in the end cap region was 110~ps. In 2015, the end cap TOF
system was upgraded with multigap resistive plate chamber
technology, improving the time resolution to
60~ps, which benefits approximately 68\% of the data used in this analysis~\cite{li2017radiat, guo2017radiat, cao2020nucl}.\\

\indent Monte Carlo (MC) simulated data samples produced with a {\sc geant}4-based software package~\cite{geant4}, which includes the geometric description of the BESIII detector and the detector response, are used to determine detection efficiencies and to estimate backgrounds. The simulation models the beam energy spread and initial state radiation (ISR) in the $e^{+}e^{-}$ annihilations with the generator {\sc kkmc}~\cite{kkmc1,kkmc2}. The inclusive MC samples include the production of open-charm processes, the ISR production of vector charmonium(like) states, and the continuum processes incorporated in {\sc kkmc}. All particle decays are modeled with {\sc evtgen}~\cite{evtgen_1,evtgen_2} using branching fractions taken from the Particle Data Group (PDG)~\cite{PDG}, when available, or otherwise estimated with {\sc lundcharm}~\cite{lundcharm}. Final-state radiation from charged final-state particles is incorporated using the {\sc photos} package~\cite{Barberio:1990ms}.

\sloppy

\section{EVENT SELECTION AND STUDY OF BACKGROUND} \label{sec:EVENT SELECTION}
\begin{figure*}[htbp]
    \begin{subfigure}{0.32\textwidth}
        \includegraphics[width=\linewidth]{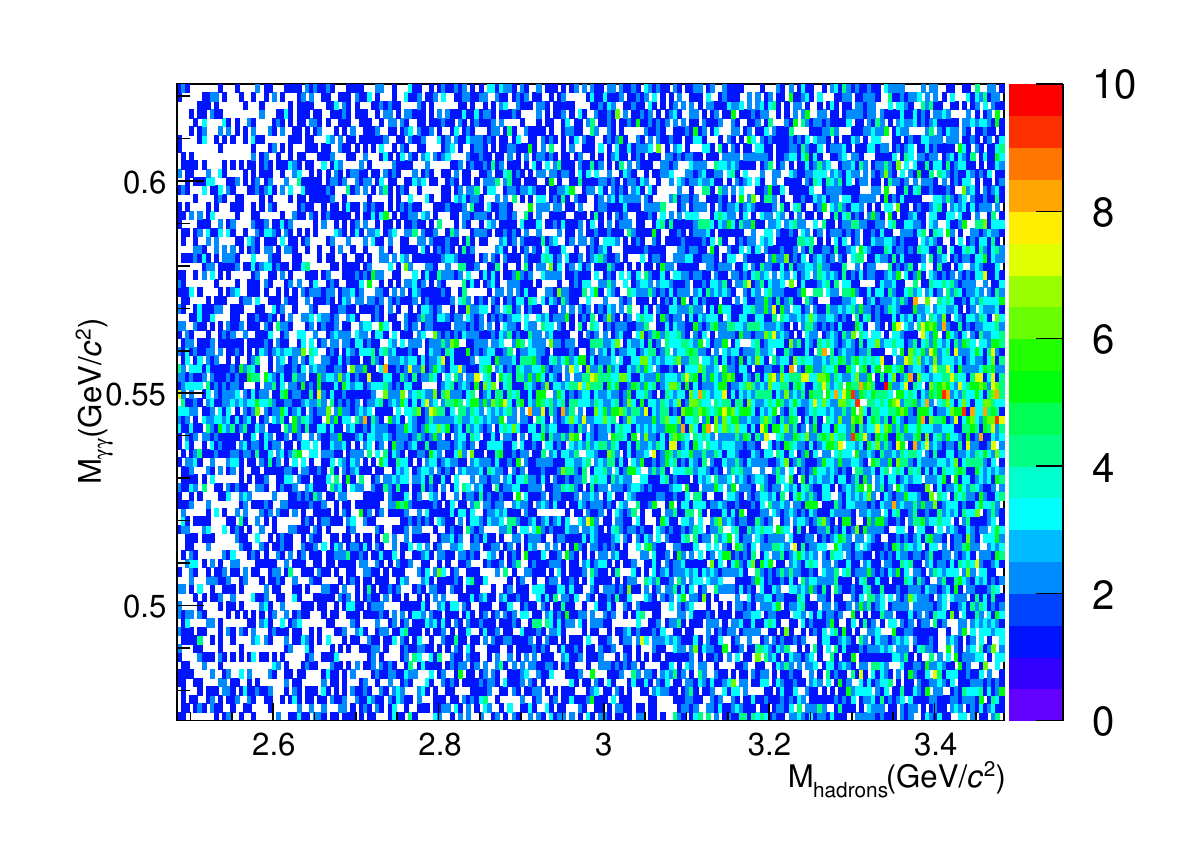}
        \captionsetup{skip=-7pt,font=normalsize}
        \caption{}
        \label{subfig:eta_2d_data}
    \end{subfigure}
    \begin{subfigure}{0.32\textwidth}
        \includegraphics[width=\linewidth]{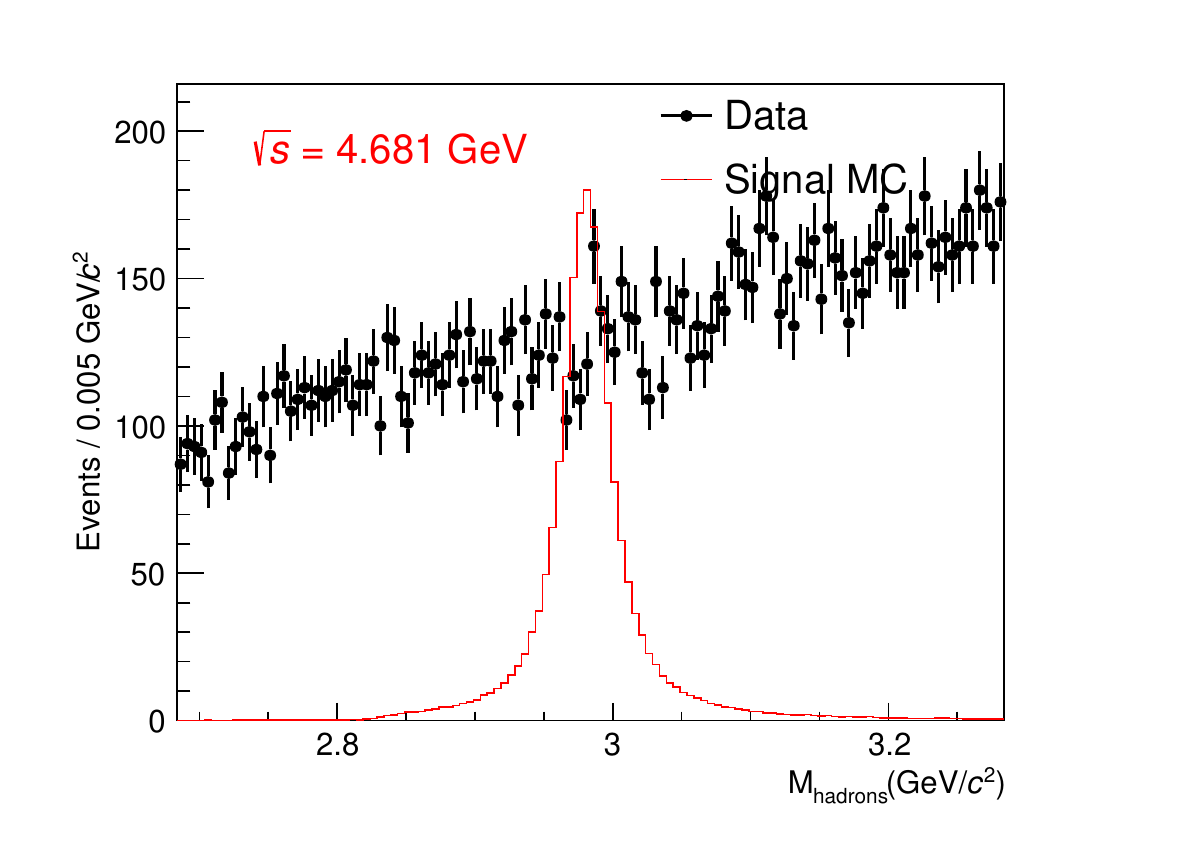}
        \captionsetup{skip=-7pt,font=normalsize}
        \caption{}
        \label{subfig:eta_2d_etac}
    \end{subfigure}
    \begin{subfigure}{0.32\textwidth}
        \includegraphics[width=\linewidth]{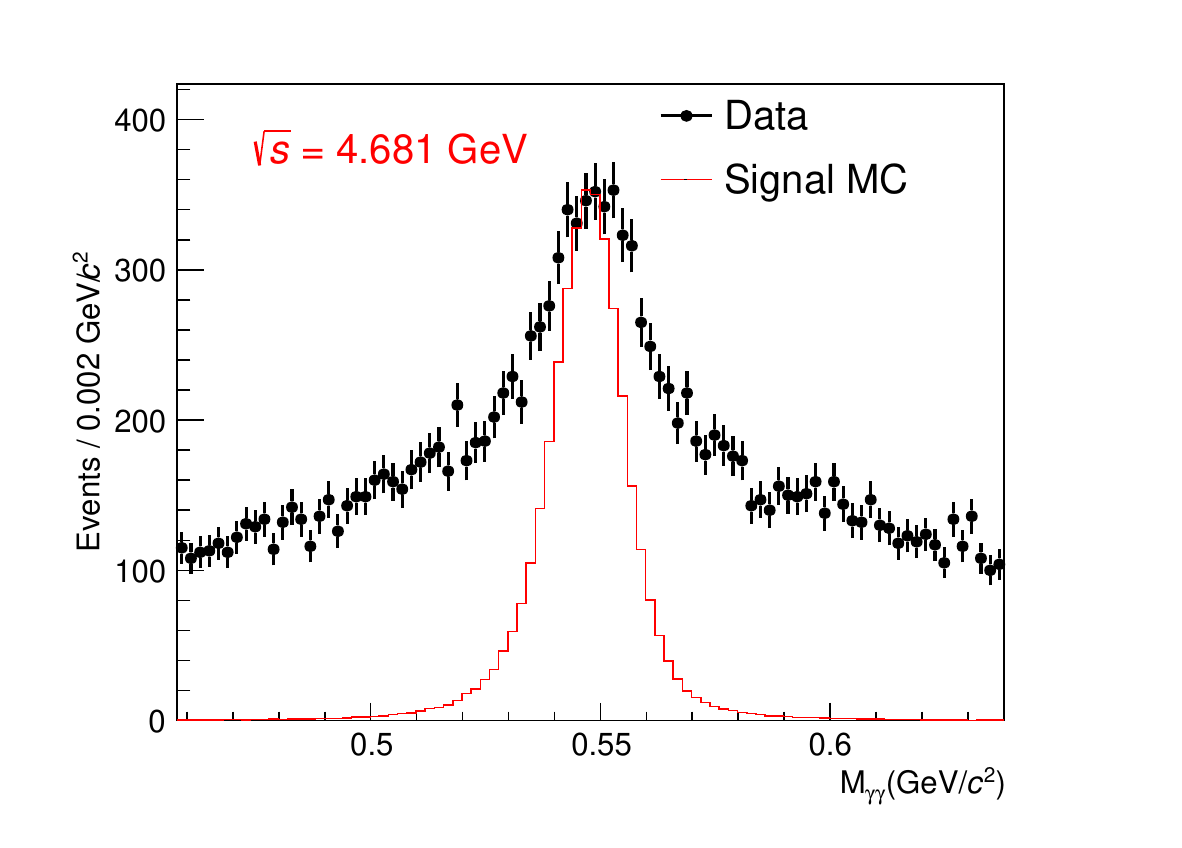}
        \captionsetup{skip=-7pt,font=normalsize}
        \caption{}
        \label{subfig:eta_2d_eta}
    \end{subfigure}
  \captionsetup{justification=raggedright}
  \caption{The distributions of (a) $M_{\gamma\gamma}$ versus $M_{hadrons}$, (b) $M_{hadrons}$, and (c) $M_{\gamma\gamma}$ at $\sqrt s=4.681$~GeV.
 For (b) and (c), the dots with error bars are data and the red histograms represent the signal MC. }
  \label{fig:dis_eta}
\end{figure*}

\begin{figure*}[htbp]
    \begin{subfigure}{0.32\textwidth}
        \includegraphics[width=\linewidth]{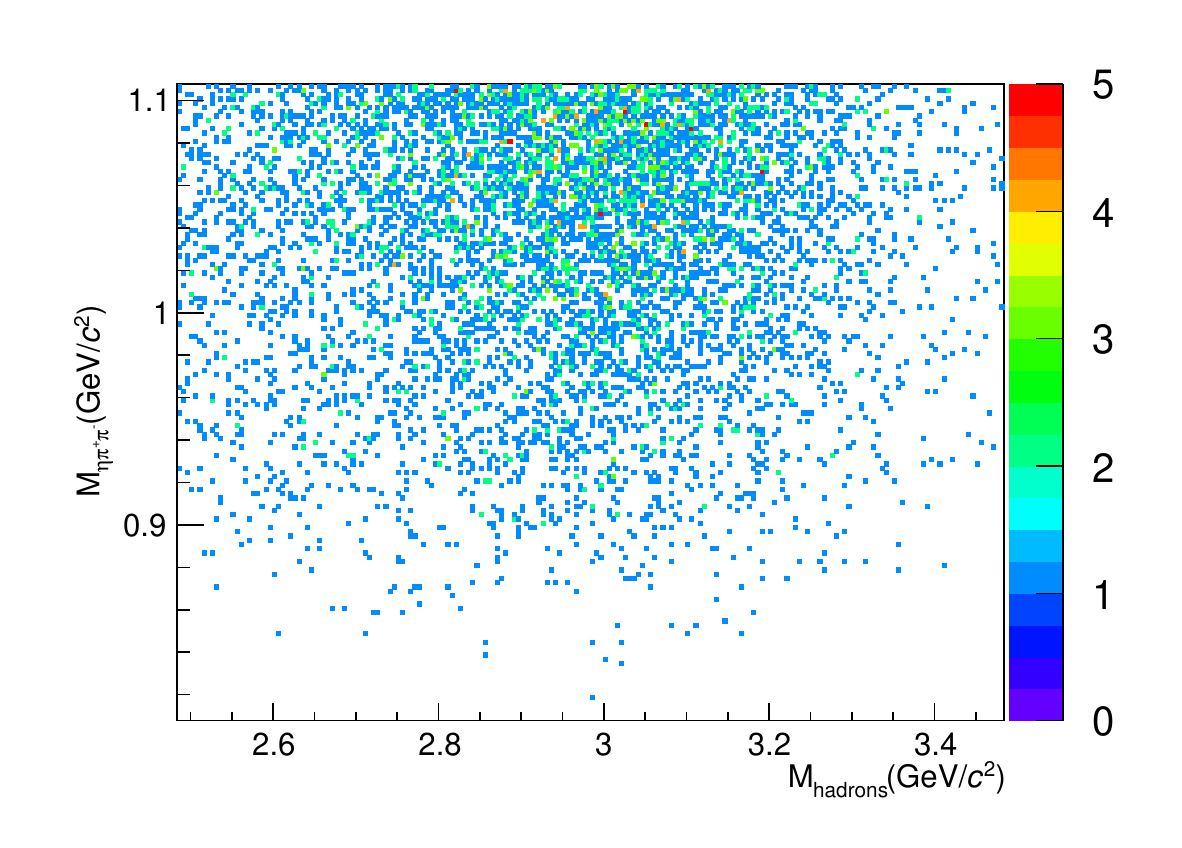}
        \captionsetup{skip=-7pt,font=normalsize}
        \caption{}
        \label{subfig:etap_2d_data}
    \end{subfigure}
    \begin{subfigure}{0.32\textwidth}
        \includegraphics[width=\linewidth]{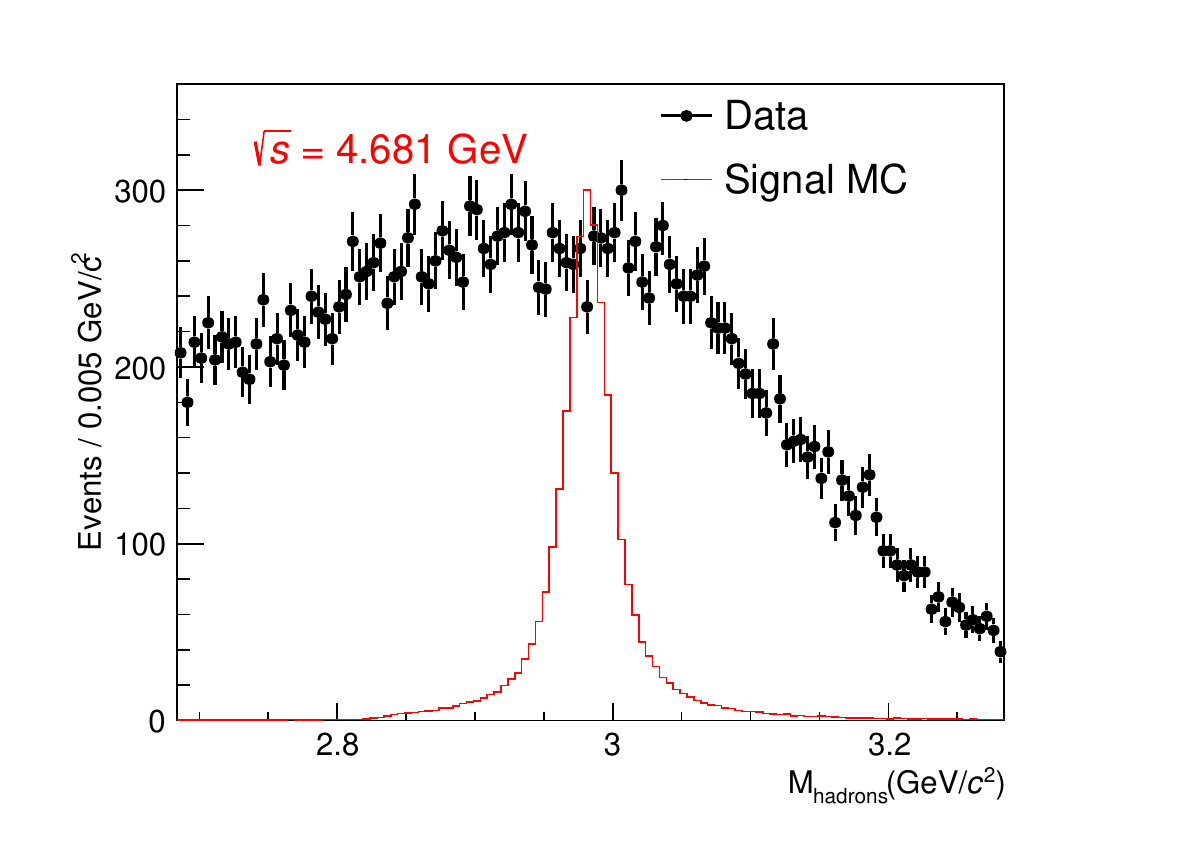}
        \captionsetup{skip=-7pt,font=normalsize}
        \caption{}
        \label{subfig:etap_2d_etac}
    \end{subfigure}
    \begin{subfigure}{0.32\textwidth}
        \includegraphics[width=\linewidth]{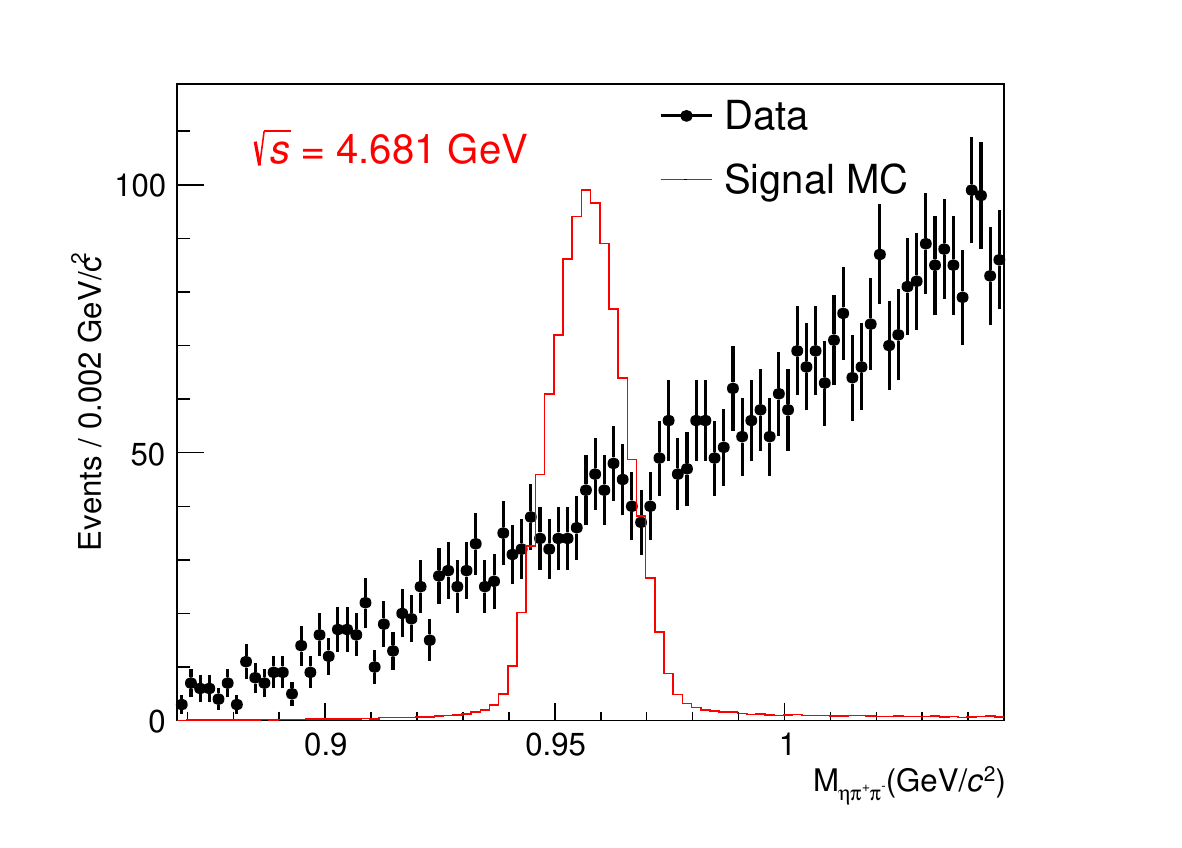}
        \captionsetup{skip=-7pt,font=normalsize}
        \caption{}
        \label{subfig:etap_2d_eta}
    \end{subfigure}
  \captionsetup{justification=raggedright}
  \caption{The distributions of (a) $M_{\eta\pi^+\pi^-}$ versus $M_{hadrons}$, (b) $M_{hadrons}$, and (c) $M_{\eta\pi^+\pi^-}$ at $\sqrt s=4.681$~GeV. 
  For (b) and (c), the dots with error bars are data and the red histograms represent the signal MC.}
  \label{fig:dis_etap}
\end{figure*}

To select candidate events for $\ee\rightarrow\gamma\eta^{(\prime)}\etac$, the $\eta^{\prime}$ is reconstructed through the final states of $\eta\pippim$, the $\eta_c$ is reconstructed through 16 hadronic final states: $p\bar{p}$, $2(\pi^{+}\pi^{-})$, $2(K^{+}K^{-})$, $K^{+}K^{-}\pi^{+}\pi^{-}$, $p\bar{p}\pi^{+}\pi^{-}$, $3(\pi^{+}\pi^{-})$, $K^{+}K^{-}2(\pi^{+}\pi^{-})$, $K^{+}K^{-}\pi^{0}$, $p\bar{p}\pi^{0}$, $K_{S}^{0}K^{\pm}\pi^{\mp}$, $K_{S}^{0}K^{\pm}\pi^{\mp}\pi^{\pm}\pi^{\mp}$, $\pi^{+}\pi^{-}\eta$, $K^{+}K^{-}\eta$, $2(\pi^{+}\pi^{-})\eta$, $\pi^{+}\pi^{-}\pi^{0}\pi^{0}$, and $2(\pi^{+}\pi^{-})\pi^{0}\pi^{0}$, in which $K_S^0$ is reconstructed from its $\pi^+\pi^-$ decay, $\pi^0$ and $\eta$ from their $\gamma\gamma$ final state.\\
\indent Charged tracks detected in the MDC must be within a polar angle ($\theta$) range of $|\cos\theta|<0.93$, where $\theta$ is defined with respect to the $z$-axis, the symmetry axis of the MDC. For charged tracks not originating from $K_S^0$ decays, the distance of closest approach to the interaction point (IP) must be less than 10\,cm along the $z$-axis, $|V_{z}|$, and less than 1\,cm in the transverse plane, $|V_{xy}|$. By combining the energy deposit d$E$/d$x$ and the TOF information, the $\chi^2_{\rm PID}(i)($ where $i=K,\pi $ or $ p)$ is calculated for each charged track under different hadron hypothesis. Both PID and kinematic fit information are used to determine the particle type of each charged track.\\
\indent Photon candidates are identified using isolated showers in the EMC.  The deposited energy of each shower must exceed 25~MeV in the barrel region ($|\cos \theta|< 0.80$) and 50~MeV in the end cap region ($0.86 <|\cos \theta|< 0.92$). To exclude showers that originate from charged tracks, the angle between the EMC shower position and the closest charged track at the EMC must exceed 10 degrees as measured from the IP. To suppress electronic noise and unrelated showers, the difference between the EMC time and the event start time must be within the range [0,700] ns.\\
\indent Each $K_{S}^0$ candidate is reconstructed from two oppositely charged tracks satisfying $|V_{z}|<$ 20~cm. The two charged tracks are assigned as $\pi^+\pi^-$ without imposing additional PID criteria. They are constrained to originate from a common vertex and are required to have an invariant mass within $|M_{\pi^{+}\pi^{-}} - m_{K_{S}^{0}}|<$ 20~MeV$/c^{2}$, where $m_{K_{S}^{0}}$ is the $K^0_{S}$ nominal mass~\cite{PDG}. The decay length of the $K^0_S$ candidate is required to be greater than
twice the vertex resolution away from the IP. The $\pi^0(\eta)$ candidates are selected with the invariant mass of $\gamma\gamma$ pair satisfying $|M_{\gamma\gamma}-m_{\pi^0(\eta)}|<15~\rm{MeV}/c^2$, where $m_{\pi^0(\eta)}$ is the nominal mass of $\pi^0(\eta)$~\cite{PDG}. To improve the energy resolution, a one-constraint (1C) kinematic fit is performed with a constraint on the $\pi^0(\eta)$ mass.\\
\indent 
After the above selections, four-constraint (4C) kinematic fits are performed for each event imposing overall energy-momentum conservation with $3\gamma(
\pippim)+hadrons$ hypothesis, where $hadrons$ represents the corresponding final states of the $\etac$ decay modes. The chi-square values of 4C kinematic fit $\chi^2_{4\mathrm{C}(3\gamma+hadrons)}$ and $\chi^2_{4\mathrm{C}(3\gamma\pippim+hadrons)}$ are required to be less than 20 and 30, respectively, to suppress background events with different final states. For each $\etac$ decay mode, if there are multiple combinations satisfying the above criteria, only the one with the minimum total chi-square $\chi^2_{\mathrm{tot}}=\chi^2_{4\mathrm{C}(3\gamma(\pippim)+hadrons)}+\chi^2_{\mathrm{PID}}+\chi^2_{1\mathrm{C}}+\chi^2_{\mathrm{vertex}}$ is retained, where $\chi^{2}_{\mathrm{PID}}$ is the sum of the $\chi^2_{\mathrm{PID}}(i)$ for each charged track in the event, $\chi^{2}_{1\mathrm{C}}$ is from the 1C kinematic fit for the $\pi^0(\eta)$ mass, and the $\chi^{2}_{\mathrm{vertex}}$ is the $\chi^{2}$ of the $K^{0}_{S}$ secondary-vertex fit. If there is no $\pi^{0}/\eta$($K^{0}_{S}$) in an event, the corresponding $\chi^{2}_{1\mathrm{C}}$($\chi^{2}_{\mathrm{vertex}}$) is set to zero. \\
\indent For the $\ee\rightarrow\gamma\eta\etac$ process, 4C kinematic fits are performed by constraining energy-momentum conservation under the hypotheses $e^{+}e^{-}\rightarrow2\gamma+hadrons$ and $e^{+}e^{-}\rightarrow4\gamma + hadrons$, to suppress the backgrounds from processes with two or four photons in the final state. The $\chi^{2}_{4\mathrm{C}(3\gamma+hadrons)}$ is required to be less than all possible $\chi^{2}_{4\mathrm{C}(2\gamma+hadrons)}$ and $\chi^{2}_{4\mathrm{C}(4\gamma+hadrons)}$.
In case of multiple $\etac$ candidates in an event, the one with the invariant mass closest to the $\eta_{c}$ nominal mass is selected to reconstruct the $\eta_{c}$ meson. Finally, the remaining $\eta$ candidate is chosen to reconstruct the $\eta$ meson from the $\ee$ collision.\\
\indent For the $\ee\rightarrow\gamma\eta^{\prime}\etac$ process, the $\etac$ and $\etap$ candidates are formed from $hadrons$ and $\eta\pippim$ combinations, respectively, with the least $\frac{(M_{hadrons} - m_{\eta_{c}})^{2}}{\sigma_{m_{\eta_{c}}}^{2}} + \frac{(M_{\eta\pi^+\pi^-} - m_{\eta'})^{2}}{\sigma_{m_{\eta'}}^{2}}$. Here, $m_{\etac}$ and $m_{\etap}$ are the nominal masses of $\etac$ and $\etap$, respectively\cite{PDG}, $M_{hadrons}$ is the invariant mass of the 16 different decay modes of $\eta_{c}$ mentioned above, $M_{\eta\pi^{+}\pi^{-}}$ is the invariant mass of $\eta\pi^{+}\pi^{-}$, $\sigma_{m_{\eta'}}$ and $\sigma_{m_{\eta_{c}}}$ are the visible widths, determined by fitting with a Gaussian function and a Breit-Wigner function the individual distributions, respectively.\\
\indent After applying the reconstructed selection criteria to all data samples, the two-dimensional distributions of $M_{\gamma\gamma}$ versus $M_{hadrons}$ and $M_{\eta\pippim}$ versus $M_{hadrons}$ in data and MC simulation at $\sqrt{s}=4.681~{\rm{GeV}}$ are shown in Figs.~\ref{subfig:eta_2d_data} and \ref{subfig:etap_2d_data}, where $M_{\gamma\gamma(\eta\pippim)}$ is the invariant mass of the reconstructed $\eta^{(\prime)}$ from $\ee$ collision.
No significant $\ee\rightarrow\gamma\eta^{(\prime)}\etac$ signals are observed in the projections of $M_{\gamma\gamma(\eta\pippim)}$ versus $M_{hadrons}$. The distributions of $M_{hadrons}$ and $M_{\gamma\gamma(\eta\pippim)}$ at $\sqrt{s}=4.681~$GeV are shown in Figs.~\ref{subfig:eta_2d_etac}, \ref{subfig:eta_2d_eta}, \ref{subfig:etap_2d_etac}, and \ref{subfig:etap_2d_eta}.\\
\indent Since the luminosity of data samples at $\sqrt{s}=4.681~{\rm{GeV}}$ is the largest, we use the inclusive MC sample at this energy point, with the luminosity 40 times larger than the data to study the background components. After applying the mentioned selection criteria to the inclusive MC sample, no significant peaking backgrounds containing both $\etac$ and $\eta^{(\prime)}$ simultaneously are observed. However, for the $\ee\rightarrow\gamma\eta\etac$ process, there are some background events containing $\eta$ mesons. These backgrounds are considered in the two-dimensional fit as shown in Sec.~\ref{sec:SIGNAL YIELD EXTRACTION}.\\

\section{SIGNAL YIELD EXTRACTION} \label{sec:SIGNAL YIELD EXTRACTION}
\begin{figure*}[htbp]
  \begin{subfigure}{0.45\textwidth}
    \centering
    \includegraphics[width=\textwidth]{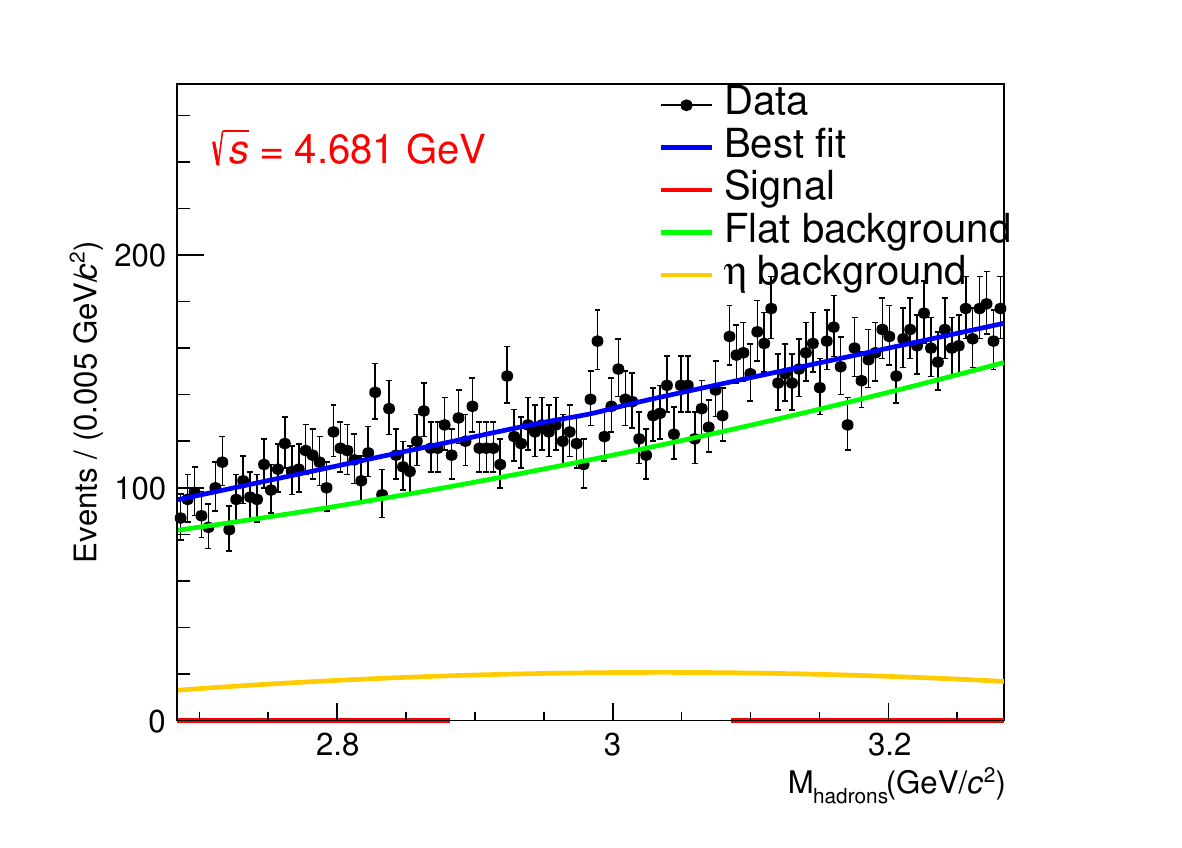}
    \captionsetup{skip=-10pt,font=large}
    \caption{}
    \label{fig:fit_eta_etac}
  \end{subfigure}
  \begin{subfigure}{0.45\textwidth}
    \centering
    \includegraphics[width=\textwidth]{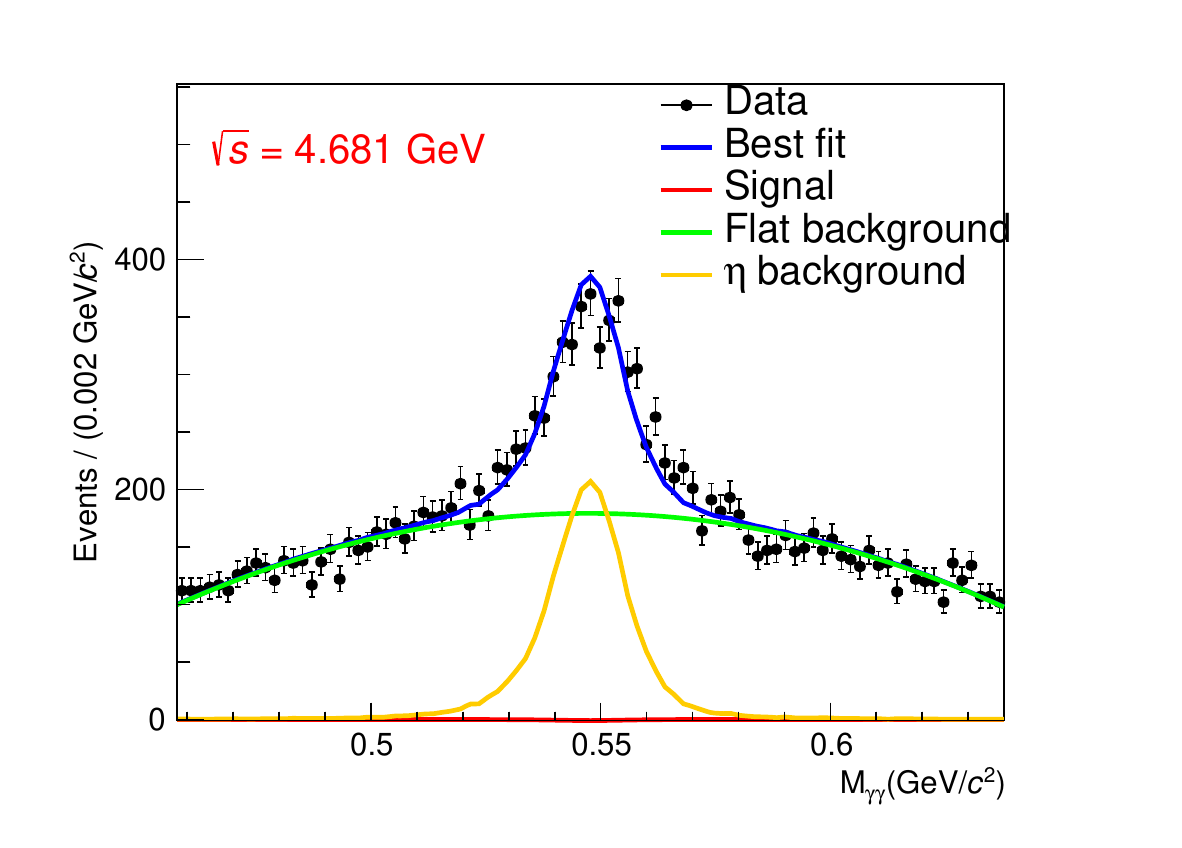}
    \captionsetup{skip=-10pt,font=large}
    \caption{}
    \label{fig:fit_eta_eta}
  \end{subfigure}
   \begin{subfigure}{0.45\textwidth}
    \centering
    \includegraphics[width=\textwidth]{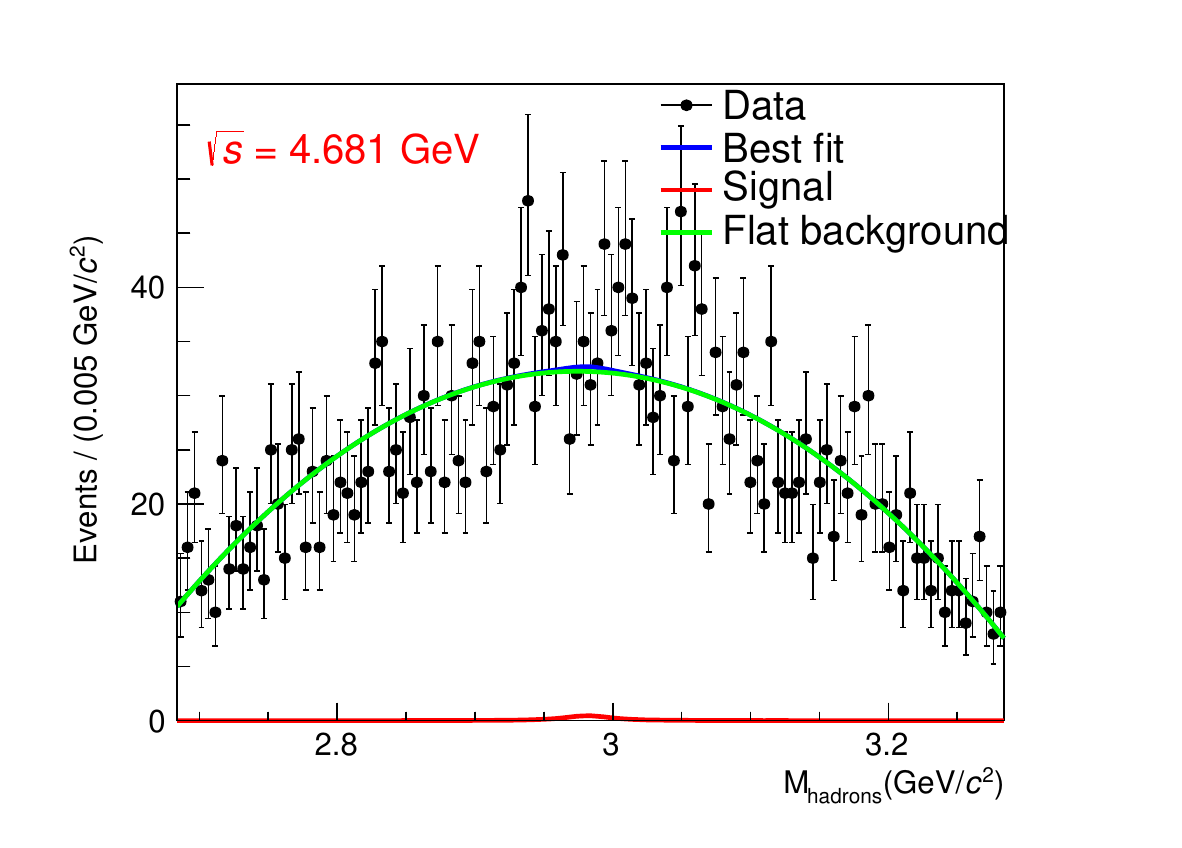}
    \captionsetup{skip=-10pt,font=large}
    \caption{}
    \label{fig:fit_etap_etac}
  \end{subfigure}
    \begin{subfigure}{0.45\textwidth}
    \centering
    \includegraphics[width=\textwidth]{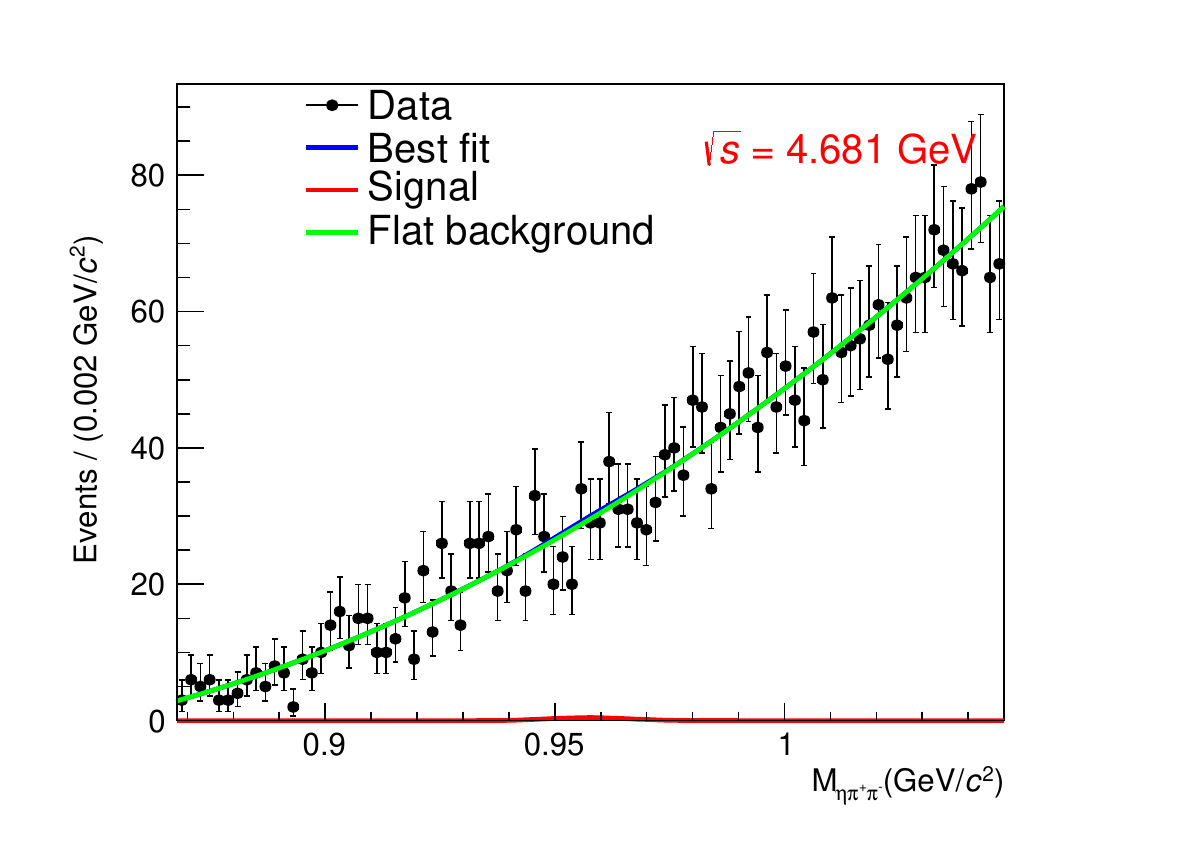}
    \captionsetup{skip=-10pt,font=large}
    \caption{}
    \label{fig:fit_etap_etap}
  \end{subfigure}
  \captionsetup{justification=raggedright}
  \caption{Fits to the invariant mass distributions of (a)(c)$M(hadrons)$, (b)$M(\gamma\gamma)$ and (d)$M(\eta\pi^{+}\pi^{-})$ at $\sqrt s=4.681$~GeV. The black dots with error bars are data, the red solid lines are the MC signals, the green solid lines are the flat background and the orange solid lines indicate the background containing $\eta$.}
  \label{fig:fit}
\end{figure*}

\indent The signal yield of the $\ee\rightarrow\gamma\eta^{(\prime)}\etac$ process is determined through an unbinned maximum likelihood fit to the two-dimensional distribution of $M_{\gamma\gamma(\eta\pippim)}$ versus $M_{hadrons}$. The fit ranges are set as follows: [0.458,0.638]~GeV$/c^2$(for $M_{\gamma\gamma}$), [0.868,1.048]~GeV$/c^2$(for $M_{\eta\pip\pim}$) and [2.684,3.284]~GeV$/c^2$(for $M_{hadrons}$).
The signal is described by a combination of 16 weighted shapes obtained from the signal MC, where the weight is $\frac{\mathcal{B}_{i}\epsilon_{i}}{\sum_{i}{\mathcal{B}_{i}\epsilon_{i}}}$. Here, $\mathcal{B}_{i}$ denotes the branching fraction of the $\eta_{c}$ decay to the $i$th decay mode and $\epsilon_i$ represents the detection efficiency obtained from MC samples. For the $\ee\rightarrow\gamma\eta\etac$ process, the background is described by two parts. A two-dimensional probability density function (PDF) (the MC shape of $\eta$ obtained from the signal MC sample in the $M_{\gamma\gamma}$ distribution  and a second-order Chebyshev polynomial function in the $M_{hadrons}$ distribution) is used to describe the background containing only $\eta$, while the remaining backgrounds are described by a two-dimensional second-order Chebyshev polynomial. For the $\ee\rightarrow\gamma\etap\etac$ process, since there is no background containing $\etap$, background is described solely by a two-dimensional second-order Chebyshev polynomial.
The fit results at $\sqrt{s}=4.681~{\rm{GeV}}$ are shown in Fig.~\ref{fig:fit}. The signal yield of $\ee\rightarrow\gamma\eta^{(\prime)}\etac$ at $\sqrt{s}=4.681~{\rm{GeV}}$ is $70.3\pm59.0(-9.9\pm13.4)$ with a statistical significance of 1.0$\sigma$(0.7$\sigma$). The significance is calculated from the change of the negative log-likelihood function ${\rm lnL}$ with and without assuming the presence of signal, while considering the change in degrees of freedom in the fits. The significance at other energy points is listed in Table~\ref{tab:summary}. \\
\\
 
\section{UPPER LIMIT ON BORN CROSS SECTIONS} \label{sec:UPPER LIMIT ON BORN CROSS SECTIONS}
\indent The Born cross sections of $\ee\rightarrow\gamma\eta^{(\prime)}\etac$ are calculated by
 \begin{equation}
  \begin{aligned}
  \sigma^{\rm{Born}}=\frac{N^{\rm{obs}}}{\mathcal{L}\cdot(1+\delta)\cdot\frac{1}{|1-\Pi|^2}\cdot\mathcal{B}\cdot\mathcal\epsilon}.
    \end{aligned}
    \label{eq:born}
\end{equation}
Here, $N^{\rm{obs}}$ is the signal yield, $\mathcal{L}$ is the integrated luminosity of the data sample
 taken at each c.m. energy and $\frac{1}{|1-\Pi|^2}$ is the vacuum polarization factor~\cite{WorkingGrouponRadiativeCorrections:2010bjp}. For $\ee\rightarrow\gamma\eta\etac$, $\mathcal{B}\cdot\epsilon=\mathcal{B}(\eta\rightarrow\gamma\gamma)\sum_{i}\mathcal{B}_{i}\mathcal\epsilon_{i}$, while for $\ee\rightarrow\gamma\etap\etac$, $\mathcal{B}\cdot\epsilon=\mathcal{B}(\eta'\rightarrow\eta\pi^{+}\pi^{-})\mathcal{B}(\eta\rightarrow\gamma\gamma)\sum_{i}\mathcal{B}_{i}\mathcal\epsilon_{i}$, where $\mathcal{B}(\eta\rightarrow\gamma\gamma)$ and $\mathcal{B}(\etap\rightarrow\eta\pippim)$ are the branching fractions of $\eta\rightarrow\gamma\gamma$ and $\etap\rightarrow\eta\pippim$ taken from the PDG~\cite{PDG}. $1+\delta$ is the radiative (ISR) correction factor, which is defined as
 \begin{equation}
     (1+\delta)=\frac{\int\sigma(s(1-x))F(x,s)dx}{\sigma(s)}.
 \end{equation}
Here, $F(x,s)$ is the radiator function, which is known from a QCD calculation with an accuracy of 0.1\%~\cite{Kuraev:1985hb}; $s$ is the square of the c.m. energy and $x=2E_{\gamma {\rm ISR}}/\sqrt{s}$, where $E_{\gamma {\rm ISR}}$ is the energy of the ISR photon. The $\sigma(s)$ is the Born cross section.\\
 \indent Since no significant $\ee\rightarrow\gamma\eta^{(\prime)}\etac$ signal is observed, we set the upper limit at the 90\% confidence level (C.L.) on $\sigma(\ee\rightarrow\gamma\eta^{(\prime)}\etac)$ incorporating the systematic uncertainty and using the Bayesian method~\cite{Conrad:2002kn}. The likelihood that produces the most conservative upper limit is selected after considering the additive systematic terms. To incorporate the multiplicative terms into the Born cross section limit, the probability of the signal yield $P(N)$ determined using a maximum likelihood fit is convolved with a probability density function of the sensitivity. The details of systematic uncertainties can be found in Sec.~\ref{sec:SYSTEMATIC UNCERTAINTY}. The upper limits of the signal yields~($N_{\rm{UL}}^{\rm{obs}}$) at the 90\%
C.L. are determined by 
\begin{equation}
    0.9\int_{0}^{\infty}P(N)dN=\int_{0}^{N_{\rm{UL}}^{\rm{obs}}}P(N)dN.
\end{equation}
Using Eq.~\ref{eq:born}, we set the upper limits on the Born cross sections for $\ee\rightarrow\gamma\eta^{(\prime)}\etac$ at 90\% C.L.. The results are summarized in Table~\ref{tab:summary}.
\section{SYSTEMATIC UNCERTAINTY} \label{sec:SYSTEMATIC UNCERTAINTY}

The sources of systematic uncertainties are categorized into two types: additive and multiplicative terms.
The additive uncertainties arise from the fit range, signal shape, background shape and the difference between MC sample and data. A conservative approach is adopted by considering all possible variations and selecting the one that yields the largest upper limit.
\begin{itemize}
    \item[(i)]The systematic uncertainty arising from the fit range is determined by varying the fit ranges of $M_{\gamma\gamma(\eta\pip\pim)}$ and $M_{hadrons}$ by $\pm 5~$MeV$/c^2$ and $\pm 10~$MeV$/c^2$.
    \item[(ii)]The nominal mass and width of $\etac$ are 2984.1~MeV$/c^2$ and 30.5~MeV, respectively~\cite{PDG}. The uncertainty associated with the $\eta_{c}$ shape is estimated by changing its mass and width parameters within their uncertainties.
    \item[(iii)]The uncertainty due to the background shape is estimated by replacing the function from the second-order Chebyshev polynomial with both the first-order and the third-order Chebyshev polynomials.
    \item[(iv)]The uncertainty due to the difference of mass resolution between data and MC simulation is estimated by convolving a Gaussian function to $\etac$ signal shape, where the mean value is varied by $\pm 1$ MeV$/c^2$, and the standard derivation is varied by 1, 2 and 3 MeV$/c^2$. For the $\ee\rightarrow\gamma\eta\etac$ process, the $\eta$ signal shape is convolved with a Gaussian function parameterized by free parameters.  

\end{itemize}

The multiplicative systematic uncertainty is considered by convolving a Gaussian function to the likelihood distribution. These uncertainties include the following sources.
\begin{itemize}
    \item[(i)]The integrated luminosity is determined using Bhabha scattering events, with a relative uncertainty of 1.0\% uniformly assigned to all the measured energy points~\cite{BESIII:2015qfd,BESIII:2022ulv}. 
    \item[(ii)]The systematic uncertainty of the ISR correction factor $1+\delta$ is estimated by comparing the difference between factors obtained by the flat line shape and $\psi(4660)$ with mass set to 4641~MeV/$c^2\pm5$, $10~{\rm MeV}/c^2$ and width to 73~MeV$\pm5$, $10~{\rm MeV}/c^2$.\\
    \item[(iii)]The part of the Born cross sections uncertainties associated with the branching fraction are calculated by $\frac{\sqrt{\sum^{16}_{i=1}(\epsilon_{i} \cdot \delta\mathcal{B}_{i})^{2}}}{\sum^{16}_{i=1}(\epsilon_{i} \cdot \mathcal{B}_{i})}$, where the $\delta\mathcal{B}_{i}$ is the total branching fraction uncertainty after error propagation. The uncertainties of the branching fractions of $\etac\rightarrow hadrons$ are obtained from the BESIII measurement result~\cite{BESIII:2012urf}, and the others are from the PDG~\cite{PDG}.
    \item[(iv)]The uncertainty from tracking or PID for charged tracks is determined to be 1.0\% per track using the control samples $J/\psi\rightarrow \pi^{+}\pi^{-}\pi^{0}$, $J/\psi\rightarrow p\bar{p}\pi^{+}\pi^{-}$ and $J/\psi\rightarrow K^{0}_{S}K^{\pm}\pi^{\mp} + c.c.$~\cite{BESIII:2011ysp}.
    \item[(v)]The systematic uncertainty due to the photon detection is assigned as 1.0\% per photon, by analyzing the control samples $J/\psi \rightarrow \rho^{0}\pi^{0}$ and $e^{+}e^{-} \rightarrow \gamma \gamma$~\cite{BESIII:2010ank}.
    \item[(vi)]The systematic uncertainty due to the $K^{0}_{S}$ reconstruction is studied using control samples $J/\psi \rightarrow K^{*\pm}K^{\mp}$ and $J/\psi \rightarrow \phi K^{0}_{S}K^{\pm}\pi^{\mp}$, and estimated to be 1.0\%~\cite{BESIII:2015jmz}.
    \item[(vii)]Using a high purity control sample of $J/\psi \rightarrow \pi^{0} p \bar{p}$  or $J/\psi \to \eta p\bar p$, the systematic uncertainty from the $\pi^{0}$ or $\eta$ reconstruction is determined to be 1.0\%~\cite{BESIII:2010tfr}.
    \item[(viii)]The systematic uncertainty associated with the kinematic fit is assigned as the difference between the detection efficiencies before and after the helix parameter corrections~\cite{BESIII:2012mpj} in the signal MC simulation.
    \item[(ix)]The contamination among the 16 decay modes of $\eta_{c}$ is calculated by using the signal MC simulation.
\end{itemize}
Table~\ref{tab:m_eta} summarizes the multiplicative systematic uncertainties. All sources are treated as uncorrelated, and the total systematic uncertainty is obtained by adding them in quadrature.

\section{SUMMARY} \label{sec:SUMMARY AND DISSCUSSION}
Based on $e^{+}e^{-}$ annihilation data corresponding to an integrated luminosity of 10.6 $\rm{fb^{-1}}$
collected with the BESIII detector at the BEPCII collider for c.m. energies in the range $\sqrt{s}=$ 4.258-4.681 GeV, 
we search for the $1^{- +}$ charmonium-like hybrid via $e^{+}e^{-}\rightarrow\gamma\eta^{(\prime)}\eta_{c}$ for the first time. 
No significant signal is observed and the upper limits at the 90\% C.L. on the Born cross sections for $e^{+}e^{-}\rightarrow\gamma\eta\eta_{c}$ and $e^{+}e^{-}\rightarrow\gamma\etap\eta_{c}$ are determined. Some theoretical calculations also predict that $\eta_{c1}$ can decay to the open-charm final states, such as $D_1\bar{D},D^{*}\bar{D}$ and $D^*\bar{D}^*$~\cite{Shi:2023sdy}. With future datasets of higher statistics, searches for the $\eta_{c1}$ in these final states can be further explored.

\begin{table*}[htbp]
    \centering
    \captionsetup{justification=raggedright}
    \caption{Data sets and results of the upper limits on Born cross sections of $e^{+}e^{-}\rightarrow \gamma\eta^{(\prime)}\eta_{c}$. The integrated luminosities $\mathcal{L}$, the upper limits of observed signal yields $N^{\rm{obs}}_{\rm{UL}}$, the ISR correction ($1+\delta$), vacuum polarization correction factor $\frac{1}{|1-\Pi|^2}$, the sum of the products of the branching fraction and efficiency $\sum\mathcal{B}_i\epsilon_{i}$, the upper limits of Born cross sections $\sigma^{\rm{Born}}_{\rm{UL}}(\rm{pb})$ and the statistical significance $\mathcal{S}$.}
    \label{tab:summary}
    \scalebox{1}{ 
        \begin{tabular}{|c|c|c|c|c|c|c|c|c|c|c|c|c|c|}
            \hline 
            \multirow{2}{*}{$\sqrt{s}({\rm{GeV}})$} &\multirow{2}{*}{$\mathcal{L}(\rm{pb^{-1}})$} & \multicolumn{6}{c|}{$e^{+}e^{-}\rightarrow \gamma\eta\eta_{c}$} &\multicolumn{6}{c|}{$e^{+}e^{-}\rightarrow \gamma\eta'\eta_{c}$} \\
            \cline{3-14}
                & &$1+\delta$& $\frac{1}{|1-\Pi|^2}$& $N^{\rm{obs}}_{\rm{UL}}$ &$\sum_{i}(\mathcal{B}_{i} \epsilon_{i})(\%)$ &$\sigma^{\rm{Born}}_{\rm{UL}}(\rm{pb})$&$\mathcal{S}(\sigma)$ &$1+\delta$ &$\frac{1}{|1-\Pi|^2}$ & $N^{\rm{obs}}_{\rm{UL}}$ &$\sum_{i}(\mathcal{B}_{i}\epsilon_{i})(\%)$ &$\sigma^{\rm{Born}}_{\rm{UL}}(\rm{pb})$&$\mathcal{S}(\sigma)$ \\

			\hline
        
                4.258& 828.4  &0.86&    1.05&  76.8 	    &    0.38    & 27.3    &$0.1$    &    0.85  & 1.05 &65.8 &0.17  &  51.2&2.8  \\
                4.267&  531.1 &0.86&    1.05&  62.3  	    &    0.37    & 35.3    &0.1	     &    0.85  & 1.05 &45.6 &0.17  &  56.4&1.5  \\
                4.278&  175.7 &0.86&    1.05&  28.2  	    &    0.36    & 49.1    &2.2	     &    0.85  & 1.05 &28.3 &0.16  & 110.1&1.9  \\
                4.288& 502.4  &0.86&    1.05&  30.6  	    &    0.37    & 18.5    &0.6	     &    0.86  & 1.05 &33.4 &0.17  &  44.5&2.5  \\
                4.308& 45.1   &0.85&    1.05&  36.2  	    &    0.37    & 240.8   &2.5      &    0.84  & 1.05 &12.6 &0.18  & 178.9&1.7  \\
                4.312& 501.2  &0.85&    1.05&  48.5  	    &    0.36    & 29.6    &$1.9$    &    0.87  & 1.05 &43.7 &0.17  &  56.5&2.8  \\
                4.337& 505.0  &0.85&    1.05&  61.3 	    &    0.37    & 26.8    &$0.9$    &    0.87  & 1.05 &44.2 &0.17  &  55.2&2.8  \\
                4.358& 543.9  &0.85&    1.05&  38.8  	    &    0.38    & 21.1    &0.6	     &    0.84  & 1.05 &65.0 &0.18  &  75.5&2.4  \\
                4.377& 522.7  &0.84&    1.05&  48.4  	    &    0.37    & 28.2    &1.5      &    0.86  & 1.05 &40.2 &0.18  &  48.5&1.0  \\
                4.387& 55.6   &0.84&    1.05&  13.7  	    &    0.38    & 73.6    &2.5      &    0.84  & 1.05 &14.6 &0.18  & 165.2&1.3  \\
                4.396& 507.8  &0.84&    1.05&  35.1 	    &    0.37    & 21.1    &2.5      &    0.86  & 1.05 &34.1 &0.18  &  42.7&0.6  \\
                4.416& 1043.9 &0.83&    1.05&  45.2  	    &    0.37    & 13.2    &0.1      &    0.84  & 1.05 &70.9 &0.18  &  41.9&2.3  \\
                4.436& 569.9  &0.83&    1.05&  91.3 	    &    0.38    & 49.3    &1.1      &    0.85  & 1.05 &32.4 &0.18  &  35.3&1.1  \\
                4.467& 111.1  &0.82&    1.05&  27.5  	    &    0.38    & 77.1    &0.8      &    0.82  & 1.05 &13.7 &0.18  &  79.3&0.9  \\
                4.527& 112.1  &0.81&    1.05&  60.0  	    &    0.38    & 38.9    &2.0      &    0.81  & 1.05 &16.2 &0.19  &  90.0&1.0  \\
                4.574& 48.9   &0.79&    1.05&  24.1         &    0.38    & 179.7   &1.5      &    0.79  & 1.05 &12.6 &0.20  & 158.7&0.4  \\
                4.600& 586.9  &0.78&    1.05&  60.6 	    &    0.38    & 32.7    &2.4      &    0.78  & 1.05 &17.4 &0.20  &  18.4&0.9  \\
                4.611& 103.65 &0.77&    1.05&  24.0  	    &    0.38    & 78.5    &0.7      &    0.77  & 1.05 &12.9 &0.20  &  78.3&0.1  \\
                4.628& 521.53 &0.77&    1.05&  77.6 	    &    0.38    & 48.8   & 1.3      &    0.77  & 1.05 &27.4 &0.20  &  33.7&1.2  \\
                4.641& 551.65 &0.76&    1.05&  28.7  	    &    0.38    & 17.3    &2.2      &    0.76  & 1.05 &25.1 &0.20  &  28.8&0.2  \\
                4.661& 529.43 &0.75&    1.05&  55.9 	    &    0.38    & 35.4    &1.2      &    0.75  & 1.05 &29.0 &0.20  &  34.4&1.5  \\
                4.681& 1667.39&0.75&    1.05&  71.8 	    &    0.38    & 14.5    &1.4      &    0.78  & 1.05 &46.1 &0.20  &  16.5&0.4 \\
      \hline                                            
            
            \hline                                      
        \end{tabular}
    }
\end{table*}
\begin{table*}[htbp]
    \caption{Relative systematic uncertainties~(\%) on $\sigma^{\rm{Born}}(e^{+}e^{-}\rightarrow\gamma\eta^{(\prime)}\eta_{c})$ at $\sqrt{s} = 4.681$ GeV.}
    \centering
    \begin{tabular}{cc|c|c}
    \hline
        \multicolumn{2}{c|}{Source}                                &$\sigma^{\rm{Born}}(e^{+}e^{-}\rightarrow\gamma\eta\eta_{c})$ &  $\sigma^{\rm{Born}}(e^{+}e^{-}\rightarrow\gamma\eta'\eta_{c})$ \\
        \hline
        \multicolumn{2}{c|}{Luminosity}                            &  1.0                                                                    & 1.0\\
        \hline
        \multicolumn{2}{c|}{ISR correction}                        &  3.3                                                                    & 0.9\\
        \hline
        \multirow{9}{*}{$\sum_{i}\mathcal{B}_{i}\epsilon_{i}$}
                                                               & \multicolumn{1}{|c|}{$\mathcal{B}_{\etac\rightarrow X_i}$} &  18.1        & 18.1\\
                                                          
                                                               & \multicolumn{1}{|c|}{Tracking}                        &  3.3         & 5.1\\
                                                               & \multicolumn{1}{|c|}{Photon selection }                          &  3.0         & 3.1\\
                                                               & \multicolumn{1}{|c|}{PID}                                  &  3.3         & 5.1\\
                                                               & \multicolumn{1}{|c|}{$K_{S}^{0}$ reconstruction}                     &  $0.1$     & $0.1$\\
                                                               & \multicolumn{1}{|c|}{$\pi^0$ reconstruction}                         &  0.4         & 0.4\\
                                                               & \multicolumn{1}{|c|}{$\eta$ reconstruction}                          &  0.3         & 0.3\\
                                                               & \multicolumn{1}{|c|}{Kinematic fit}                        &  1.1         & 1.0\\
                                                               & \multicolumn{1}{|c|}{Contamination}                           &  1.2         & 0.6\\
        \hline

        \multicolumn{2}{c|}{Total}                          &  19.3                                                                 & 19.9\\
        \hline
        \label{tab:m_eta}
    \end{tabular}
\end{table*}

\section{APPENDIX} \label{sec:APPENDIX}
\begin{figure*}[htbp]
    \begin{subfigure}{0.32\textwidth}
        \includegraphics[width=\linewidth]{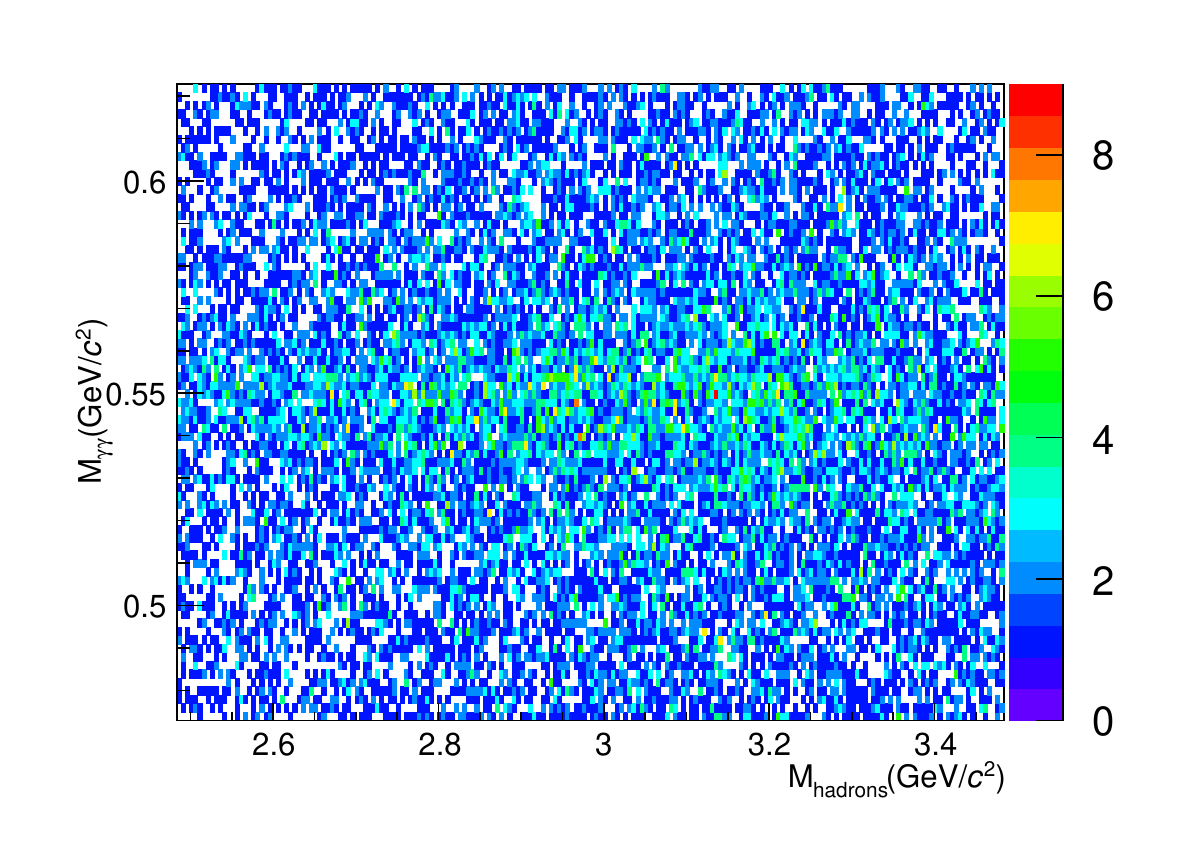}
        \captionsetup{skip=-7pt,font=normalsize}
    \end{subfigure}
    \begin{subfigure}{0.32\textwidth}
        \includegraphics[width=\linewidth]{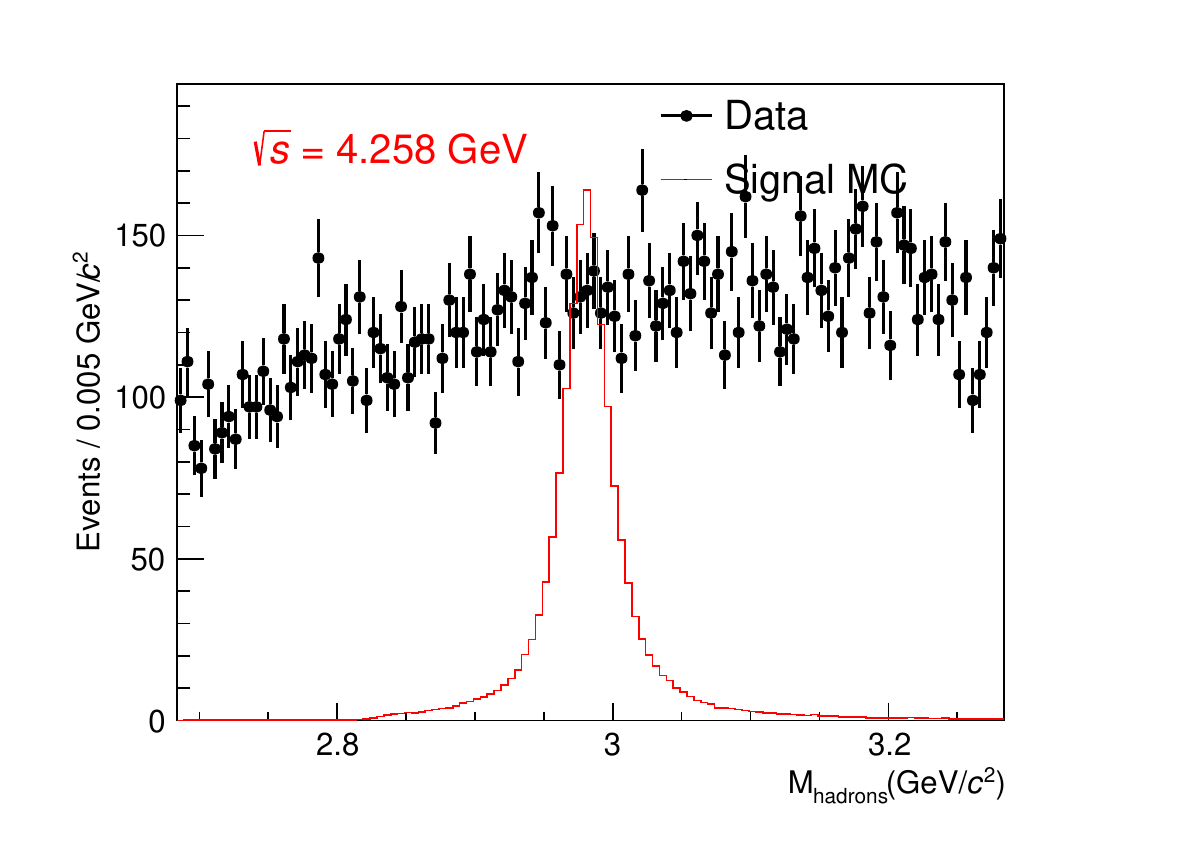}
        \captionsetup{skip=-7pt,font=normalsize}
    \end{subfigure}
    \begin{subfigure}{0.32\textwidth}
        \includegraphics[width=\linewidth]{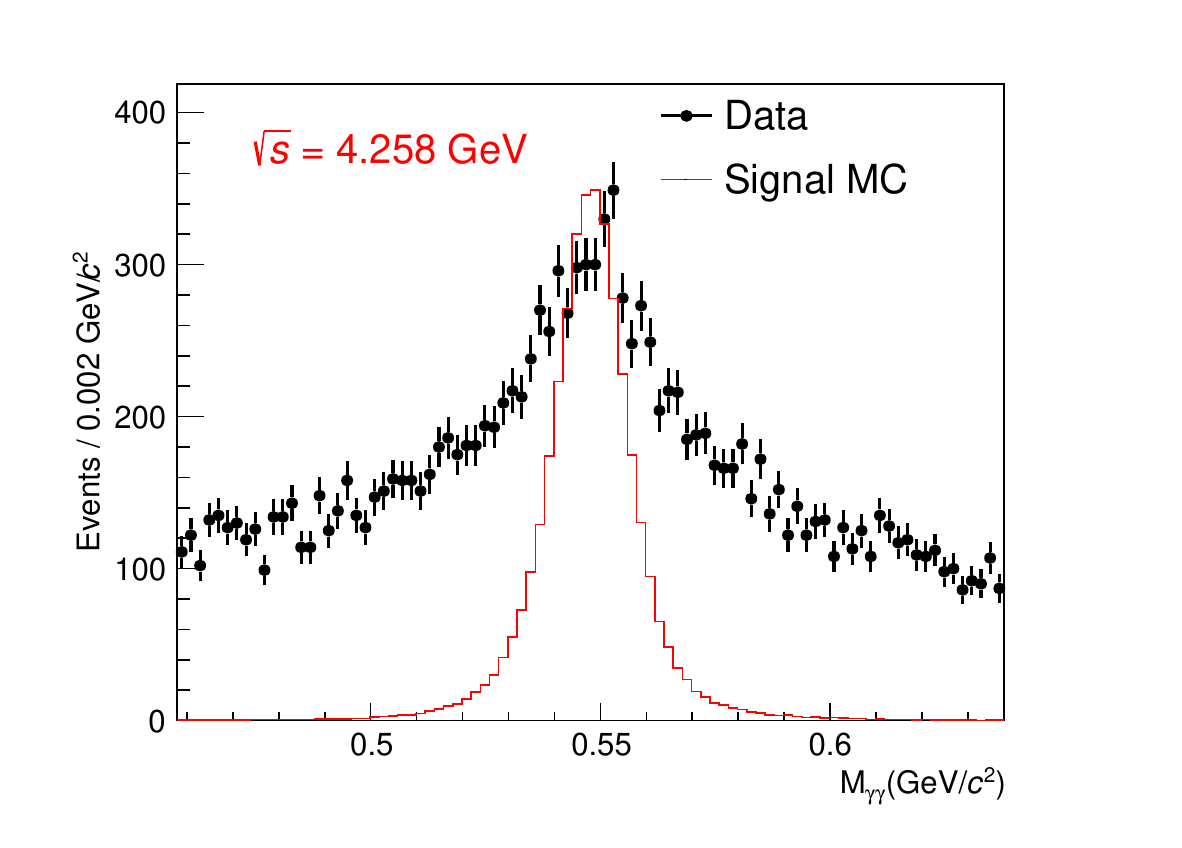}
        \captionsetup{skip=-7pt,font=normalsize}
    \end{subfigure}
    \begin{subfigure}{0.32\textwidth}
        \includegraphics[width=\linewidth]{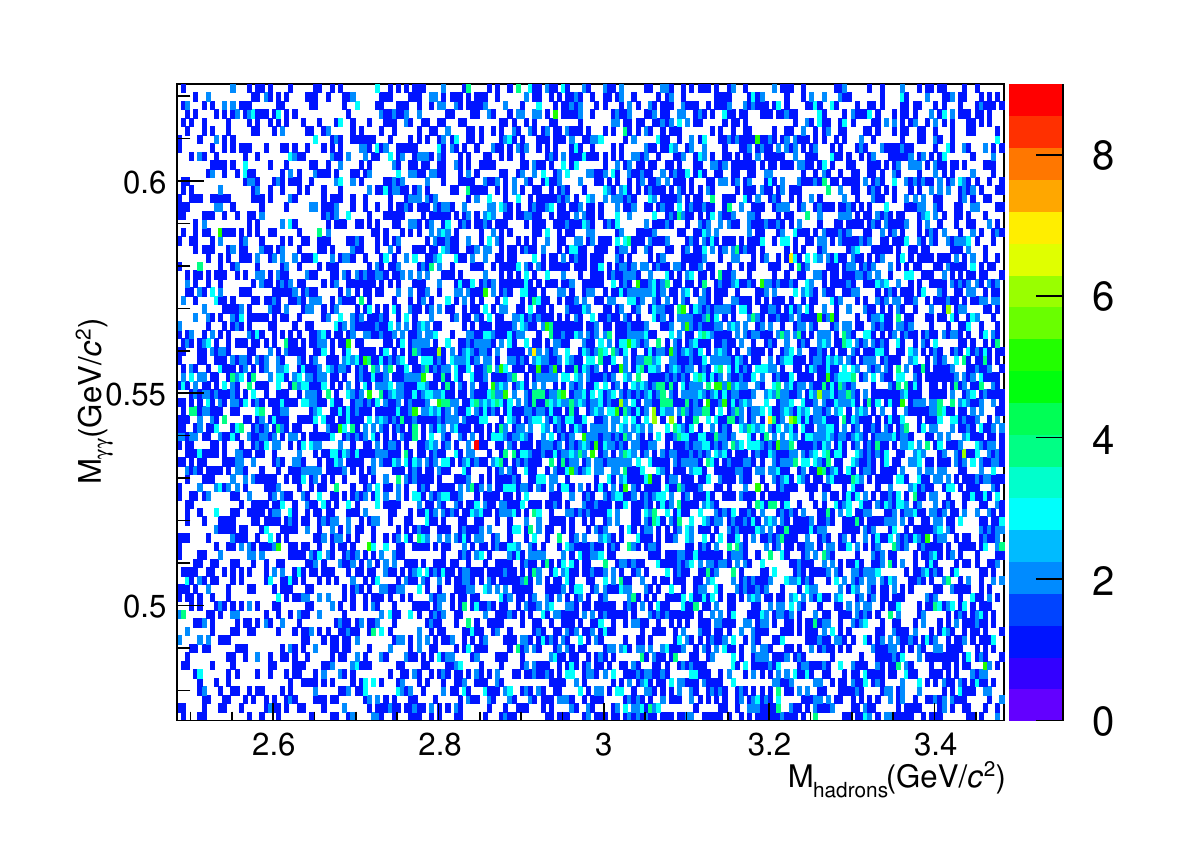}
        \captionsetup{skip=-7pt,font=normalsize}
    \end{subfigure}
    \begin{subfigure}{0.32\textwidth}
        \includegraphics[width=\linewidth]{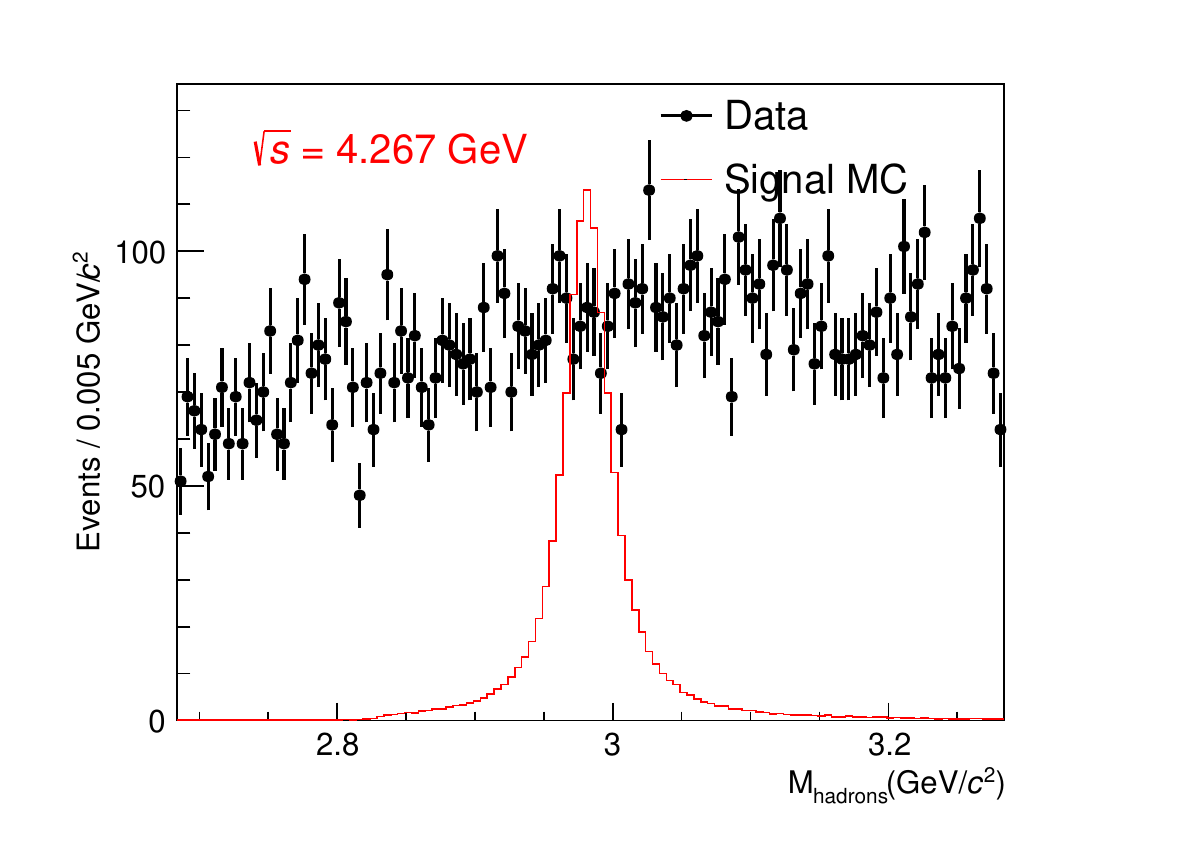}
        \captionsetup{skip=-7pt,font=normalsize}
    \end{subfigure}
    \begin{subfigure}{0.32\textwidth}
        \includegraphics[width=\linewidth]{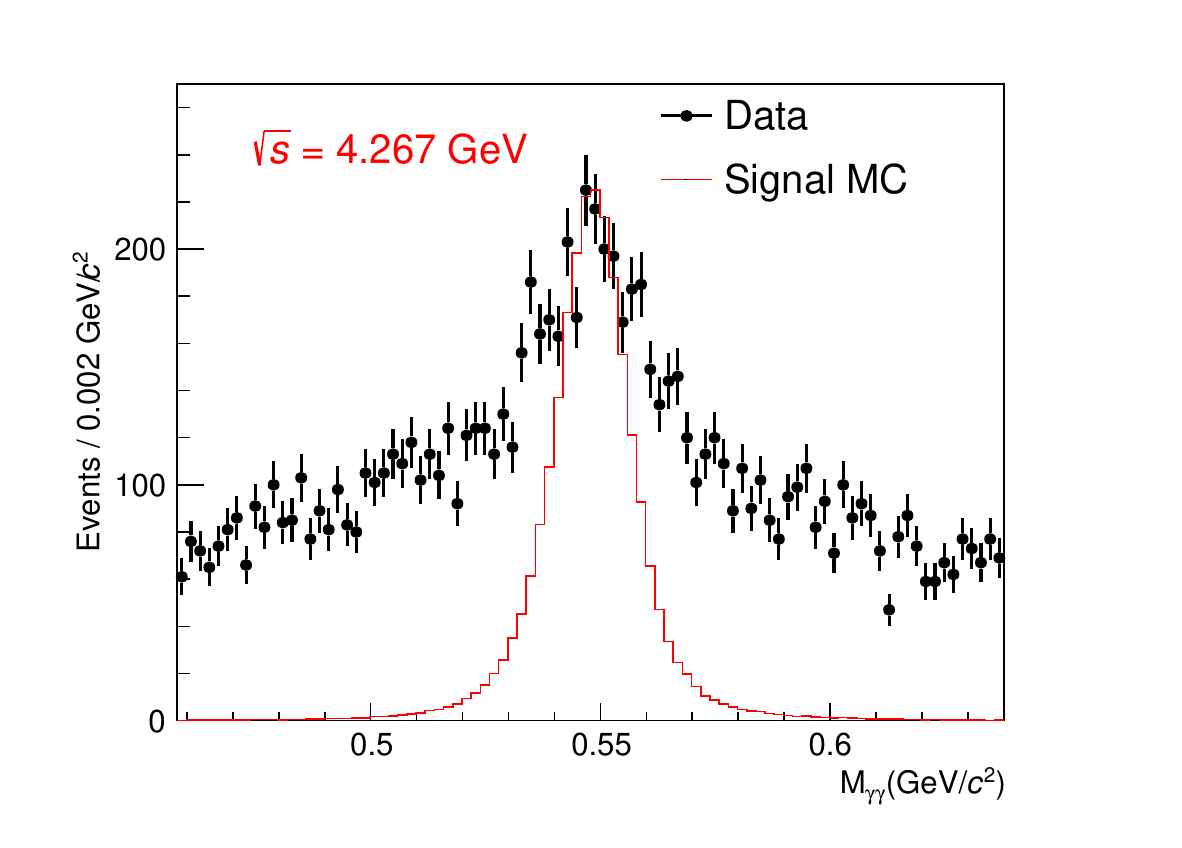}
        \captionsetup{skip=-7pt,font=normalsize}
    \end{subfigure}
    \begin{subfigure}{0.32\textwidth}
        \includegraphics[width=\linewidth]{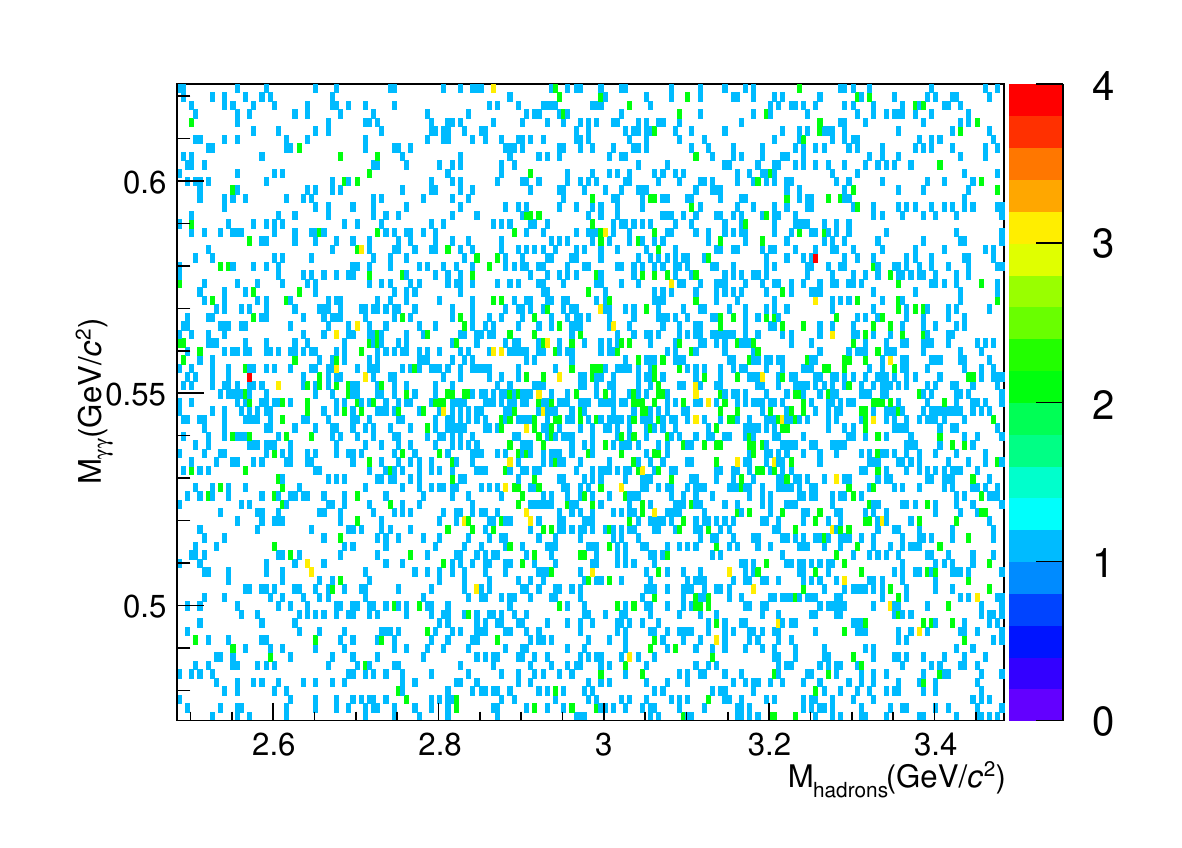}
        \captionsetup{skip=-7pt,font=normalsize}
    \end{subfigure}
    \begin{subfigure}{0.32\textwidth}
        \includegraphics[width=\linewidth]{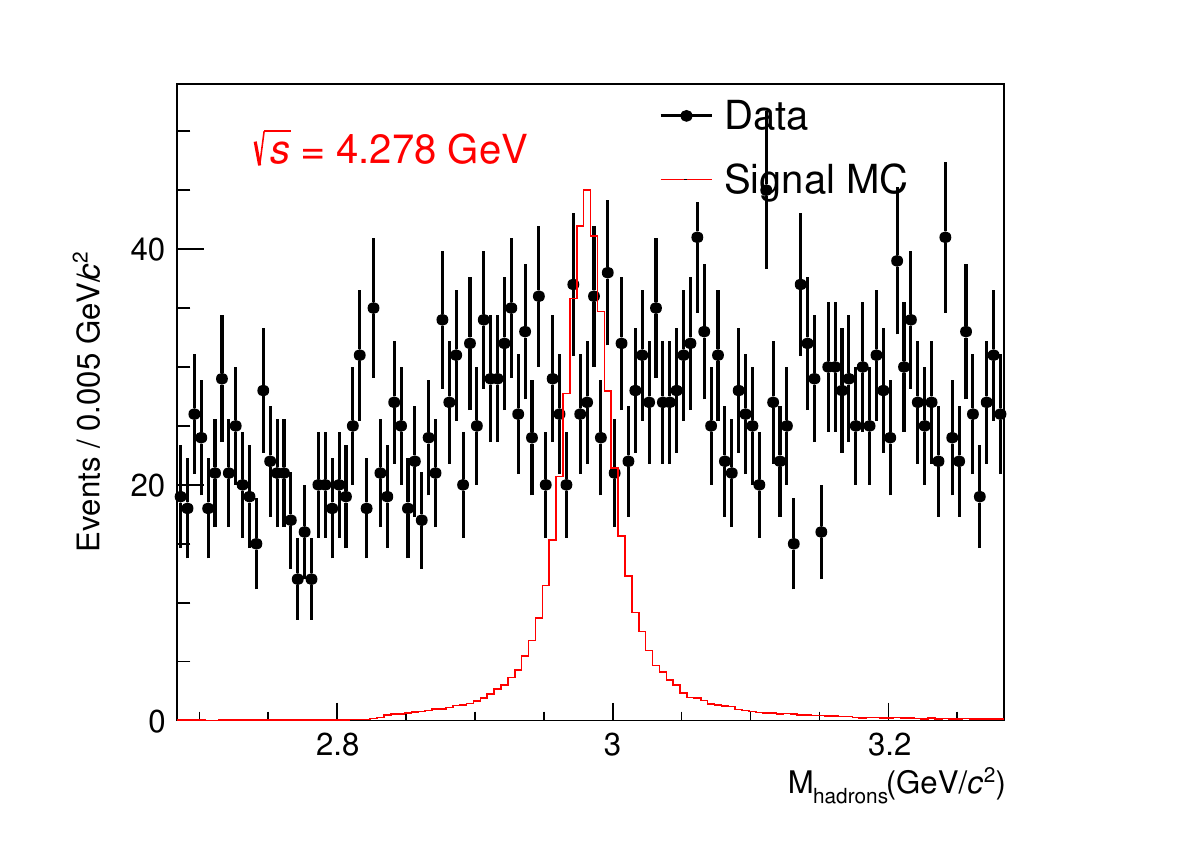}
        \captionsetup{skip=-7pt,font=normalsize}
    \end{subfigure}
    \begin{subfigure}{0.32\textwidth}
        \includegraphics[width=\linewidth]{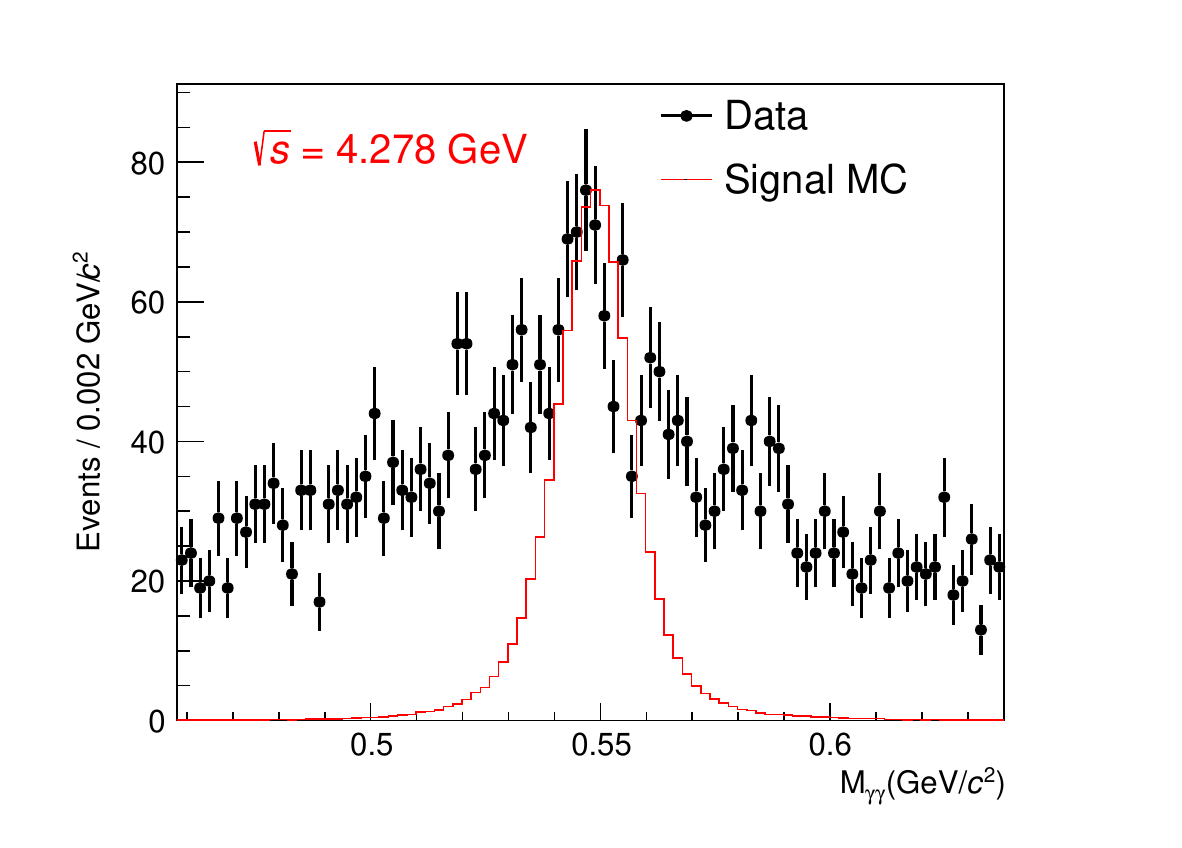}
        \captionsetup{skip=-7pt,font=normalsize}
    \end{subfigure}
    \begin{subfigure}{0.32\textwidth}
        \includegraphics[width=\linewidth]{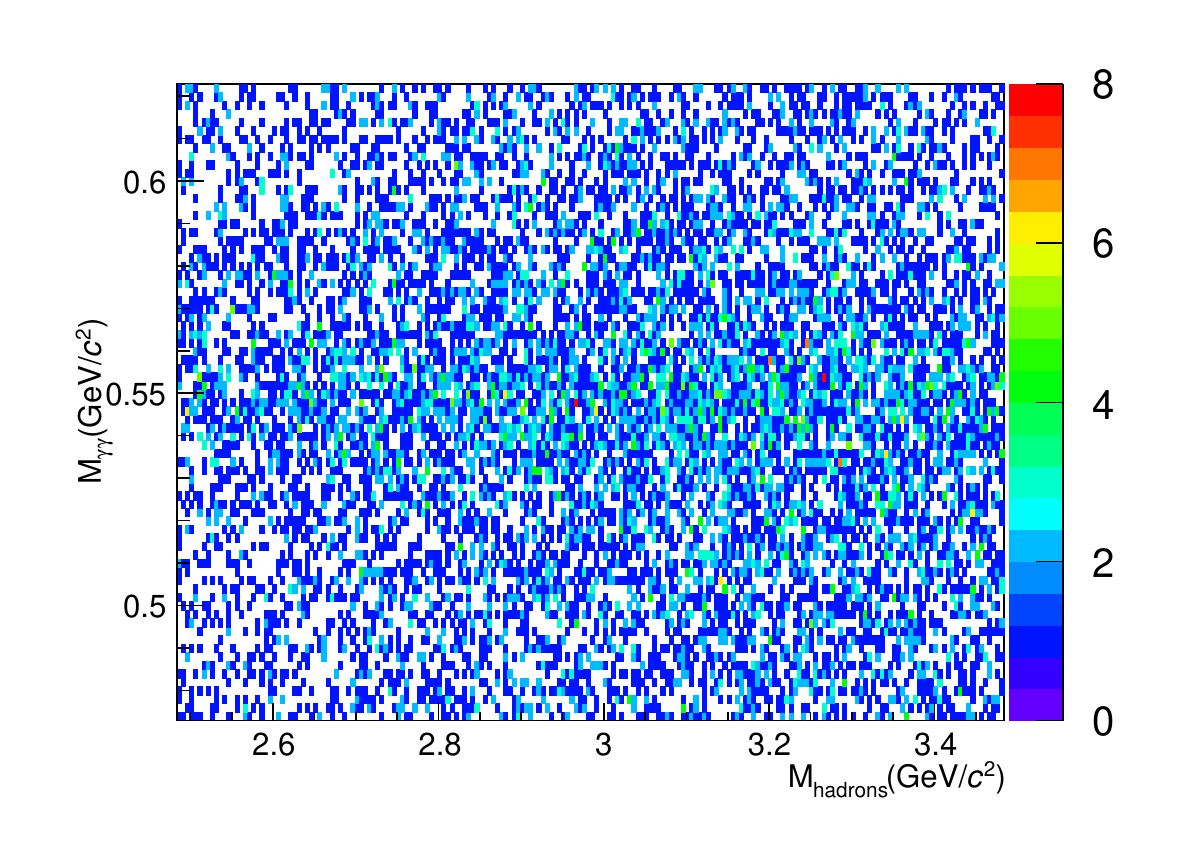}
        \captionsetup{skip=-7pt,font=normalsize}
    \end{subfigure}
    \begin{subfigure}{0.32\textwidth}
        \includegraphics[width=\linewidth]{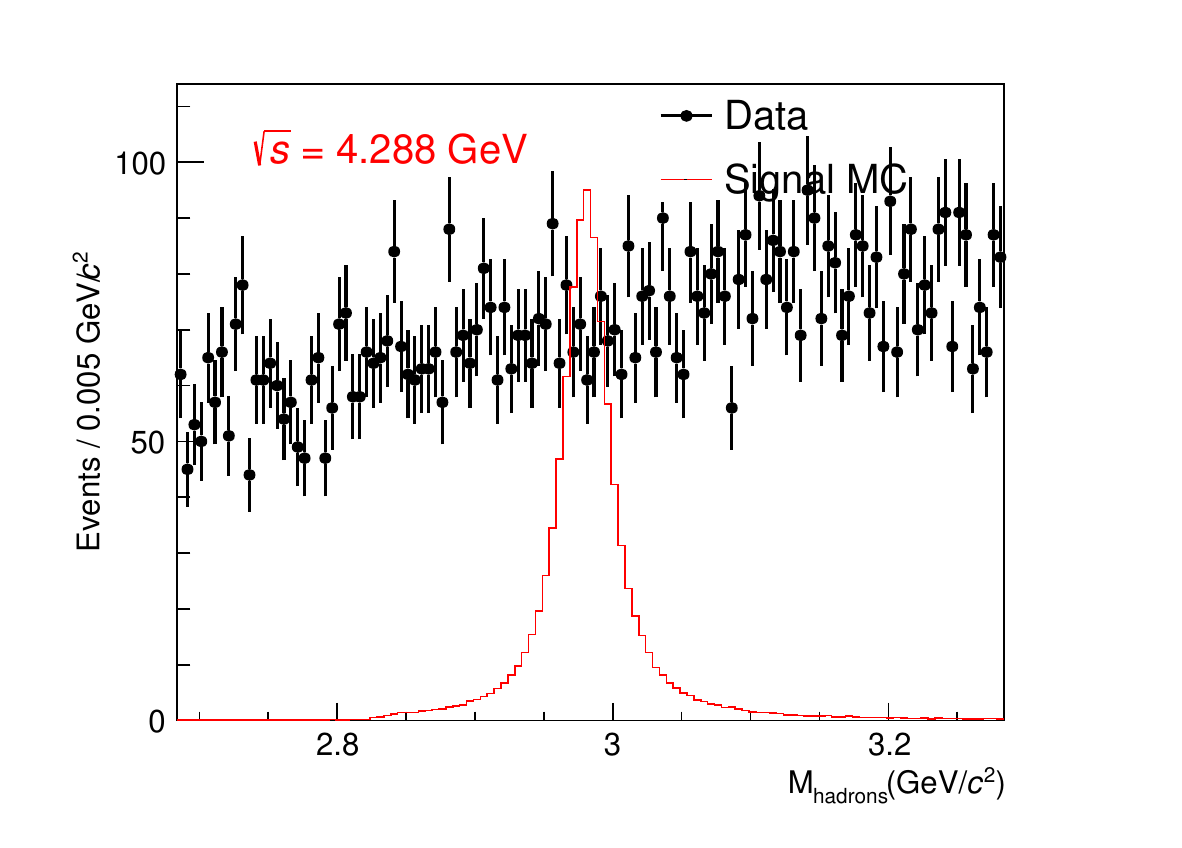}
        \captionsetup{skip=-7pt,font=normalsize}
    \end{subfigure}
    \begin{subfigure}{0.32\textwidth}
        \includegraphics[width=\linewidth]{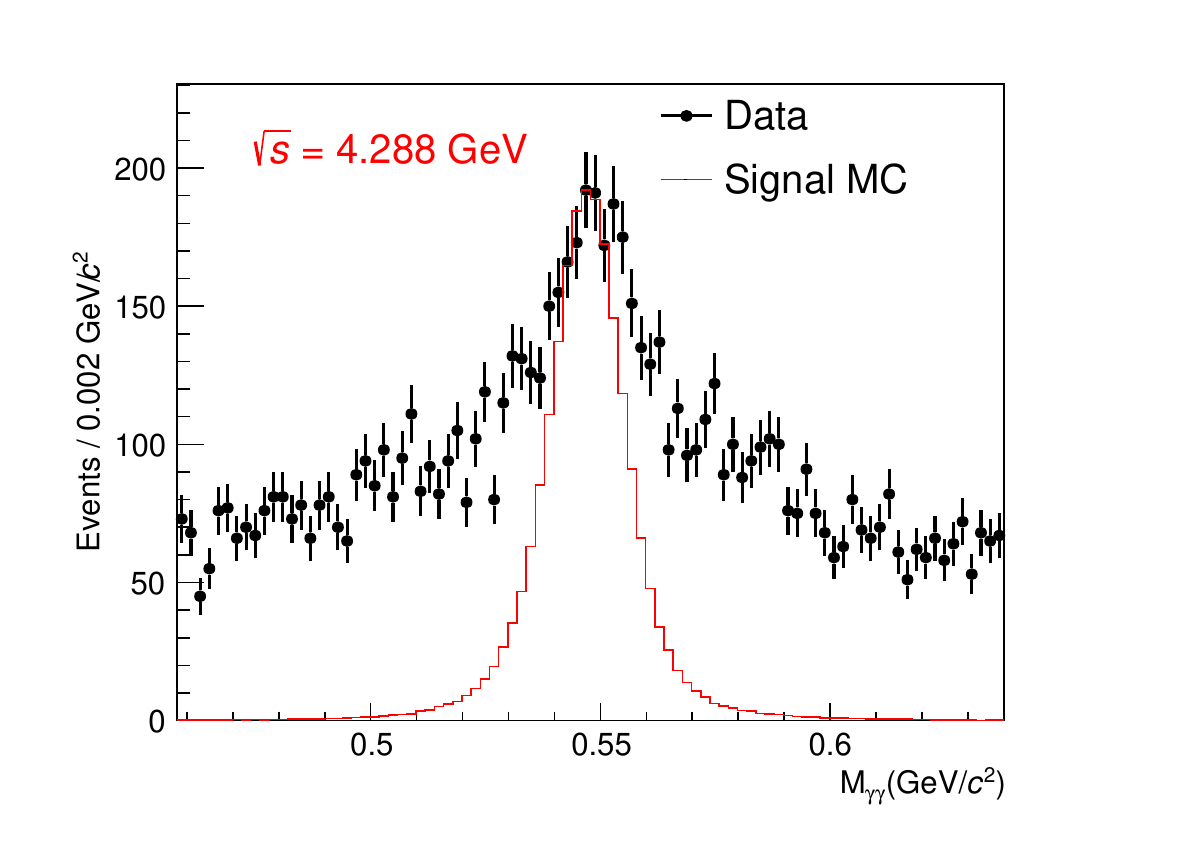}
        \captionsetup{skip=-7pt,font=normalsize}
    \end{subfigure}
\captionsetup{justification=raggedright}
\caption{The distributions of (Left) $M_{hadrons}$ versus $M_{\gamma\gamma}$, (Middle) $M_{hadrons}$, and (Right) $M_{\gamma\gamma}$ at $\sqrt s=4.258-4.288$~GeV.}
\label{fig:normal1}
\end{figure*}

\begin{figure*}[htbp]
    \begin{subfigure}{0.32\textwidth}
        \includegraphics[width=\linewidth]{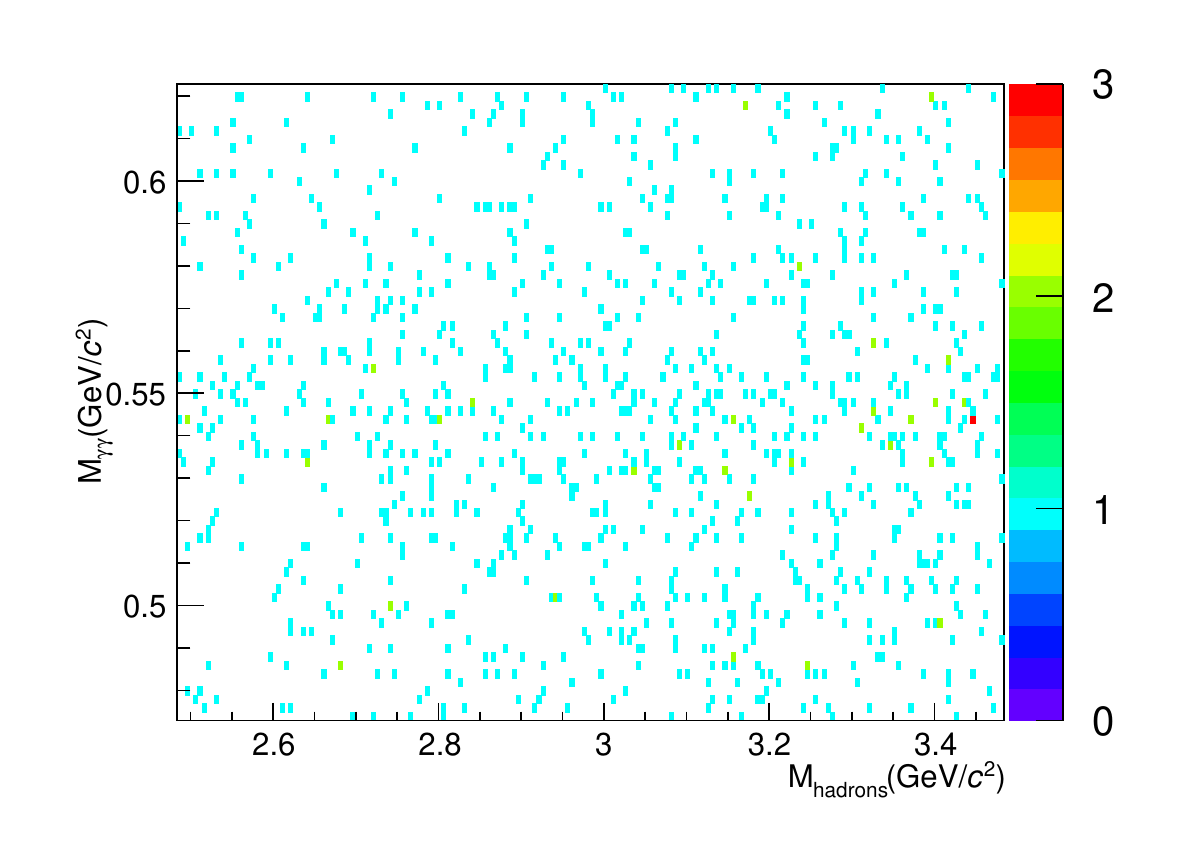}
        \captionsetup{skip=-7pt,font=normalsize}
    \end{subfigure}
    \begin{subfigure}{0.32\textwidth}
        \includegraphics[width=\linewidth]{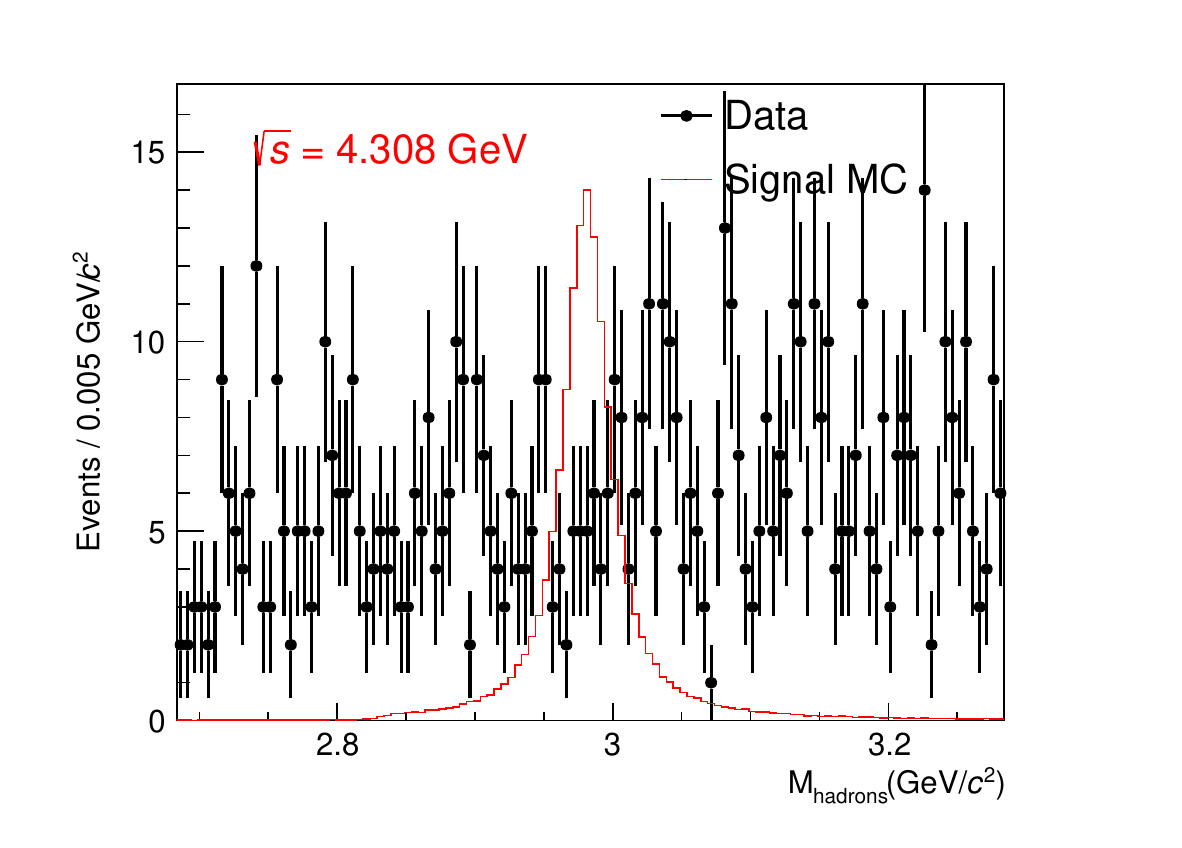}
        \captionsetup{skip=-7pt,font=normalsize}
    \end{subfigure}
    \begin{subfigure}{0.32\textwidth}
        \includegraphics[width=\linewidth]{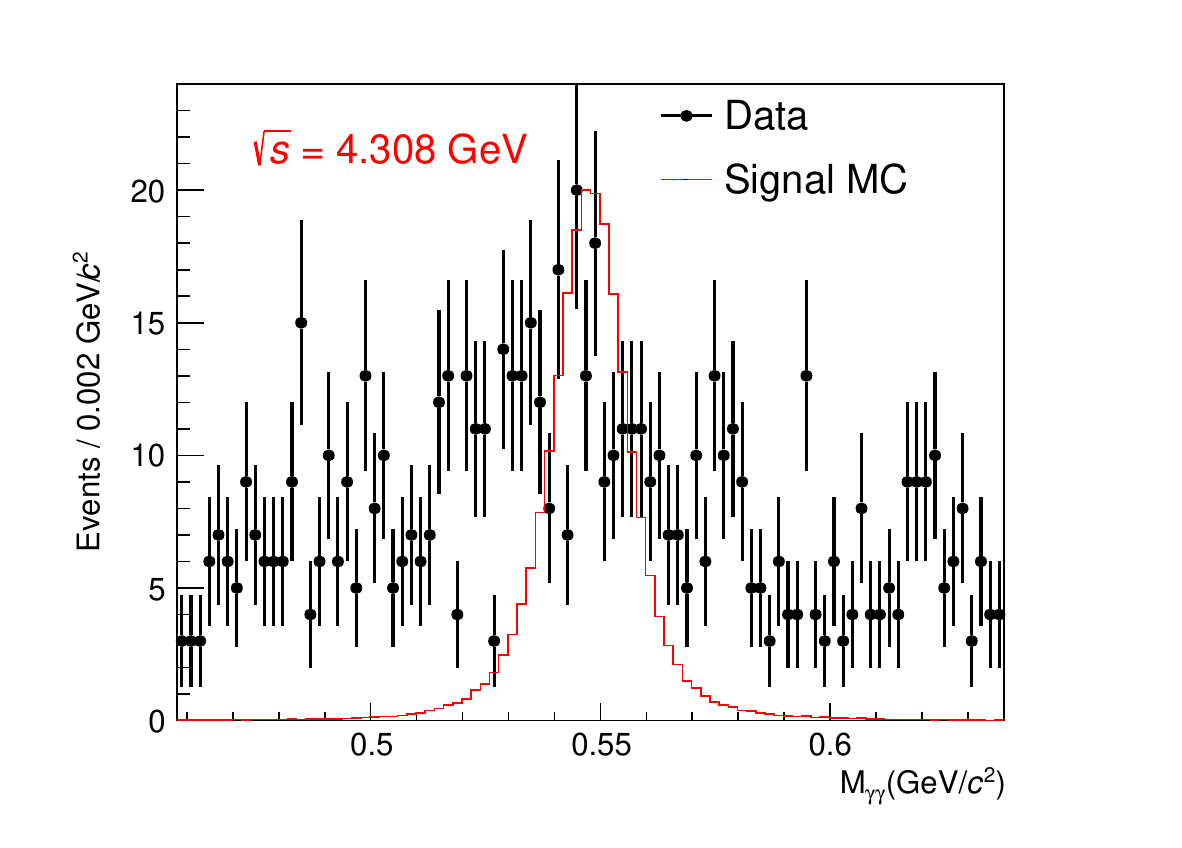}
        \captionsetup{skip=-7pt,font=normalsize}
    \end{subfigure}
    \begin{subfigure}{0.32\textwidth}
        \includegraphics[width=\linewidth]{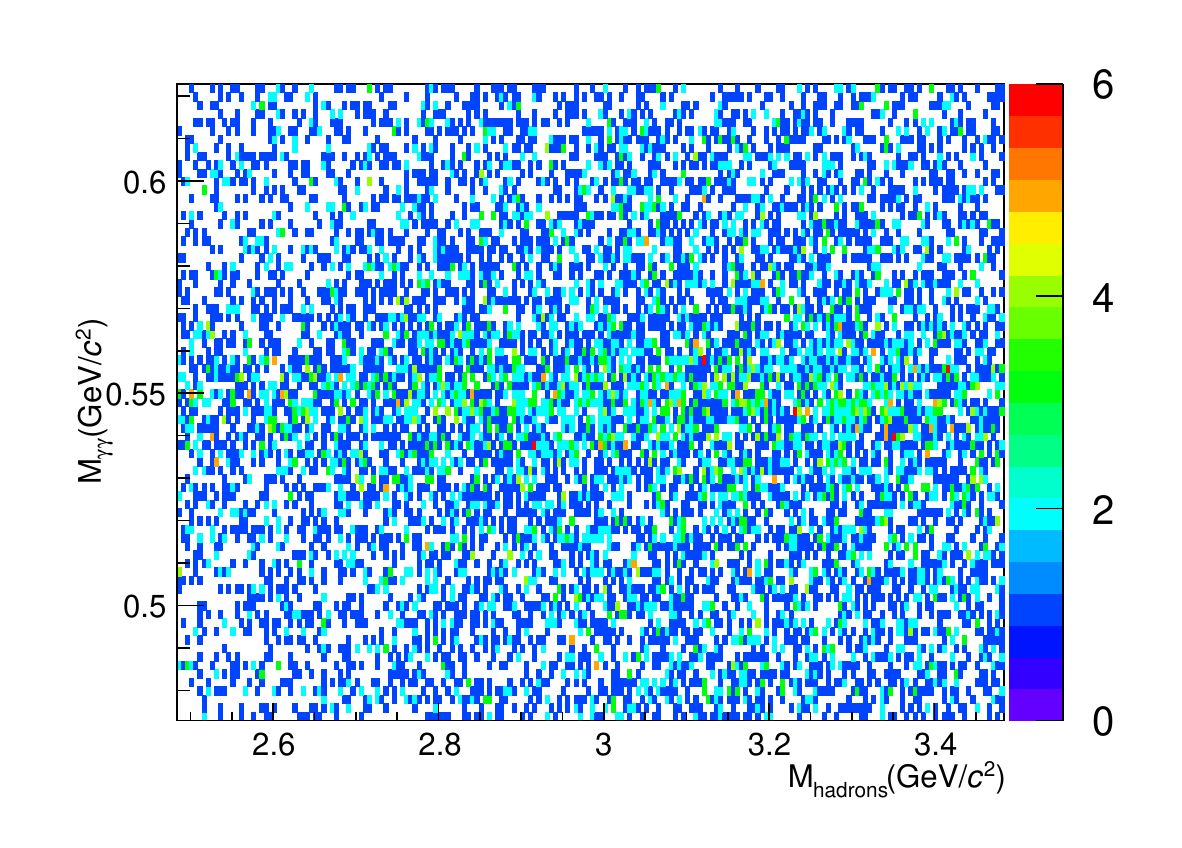}
        \captionsetup{skip=-7pt,font=normalsize}
    \end{subfigure}
    \begin{subfigure}{0.32\textwidth}
        \includegraphics[width=\linewidth]{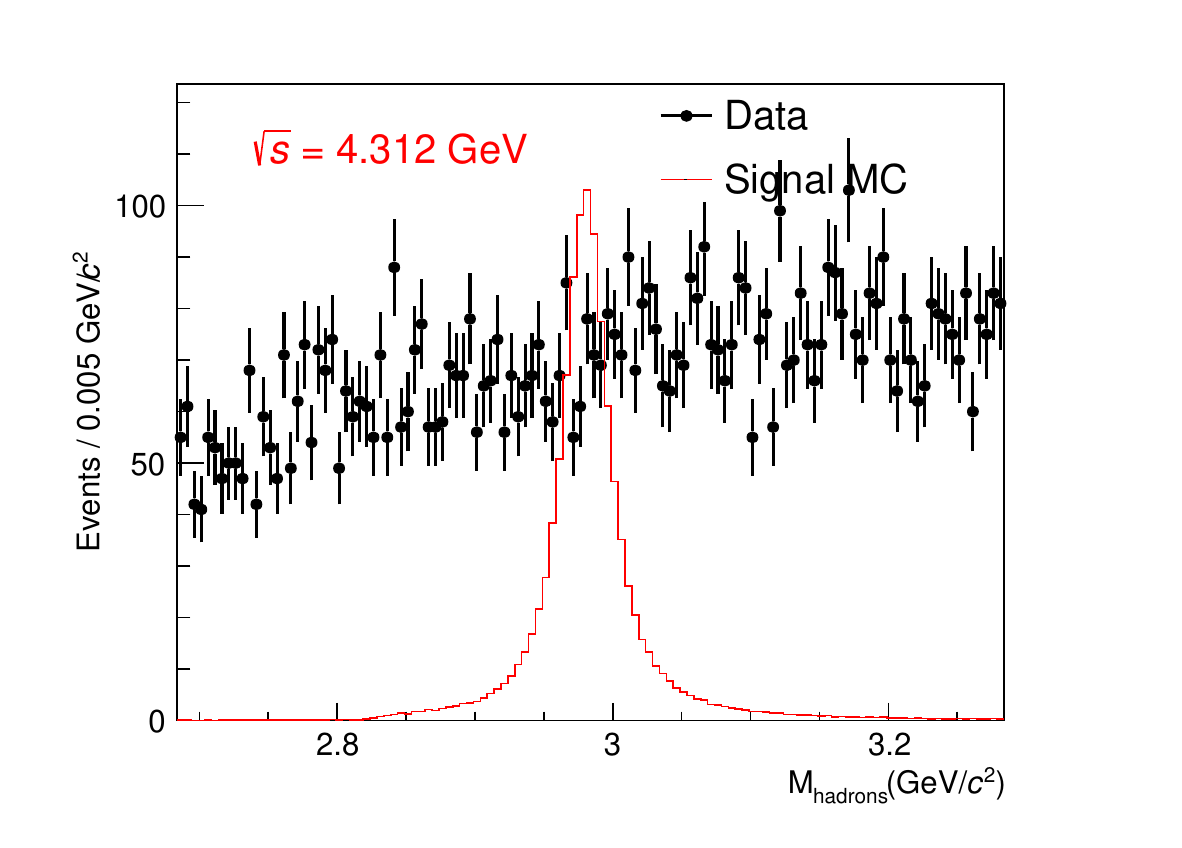}
        \captionsetup{skip=-7pt,font=normalsize}
    \end{subfigure}
    \begin{subfigure}{0.32\textwidth}
        \includegraphics[width=\linewidth]{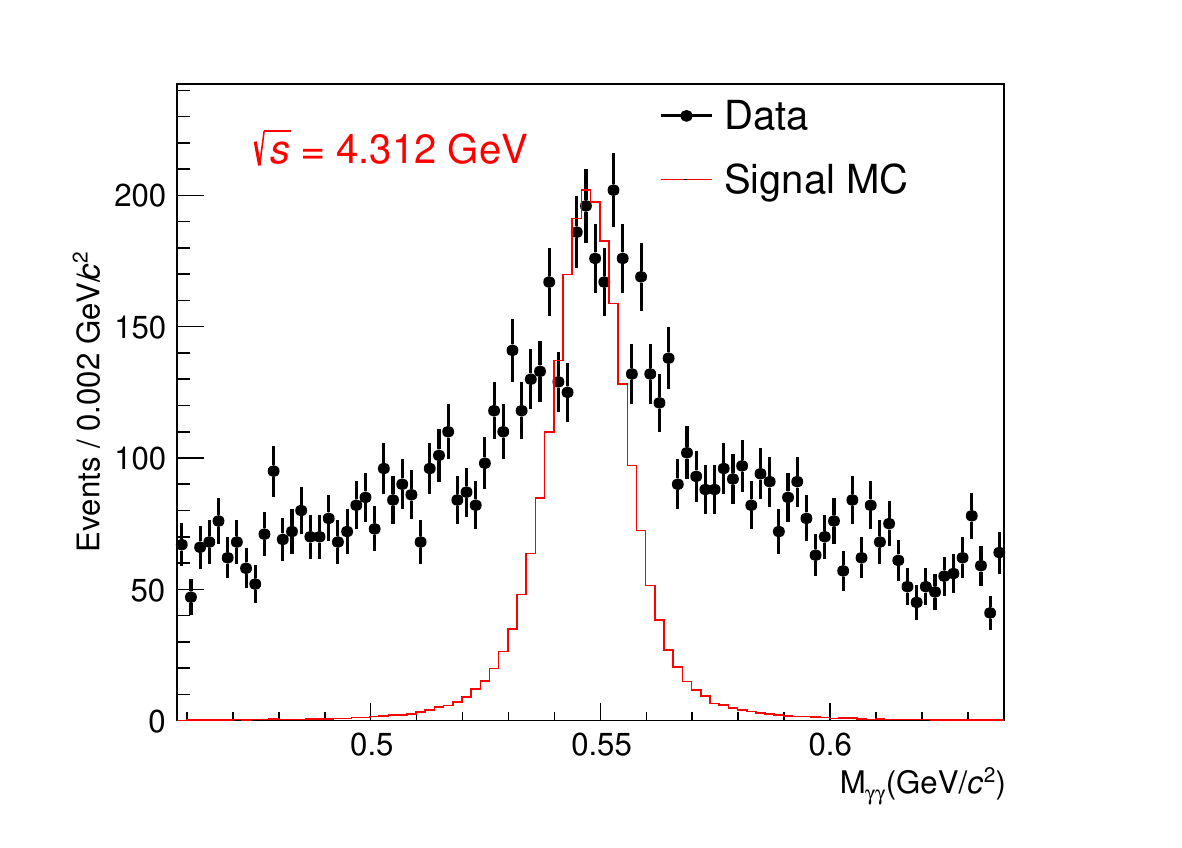}
        \captionsetup{skip=-7pt,font=normalsize}
    \end{subfigure}
    \begin{subfigure}{0.32\textwidth}
        \includegraphics[width=\linewidth]{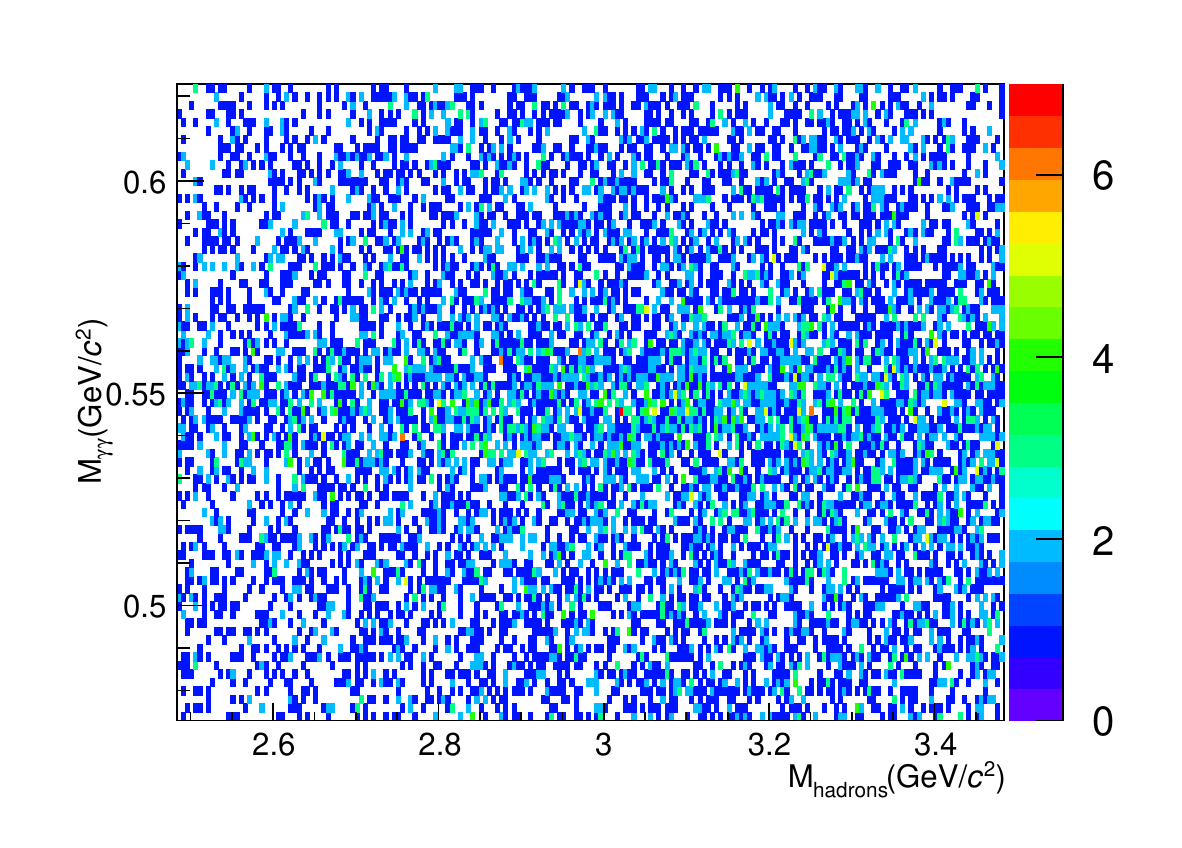}
        \captionsetup{skip=-7pt,font=normalsize}
    \end{subfigure}
    \begin{subfigure}{0.32\textwidth}
        \includegraphics[width=\linewidth]{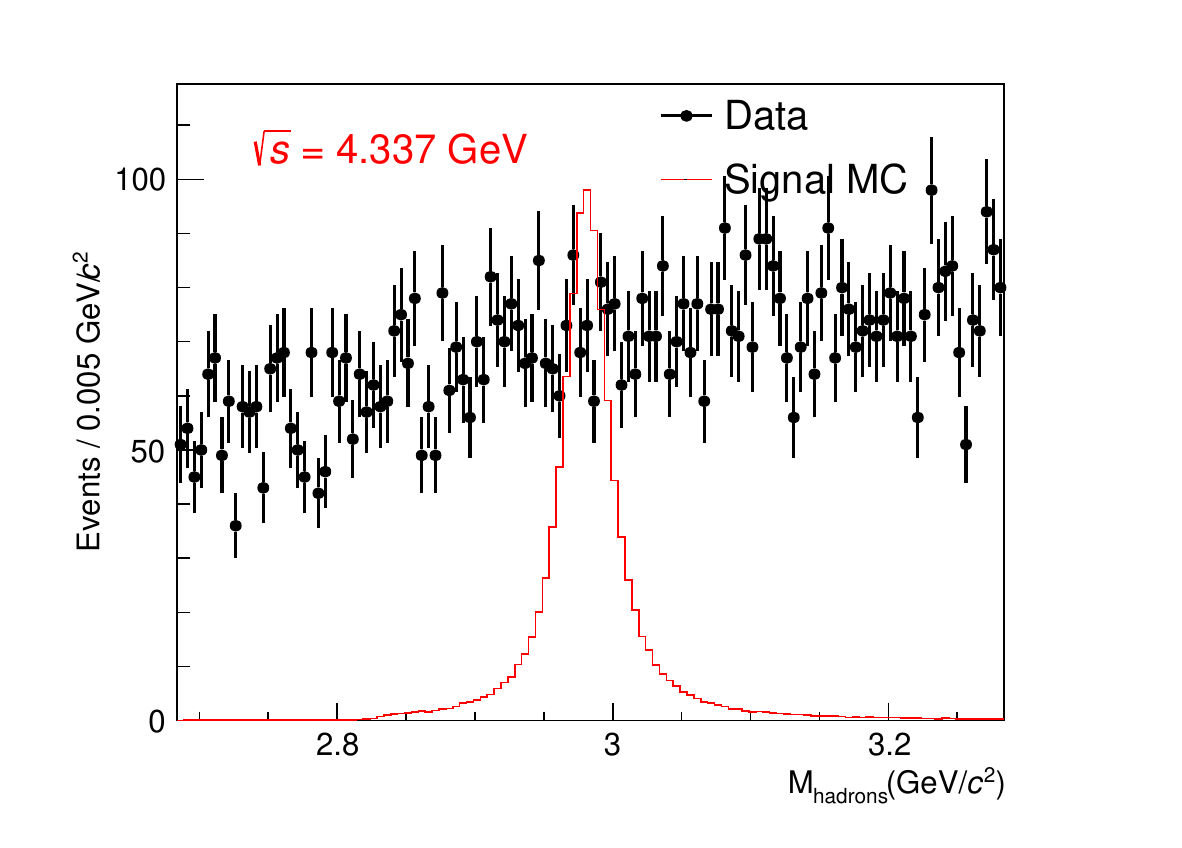}
        \captionsetup{skip=-7pt,font=normalsize}
    \end{subfigure}
    \begin{subfigure}{0.32\textwidth}
        \includegraphics[width=\linewidth]{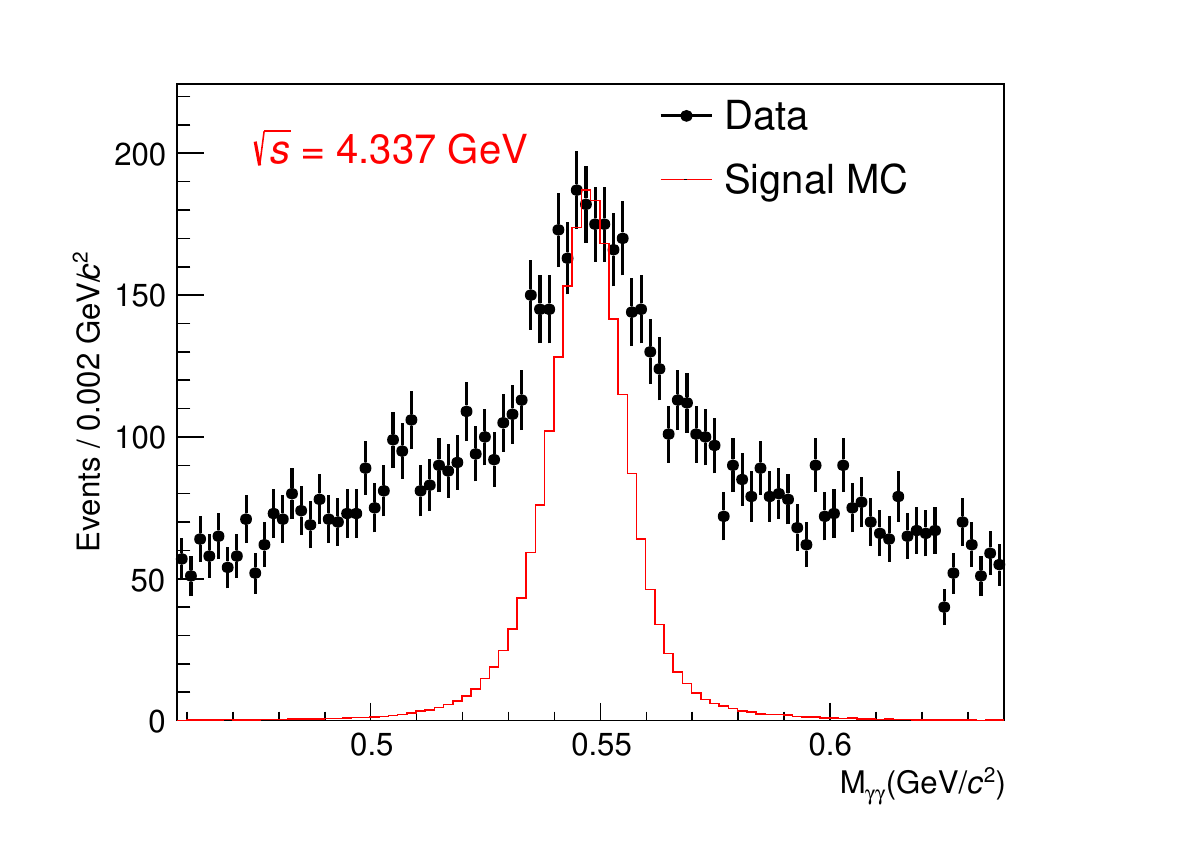}
        \captionsetup{skip=-7pt,font=normalsize}
    \end{subfigure}
    \begin{subfigure}{0.32\textwidth}
        \includegraphics[width=\linewidth]{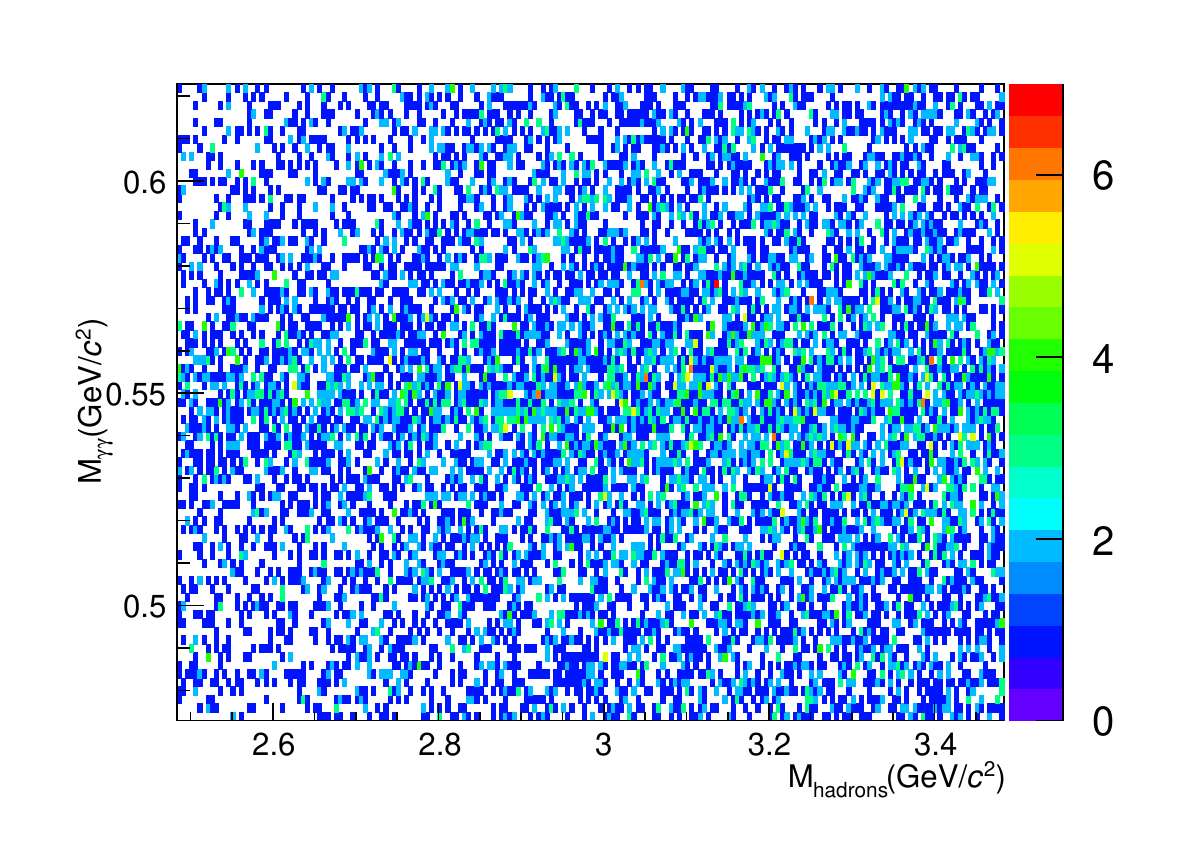}
        \captionsetup{skip=-7pt,font=normalsize}
    \end{subfigure}
    \begin{subfigure}{0.32\textwidth}
        \includegraphics[width=\linewidth]{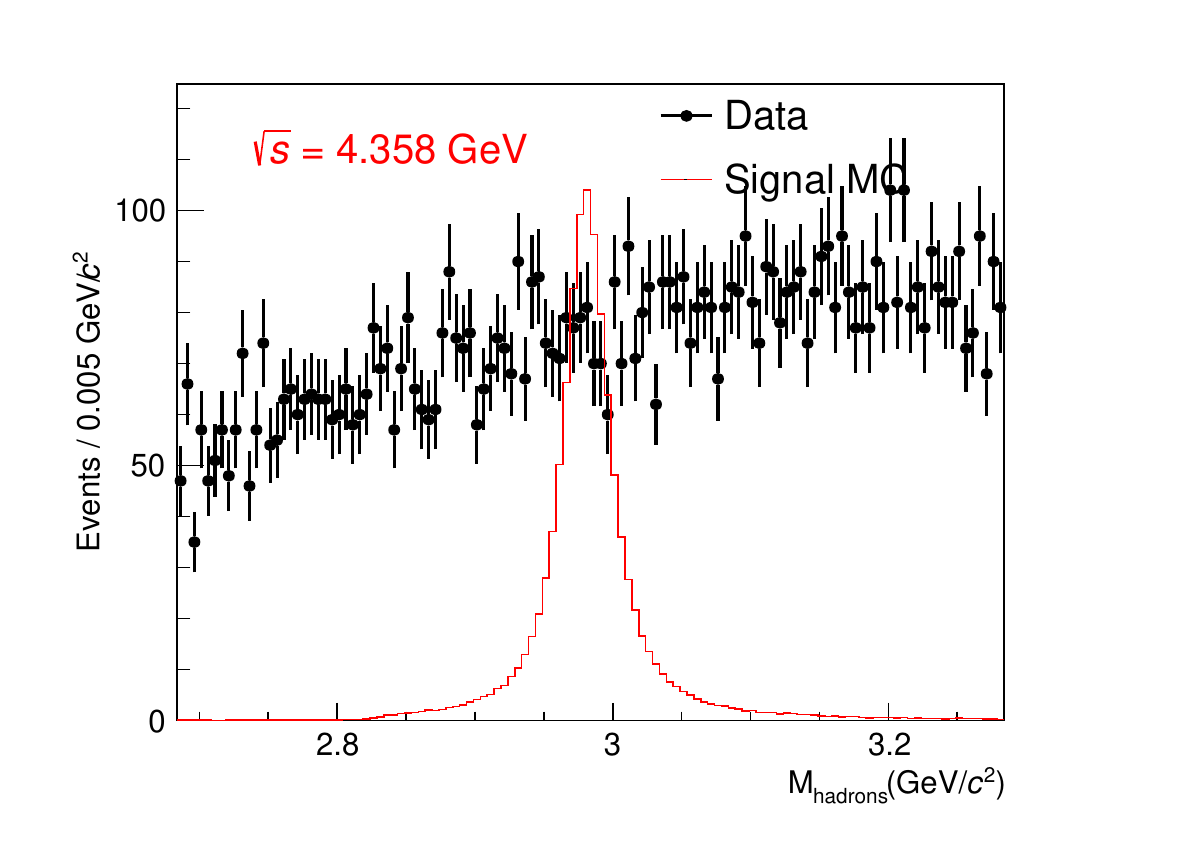}
        \captionsetup{skip=-7pt,font=normalsize}
    \end{subfigure}
    \begin{subfigure}{0.32\textwidth}
        \includegraphics[width=\linewidth]{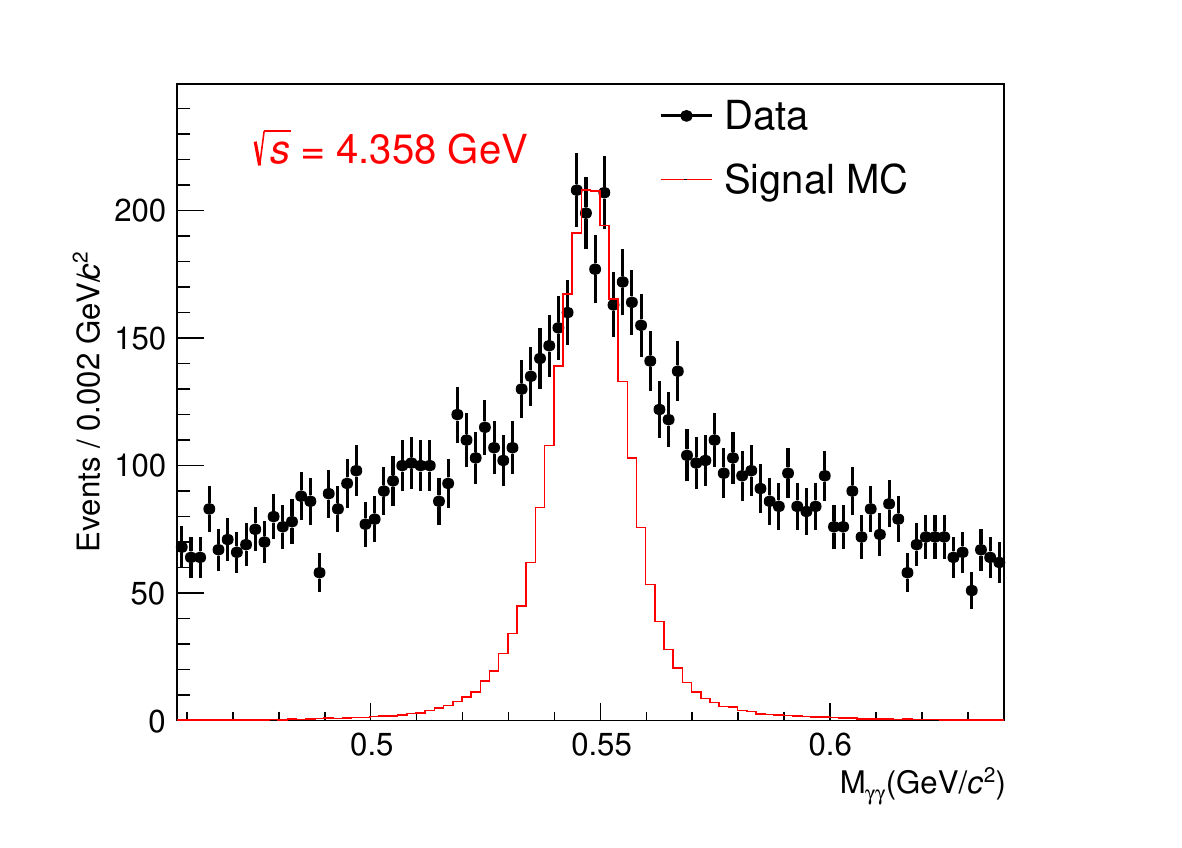}
        \captionsetup{skip=-7pt,font=normalsize}
    \end{subfigure}
\captionsetup{justification=raggedright}
\caption{The distributions of (Left) $M_{hadrons}$ versus $M_{\gamma\gamma}$, (Middle) $M_{hadrons}$, and (Right) $M_{\gamma\gamma}$ at $\sqrt s=4.308-4.358$~GeV.}
\label{fig:normal2}
\end{figure*}

\begin{figure*}[htbp]
    \begin{subfigure}{0.32\textwidth}
        \includegraphics[width=\linewidth]{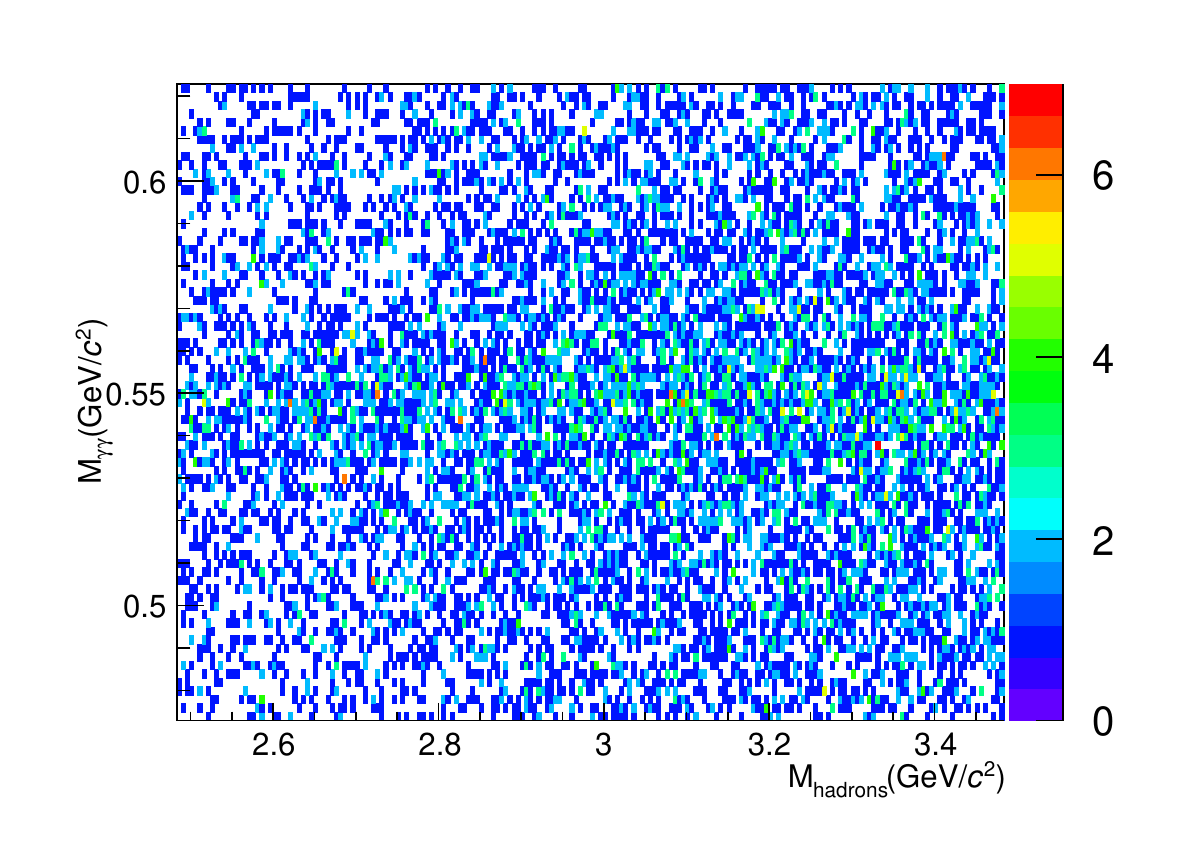}
        \captionsetup{skip=-7pt,font=normalsize}
    \end{subfigure}
    \begin{subfigure}{0.32\textwidth}
        \includegraphics[width=\linewidth]{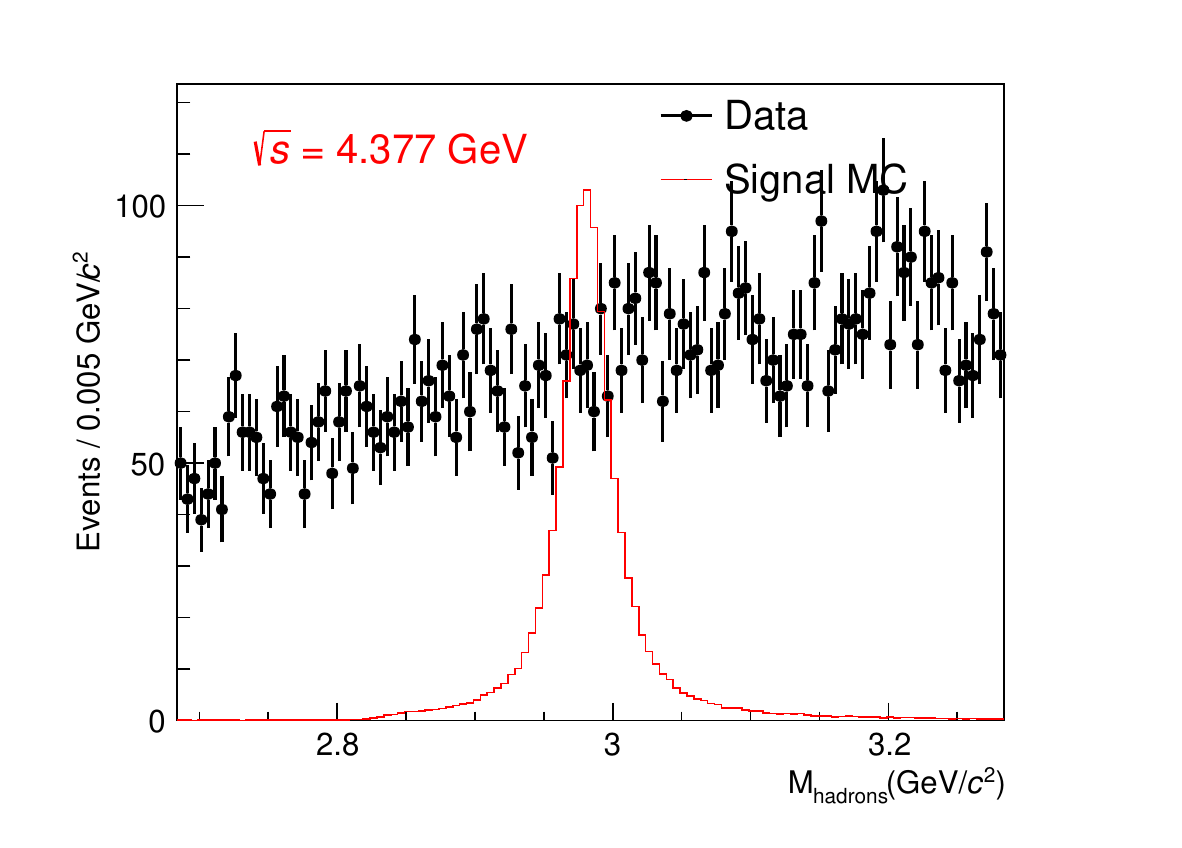}
        \captionsetup{skip=-7pt,font=normalsize}
    \end{subfigure}
    \begin{subfigure}{0.32\textwidth}
        \includegraphics[width=\linewidth]{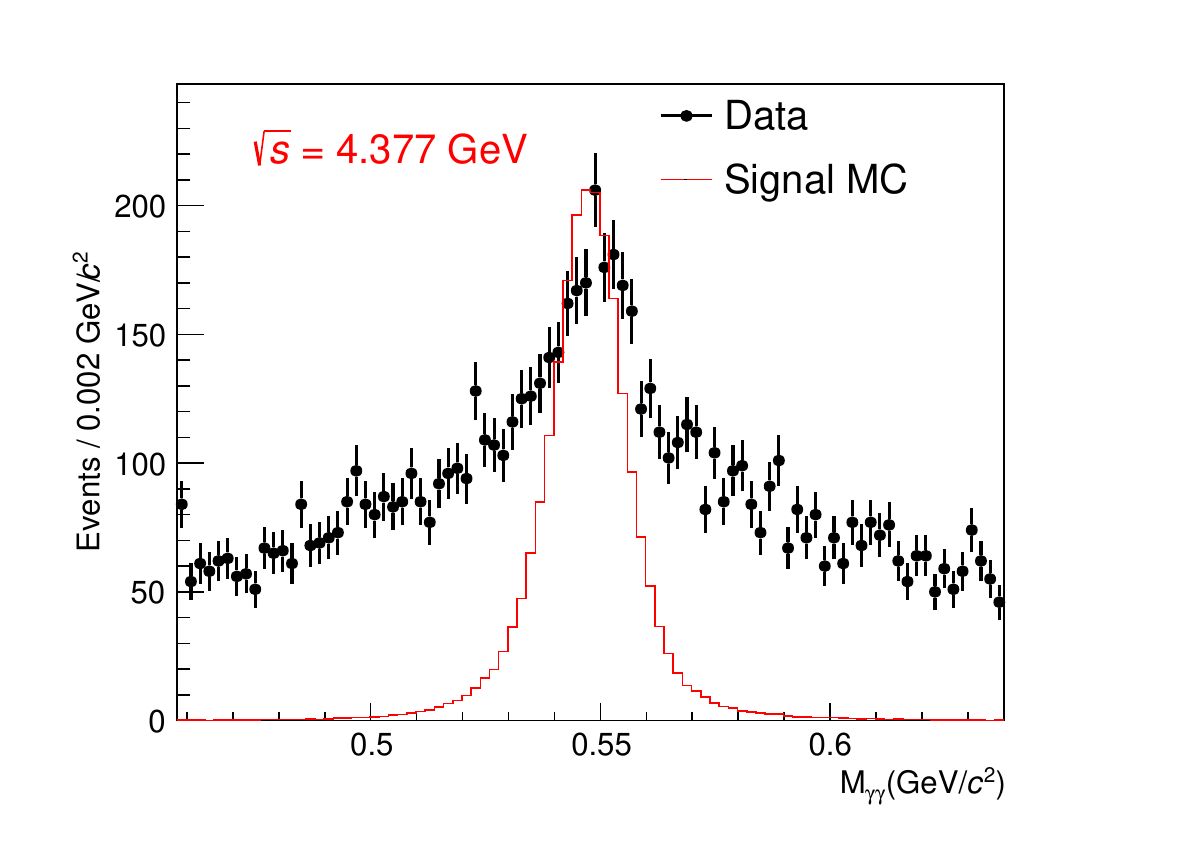}
        \captionsetup{skip=-7pt,font=normalsize}
    \end{subfigure}
    \begin{subfigure}{0.32\textwidth}
        \includegraphics[width=\linewidth]{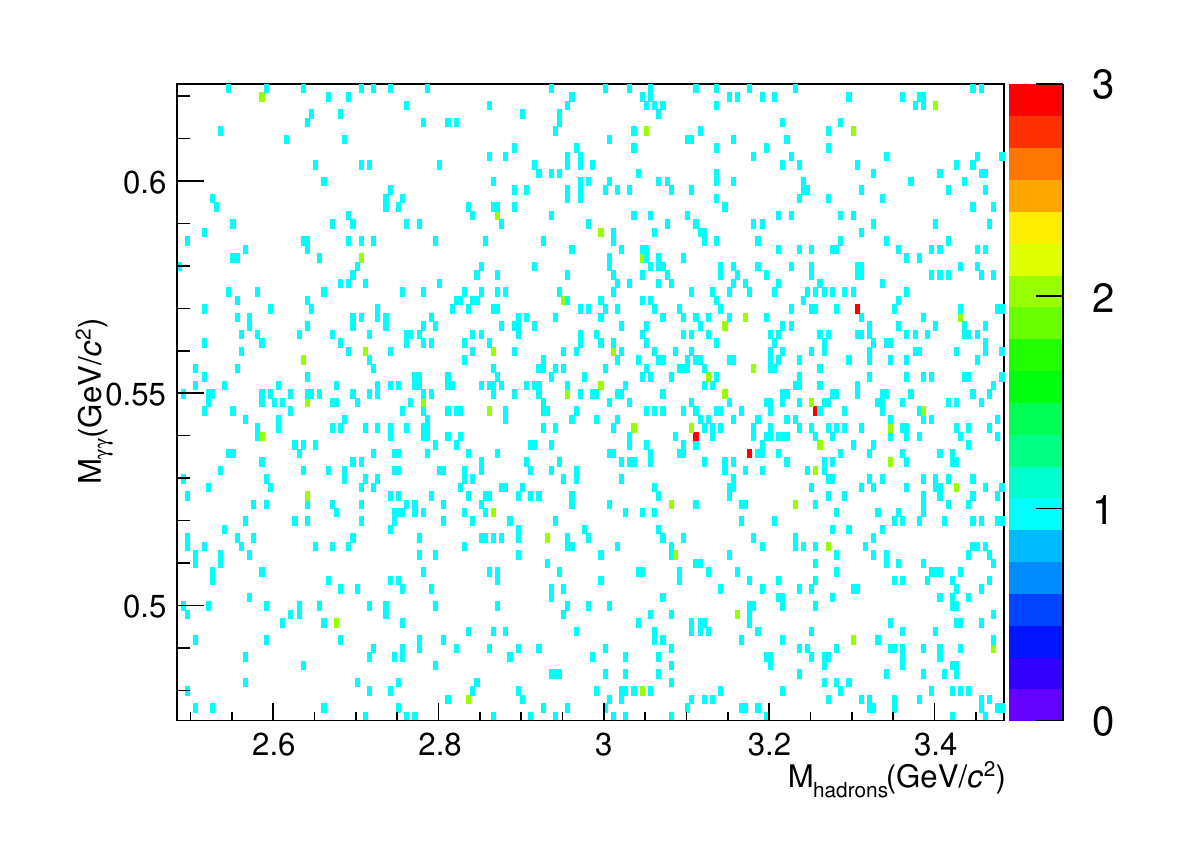}
        \captionsetup{skip=-7pt,font=normalsize}
    \end{subfigure}
    \begin{subfigure}{0.32\textwidth}
        \includegraphics[width=\linewidth]{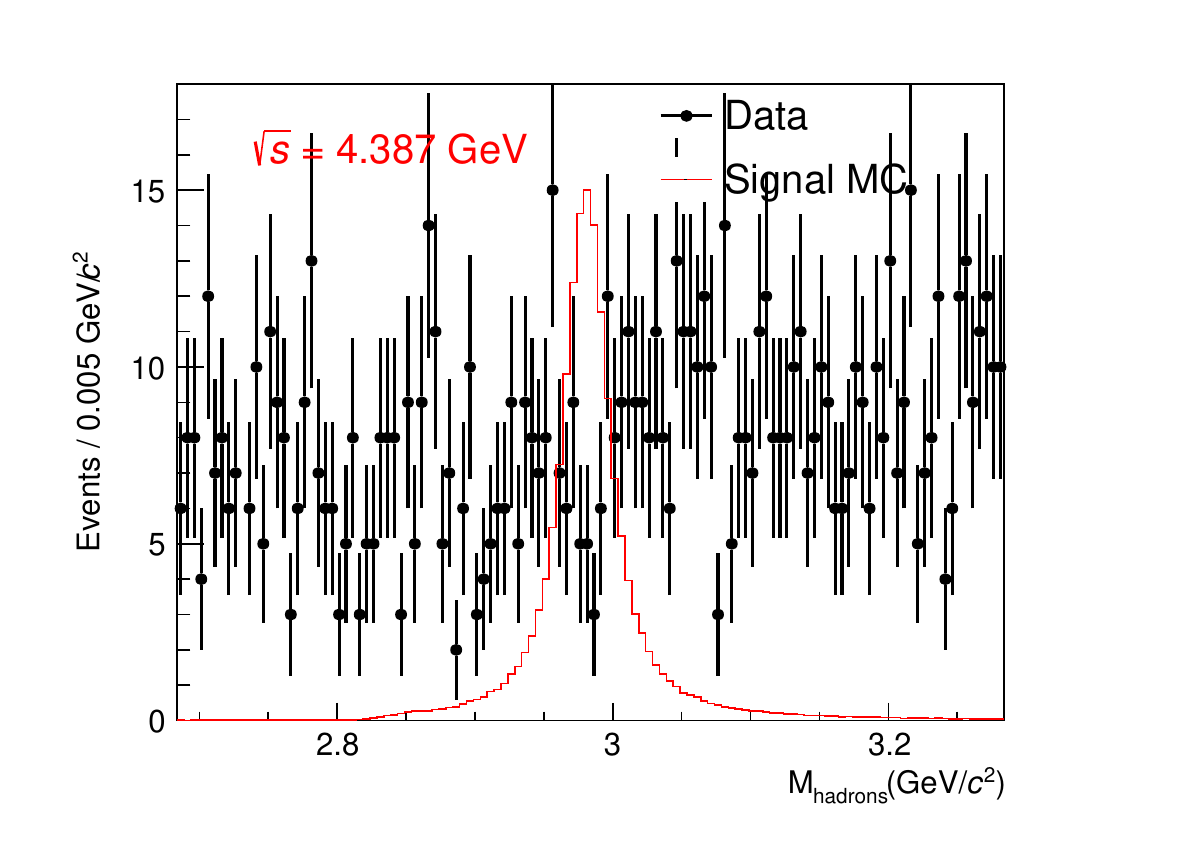}
        \captionsetup{skip=-7pt,font=normalsize}
    \end{subfigure}
    \begin{subfigure}{0.32\textwidth}
        \includegraphics[width=\linewidth]{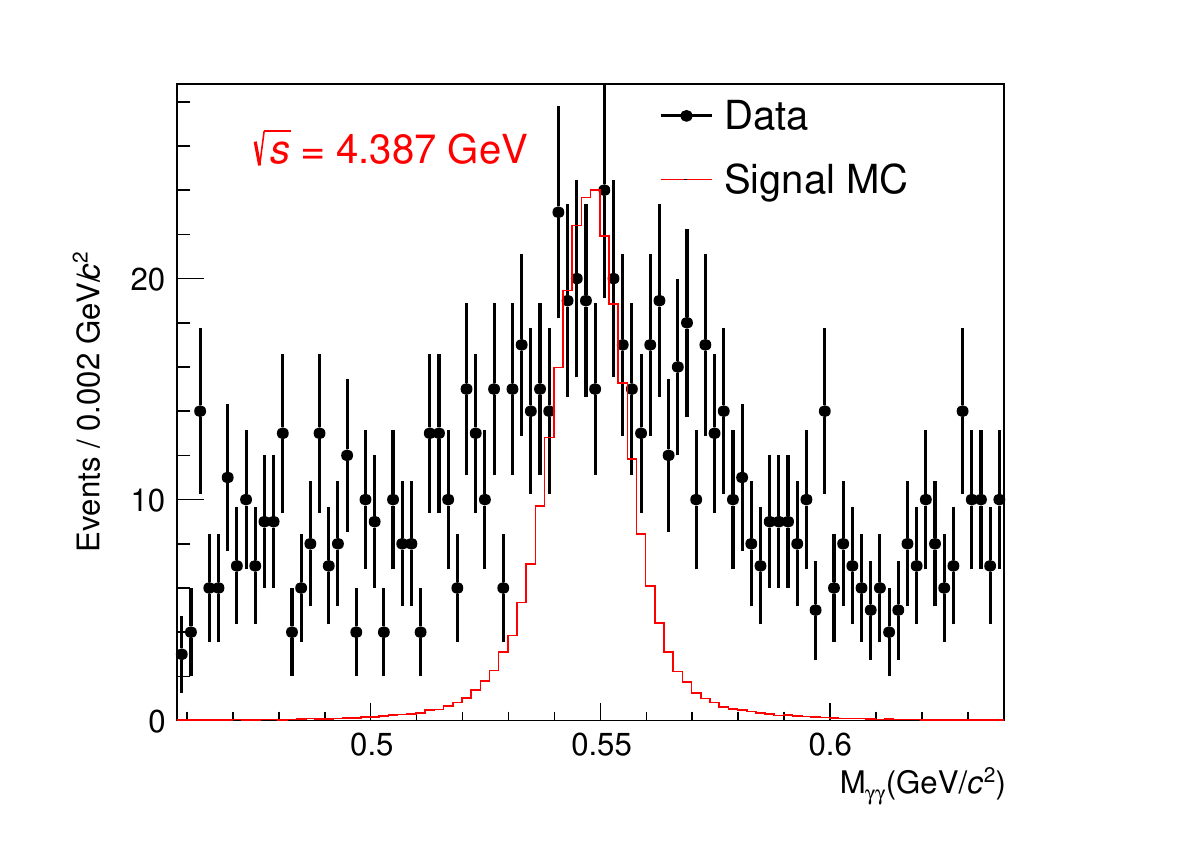}
        \captionsetup{skip=-7pt,font=normalsize}
    \end{subfigure}
    \begin{subfigure}{0.32\textwidth}
        \includegraphics[width=\linewidth]{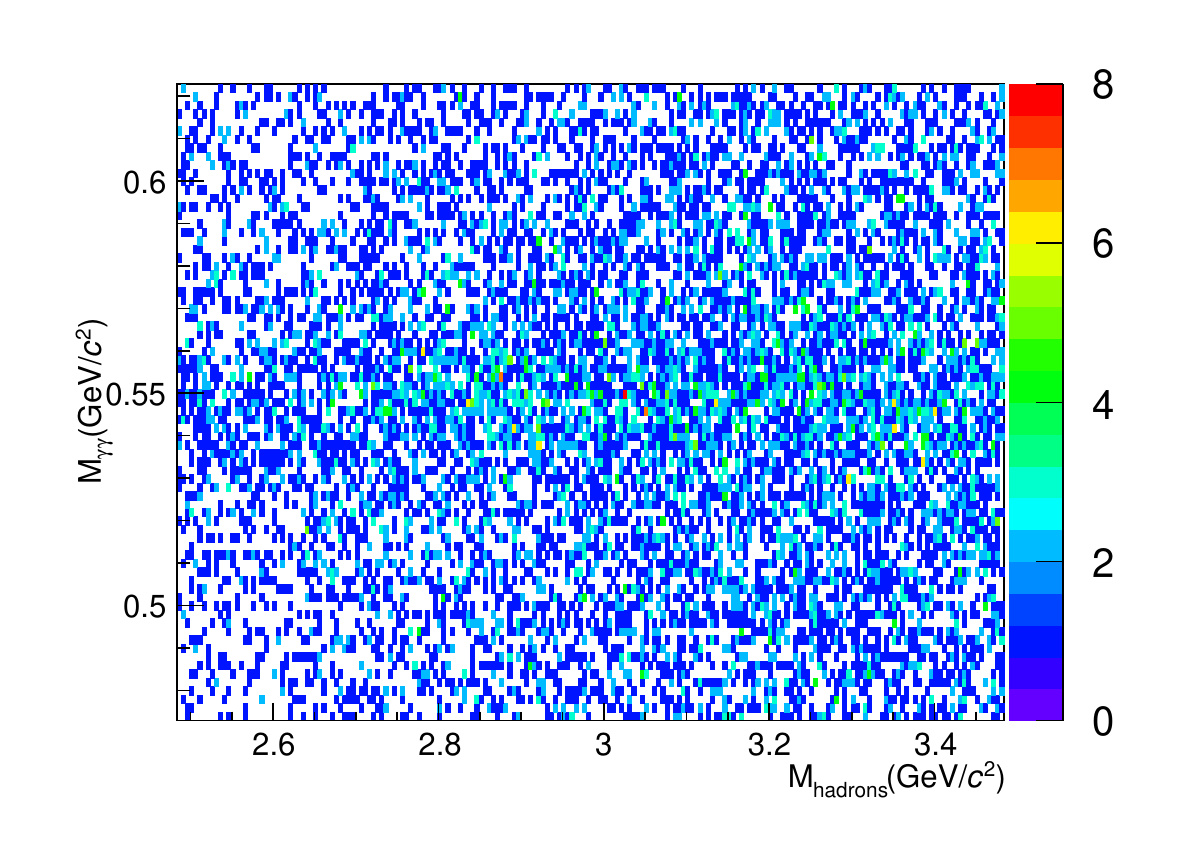}
        \captionsetup{skip=-7pt,font=normalsize}
    \end{subfigure}
    \begin{subfigure}{0.32\textwidth}
        \includegraphics[width=\linewidth]{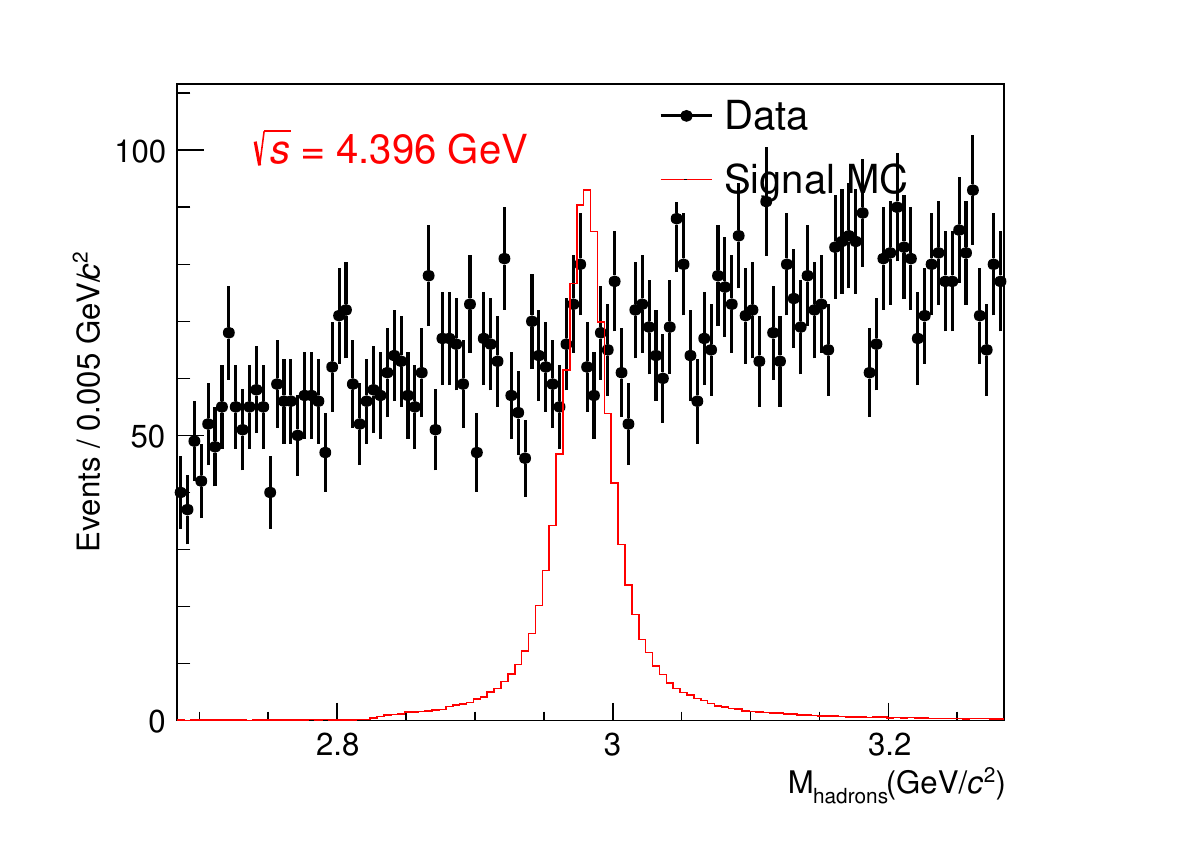}
        \captionsetup{skip=-7pt,font=normalsize}
    \end{subfigure}
    \begin{subfigure}{0.32\textwidth}
        \includegraphics[width=\linewidth]{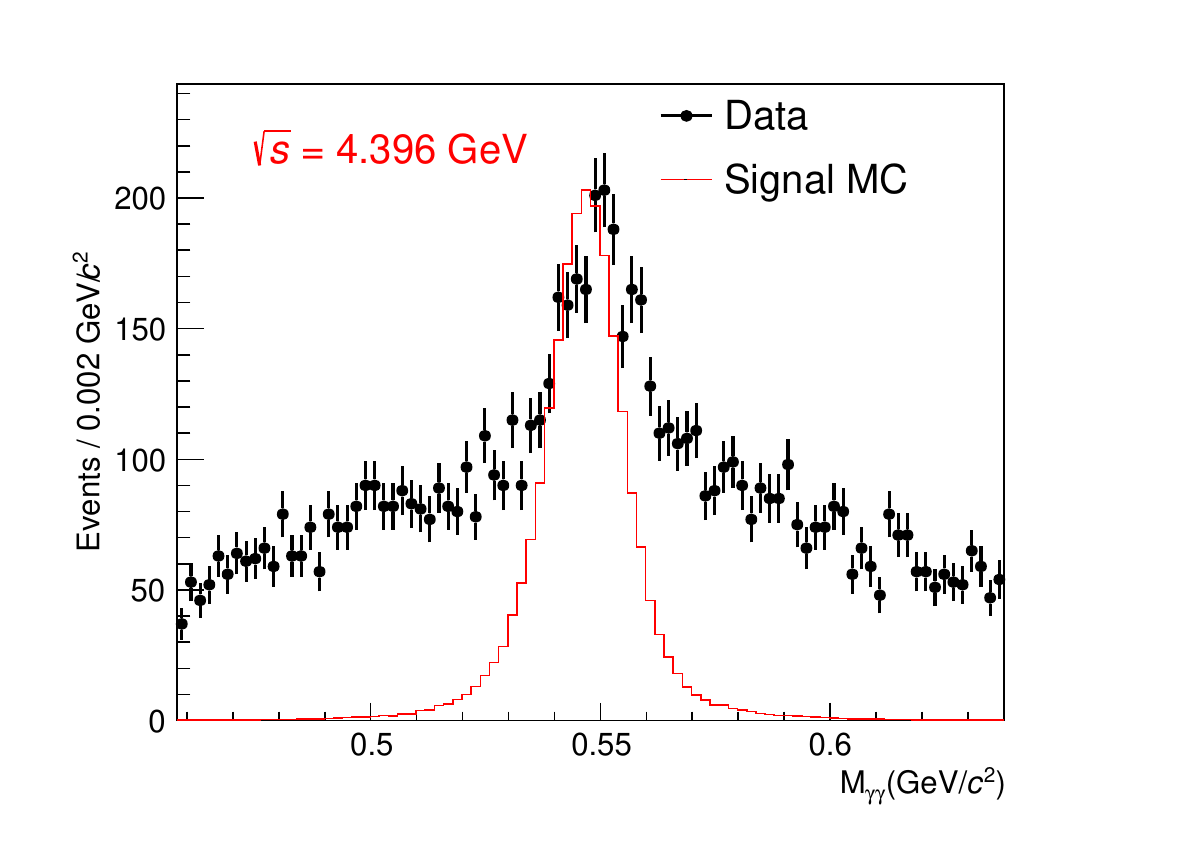}
        \captionsetup{skip=-7pt,font=normalsize}
    \end{subfigure}
    \begin{subfigure}{0.32\textwidth}
        \includegraphics[width=\linewidth]{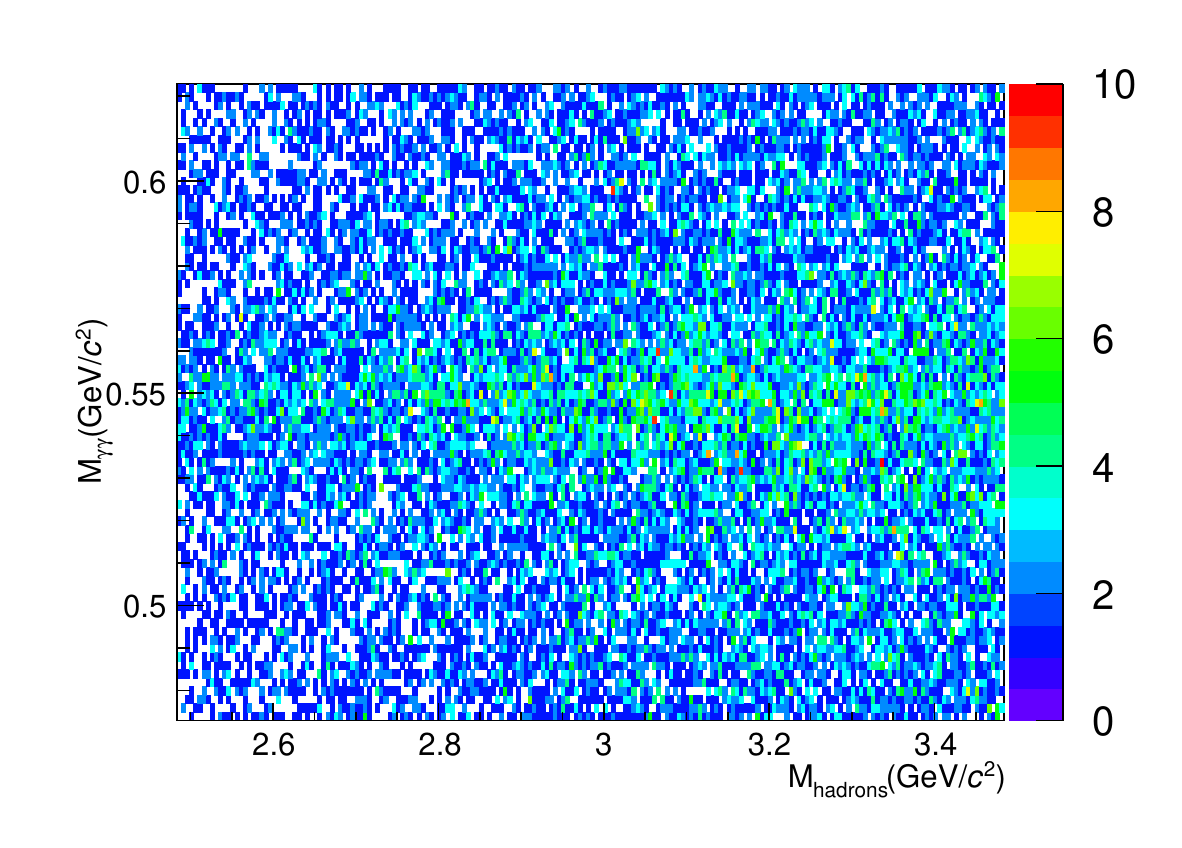}
        \captionsetup{skip=-7pt,font=normalsize}
    \end{subfigure}
    \begin{subfigure}{0.32\textwidth}
        \includegraphics[width=\linewidth]{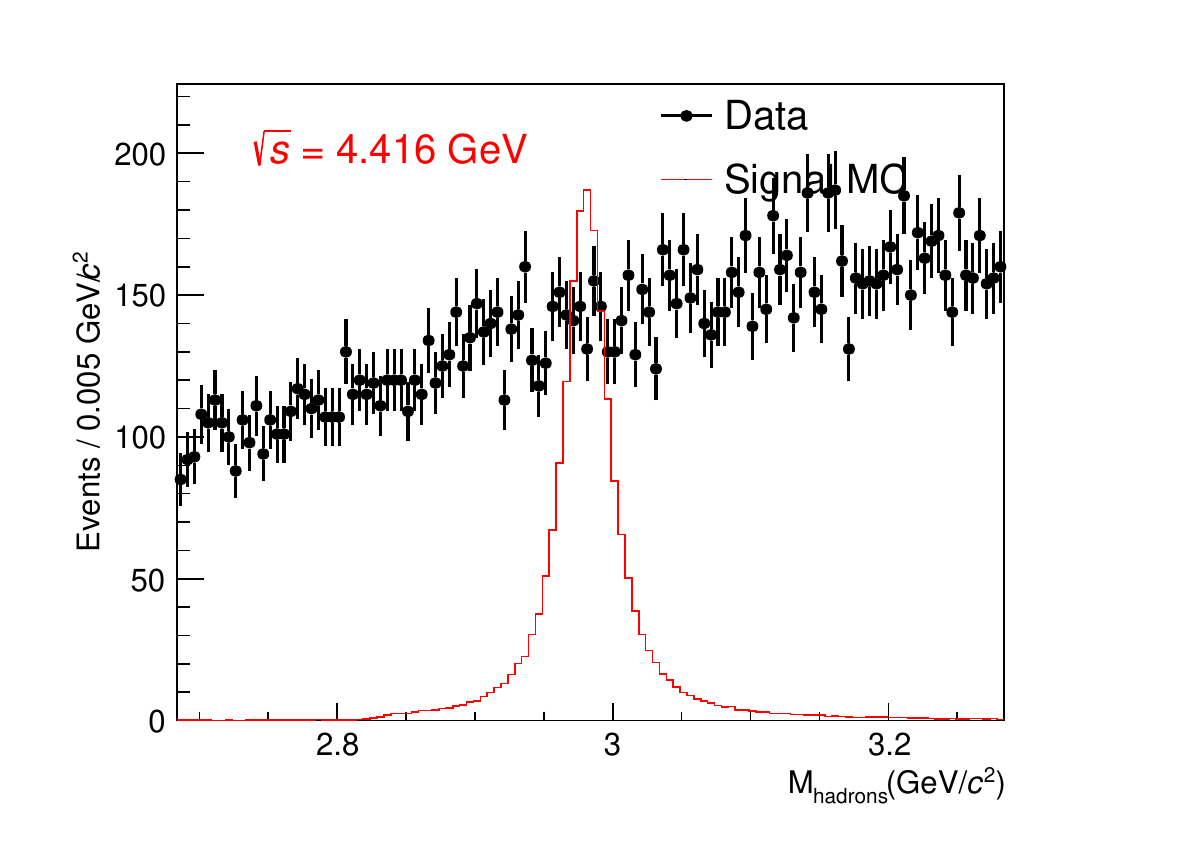}
        \captionsetup{skip=-7pt,font=normalsize}
    \end{subfigure}
    \begin{subfigure}{0.32\textwidth}
        \includegraphics[width=\linewidth]{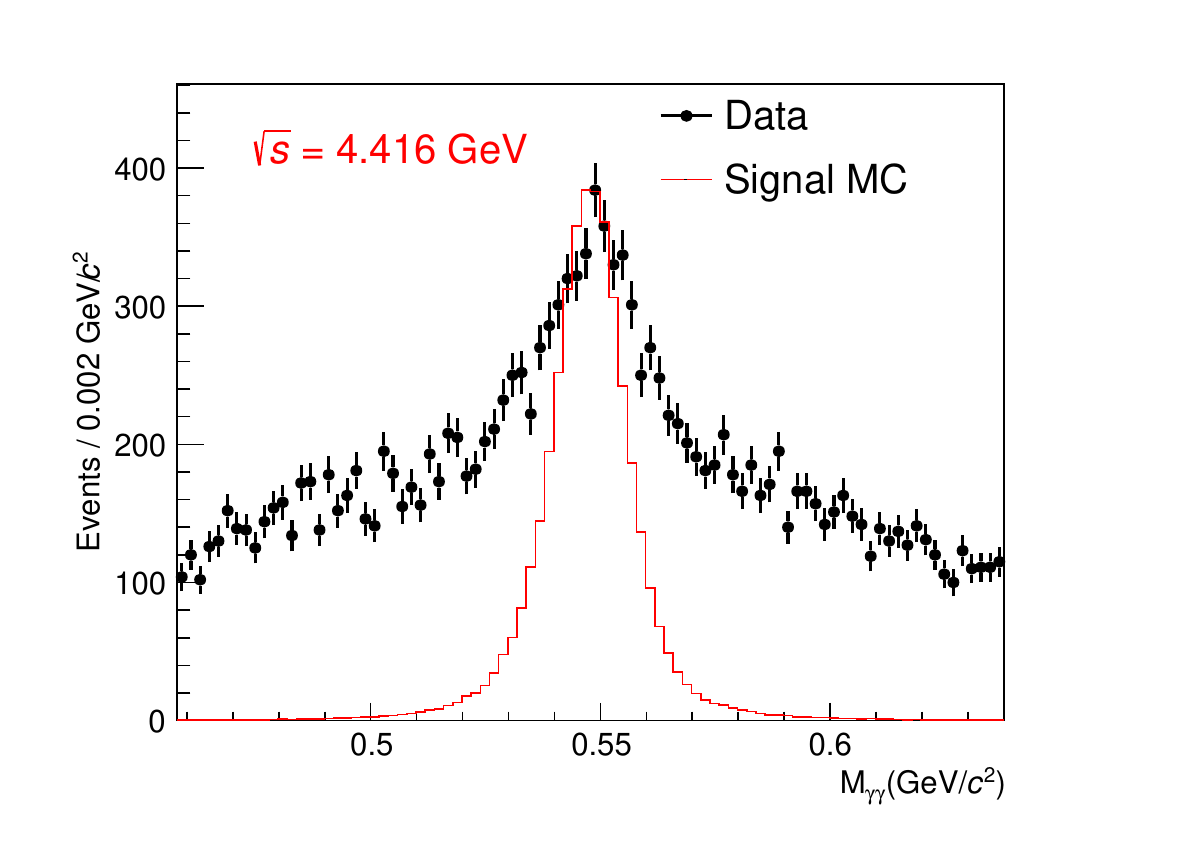}
        \captionsetup{skip=-7pt,font=normalsize}
    \end{subfigure}
\captionsetup{justification=raggedright}
\caption{The distributions of (Left) $M_{hadrons}$ versus $M_{\gamma\gamma}$, (Middle) $M_{hadrons}$, and (Right) $M_{\gamma\gamma}$ at $\sqrt s=4.377-4.416$~GeV.}
\label{fig:normal3}
\end{figure*}

\begin{figure*}[htbp]
    \begin{subfigure}{0.32\textwidth}
        \includegraphics[width=\linewidth]{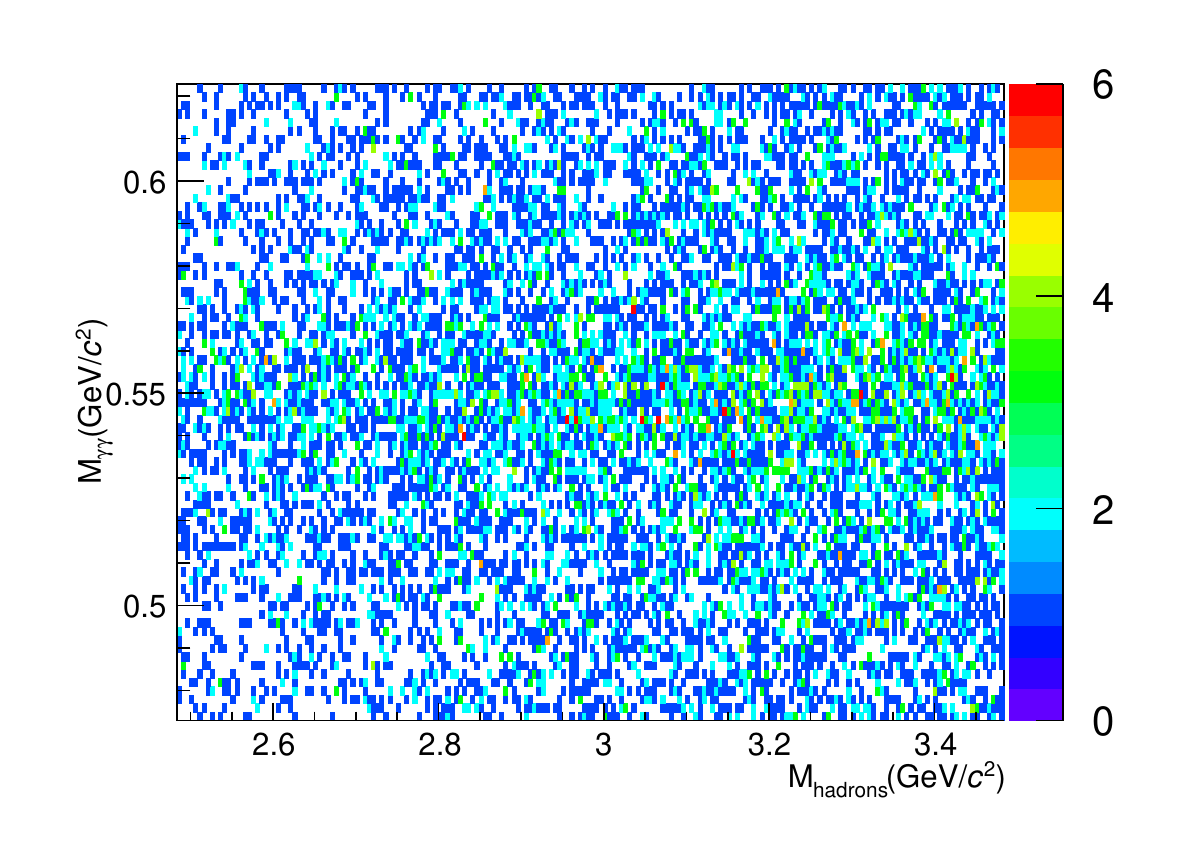}
        \captionsetup{skip=-7pt,font=normalsize}
    \end{subfigure}
    \begin{subfigure}{0.32\textwidth}
        \includegraphics[width=\linewidth]{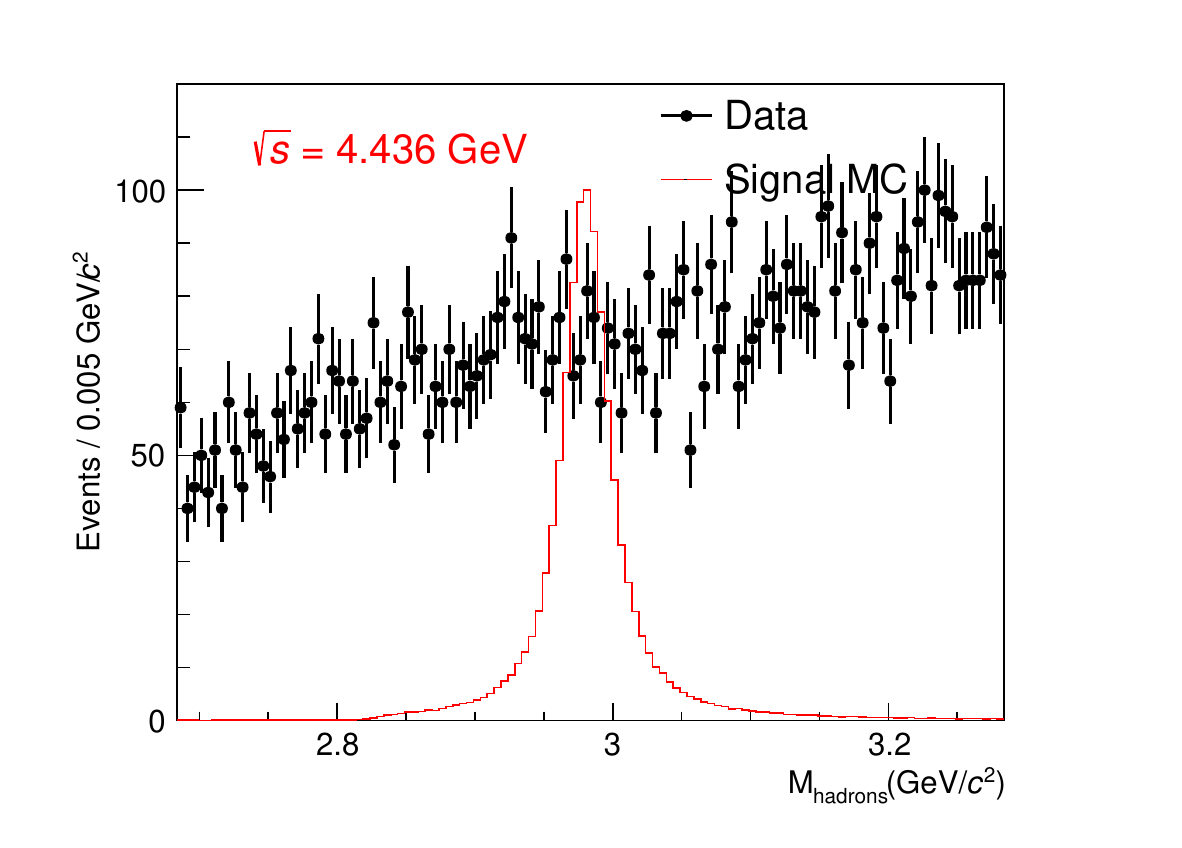}
        \captionsetup{skip=-7pt,font=normalsize}
    \end{subfigure}
    \begin{subfigure}{0.32\textwidth}
        \includegraphics[width=\linewidth]{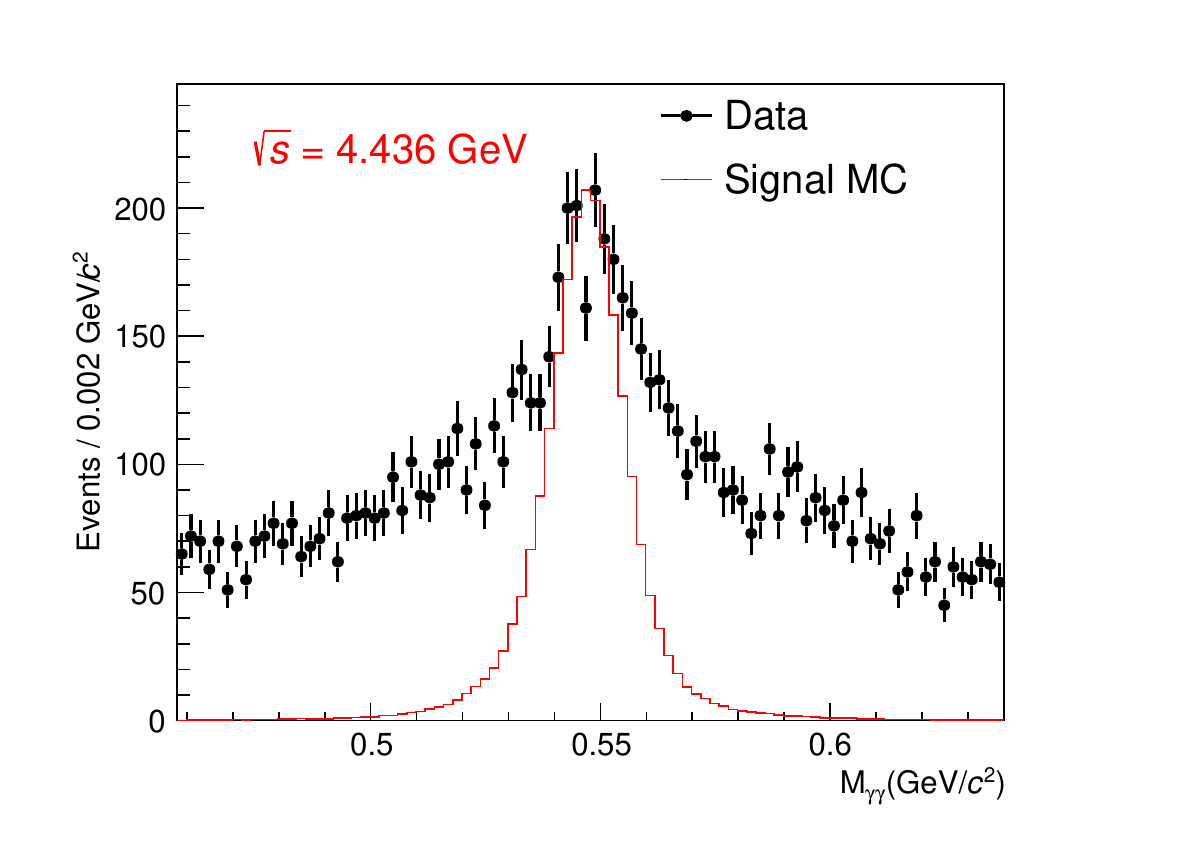}
        \captionsetup{skip=-7pt,font=normalsize}
    \end{subfigure}
    \begin{subfigure}{0.32\textwidth}
        \includegraphics[width=\linewidth]{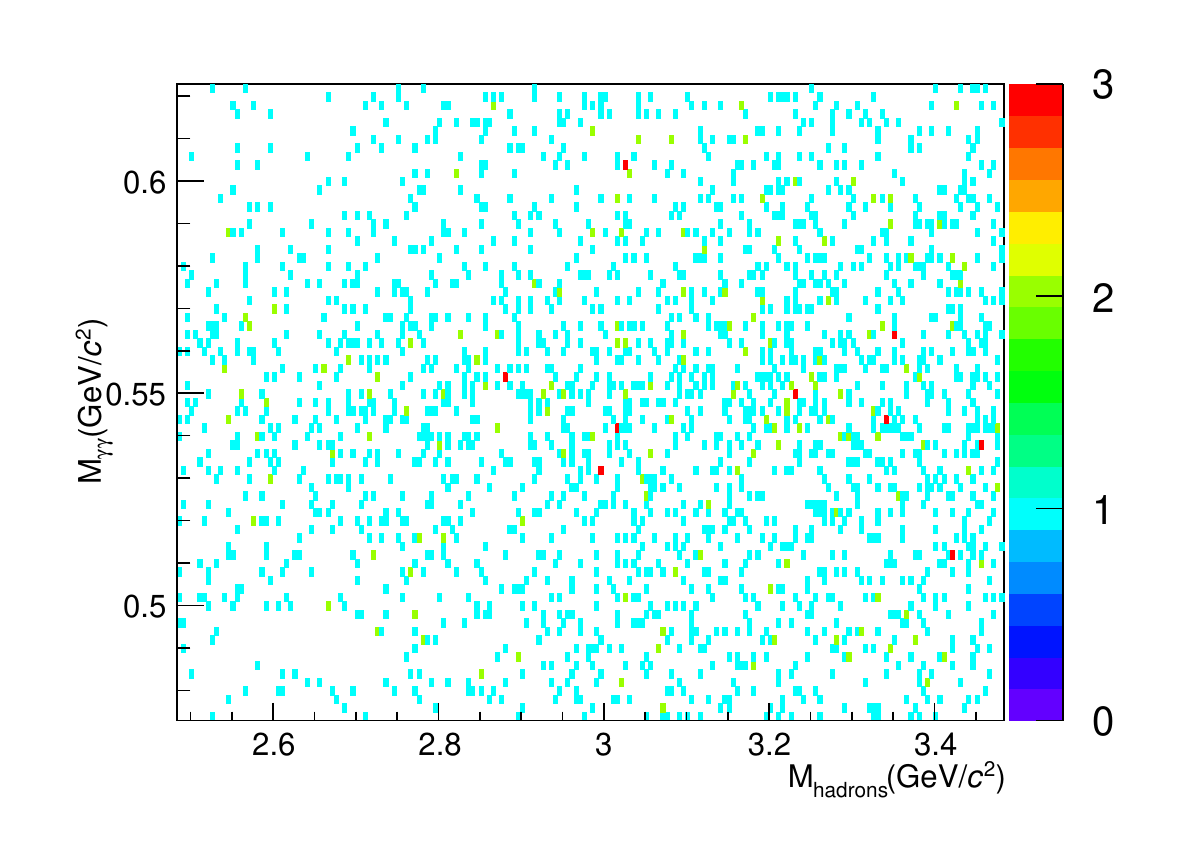}
        \captionsetup{skip=-7pt,font=normalsize}
    \end{subfigure}
    \begin{subfigure}{0.32\textwidth}
        \includegraphics[width=\linewidth]{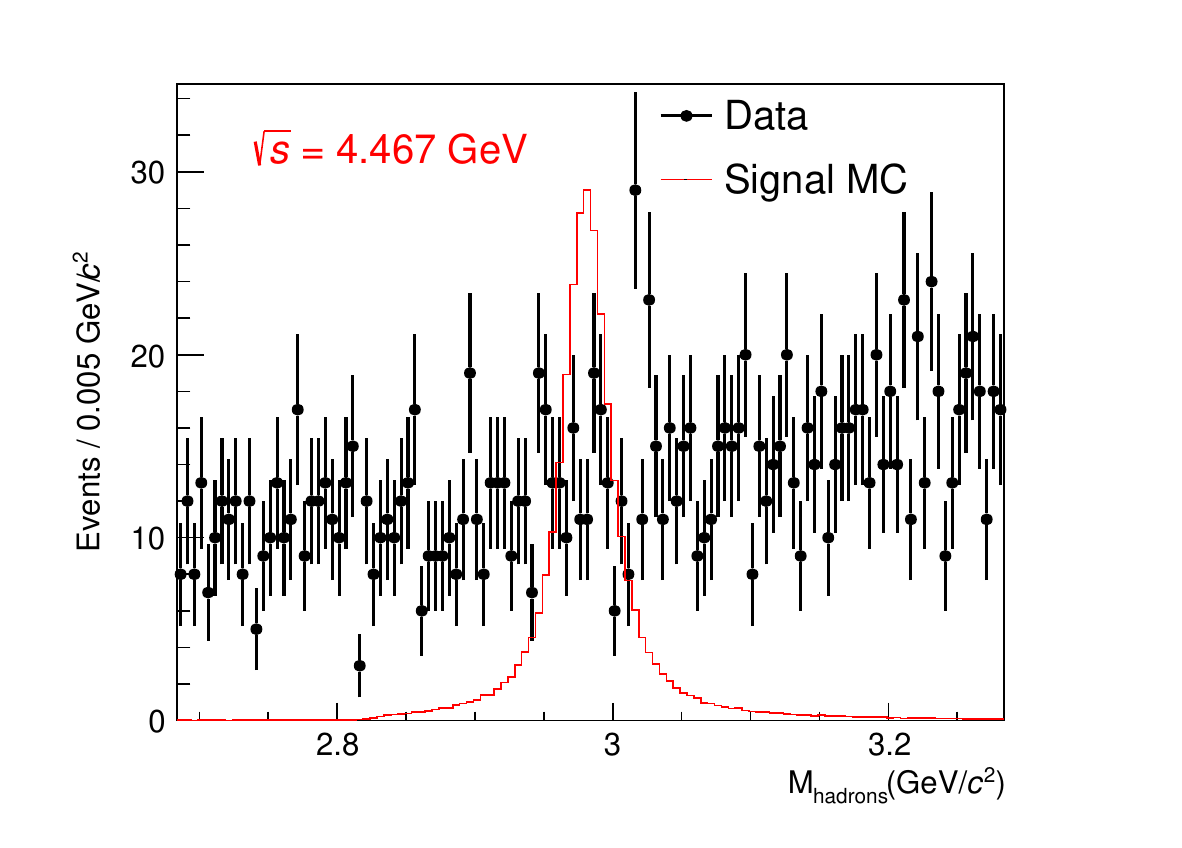}
        \captionsetup{skip=-7pt,font=normalsize}
    \end{subfigure}
    \begin{subfigure}{0.32\textwidth}
        \includegraphics[width=\linewidth]{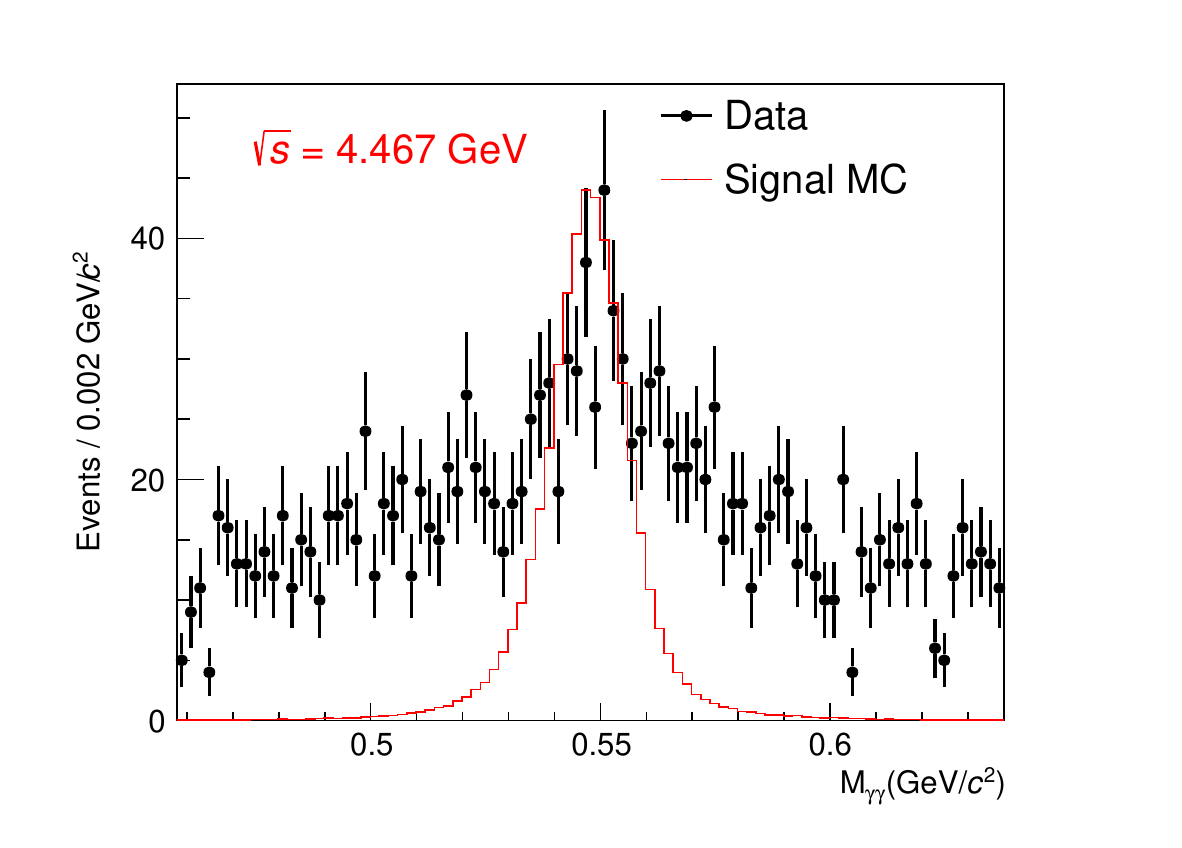}
        \captionsetup{skip=-7pt,font=normalsize}
    \end{subfigure}
    \begin{subfigure}{0.32\textwidth}
        \includegraphics[width=\linewidth]{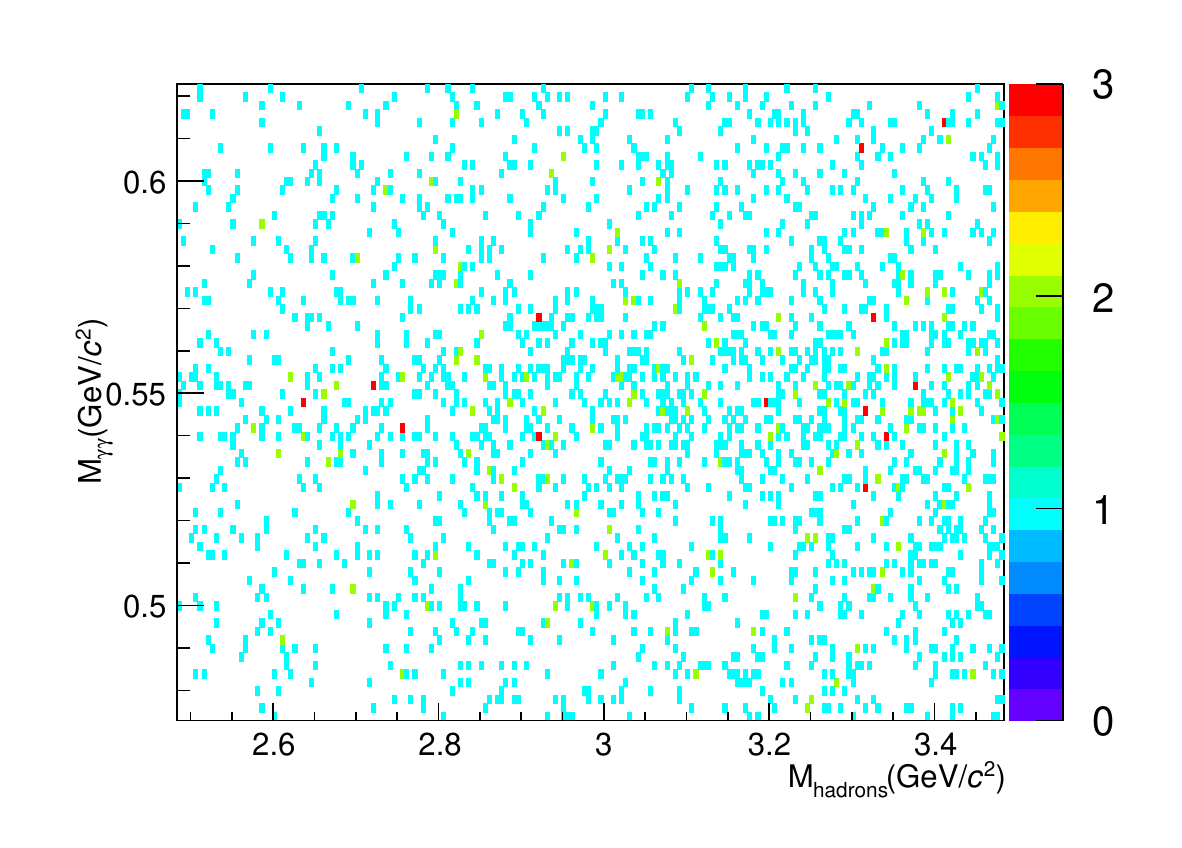}
        \captionsetup{skip=-7pt,font=normalsize}
    \end{subfigure}
    \begin{subfigure}{0.32\textwidth}
        \includegraphics[width=\linewidth]{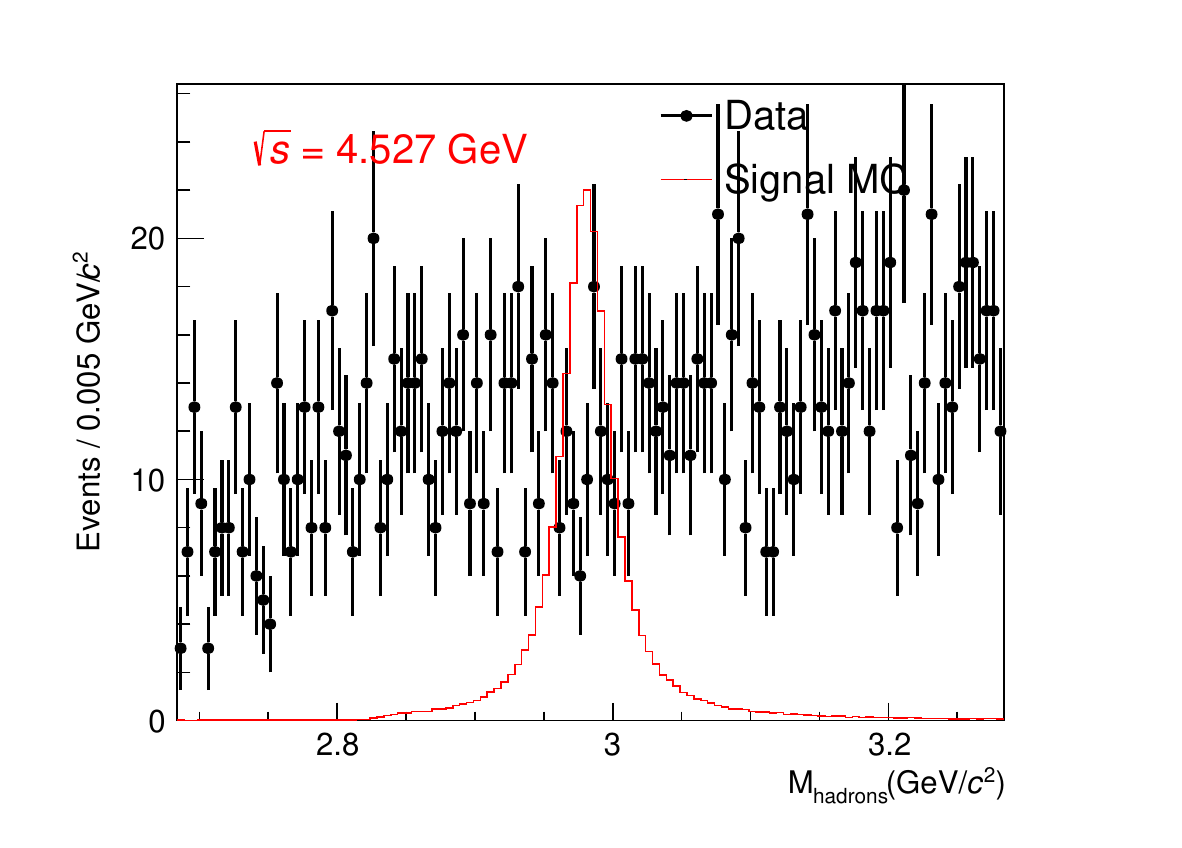}
        \captionsetup{skip=-7pt,font=normalsize}
    \end{subfigure}
    \begin{subfigure}{0.32\textwidth}
        \includegraphics[width=\linewidth]{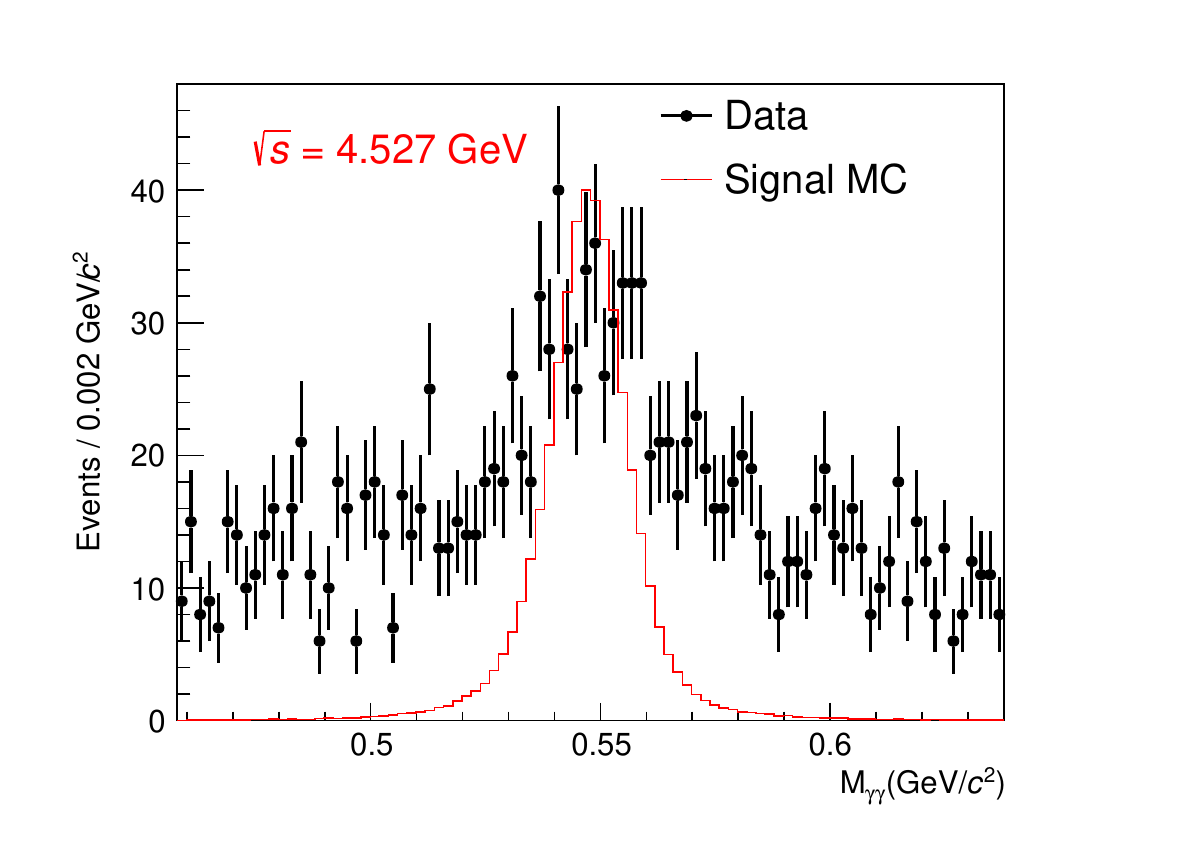}
        \captionsetup{skip=-7pt,font=normalsize}
    \end{subfigure}
    \begin{subfigure}{0.32\textwidth}
        \includegraphics[width=\linewidth]{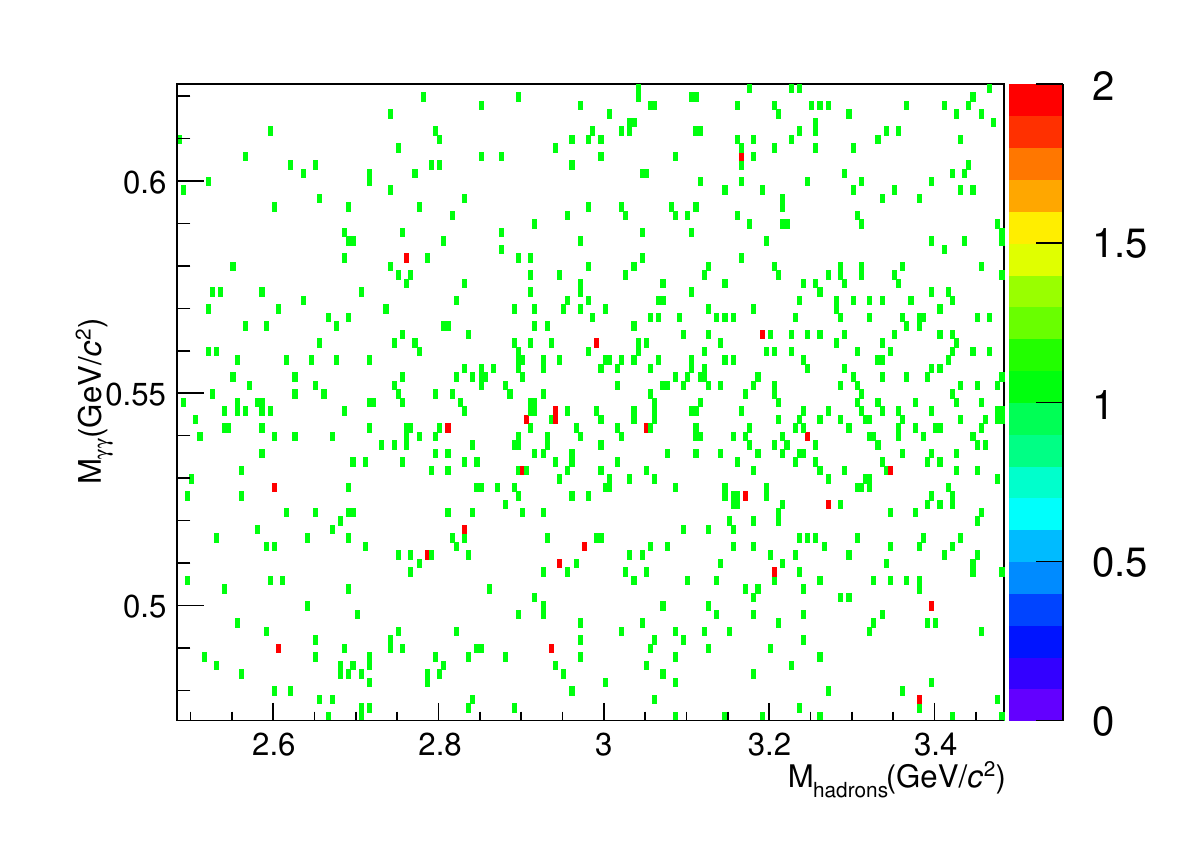}
        \captionsetup{skip=-7pt,font=normalsize}
    \end{subfigure}
    \begin{subfigure}{0.32\textwidth}
        \includegraphics[width=\linewidth]{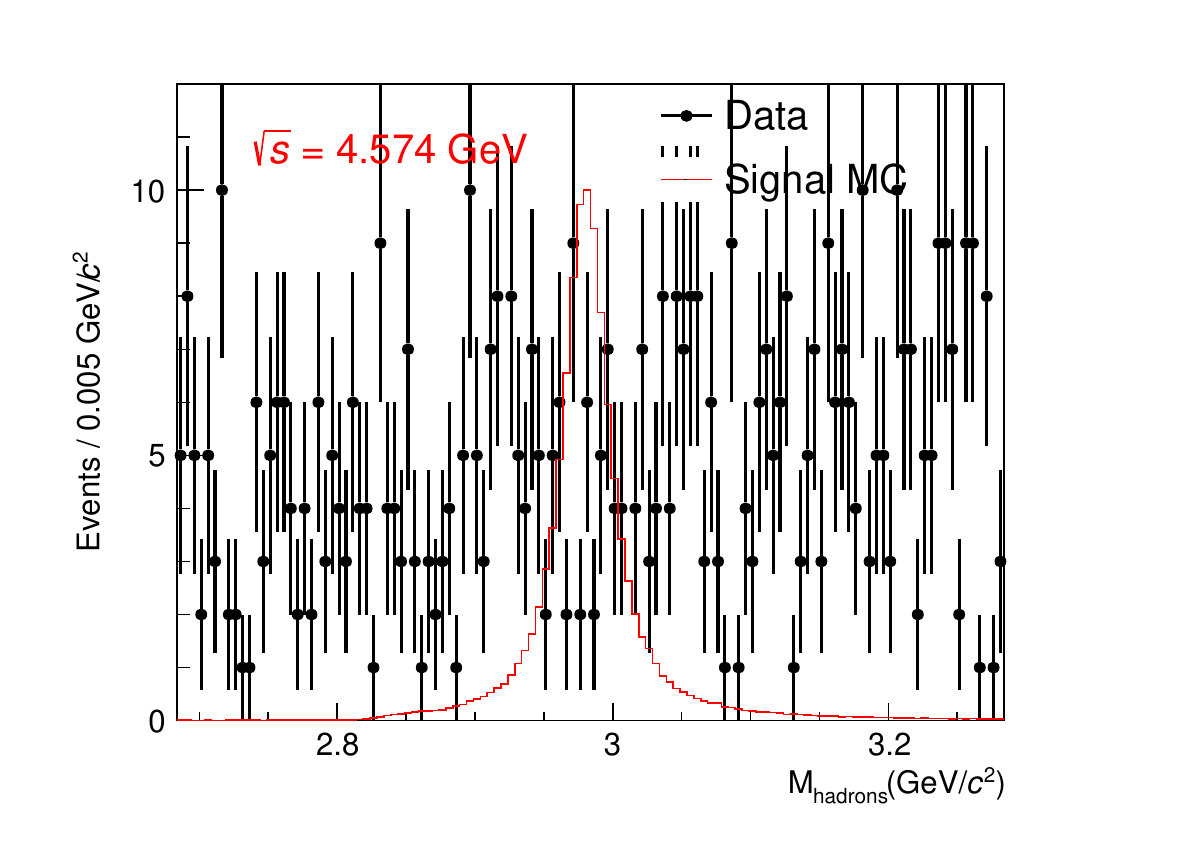}
        \captionsetup{skip=-7pt,font=normalsize}
    \end{subfigure}
    \begin{subfigure}{0.32\textwidth}
        \includegraphics[width=\linewidth]{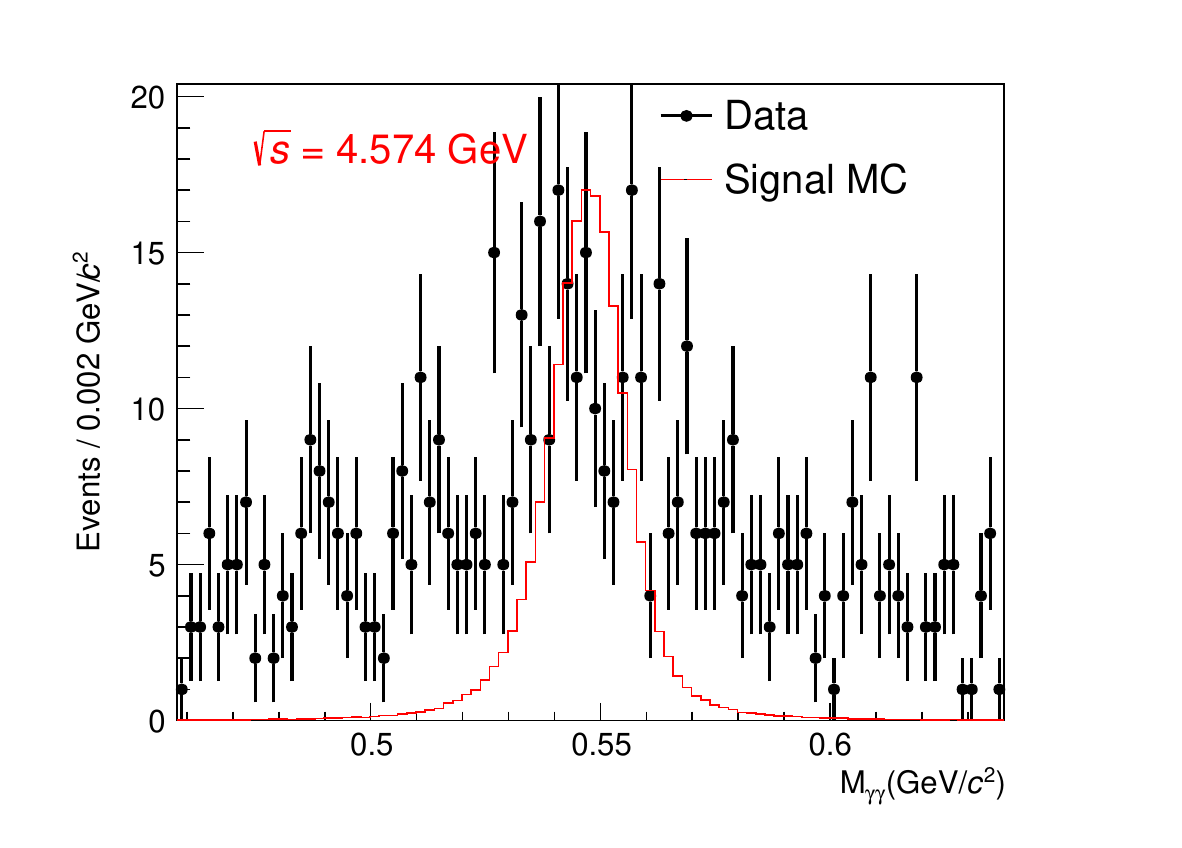}
        \captionsetup{skip=-7pt,font=normalsize}
    \end{subfigure}
\captionsetup{justification=raggedright}
\caption{The distributions of (Left) $M_{hadrons}$ versus $M_{\gamma\gamma}$, (Middle) $M_{hadrons}$, and (Right) $M_{\gamma\gamma}$ at $\sqrt s=4.436-4.574$~GeV.}
\label{fig:normal4}
\end{figure*}

\begin{figure*}[htbp]
    \begin{subfigure}{0.32\textwidth}
        \includegraphics[width=\linewidth]{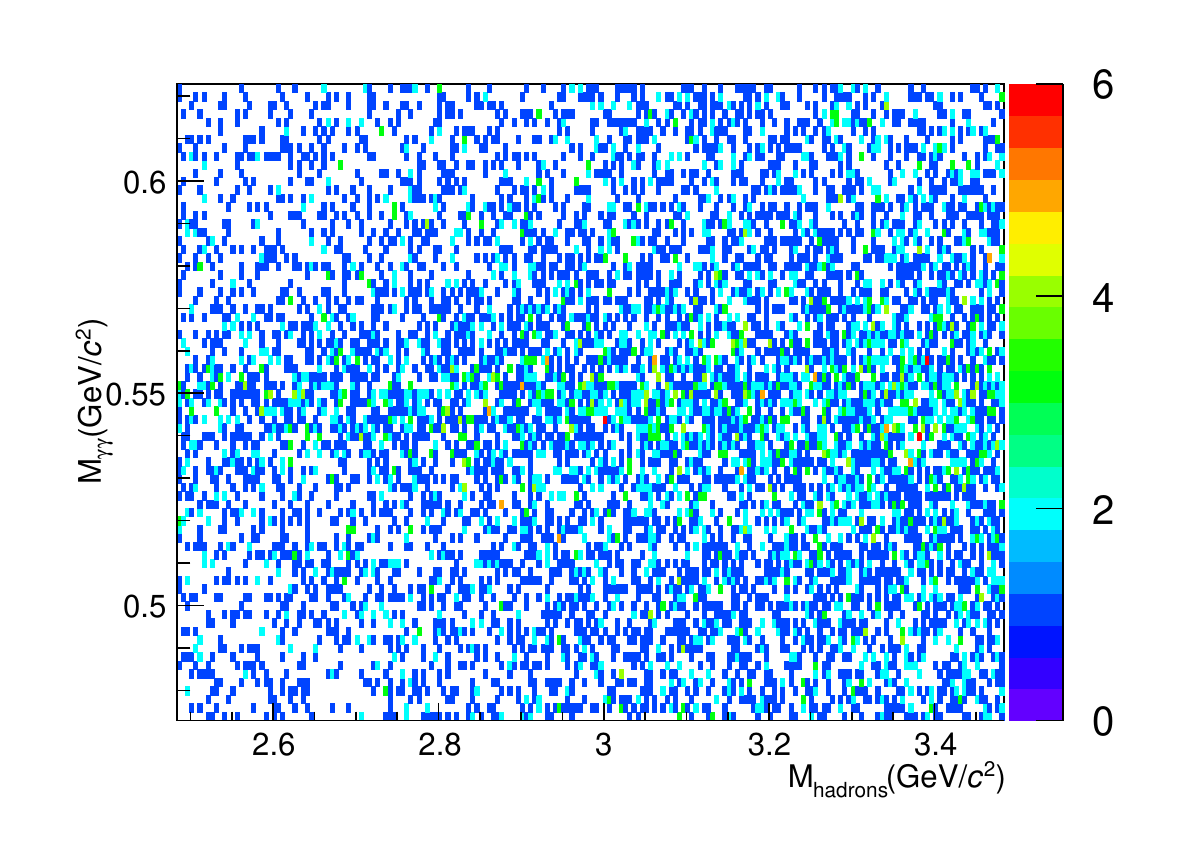}
        \captionsetup{skip=-7pt,font=normalsize}
    \end{subfigure}
    \begin{subfigure}{0.32\textwidth}
        \includegraphics[width=\linewidth]{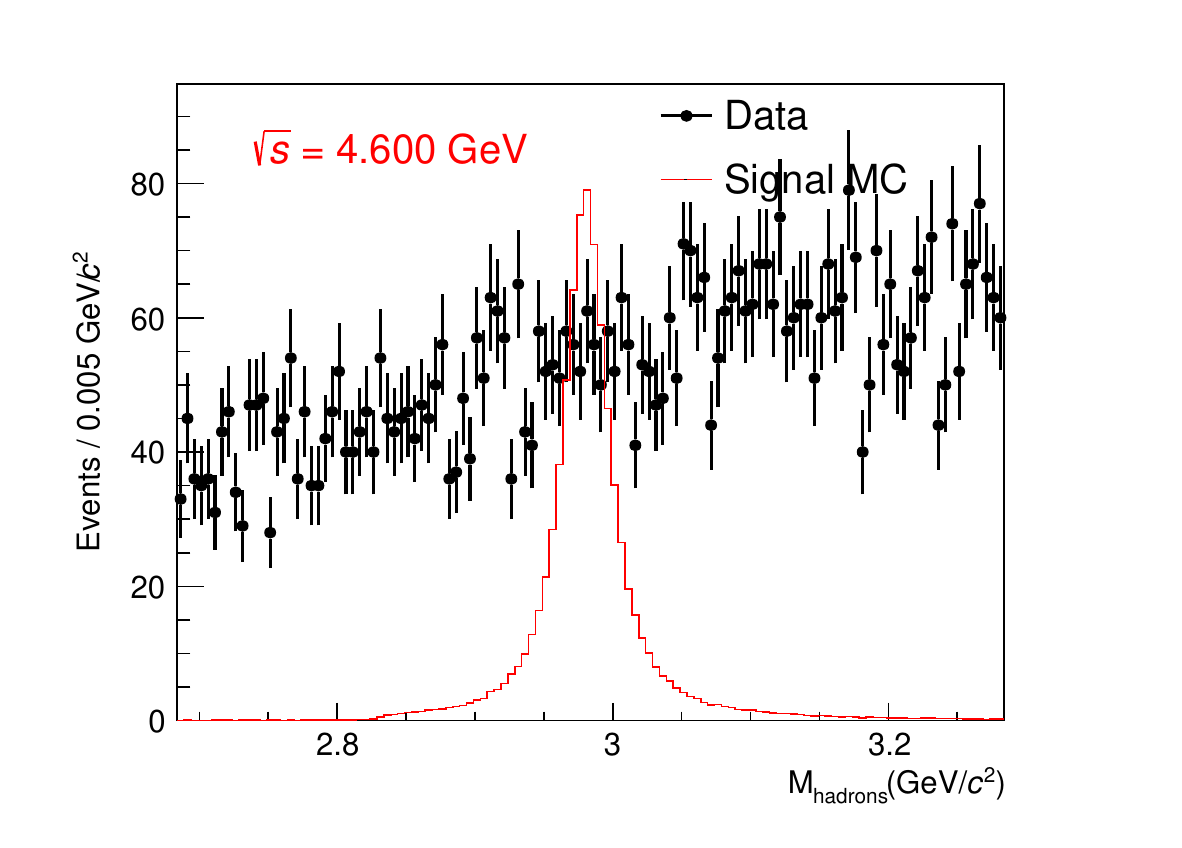}
        \captionsetup{skip=-7pt,font=normalsize}
    \end{subfigure}
    \begin{subfigure}{0.32\textwidth}
        \includegraphics[width=\linewidth]{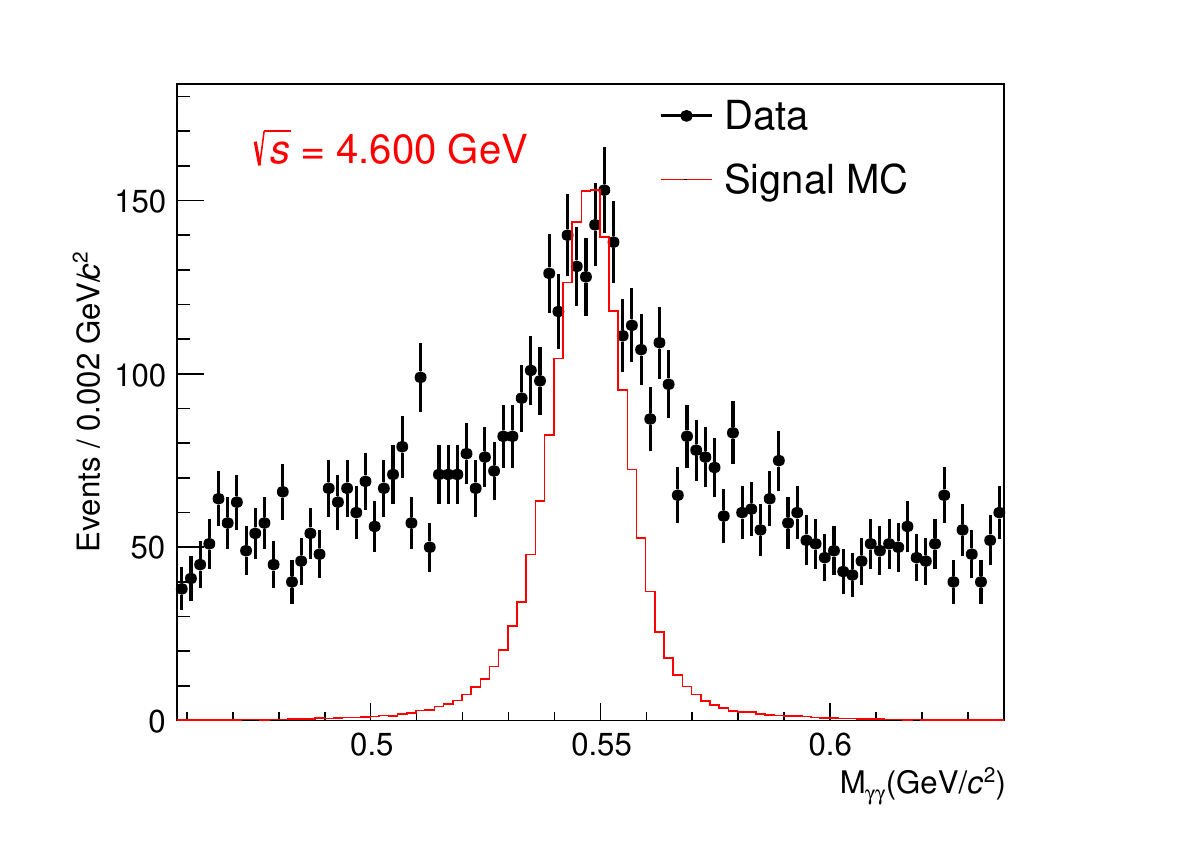}
        \captionsetup{skip=-7pt,font=normalsize}
    \end{subfigure}
    \begin{subfigure}{0.32\textwidth}
        \includegraphics[width=\linewidth]{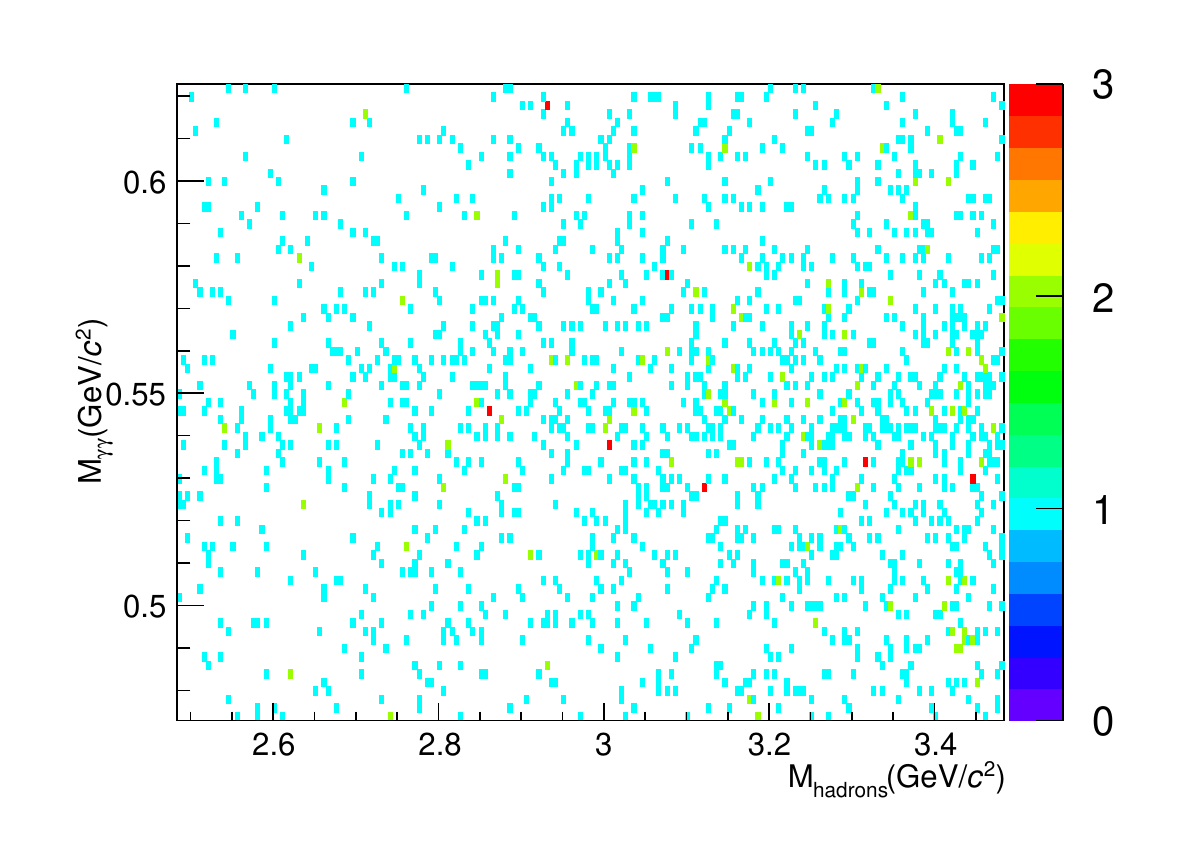}
        \captionsetup{skip=-7pt,font=normalsize}
    \end{subfigure}
    \begin{subfigure}{0.32\textwidth}
        \includegraphics[width=\linewidth]{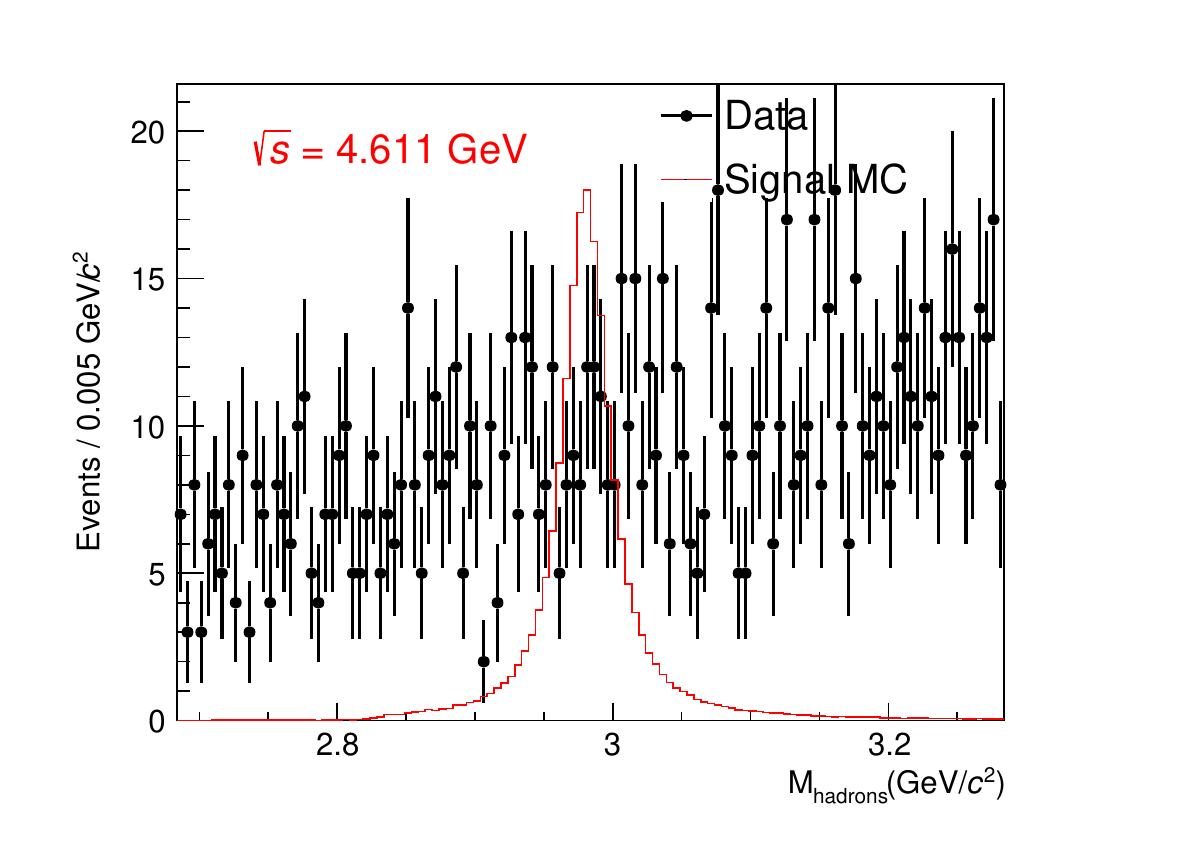}
        \captionsetup{skip=-7pt,font=normalsize}
    \end{subfigure}
    \begin{subfigure}{0.32\textwidth}
        \includegraphics[width=\linewidth]{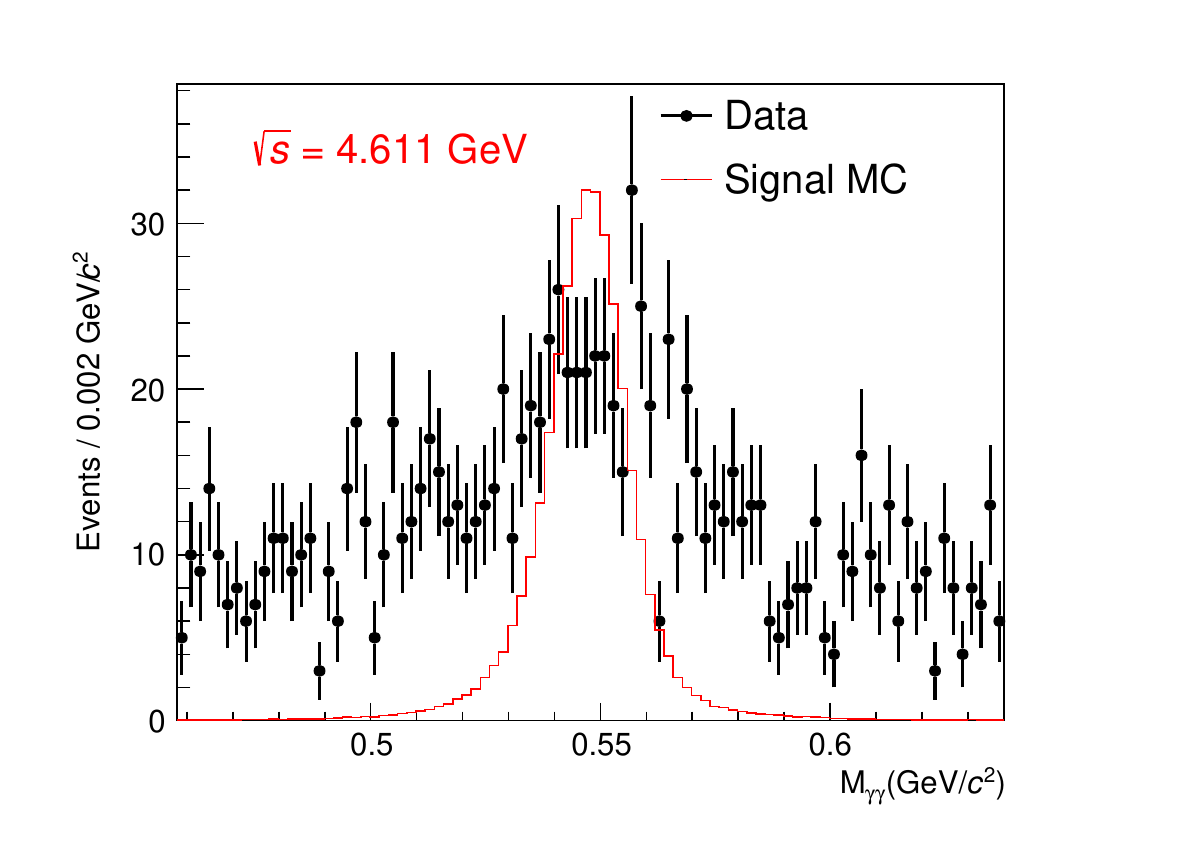}
        \captionsetup{skip=-7pt,font=normalsize}
    \end{subfigure}
    \begin{subfigure}{0.32\textwidth}
        \includegraphics[width=\linewidth]{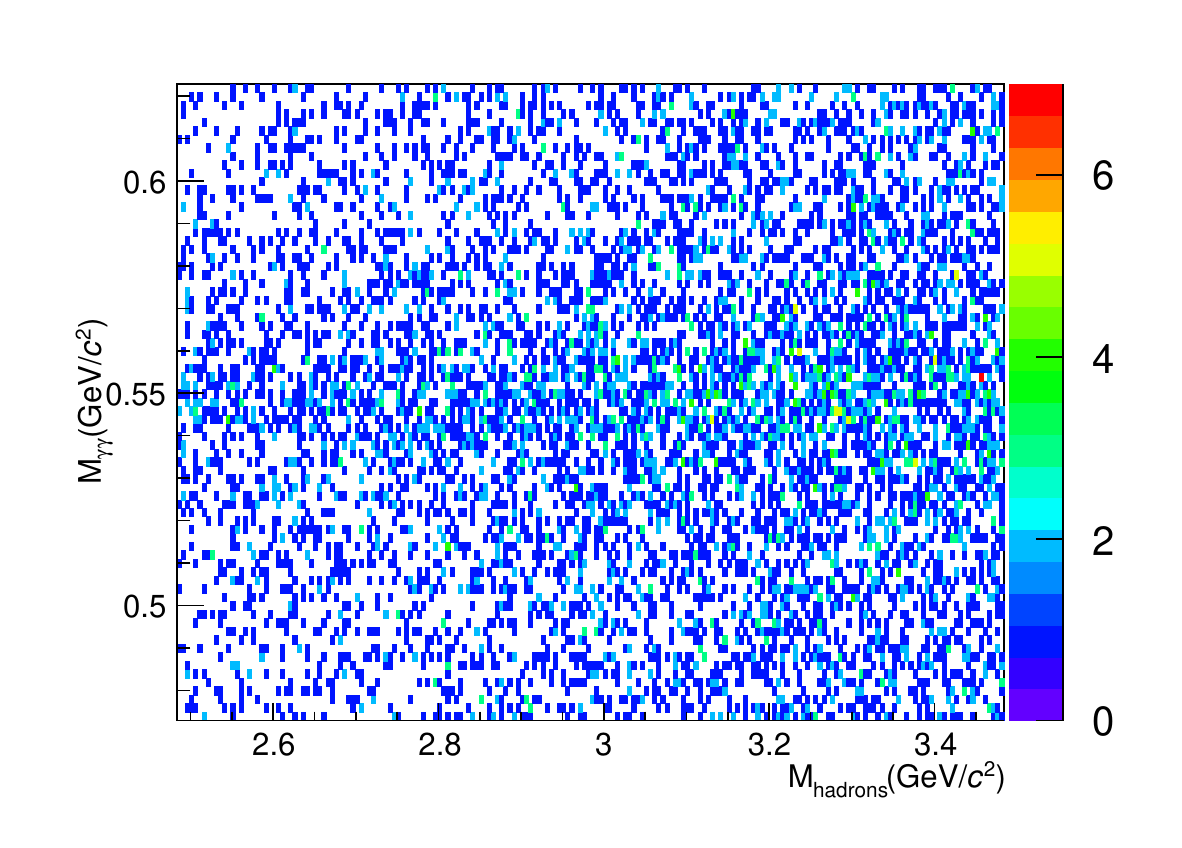}
        \captionsetup{skip=-7pt,font=normalsize}
    \end{subfigure}
    \begin{subfigure}{0.32\textwidth}
        \includegraphics[width=\linewidth]{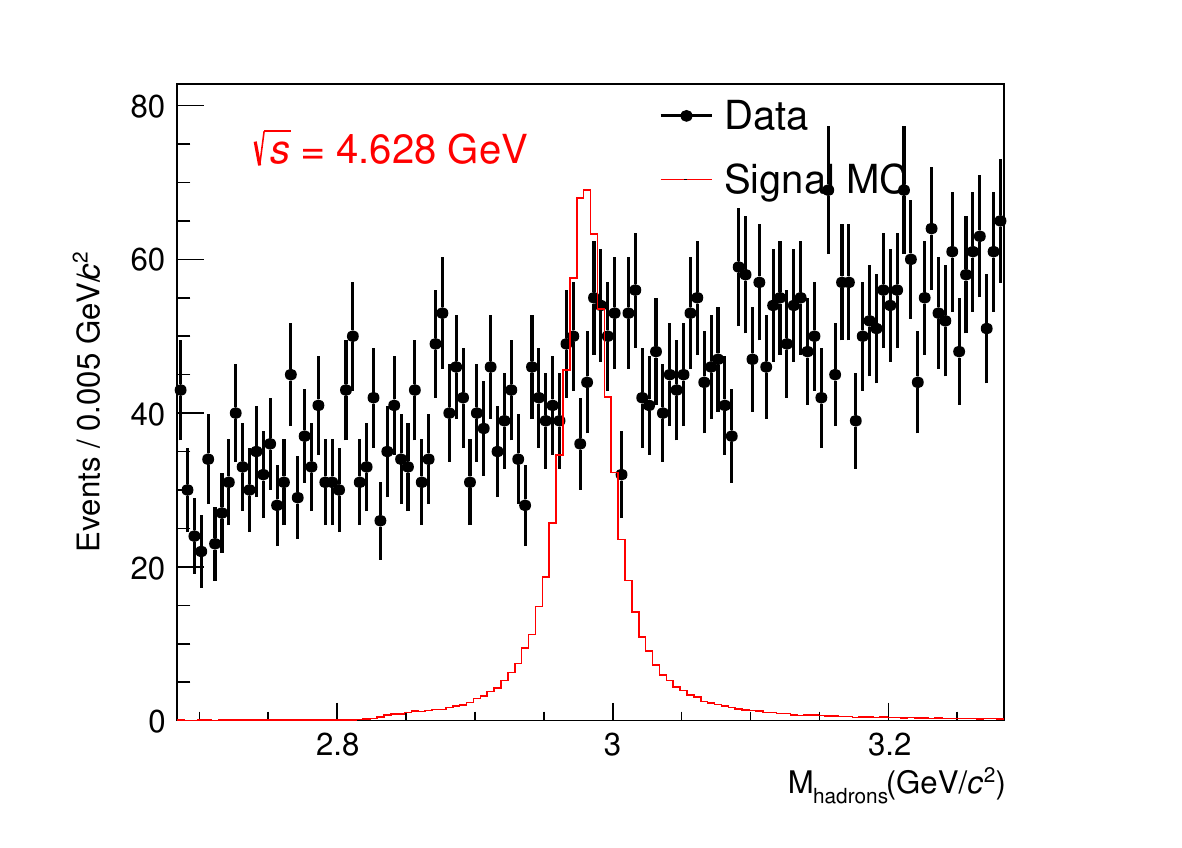}
        \captionsetup{skip=-7pt,font=normalsize}
    \end{subfigure}
    \begin{subfigure}{0.32\textwidth}
        \includegraphics[width=\linewidth]{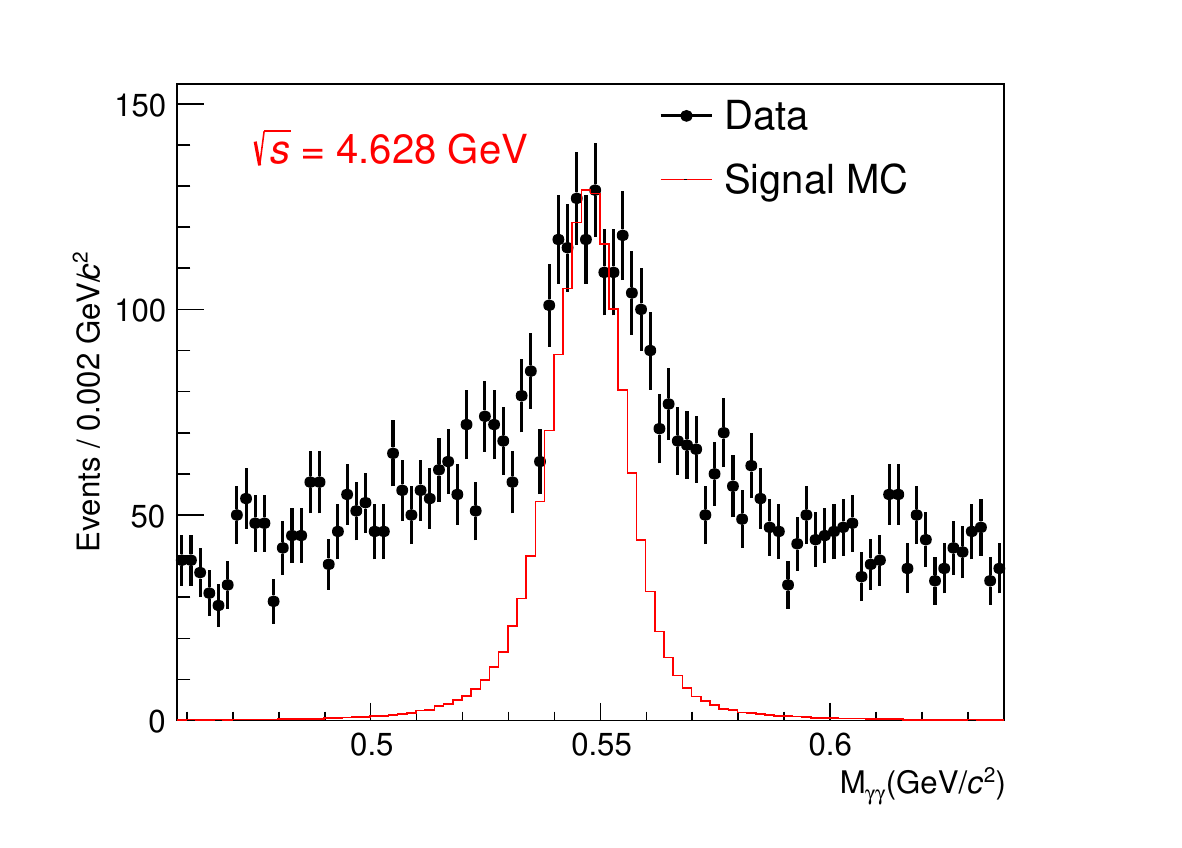}
        \captionsetup{skip=-7pt,font=normalsize}
    \end{subfigure}
    \begin{subfigure}{0.32\textwidth}
        \includegraphics[width=\linewidth]{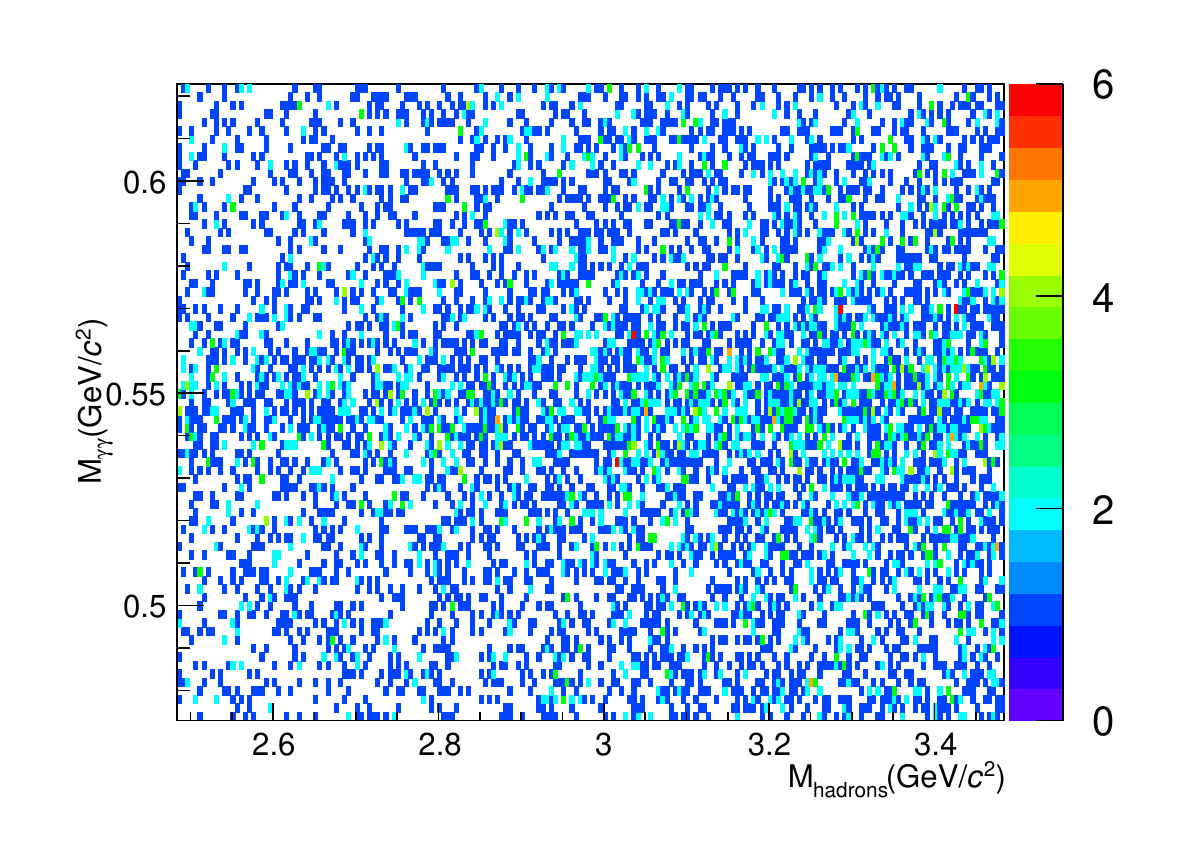}
        \captionsetup{skip=-7pt,font=normalsize}
    \end{subfigure}
    \begin{subfigure}{0.32\textwidth}
        \includegraphics[width=\linewidth]{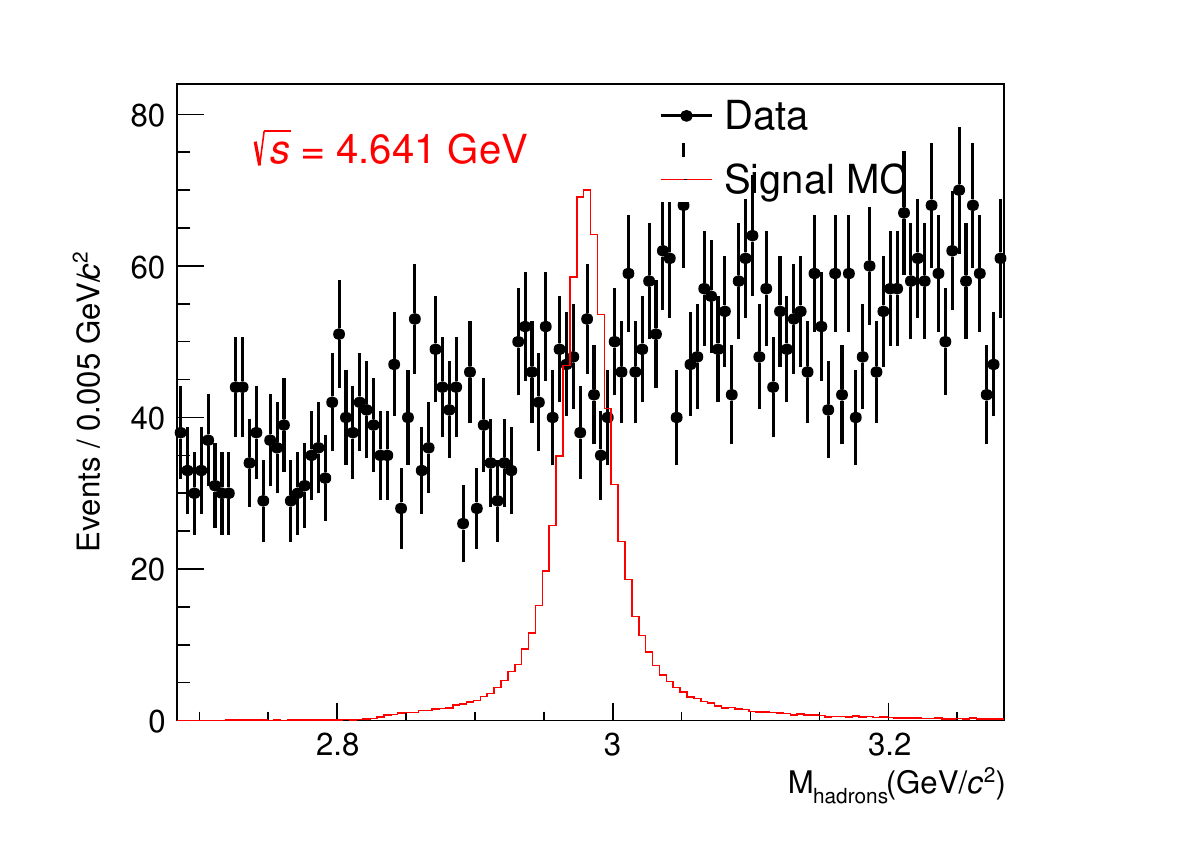}
        \captionsetup{skip=-7pt,font=normalsize}
    \end{subfigure}
    \begin{subfigure}{0.32\textwidth}
        \includegraphics[width=\linewidth]{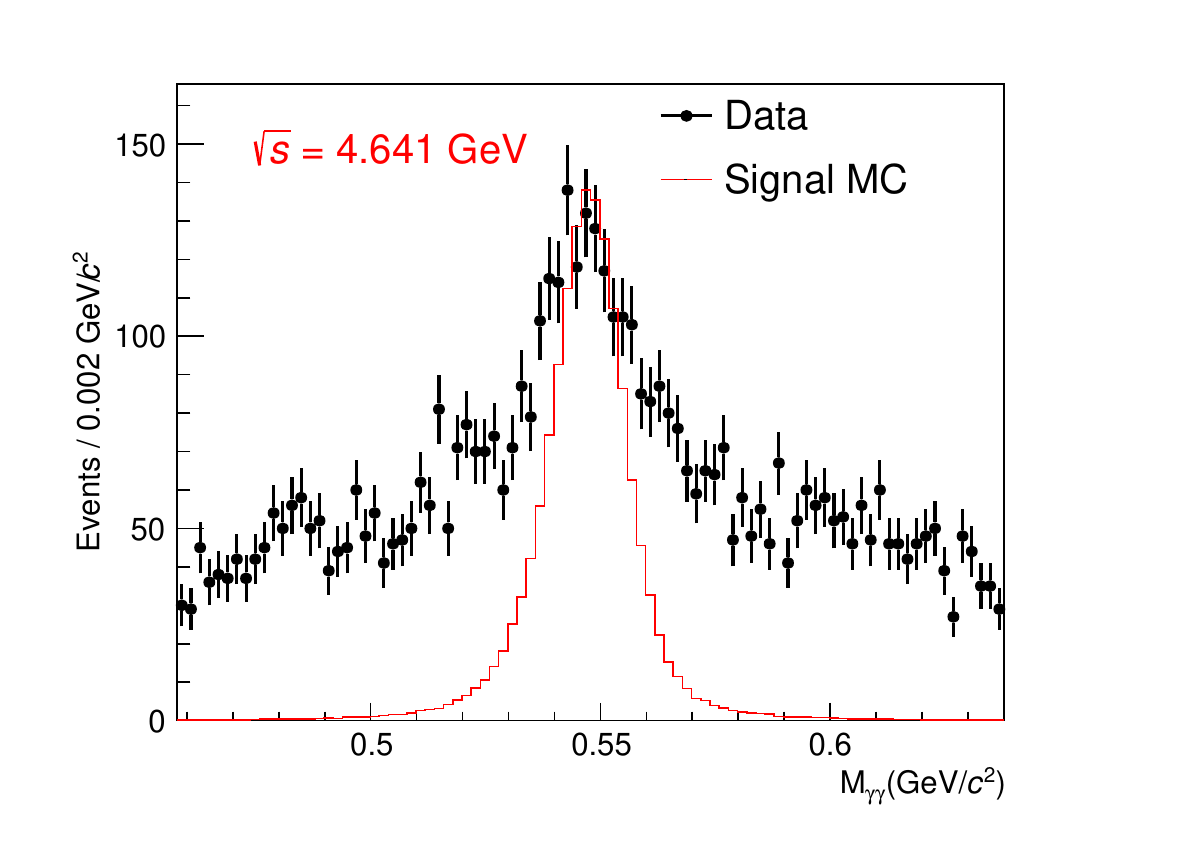}
        \captionsetup{skip=-7pt,font=normalsize}
    \end{subfigure}
    \begin{subfigure}{0.32\textwidth}
        \includegraphics[width=\linewidth]{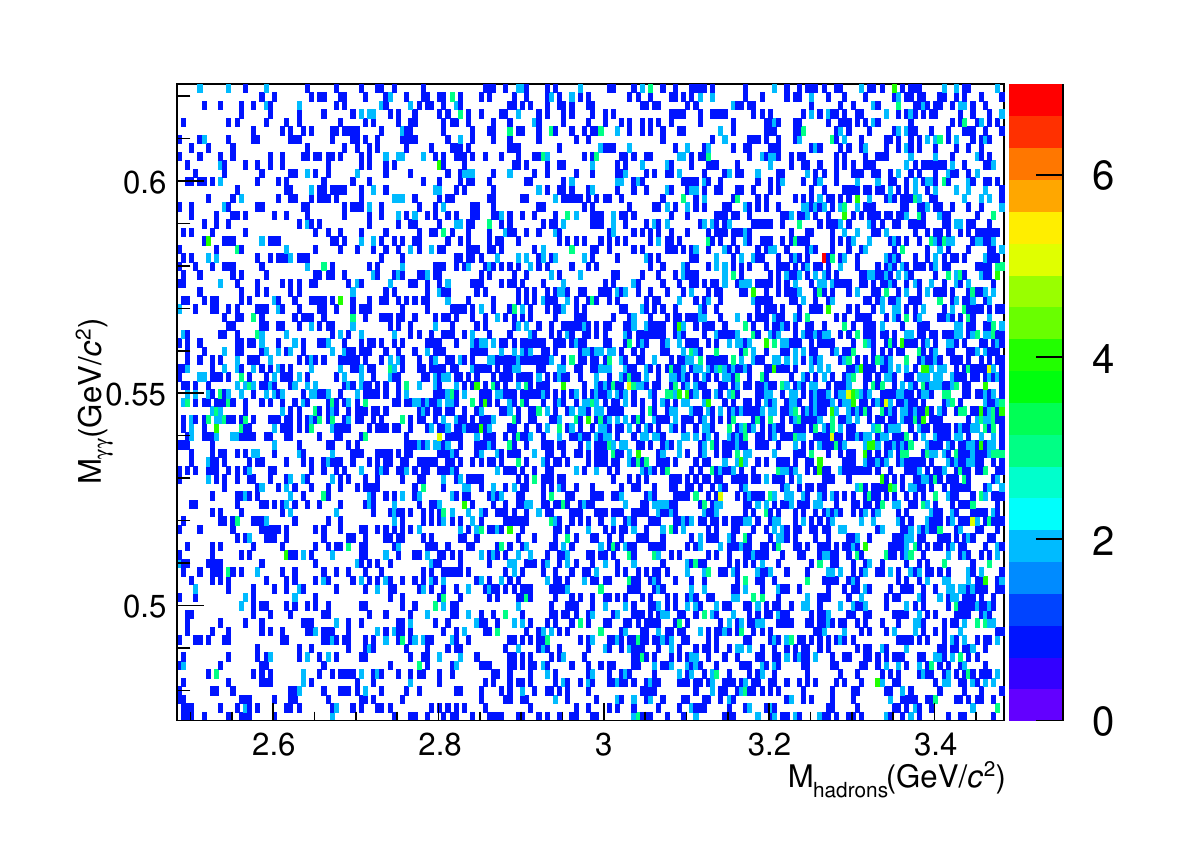}
        \captionsetup{skip=-7pt,font=normalsize}
    \end{subfigure}
    \begin{subfigure}{0.32\textwidth}
        \includegraphics[width=\linewidth]{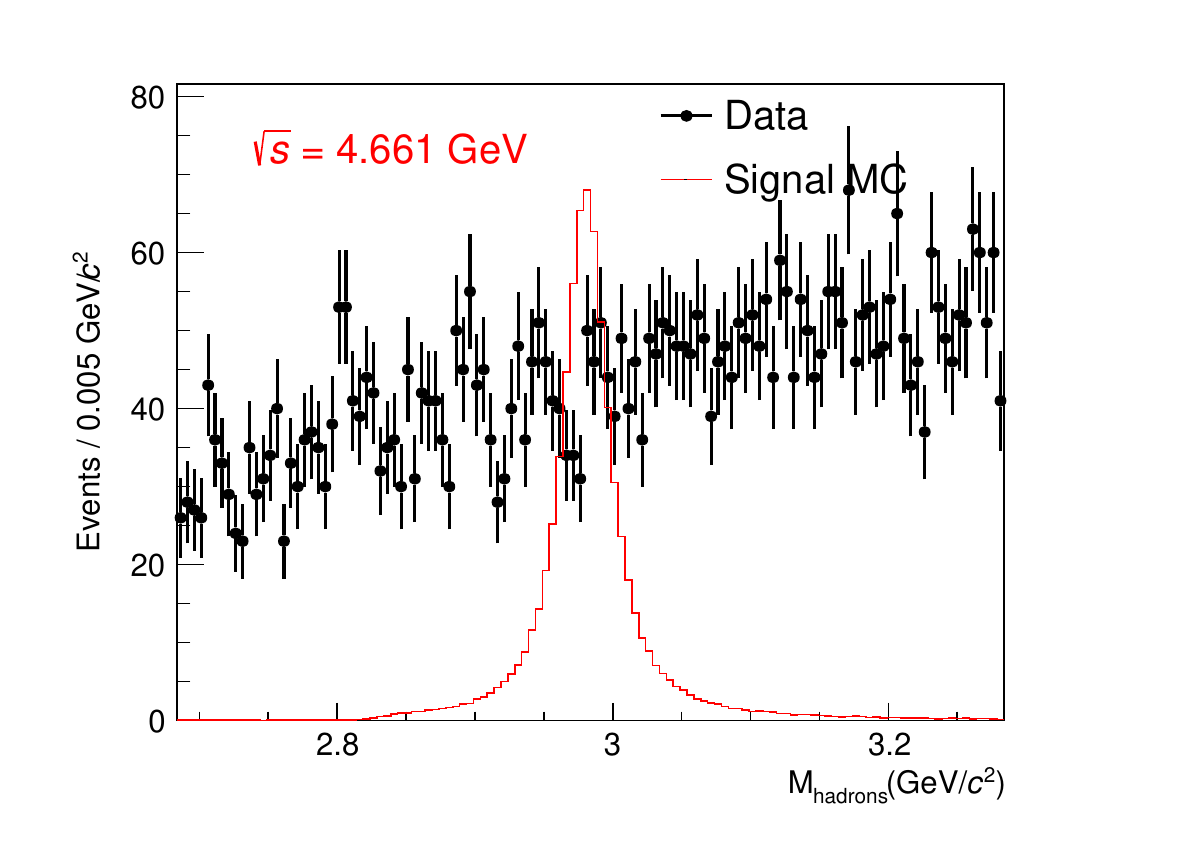}
        \captionsetup{skip=-7pt,font=normalsize}
    \end{subfigure}
    \begin{subfigure}{0.32\textwidth}
        \includegraphics[width=\linewidth]{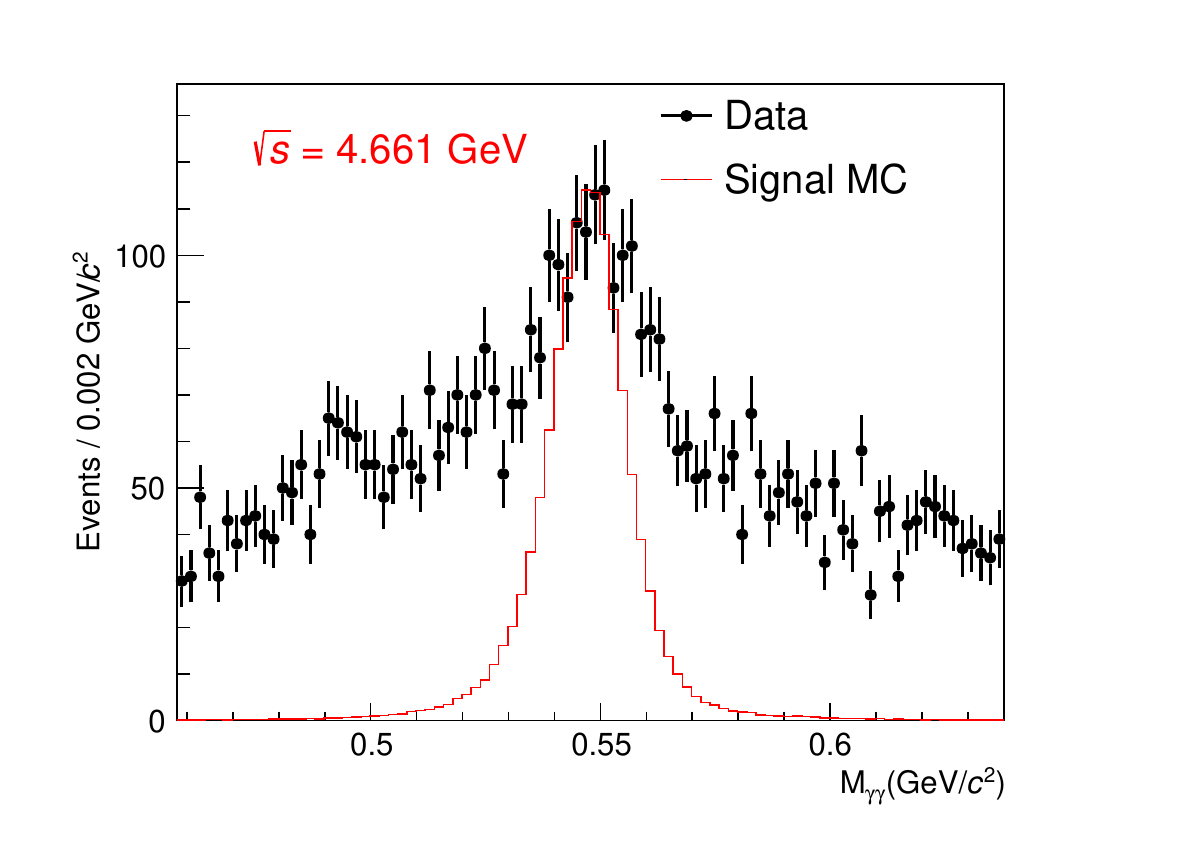}
        \captionsetup{skip=-7pt,font=normalsize}
    \end{subfigure}
\captionsetup{justification=raggedright}
\caption{The distributions of (Left) $M_{hadrons}$ versus $M_{\gamma\gamma}$, (Middle) $M_{hadrons}$, and (Right) $M_{\gamma\gamma}$ at $\sqrt s=4.600-4.661$~GeV.}
\label{fig:normal6}
\end{figure*}

\begin{figure*}[htbp]
    \begin{subfigure}{0.32\textwidth}
        \includegraphics[width=\linewidth]{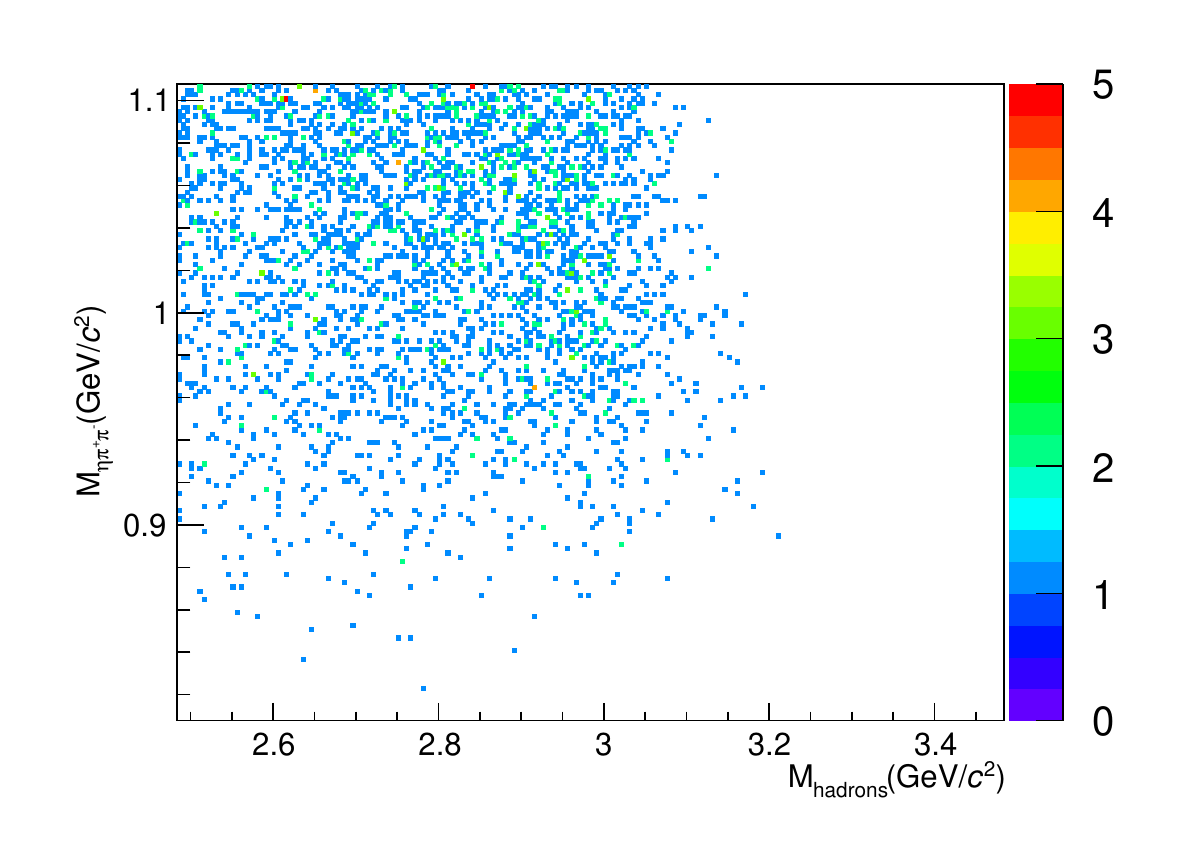}
        \captionsetup{skip=-7pt,font=normalsize}
    \end{subfigure}
    \begin{subfigure}{0.32\textwidth}
        \includegraphics[width=\linewidth]{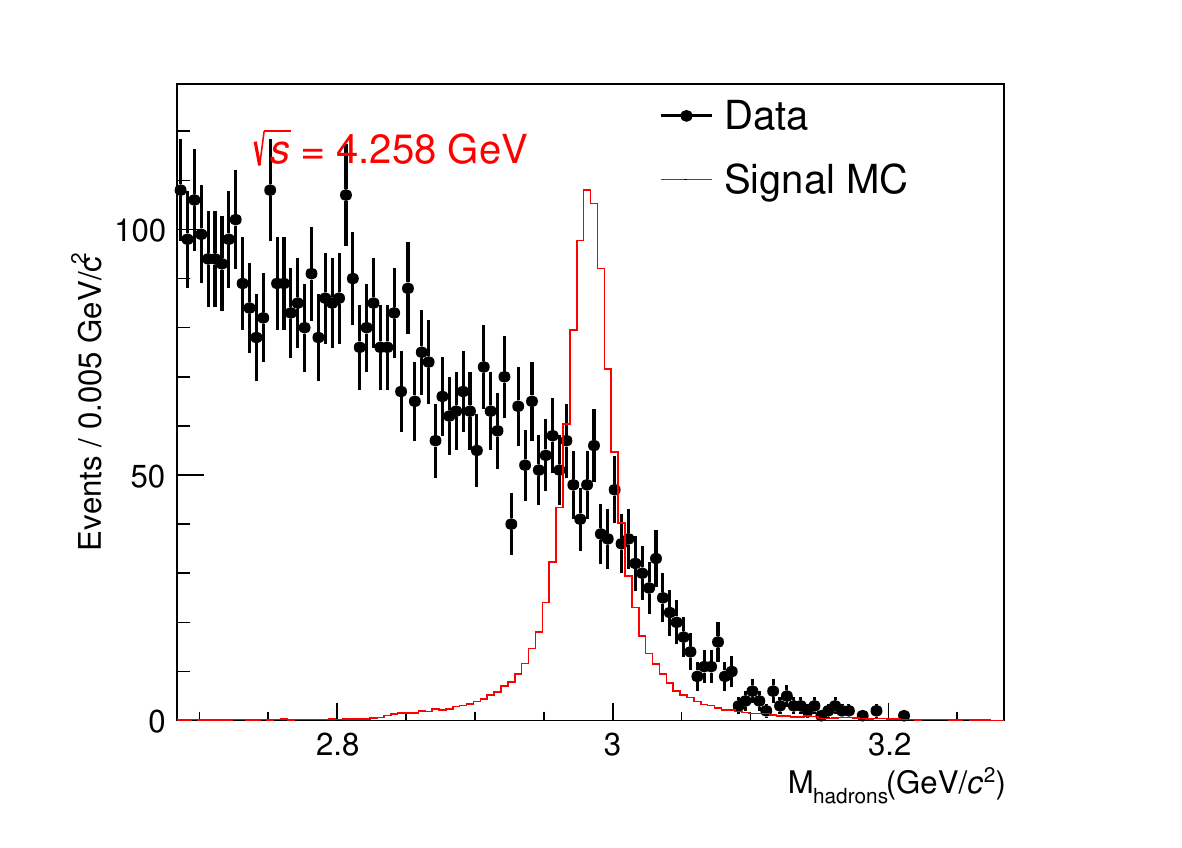}
        \captionsetup{skip=-7pt,font=normalsize}
    \end{subfigure}
    \begin{subfigure}{0.32\textwidth}
        \includegraphics[width=\linewidth]{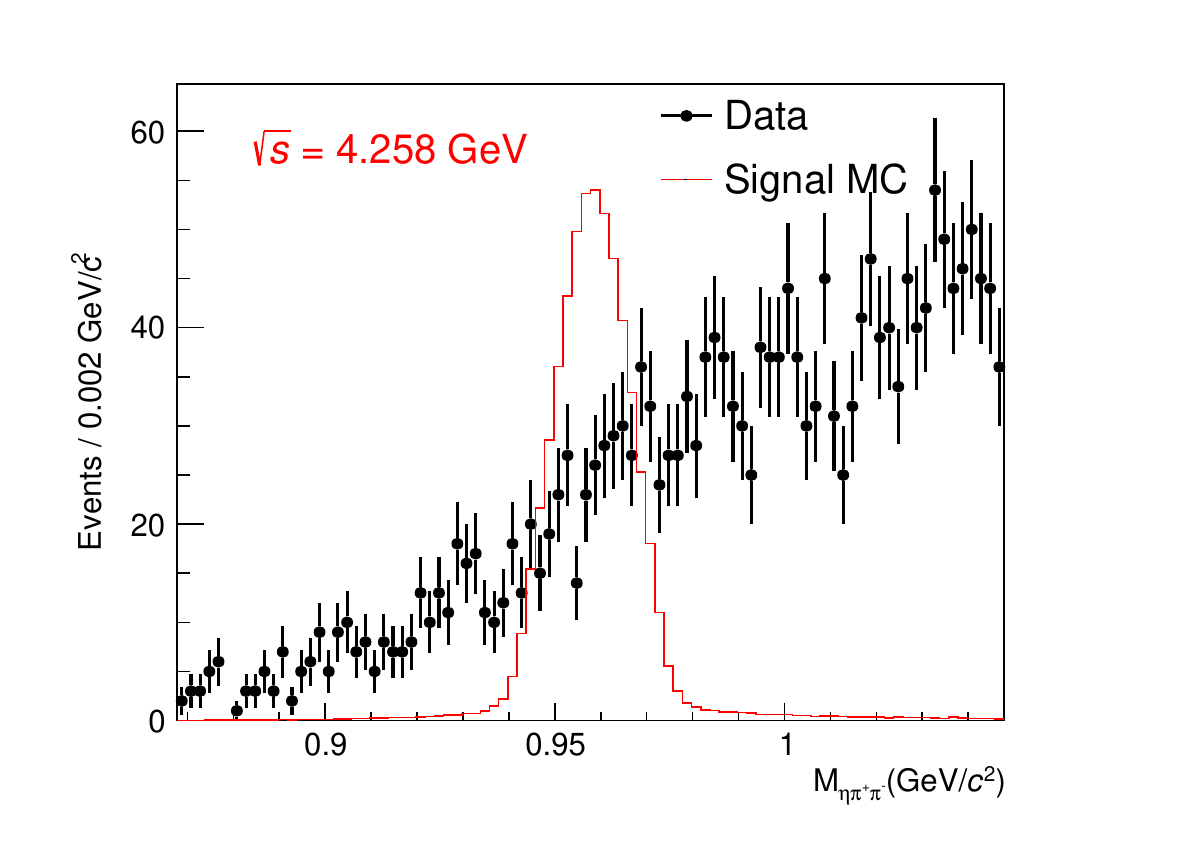}
        \captionsetup{skip=-7pt,font=normalsize}
    \end{subfigure}
    \begin{subfigure}{0.32\textwidth}
        \includegraphics[width=\linewidth]{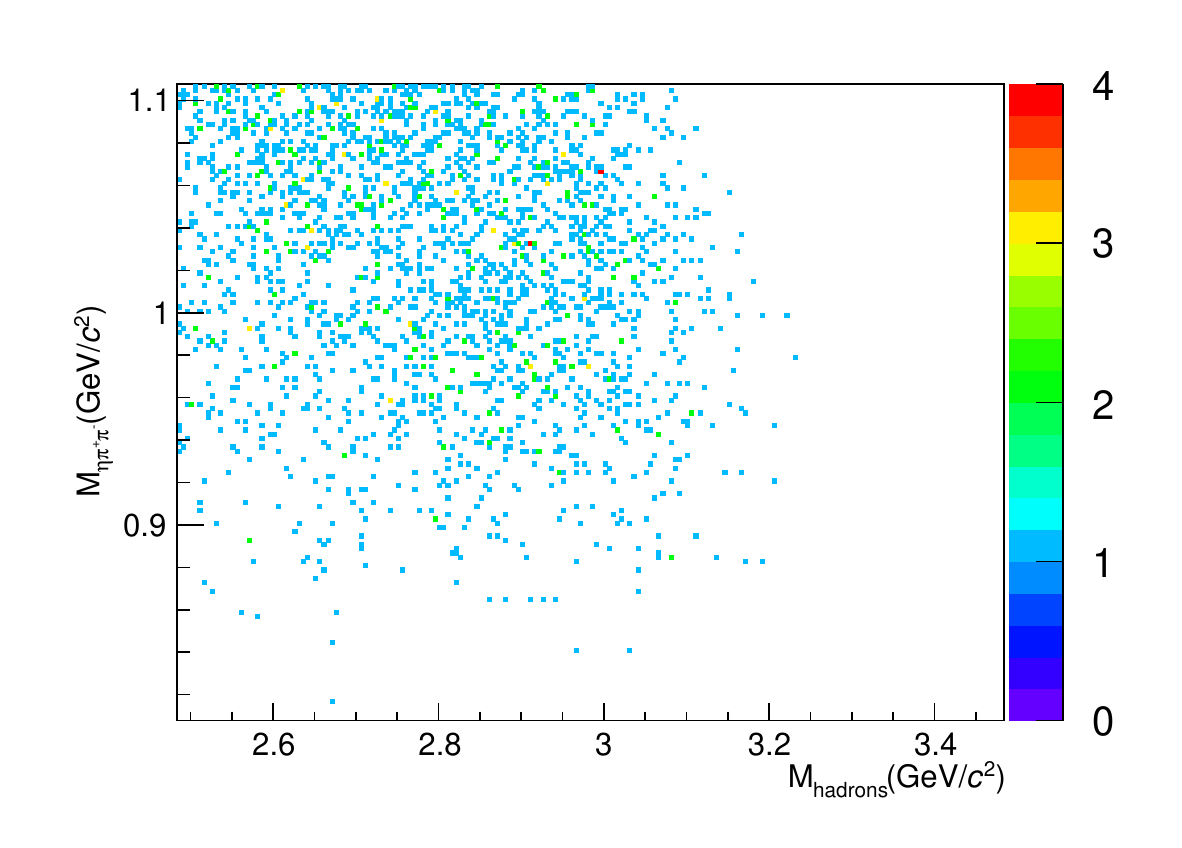}
        \captionsetup{skip=-7pt,font=normalsize}
    \end{subfigure}
    \begin{subfigure}{0.32\textwidth}
        \includegraphics[width=\linewidth]{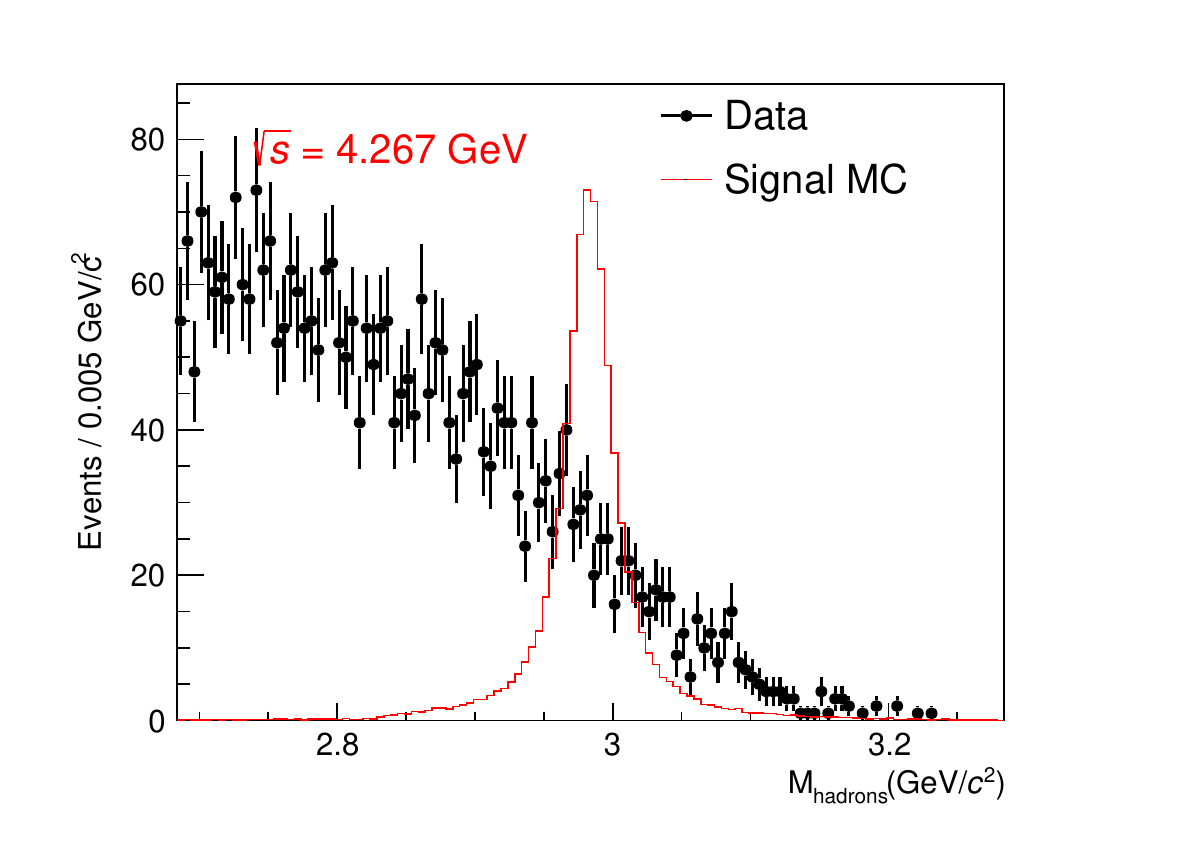}
        \captionsetup{skip=-7pt,font=normalsize}
    \end{subfigure}
    \begin{subfigure}{0.32\textwidth}
        \includegraphics[width=\linewidth]{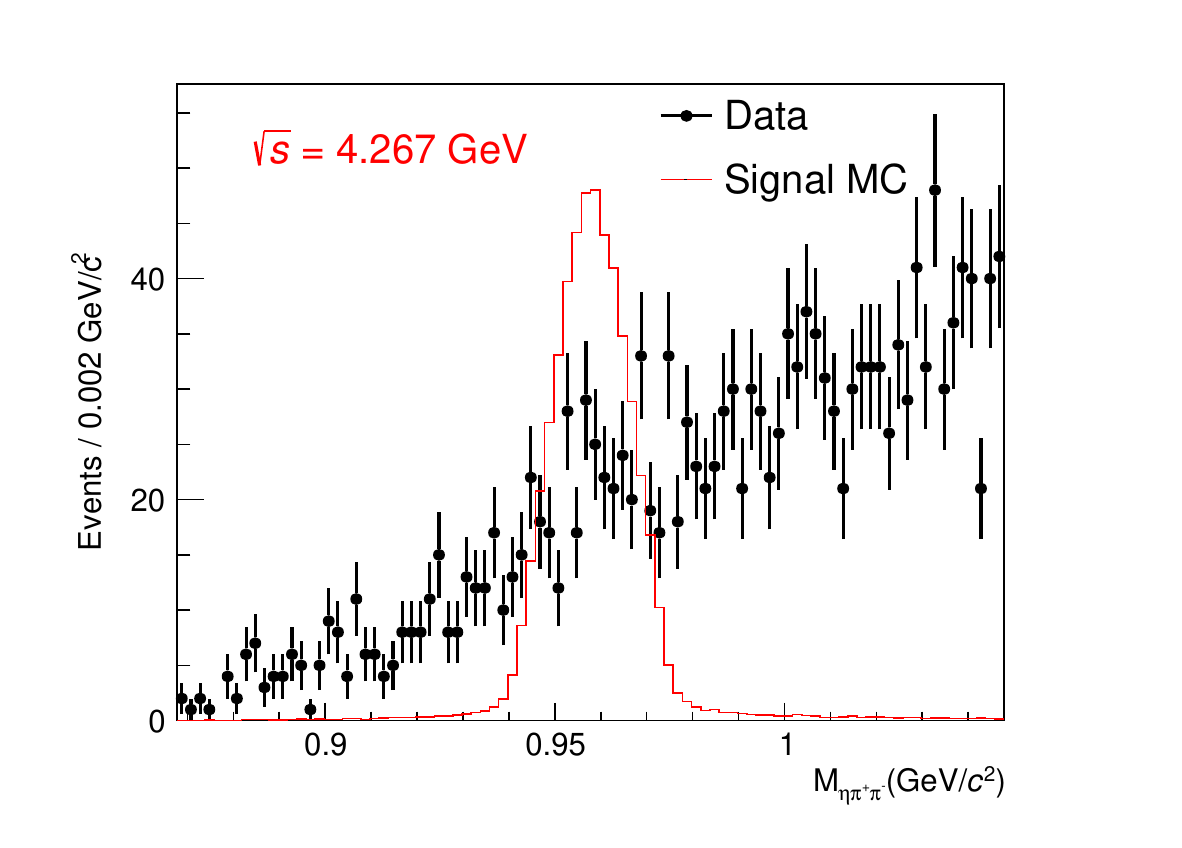}
        \captionsetup{skip=-7pt,font=normalsize}
    \end{subfigure}
    \begin{subfigure}{0.32\textwidth}
        \includegraphics[width=\linewidth]{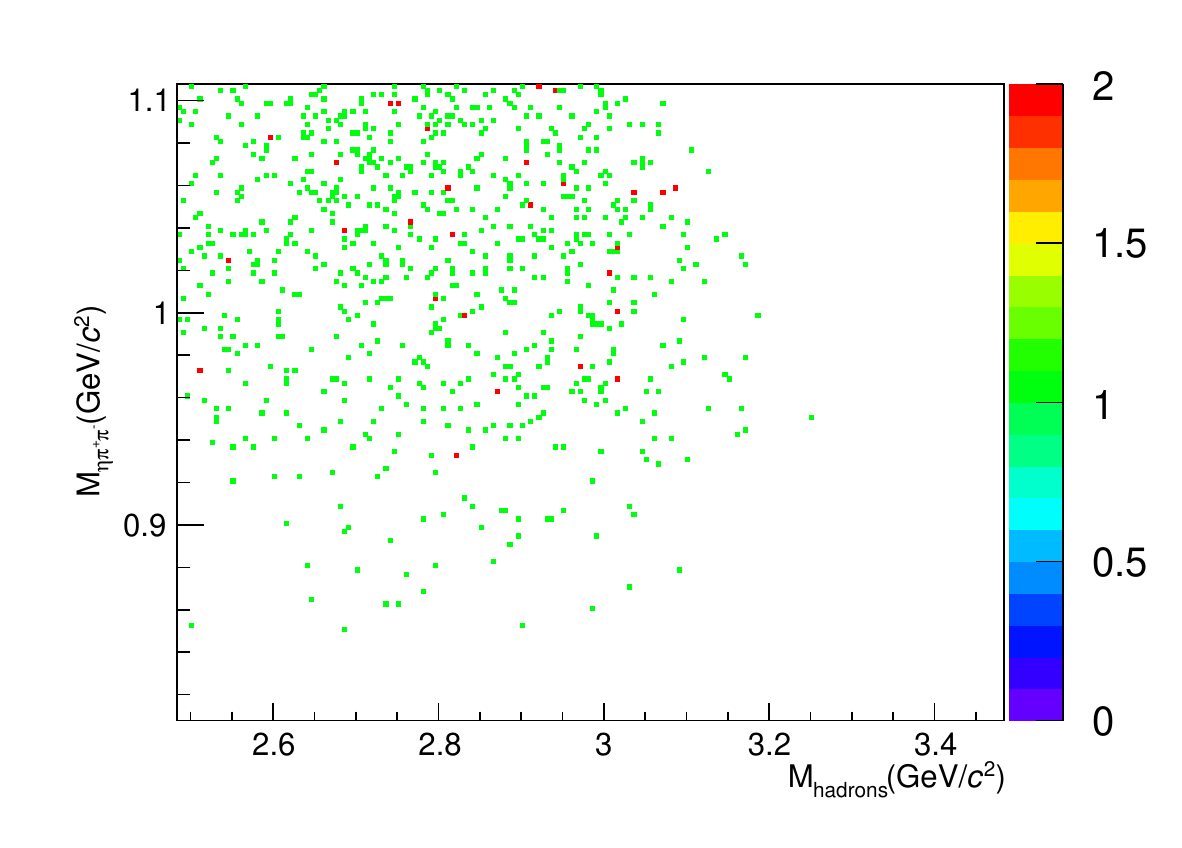}
        \captionsetup{skip=-7pt,font=normalsize}
    \end{subfigure}
    \begin{subfigure}{0.32\textwidth}
        \includegraphics[width=\linewidth]{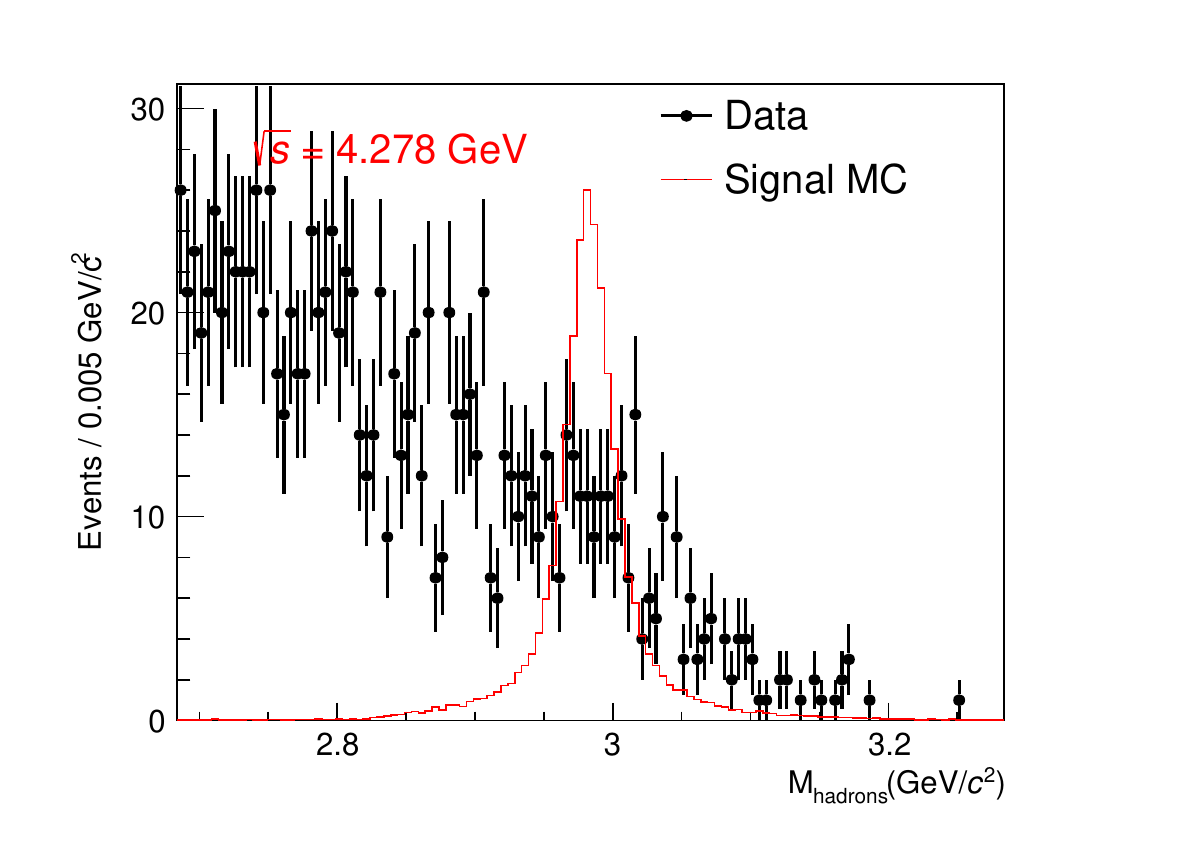}
        \captionsetup{skip=-7pt,font=normalsize}
    \end{subfigure}
    \begin{subfigure}{0.32\textwidth}
        \includegraphics[width=\linewidth]{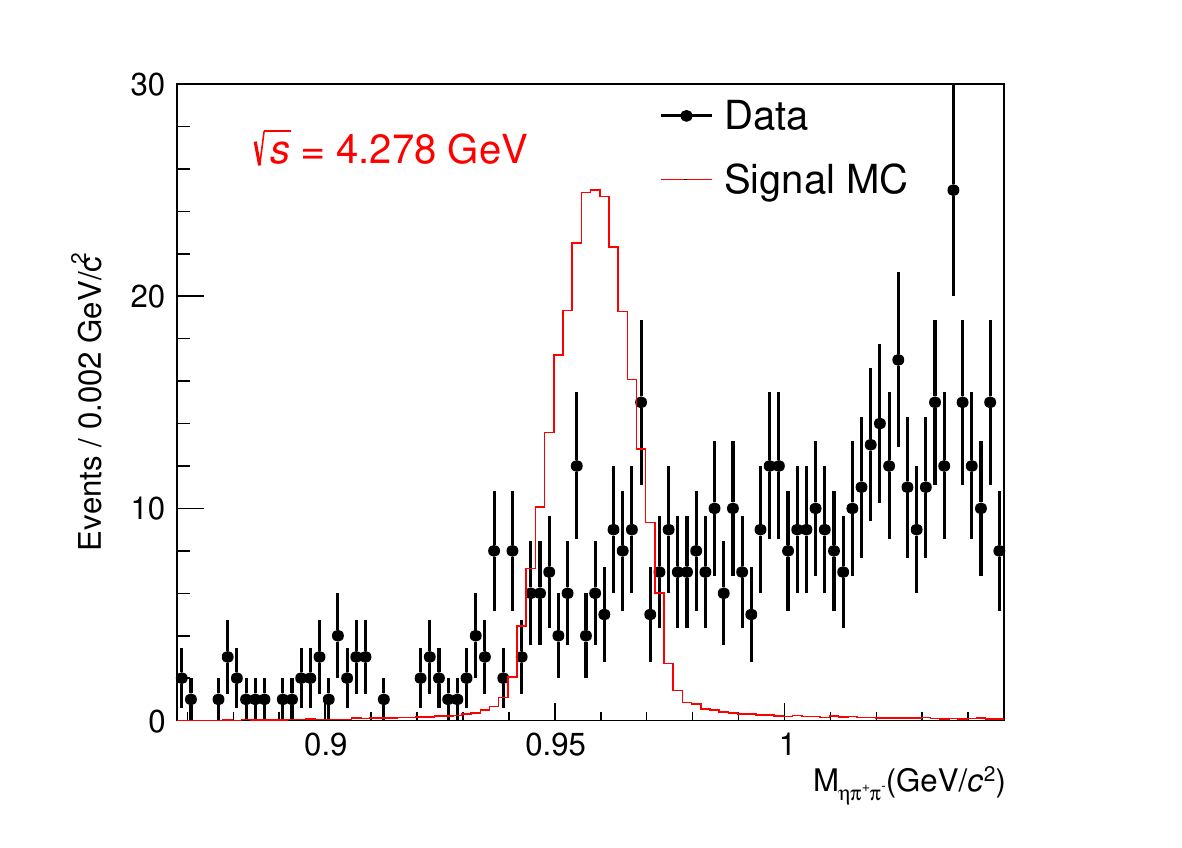}
        \captionsetup{skip=-7pt,font=normalsize}
    \end{subfigure}
    \begin{subfigure}{0.32\textwidth}
        \includegraphics[width=\linewidth]{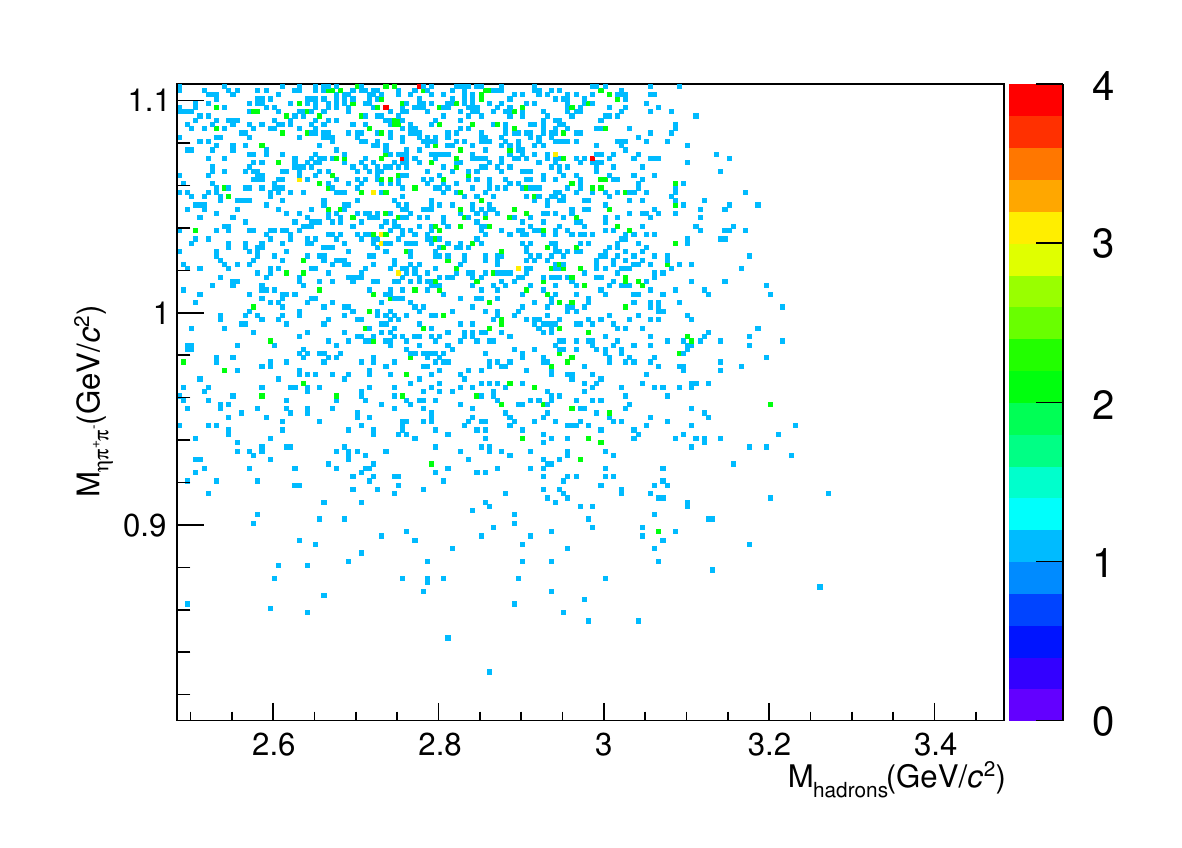}
        \captionsetup{skip=-7pt,font=normalsize}
    \end{subfigure}
    \begin{subfigure}{0.32\textwidth}
        \includegraphics[width=\linewidth]{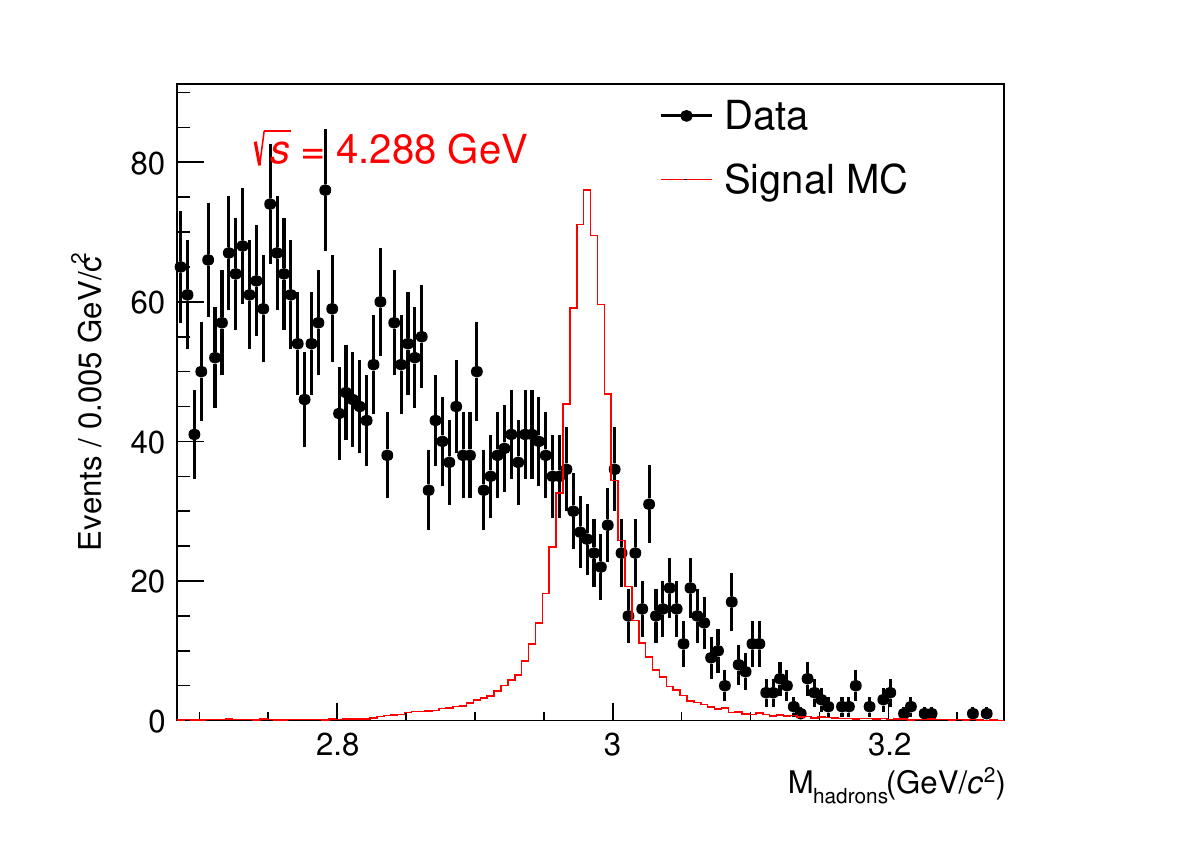}
        \captionsetup{skip=-7pt,font=normalsize}
    \end{subfigure}
    \begin{subfigure}{0.32\textwidth}
        \includegraphics[width=\linewidth]{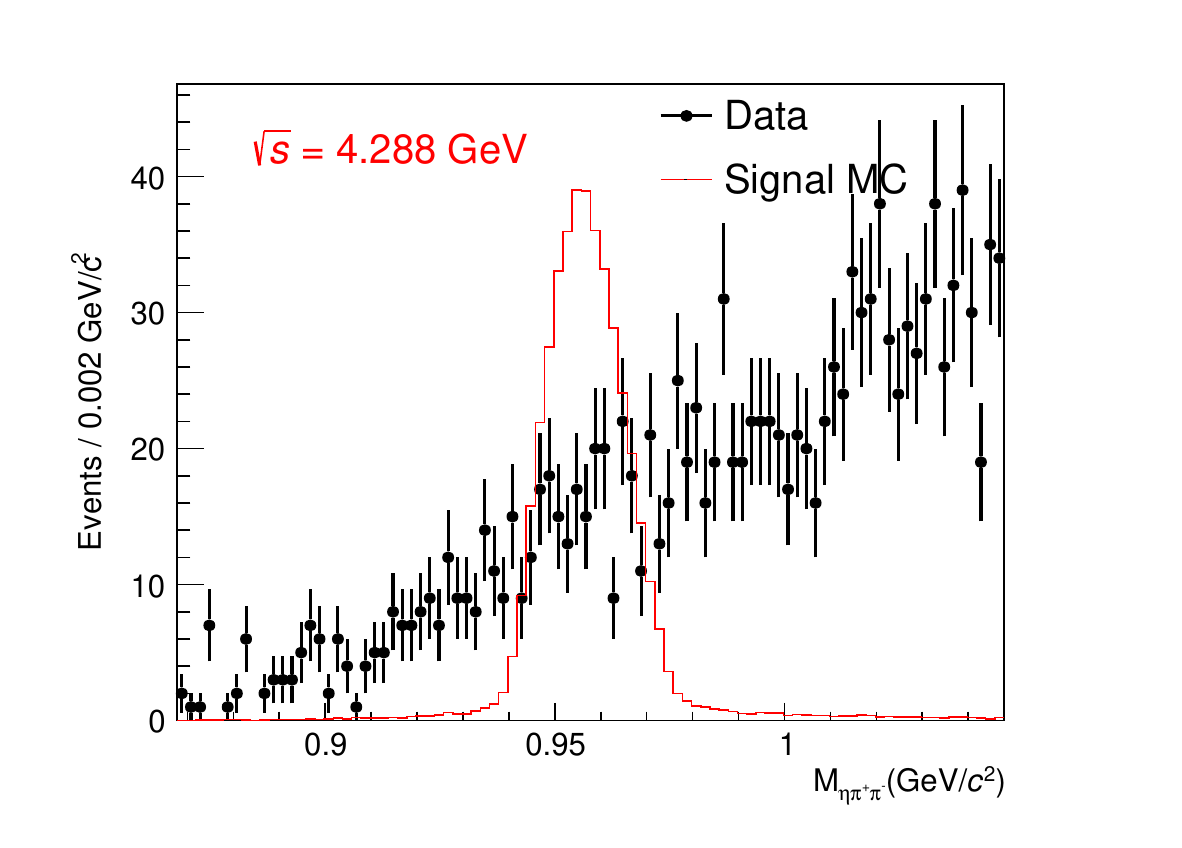}
        \captionsetup{skip=-7pt,font=normalsize}
    \end{subfigure}
\captionsetup{justification=raggedright}
\caption{The distributions of (Left) $M_{hadrons}$ versus $M_{\gamma\gamma}$, (Middle) $M_{hadrons}$, and (Right) $M_{\gamma\gamma}$ at $\sqrt s=4.258-4.288$~GeV.}
\label{fig:normal7}
\end{figure*}

\begin{figure*}[htbp]
    \begin{subfigure}{0.32\textwidth}
        \includegraphics[width=\linewidth]{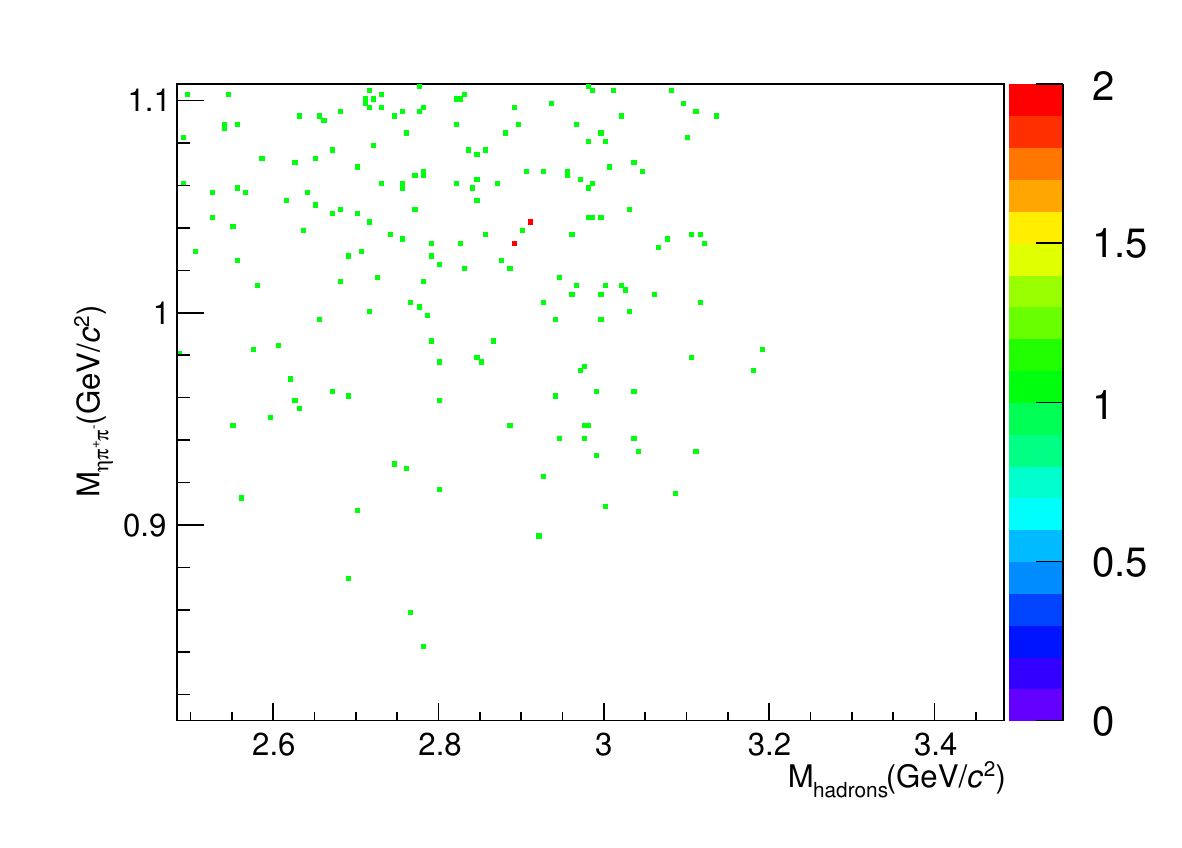}
        \captionsetup{skip=-7pt,font=normalsize}
    \end{subfigure}
    \begin{subfigure}{0.32\textwidth}
        \includegraphics[width=\linewidth]{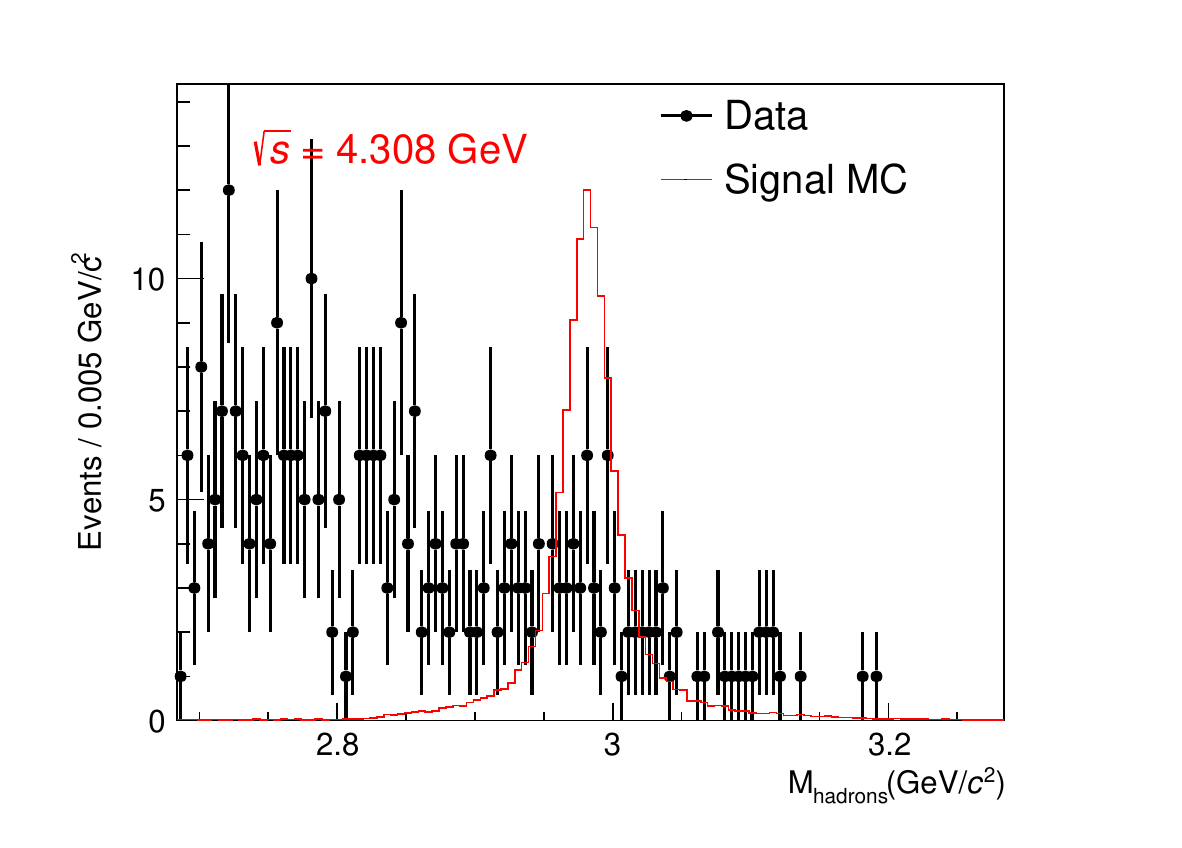}
        \captionsetup{skip=-7pt,font=normalsize}
    \end{subfigure}
    \begin{subfigure}{0.32\textwidth}
        \includegraphics[width=\linewidth]{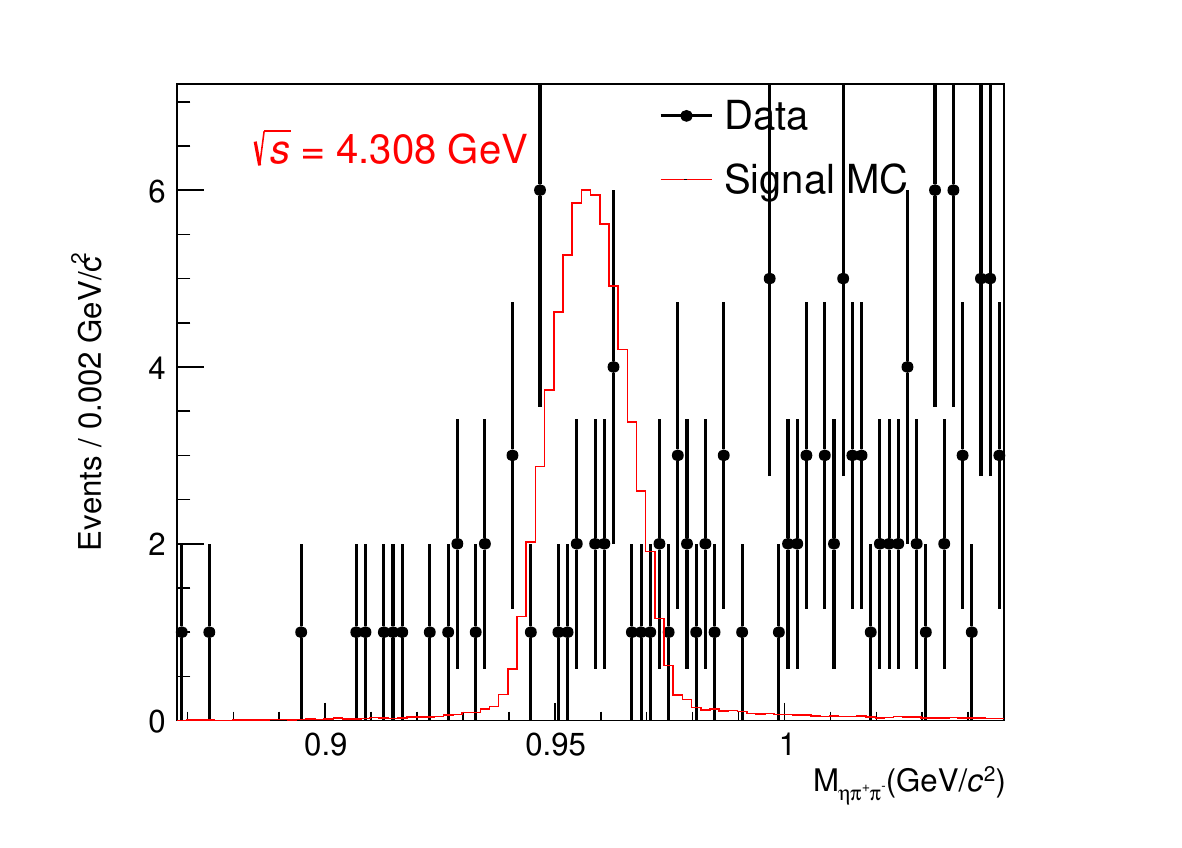}
        \captionsetup{skip=-7pt,font=normalsize}
    \end{subfigure}
    \begin{subfigure}{0.32\textwidth}
        \includegraphics[width=\linewidth]{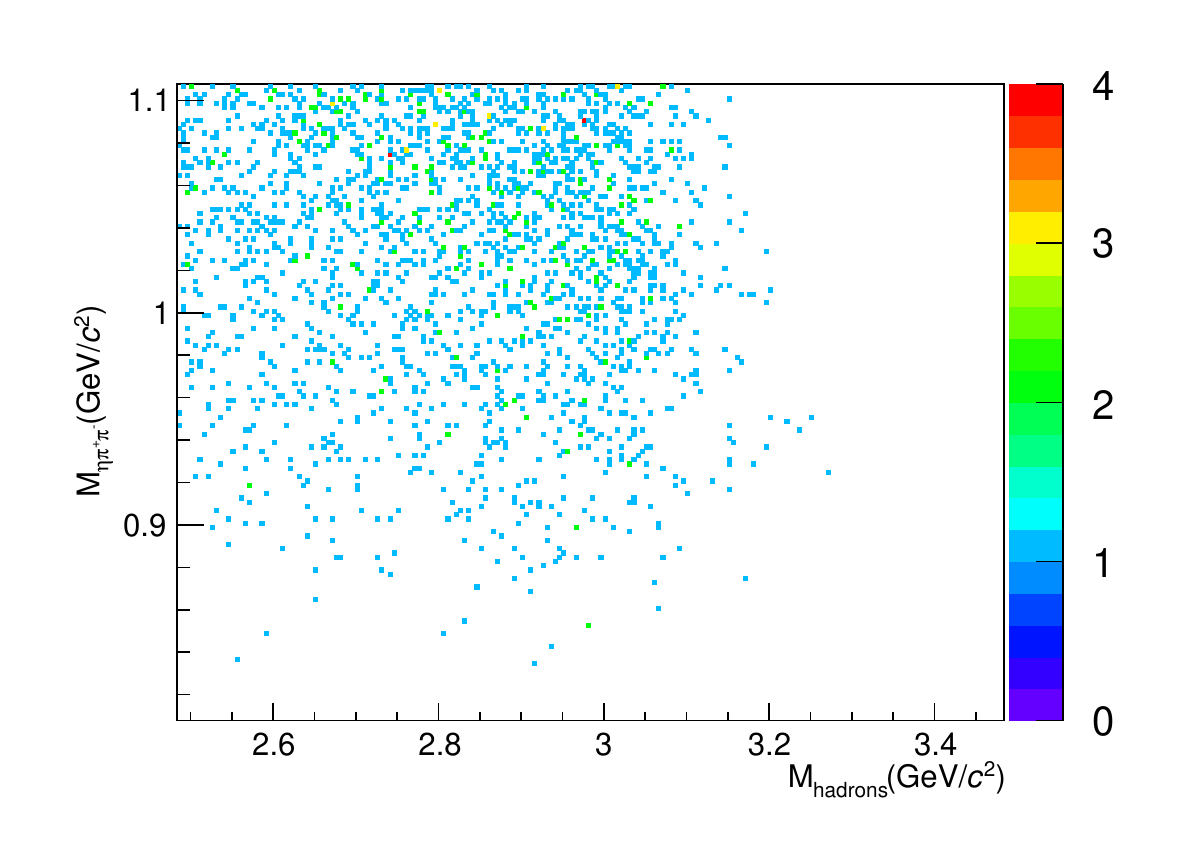}
        \captionsetup{skip=-7pt,font=normalsize}
    \end{subfigure}
    \begin{subfigure}{0.32\textwidth}
        \includegraphics[width=\linewidth]{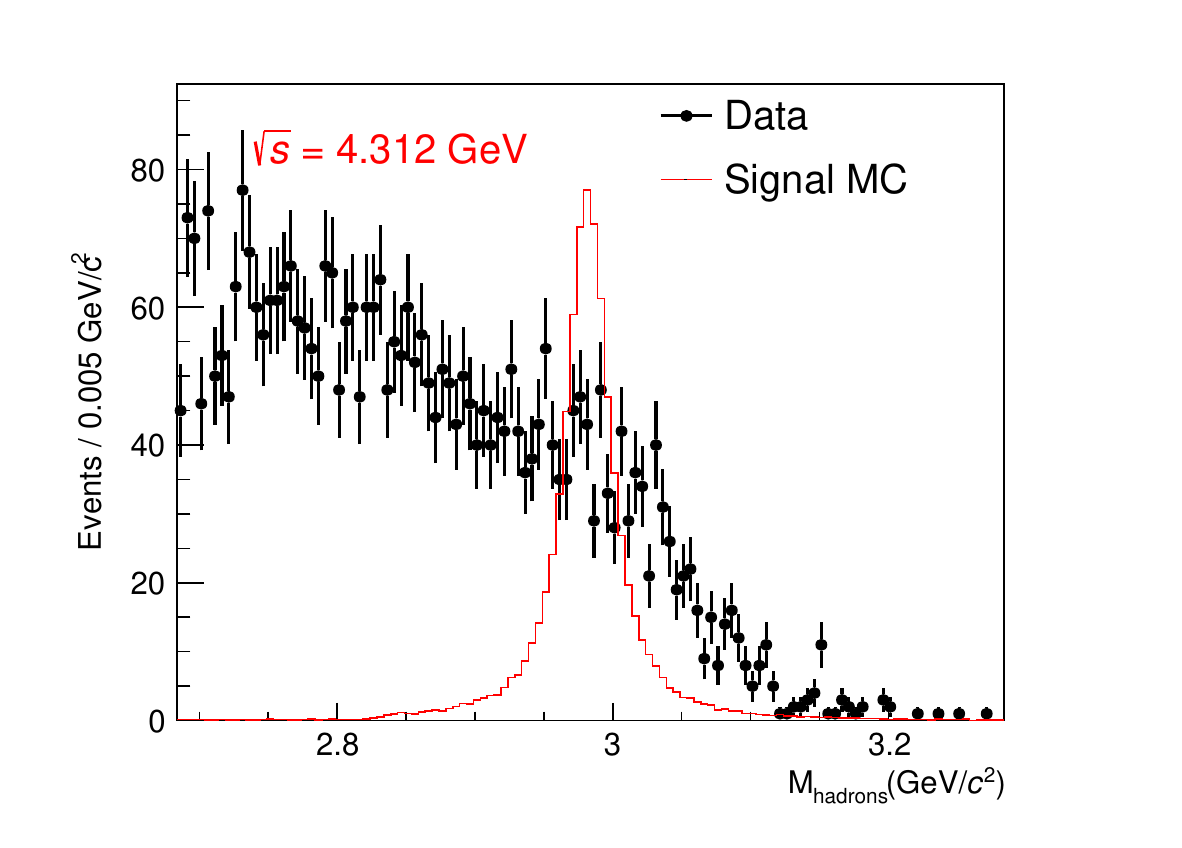}
        \captionsetup{skip=-7pt,font=normalsize}
    \end{subfigure}
    \begin{subfigure}{0.32\textwidth}
        \includegraphics[width=\linewidth]{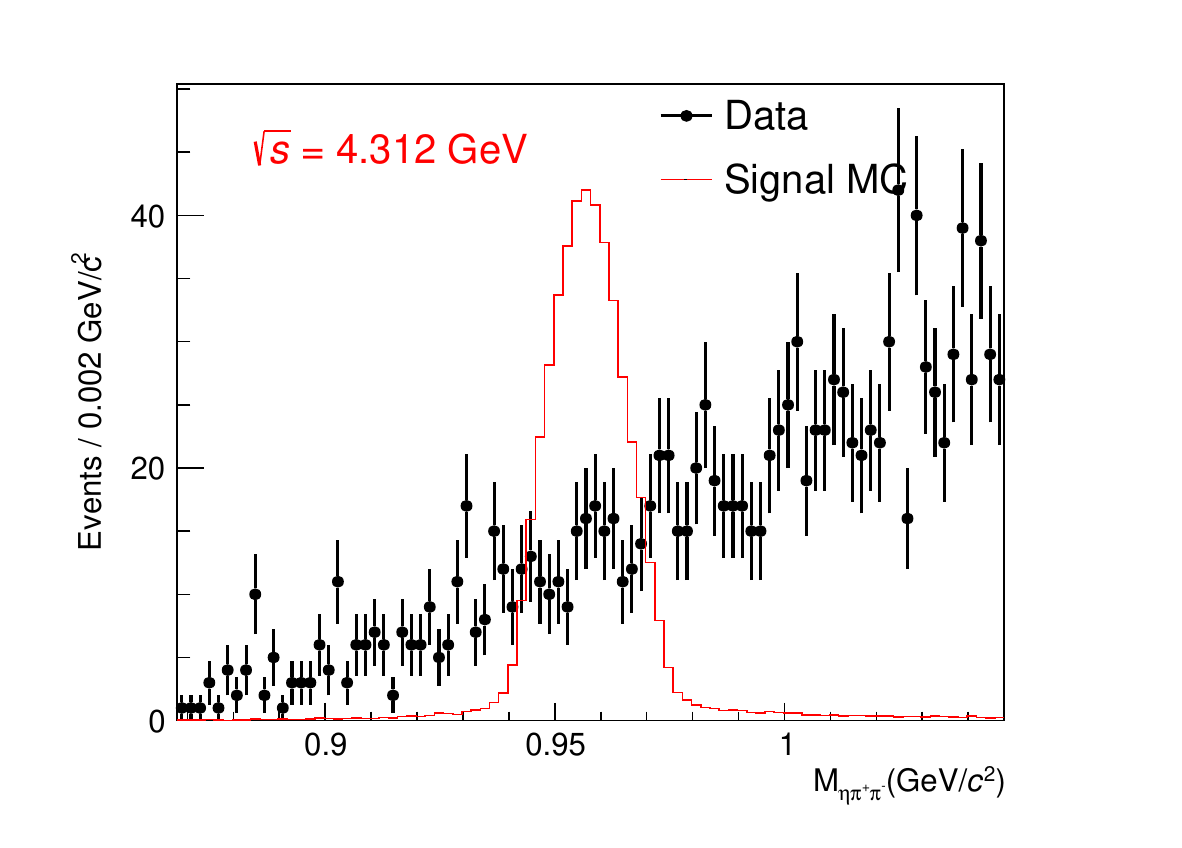}
        \captionsetup{skip=-7pt,font=normalsize}
    \end{subfigure}
    \begin{subfigure}{0.32\textwidth}
        \includegraphics[width=\linewidth]{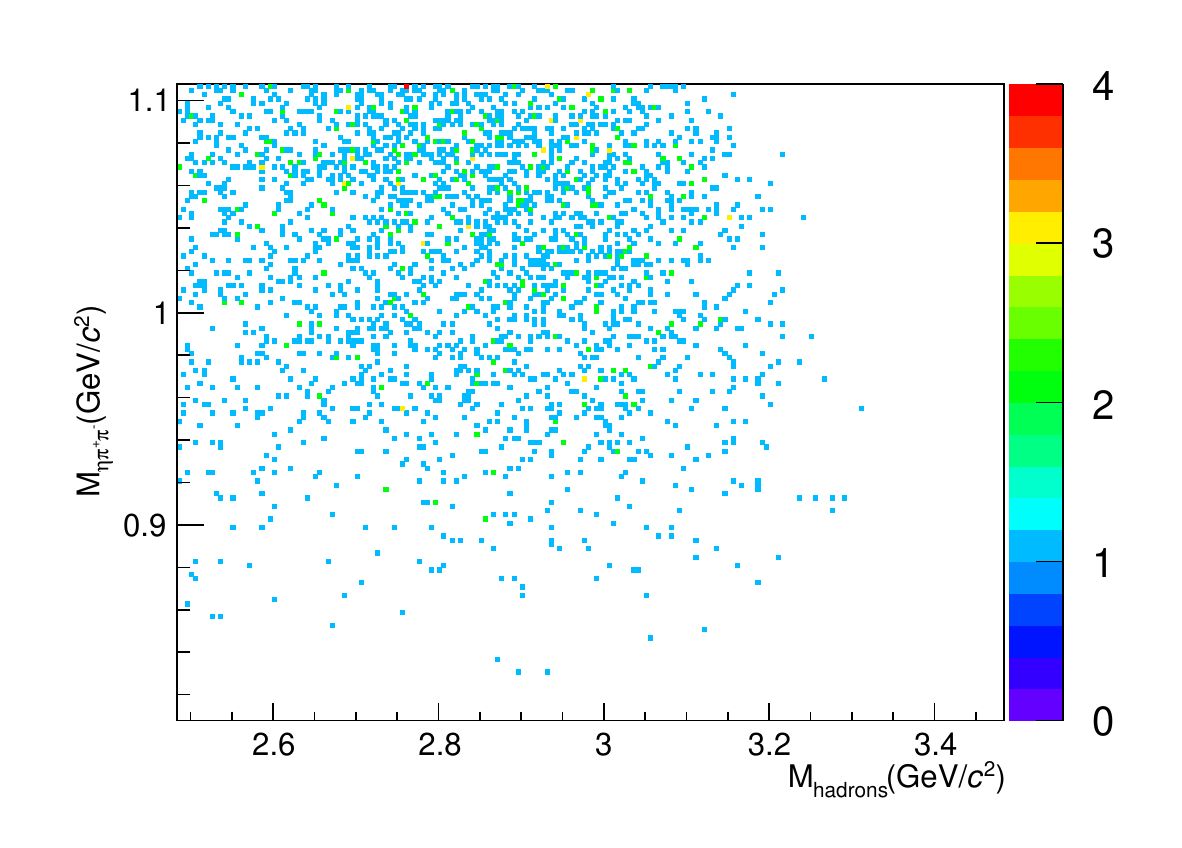}
        \captionsetup{skip=-7pt,font=normalsize}
    \end{subfigure}
    \begin{subfigure}{0.32\textwidth}
        \includegraphics[width=\linewidth]{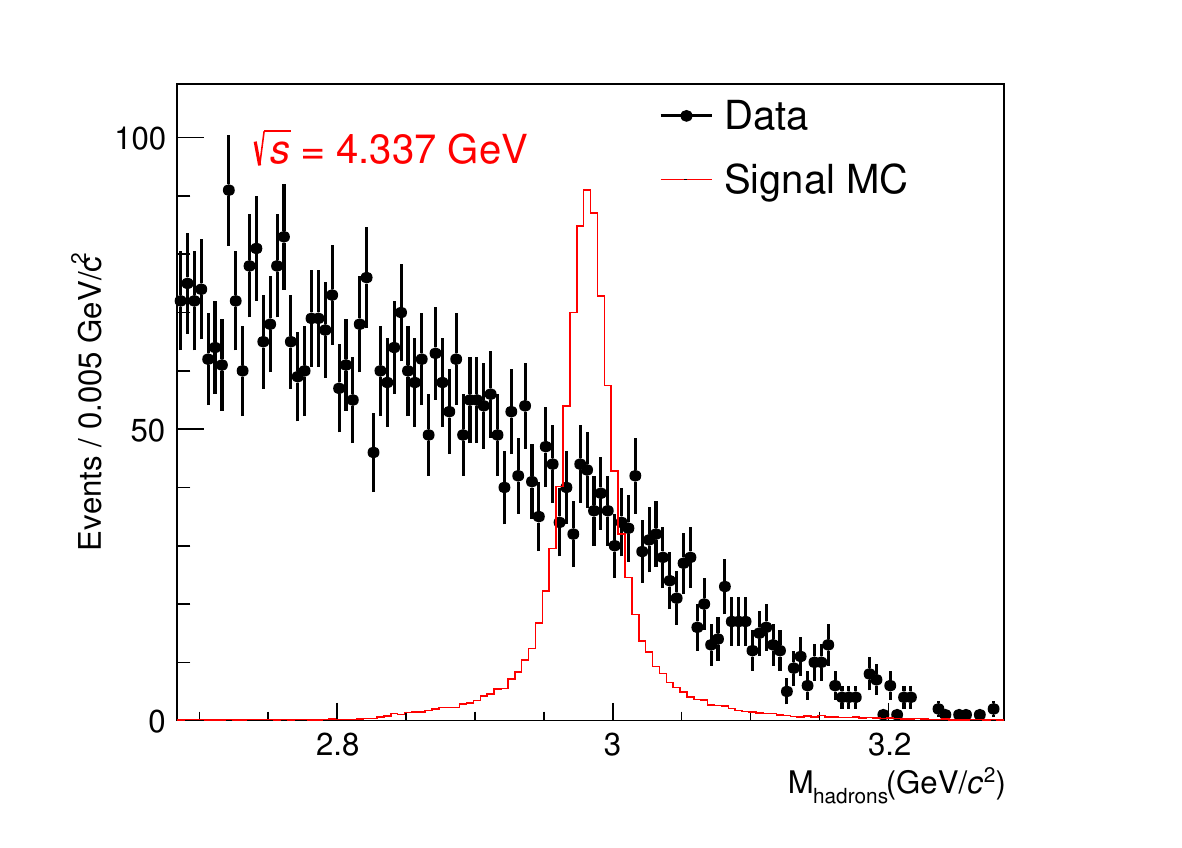}
        \captionsetup{skip=-7pt,font=normalsize}
    \end{subfigure}
    \begin{subfigure}{0.32\textwidth}
        \includegraphics[width=\linewidth]{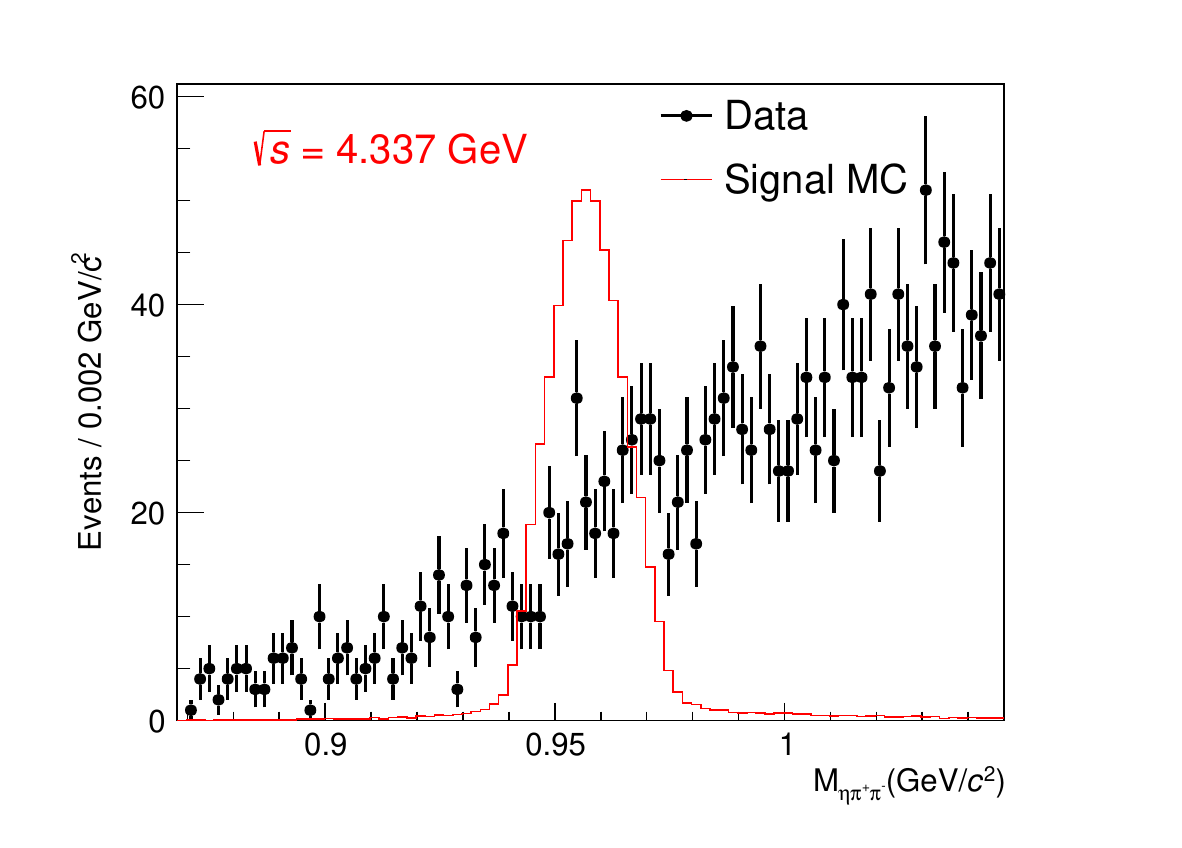}
        \captionsetup{skip=-7pt,font=normalsize}
    \end{subfigure}
    \begin{subfigure}{0.32\textwidth}
        \includegraphics[width=\linewidth]{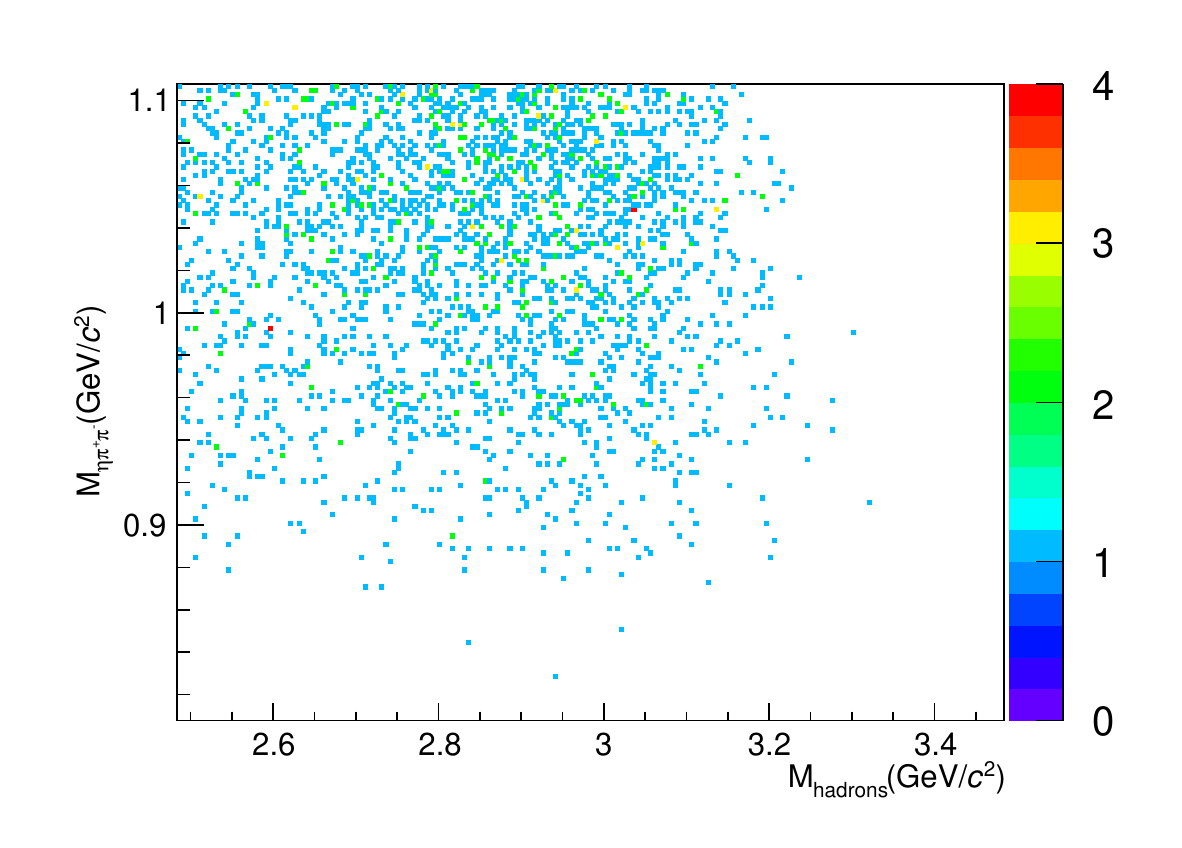}
        \captionsetup{skip=-7pt,font=normalsize}
    \end{subfigure}
    \begin{subfigure}{0.32\textwidth}
        \includegraphics[width=\linewidth]{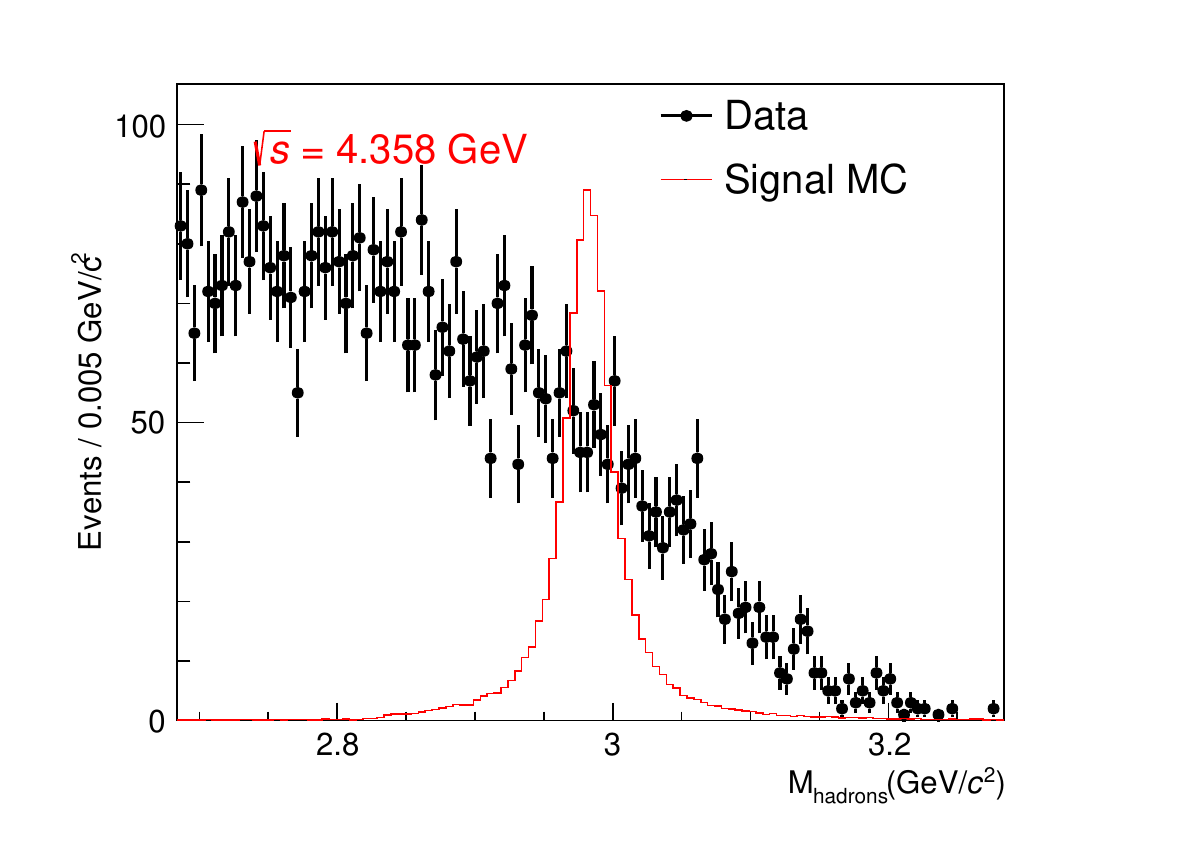}
        \captionsetup{skip=-7pt,font=normalsize}
    \end{subfigure}
    \begin{subfigure}{0.32\textwidth}
        \includegraphics[width=\linewidth]{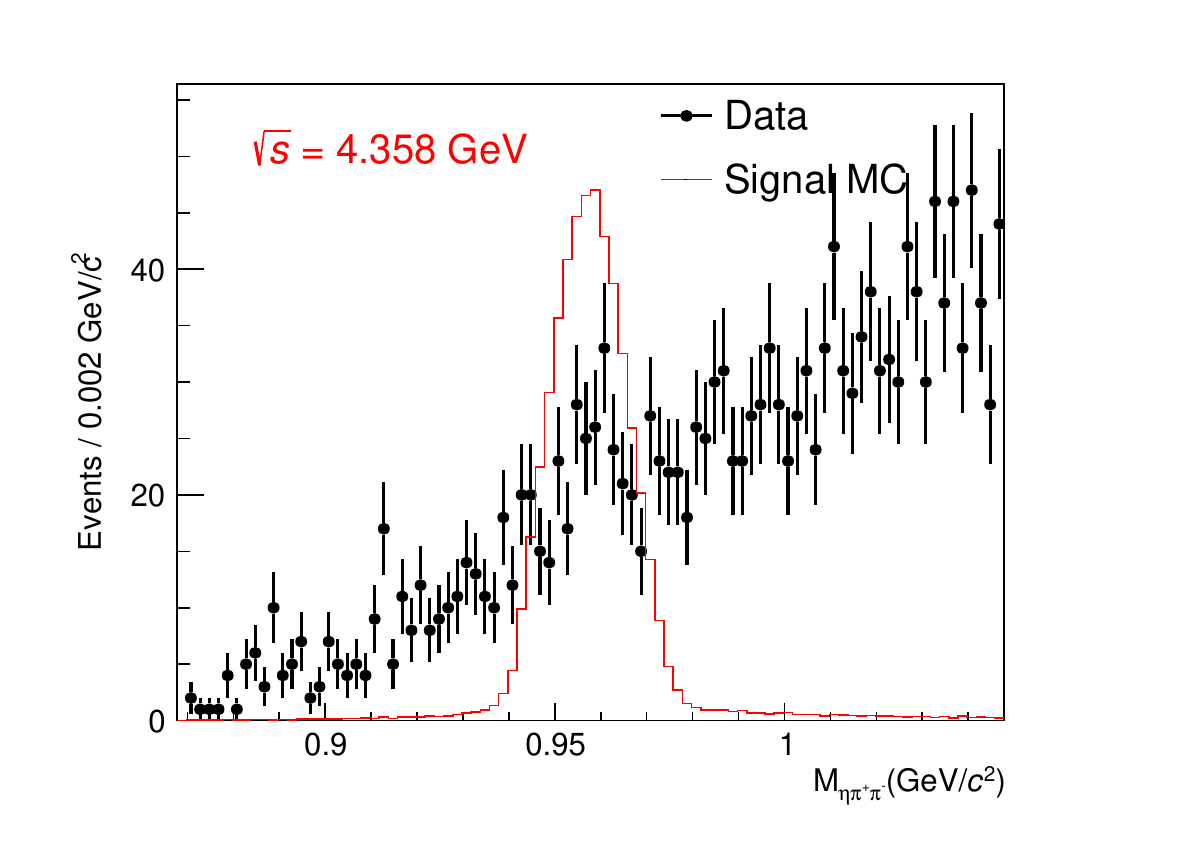}
        \captionsetup{skip=-7pt,font=normalsize}
    \end{subfigure}
\captionsetup{justification=raggedright}
\caption{The distributions of (Left) $M_{hadrons}$ versus $M_{\gamma\gamma}$, (Middle) $M_{hadrons}$, and (Right) $M_{\gamma\gamma}$ at $\sqrt s=4.308-4.358$~GeV.}
\label{fig:normal8}
\end{figure*}

\begin{figure*}[htbp]
    \begin{subfigure}{0.32\textwidth}
        \includegraphics[width=\linewidth]{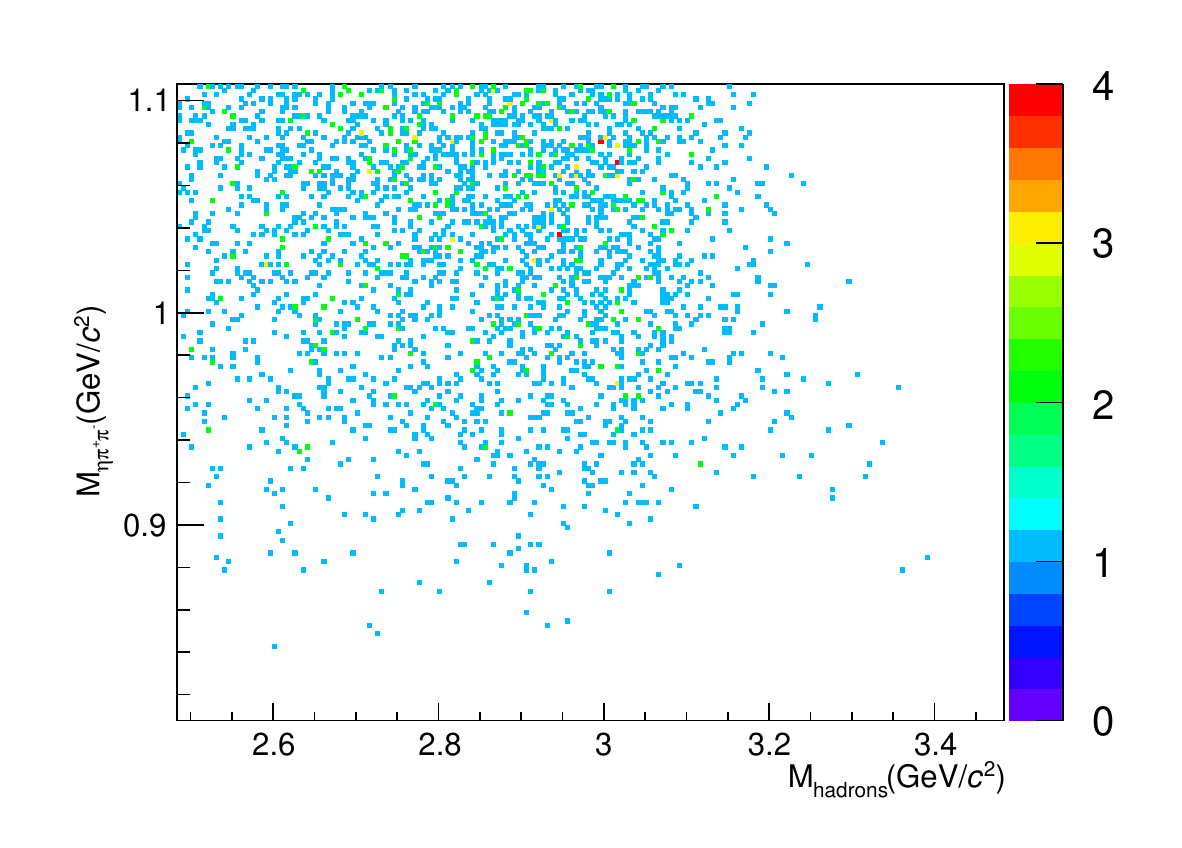}
        \captionsetup{skip=-7pt,font=normalsize}
    \end{subfigure}
    \begin{subfigure}{0.32\textwidth}
        \includegraphics[width=\linewidth]{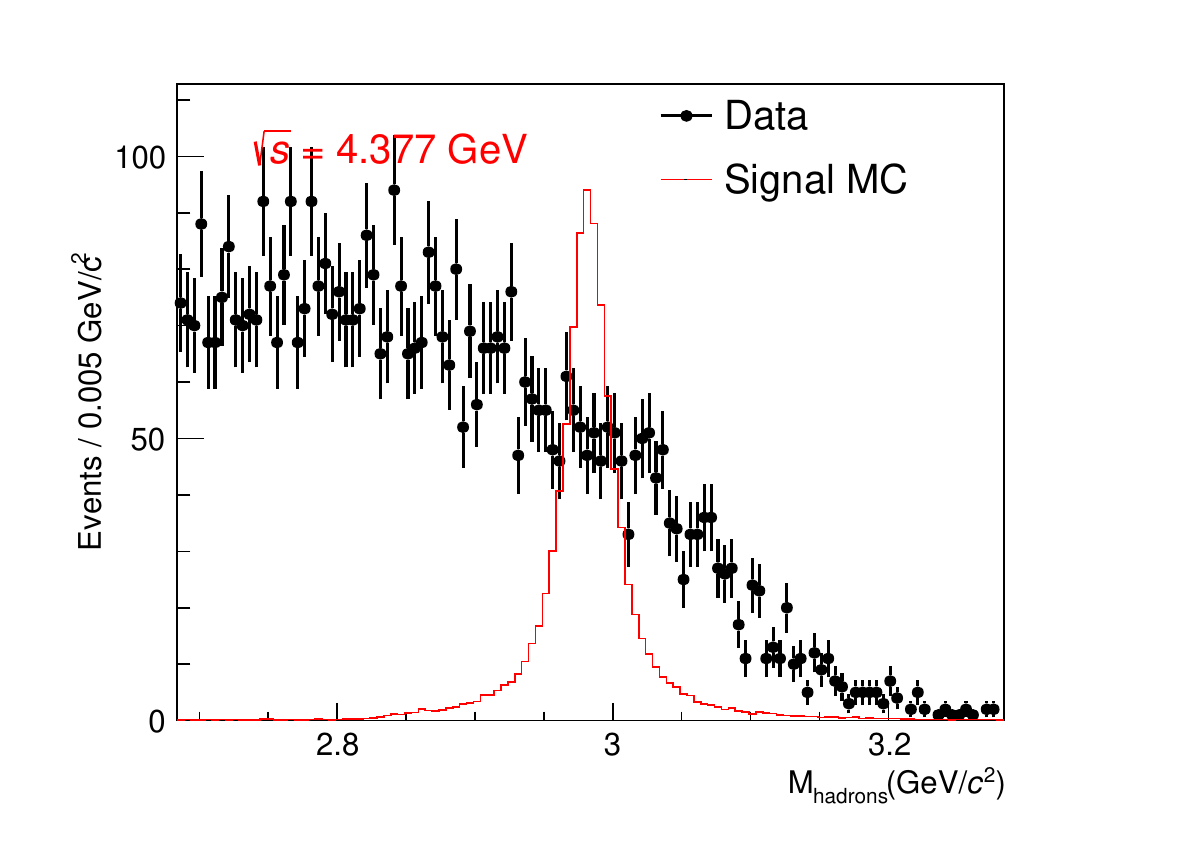}
        \captionsetup{skip=-7pt,font=normalsize}
    \end{subfigure}
    \begin{subfigure}{0.32\textwidth}
        \includegraphics[width=\linewidth]{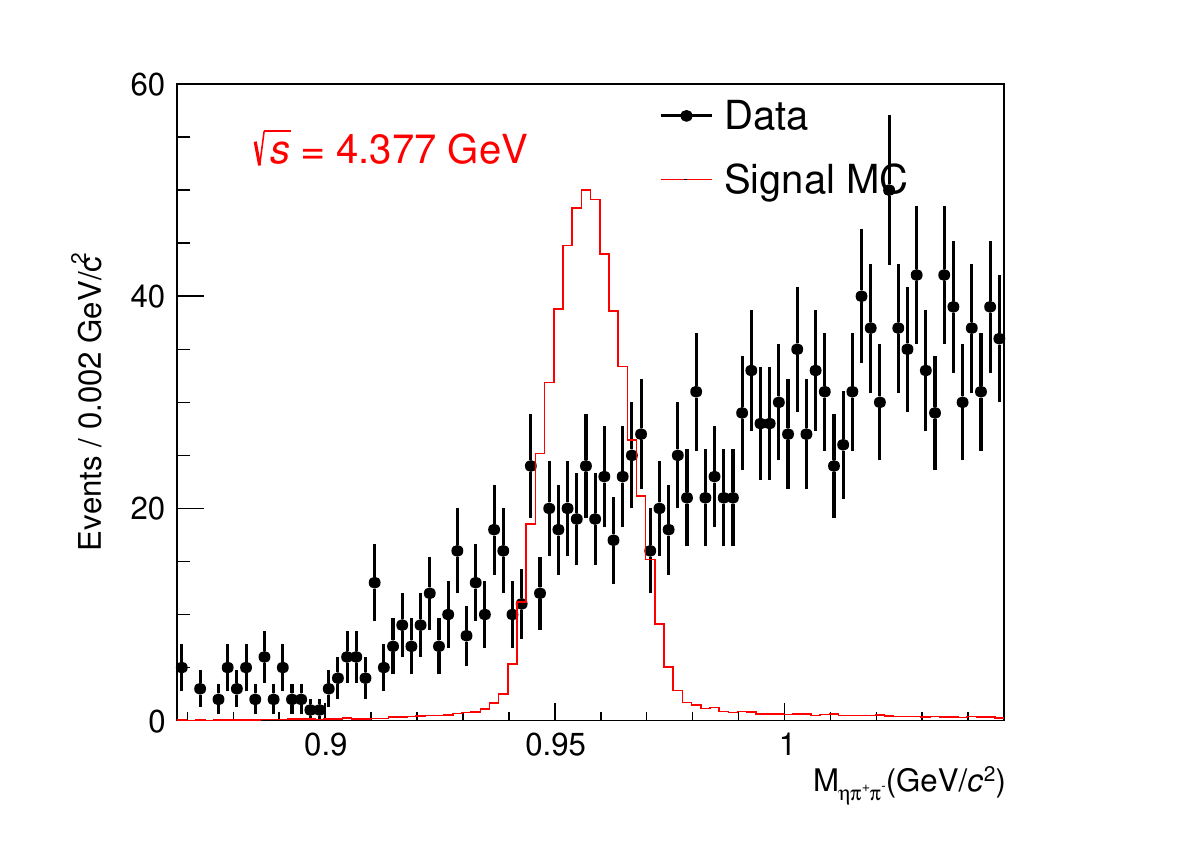}
        \captionsetup{skip=-7pt,font=normalsize}
    \end{subfigure}
    \begin{subfigure}{0.32\textwidth}
        \includegraphics[width=\linewidth]{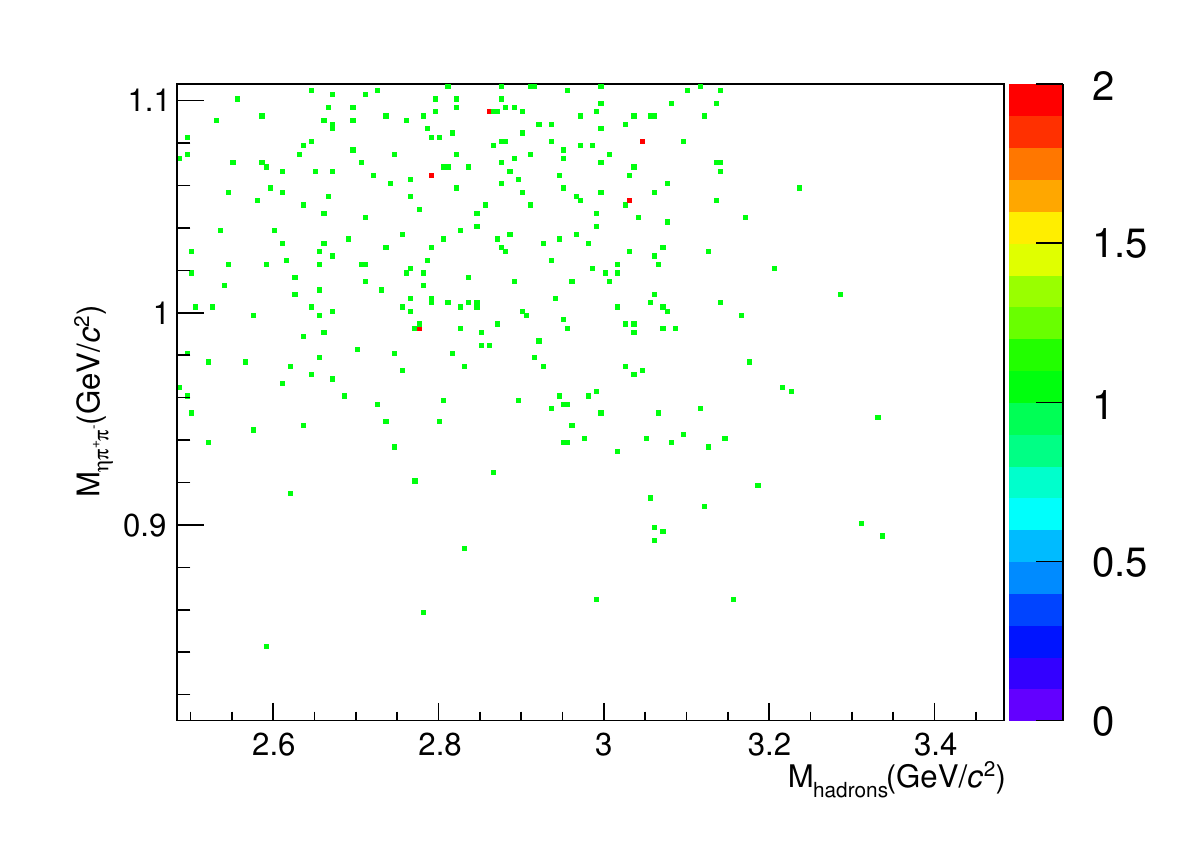}
        \captionsetup{skip=-7pt,font=normalsize}
    \end{subfigure}
    \begin{subfigure}{0.32\textwidth}
        \includegraphics[width=\linewidth]{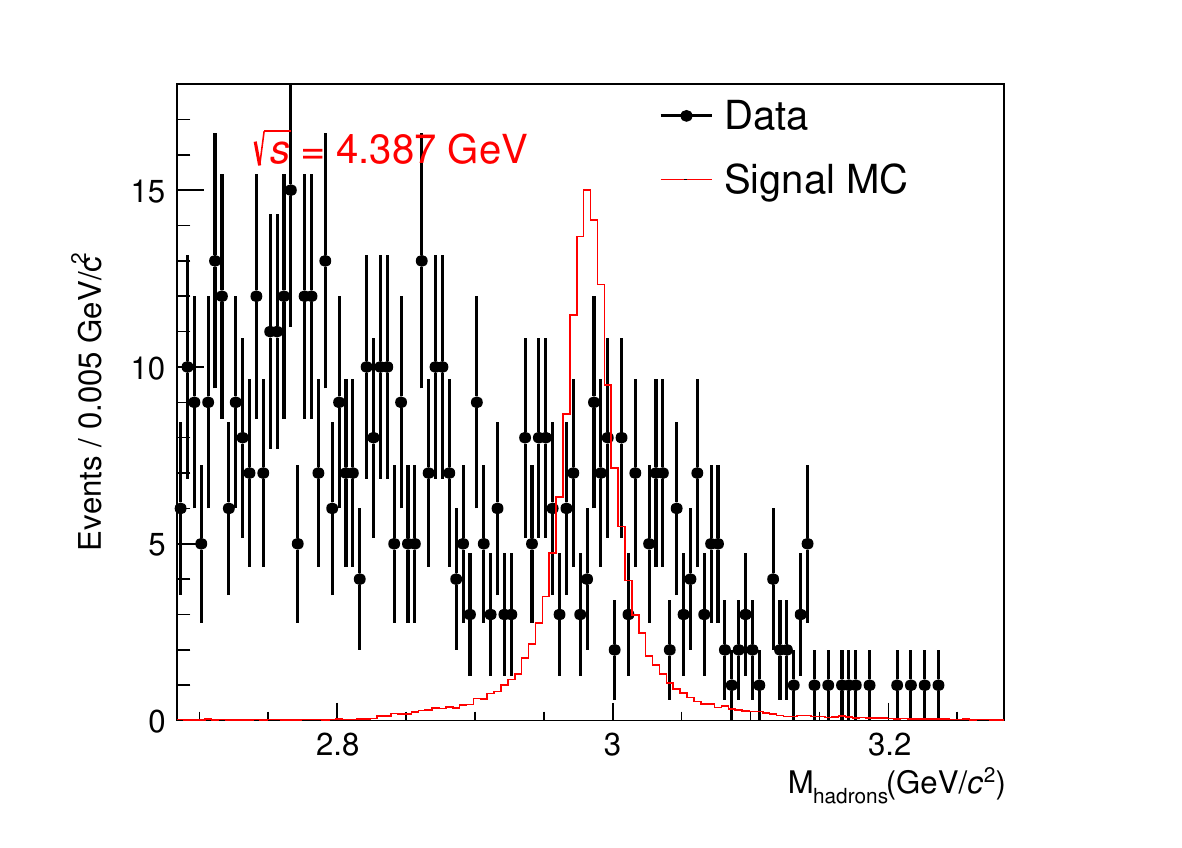}
        \captionsetup{skip=-7pt,font=normalsize}
    \end{subfigure}
    \begin{subfigure}{0.32\textwidth}
        \includegraphics[width=\linewidth]{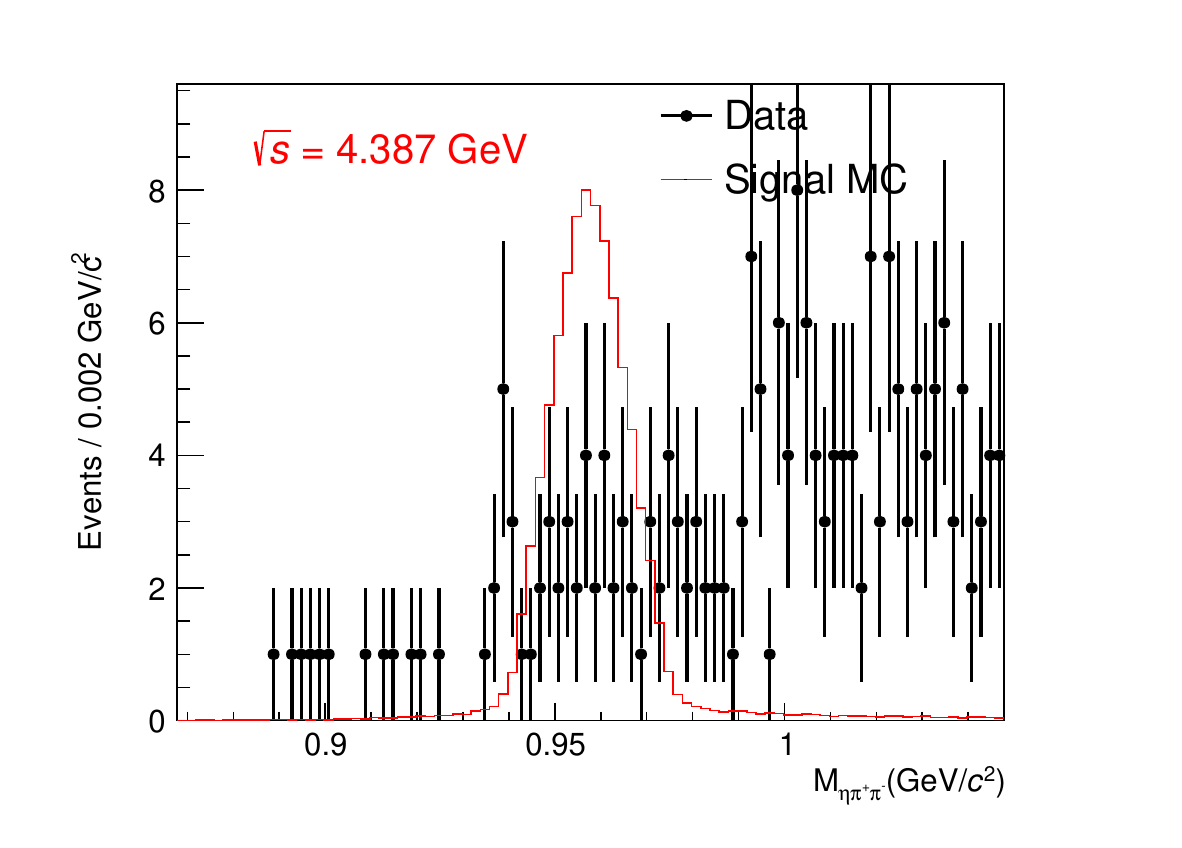}
        \captionsetup{skip=-7pt,font=normalsize}
    \end{subfigure}
    \begin{subfigure}{0.32\textwidth}
        \includegraphics[width=\linewidth]{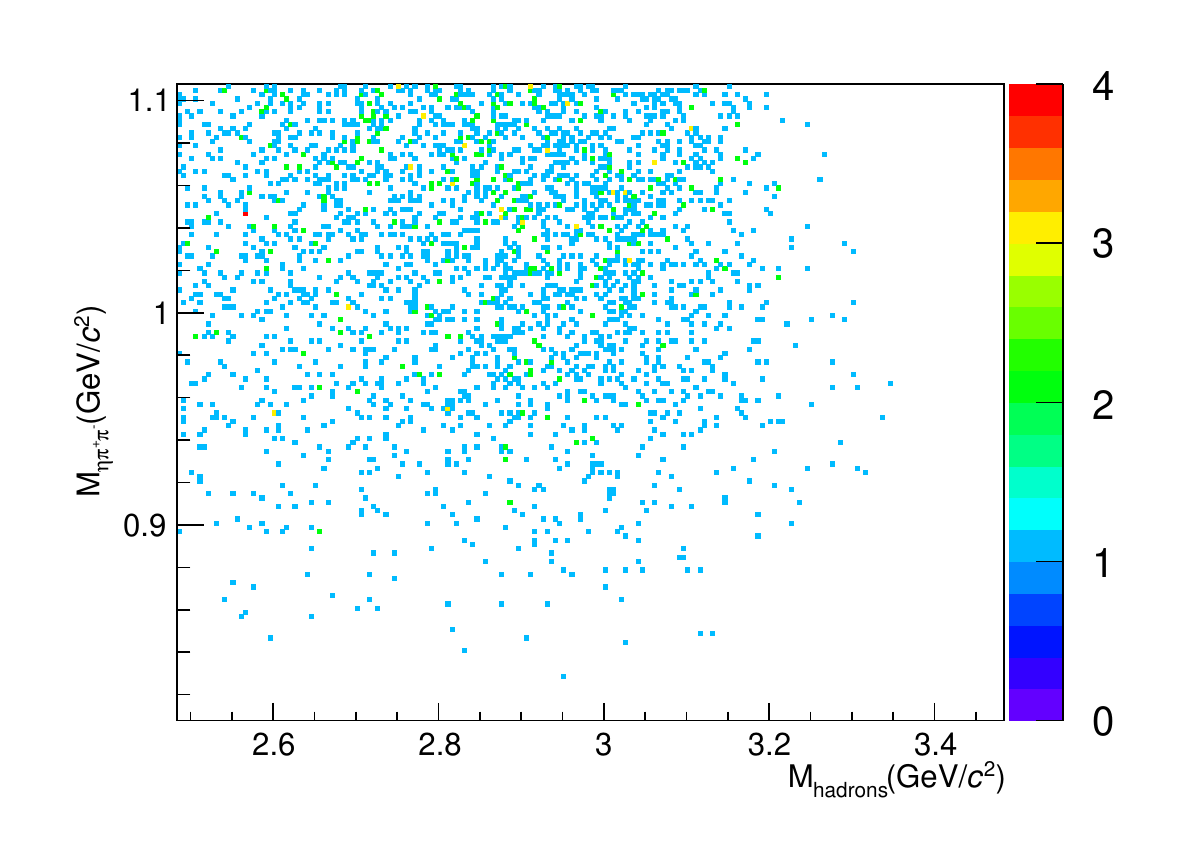}
        \captionsetup{skip=-7pt,font=normalsize}
    \end{subfigure}
    \begin{subfigure}{0.32\textwidth}
        \includegraphics[width=\linewidth]{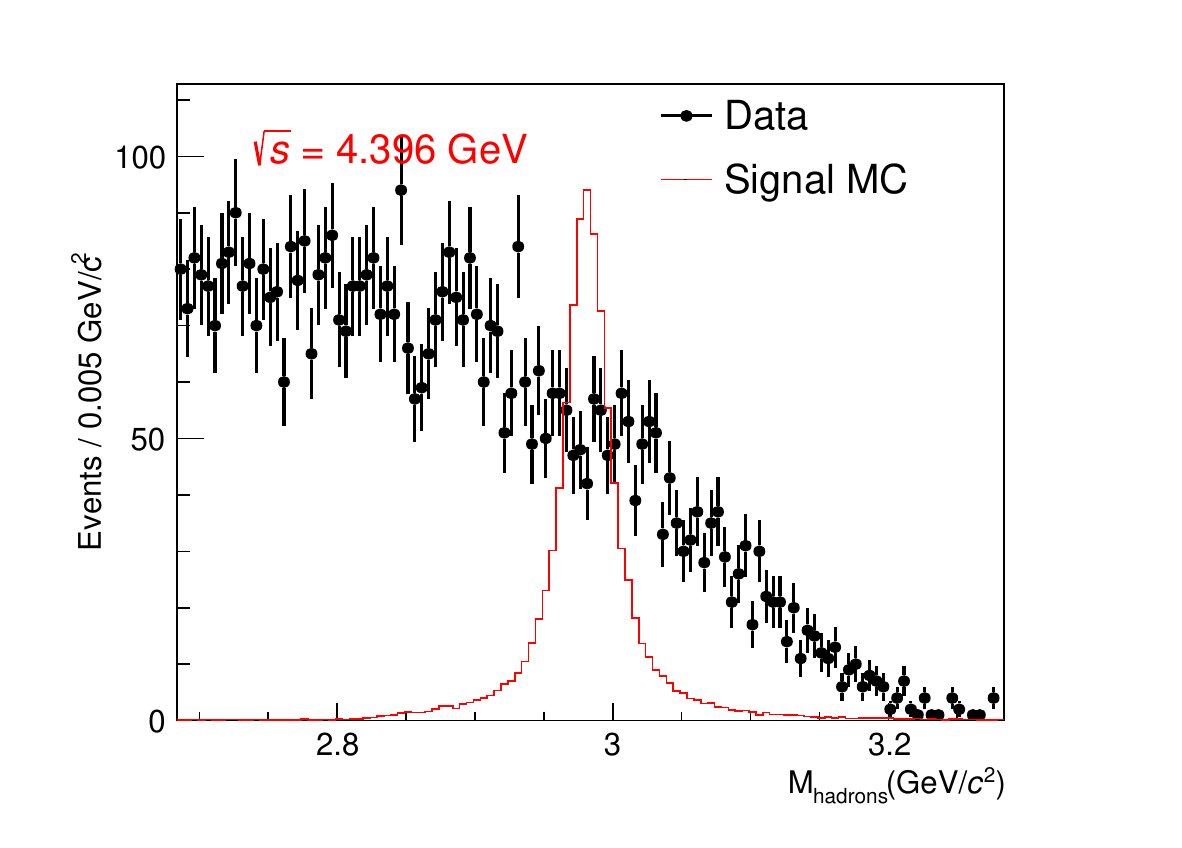}
        \captionsetup{skip=-7pt,font=normalsize}
    \end{subfigure}
    \begin{subfigure}{0.32\textwidth}
        \includegraphics[width=\linewidth]{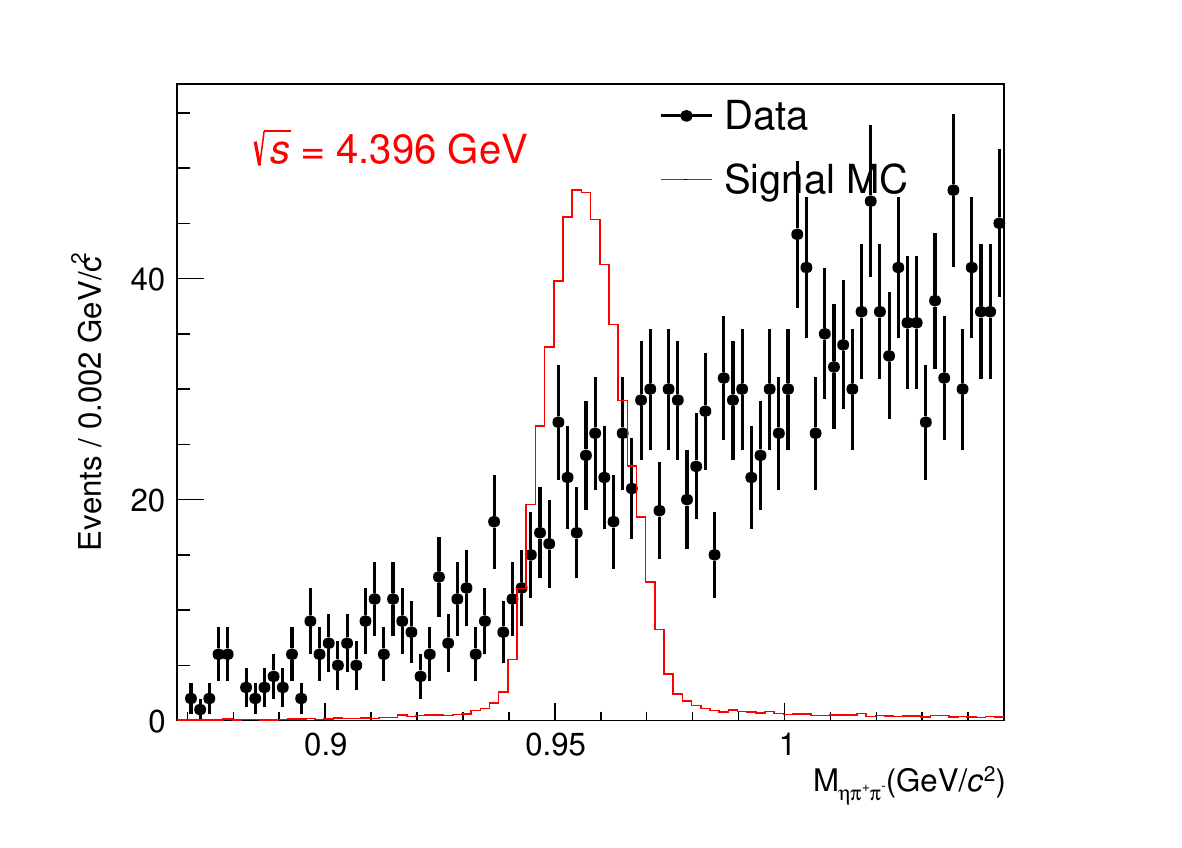}
        \captionsetup{skip=-7pt,font=normalsize}
    \end{subfigure}
    \begin{subfigure}{0.32\textwidth}
        \includegraphics[width=\linewidth]{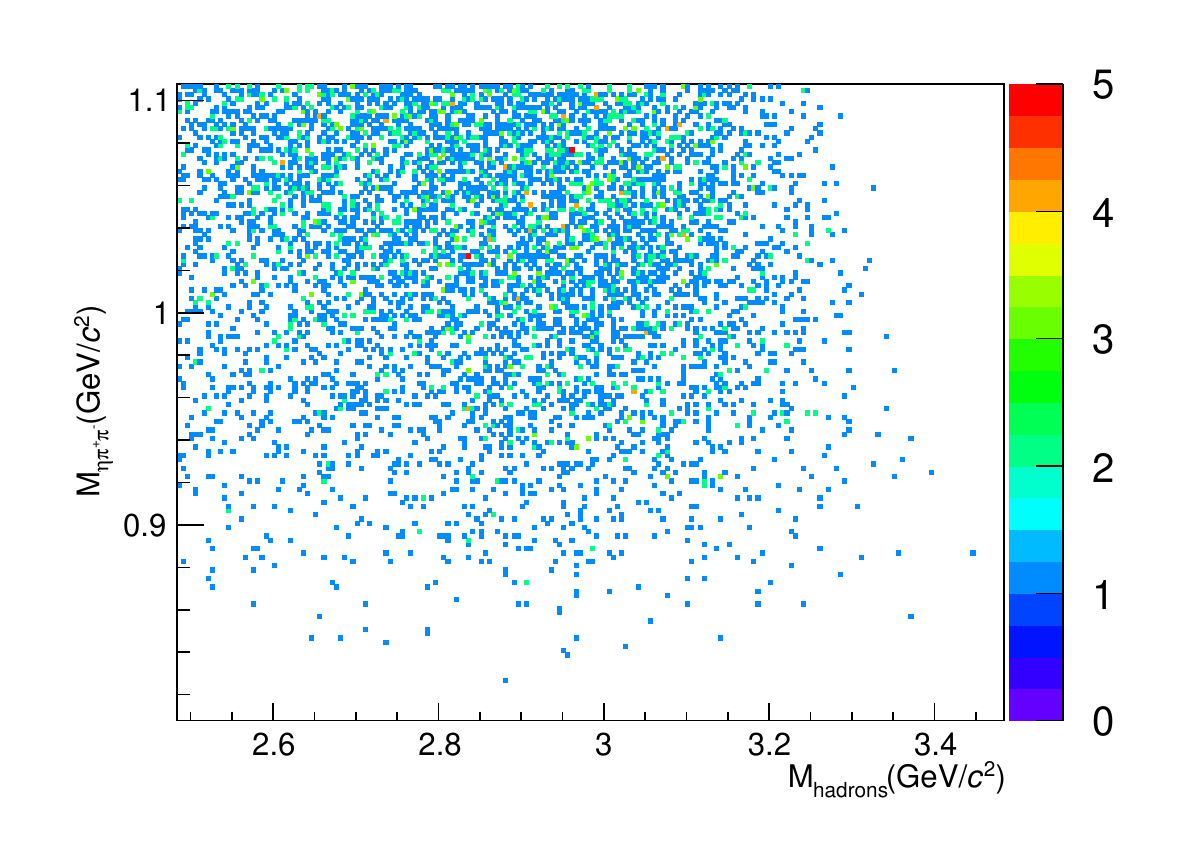}
        \captionsetup{skip=-7pt,font=normalsize}
    \end{subfigure}
    \begin{subfigure}{0.32\textwidth}
        \includegraphics[width=\linewidth]{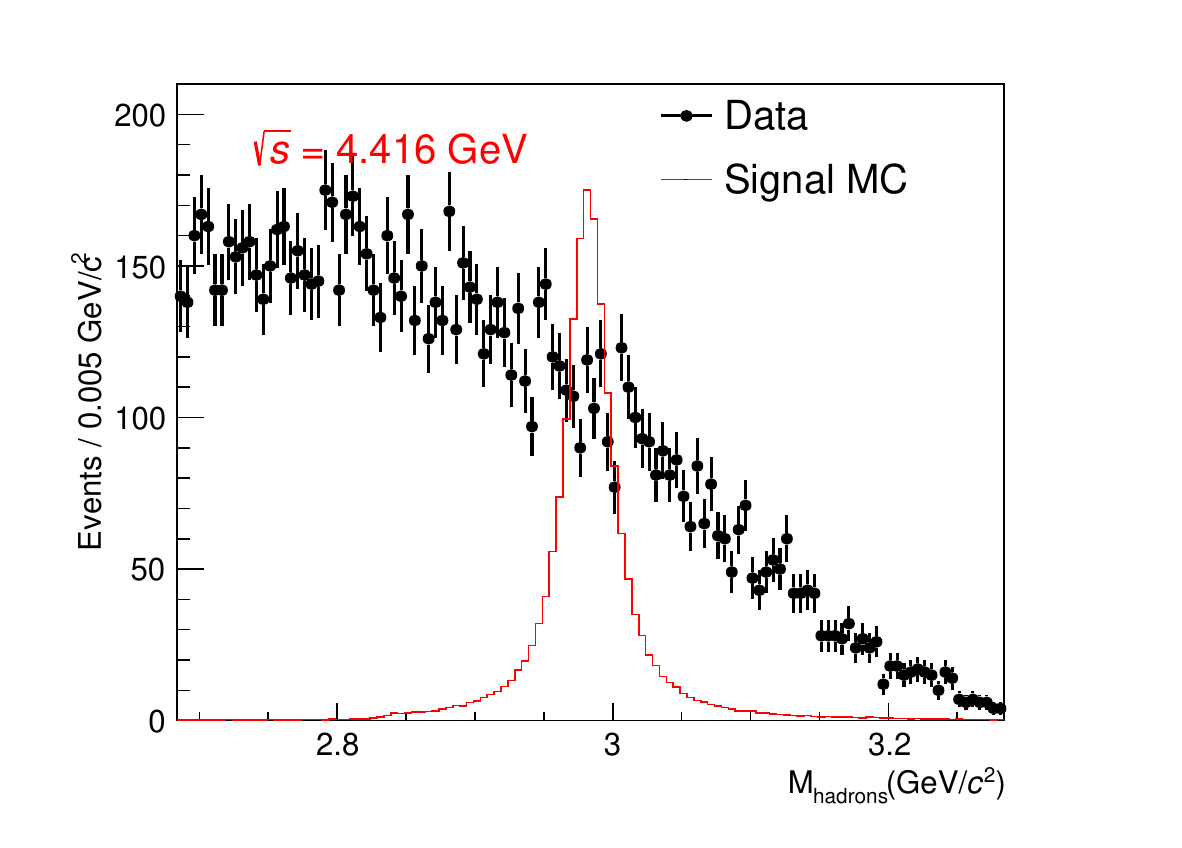}
        \captionsetup{skip=-7pt,font=normalsize}
    \end{subfigure}
    \begin{subfigure}{0.32\textwidth}
        \includegraphics[width=\linewidth]{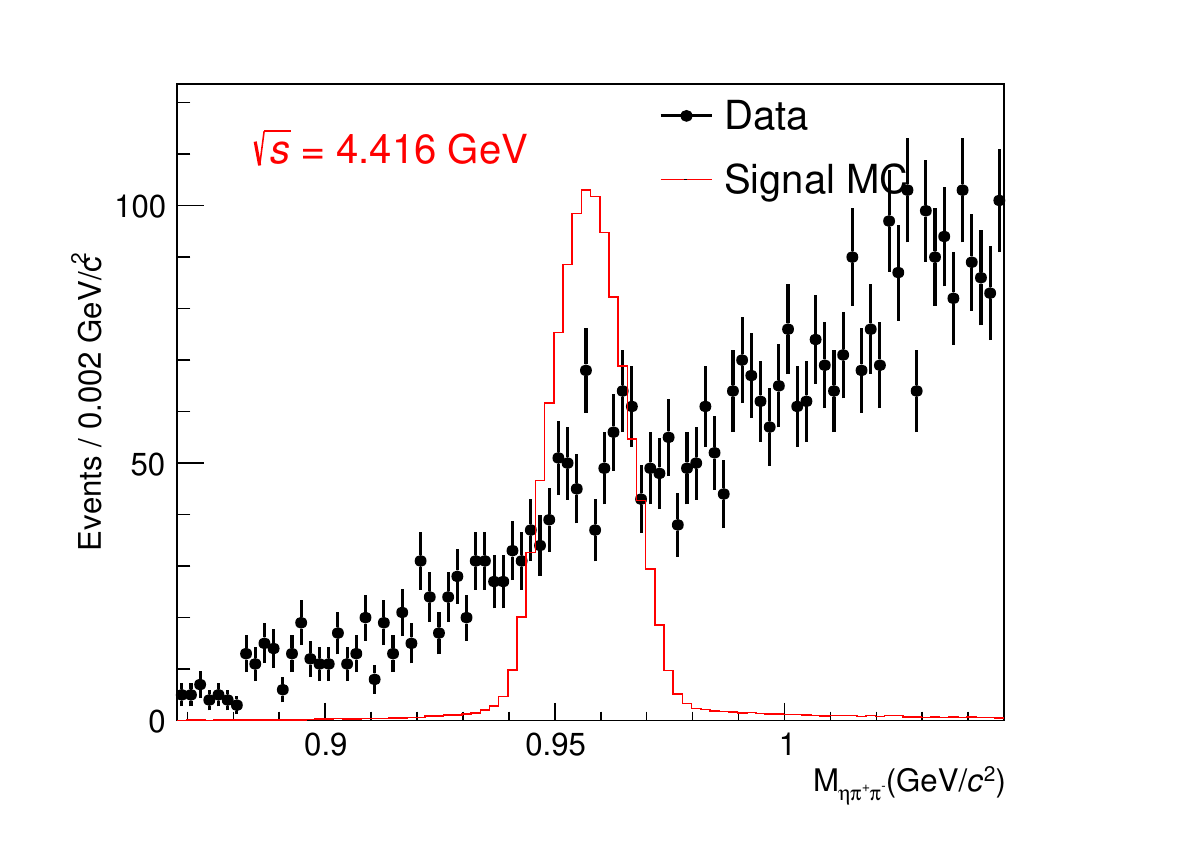}
        \captionsetup{skip=-7pt,font=normalsize}
    \end{subfigure}
\captionsetup{justification=raggedright}
\caption{The distributions of (Left) $M_{hadrons}$ versus $M_{\gamma\gamma}$, (Middle) $M_{hadrons}$, and (Right) $M_{\gamma\gamma}$ at $\sqrt s=4.377-4.416$~GeV.}
\label{fig:normal9}
\end{figure*}

\begin{figure*}[htbp]
    \begin{subfigure}{0.32\textwidth}
        \includegraphics[width=\linewidth]{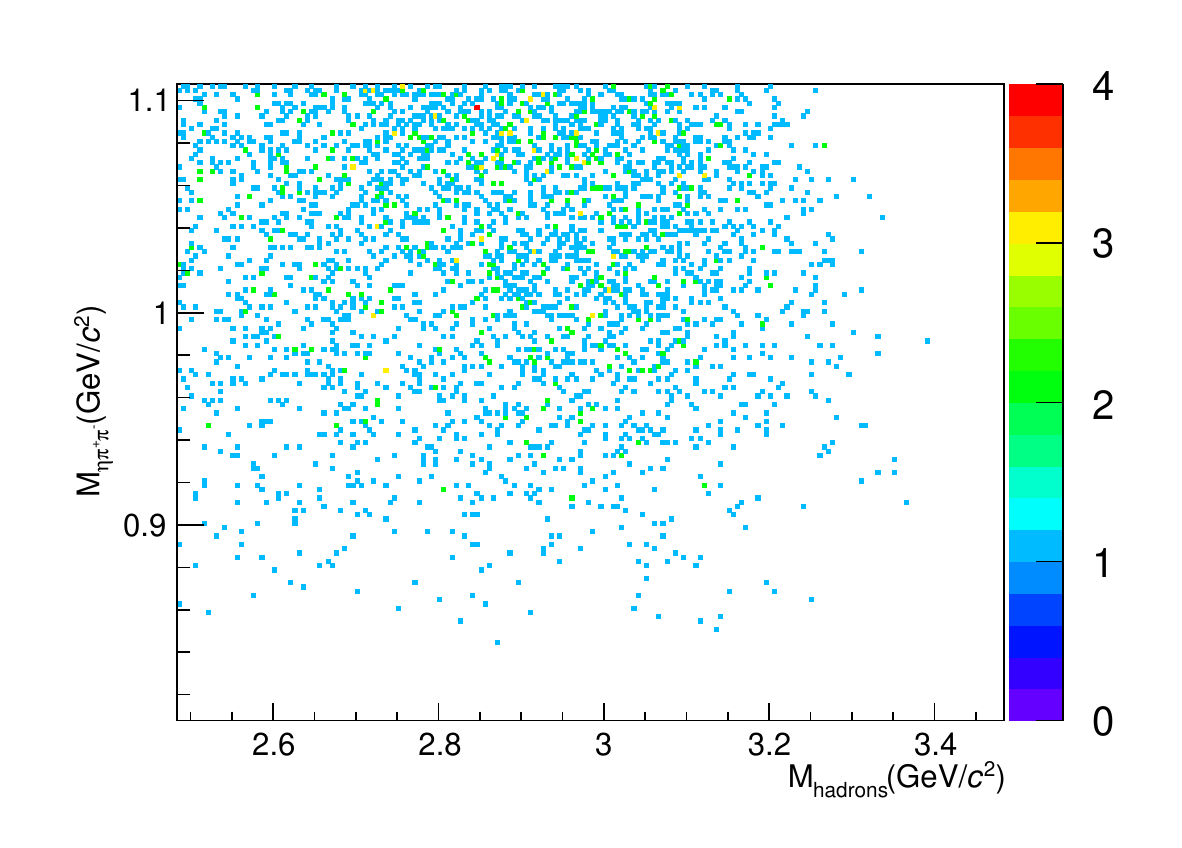}
        \captionsetup{skip=-7pt,font=normalsize}
    \end{subfigure}
    \begin{subfigure}{0.32\textwidth}
        \includegraphics[width=\linewidth]{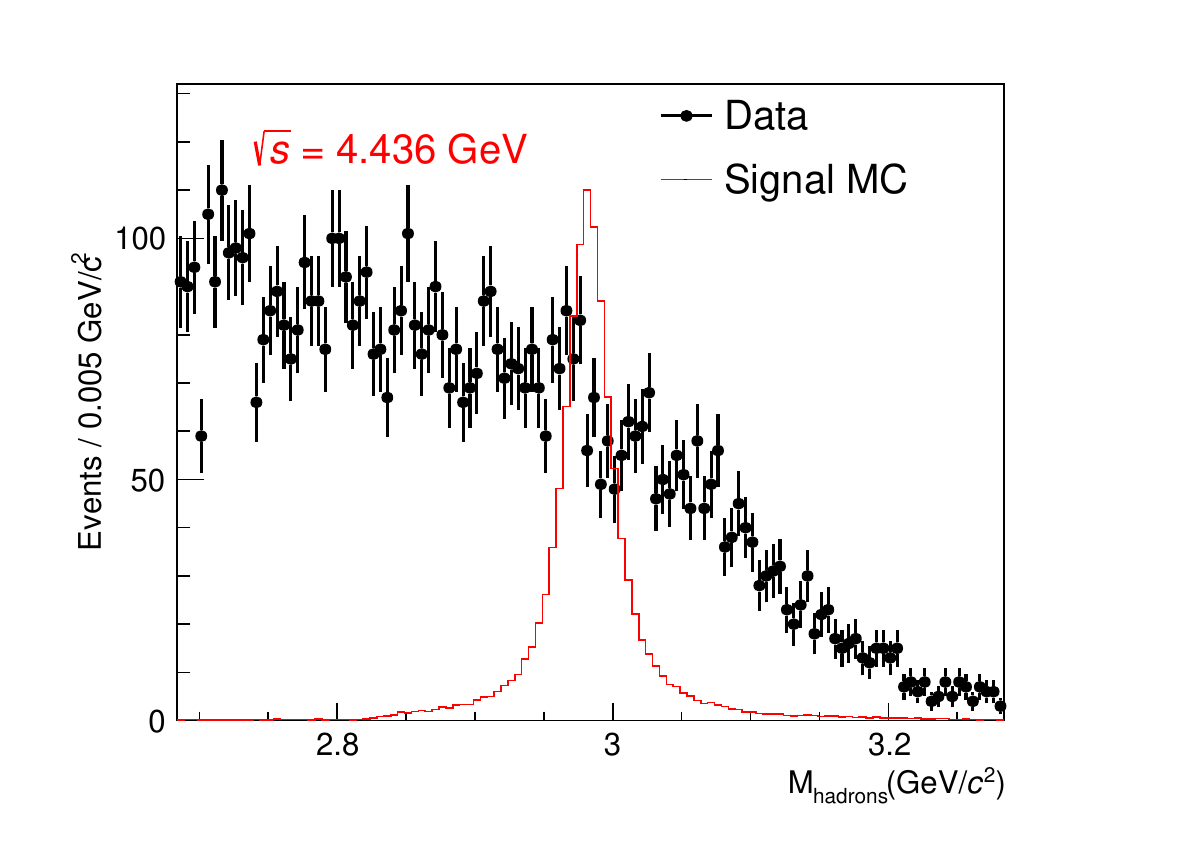}
        \captionsetup{skip=-7pt,font=normalsize}
    \end{subfigure}
    \begin{subfigure}{0.32\textwidth}
        \includegraphics[width=\linewidth]{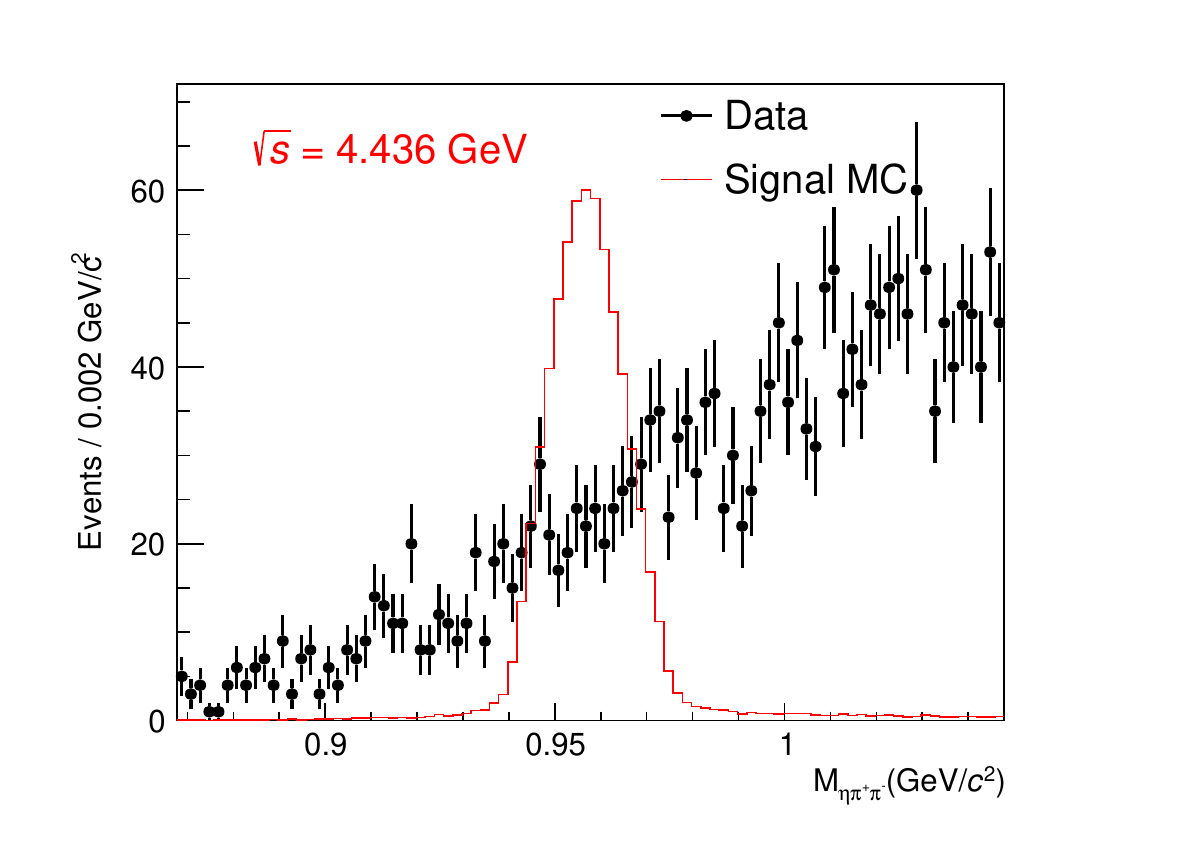}
        \captionsetup{skip=-7pt,font=normalsize}
    \end{subfigure}
    \begin{subfigure}{0.32\textwidth}
        \includegraphics[width=\linewidth]{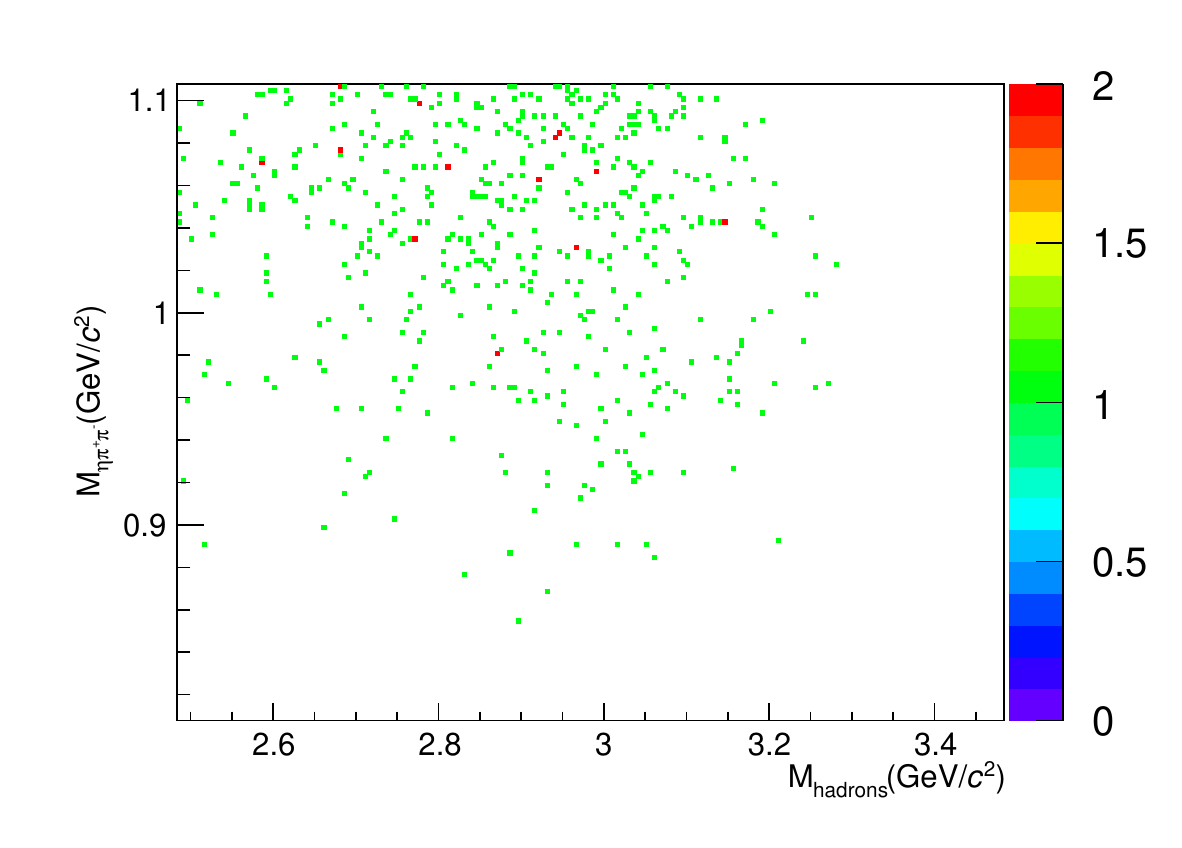}
        \captionsetup{skip=-7pt,font=normalsize}
    \end{subfigure}
    \begin{subfigure}{0.32\textwidth}
        \includegraphics[width=\linewidth]{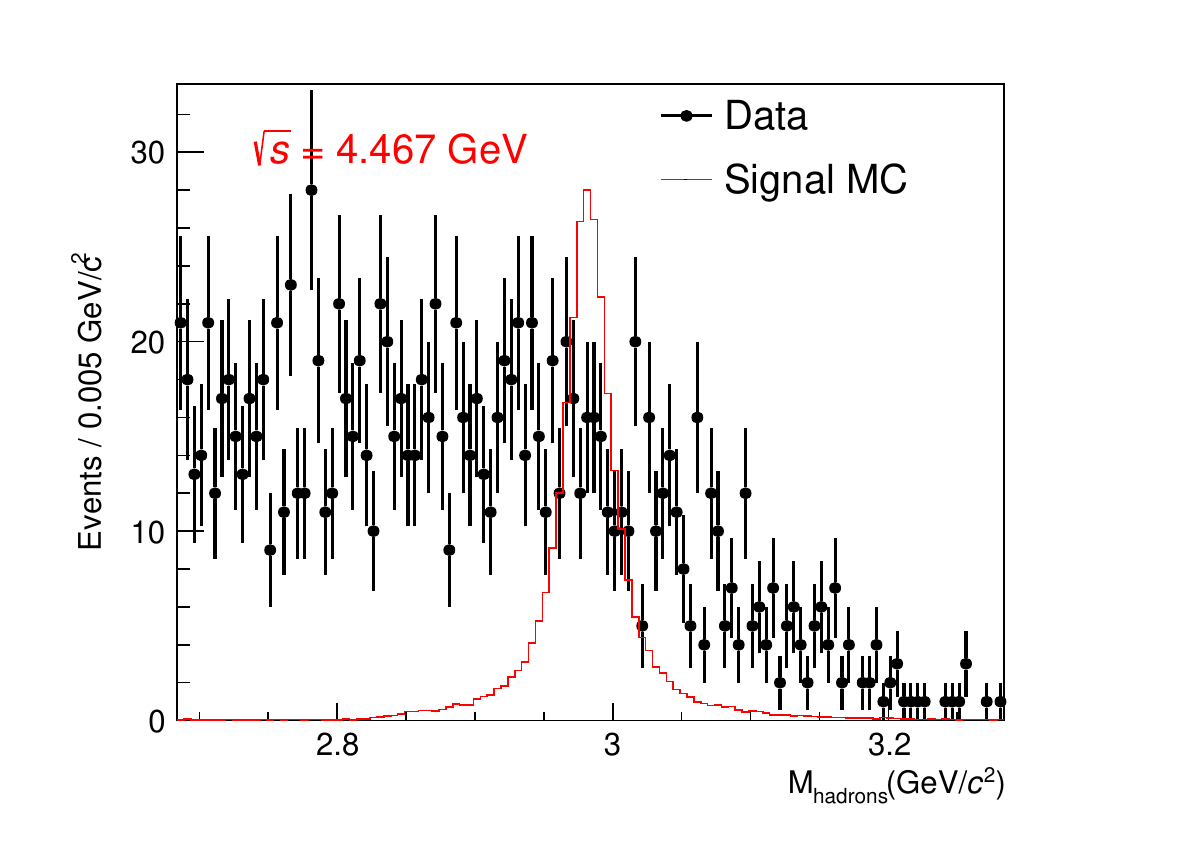}
        \captionsetup{skip=-7pt,font=normalsize}
    \end{subfigure}
    \begin{subfigure}{0.32\textwidth}
        \includegraphics[width=\linewidth]{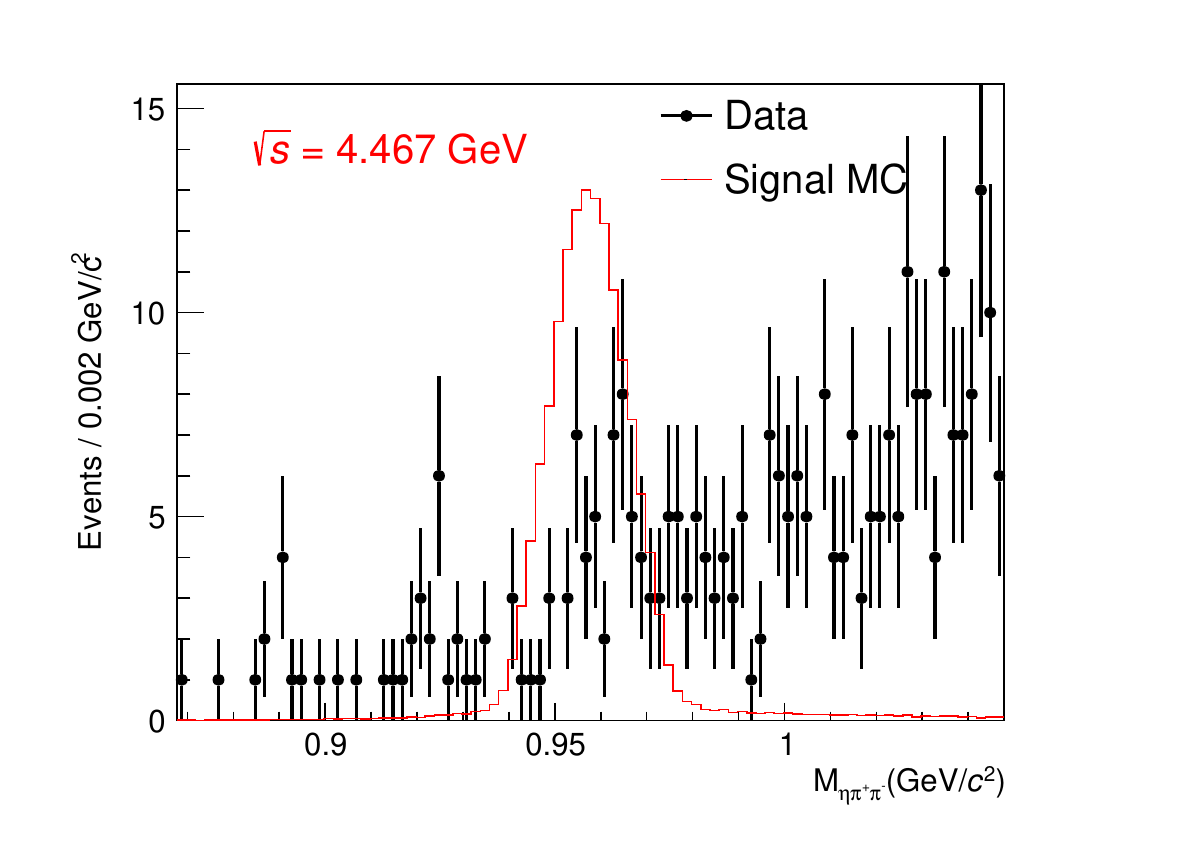}
        \captionsetup{skip=-7pt,font=normalsize}
    \end{subfigure}
    \begin{subfigure}{0.32\textwidth}
        \includegraphics[width=\linewidth]{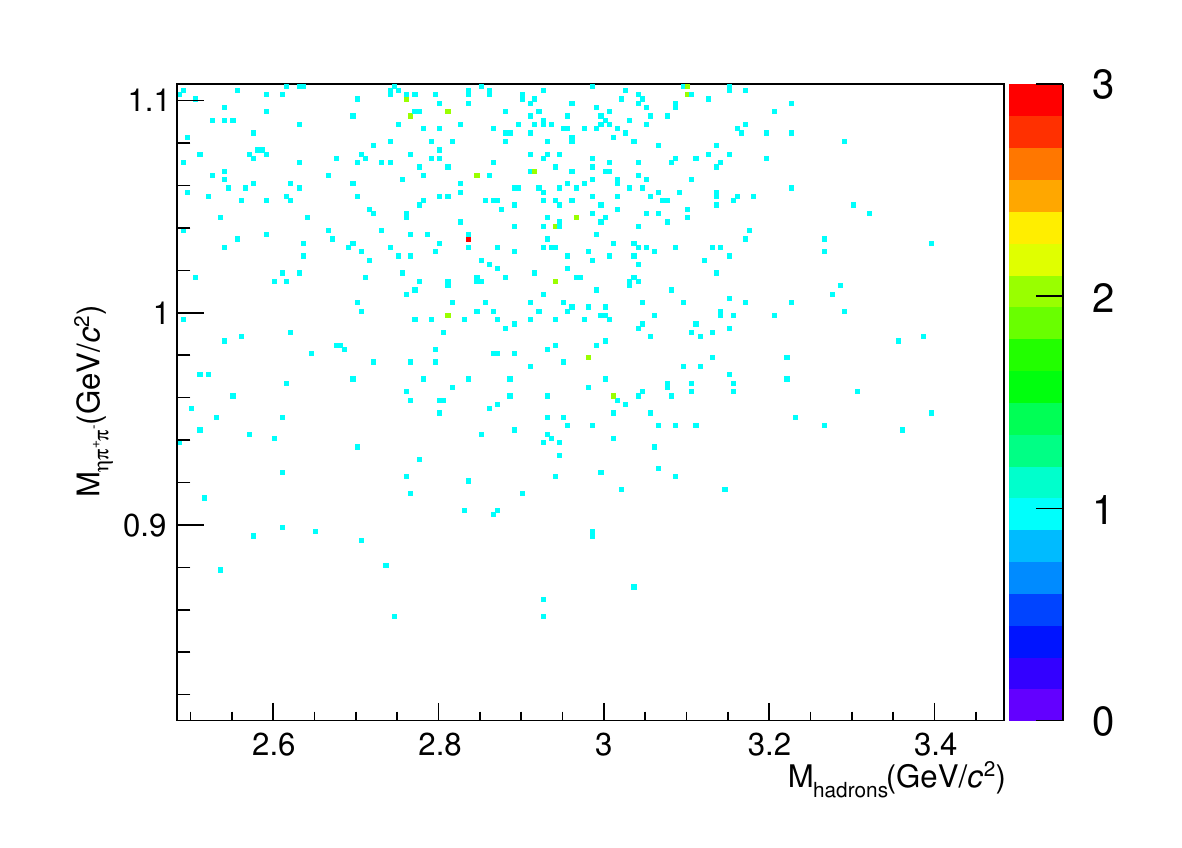}
        \captionsetup{skip=-7pt,font=normalsize}
    \end{subfigure}
    \begin{subfigure}{0.32\textwidth}
        \includegraphics[width=\linewidth]{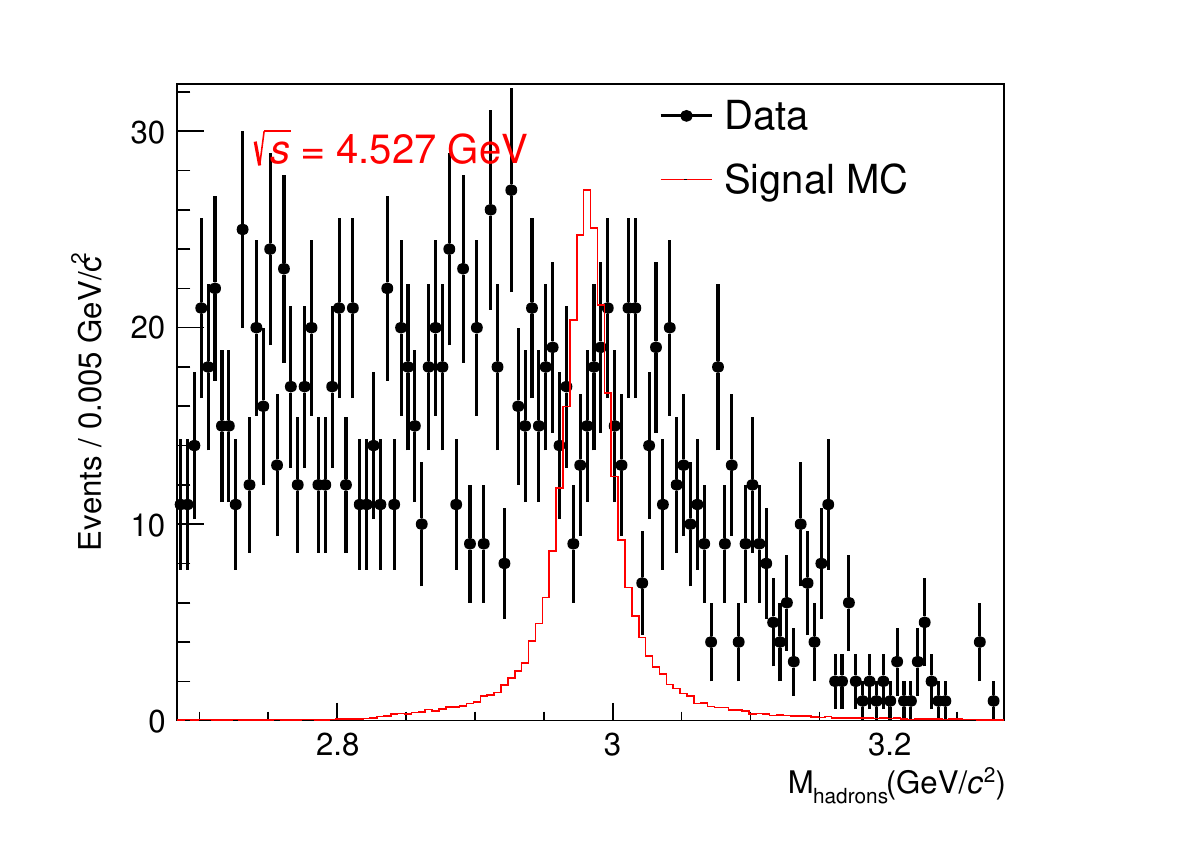}
        \captionsetup{skip=-7pt,font=normalsize}
    \end{subfigure}
    \begin{subfigure}{0.32\textwidth}
        \includegraphics[width=\linewidth]{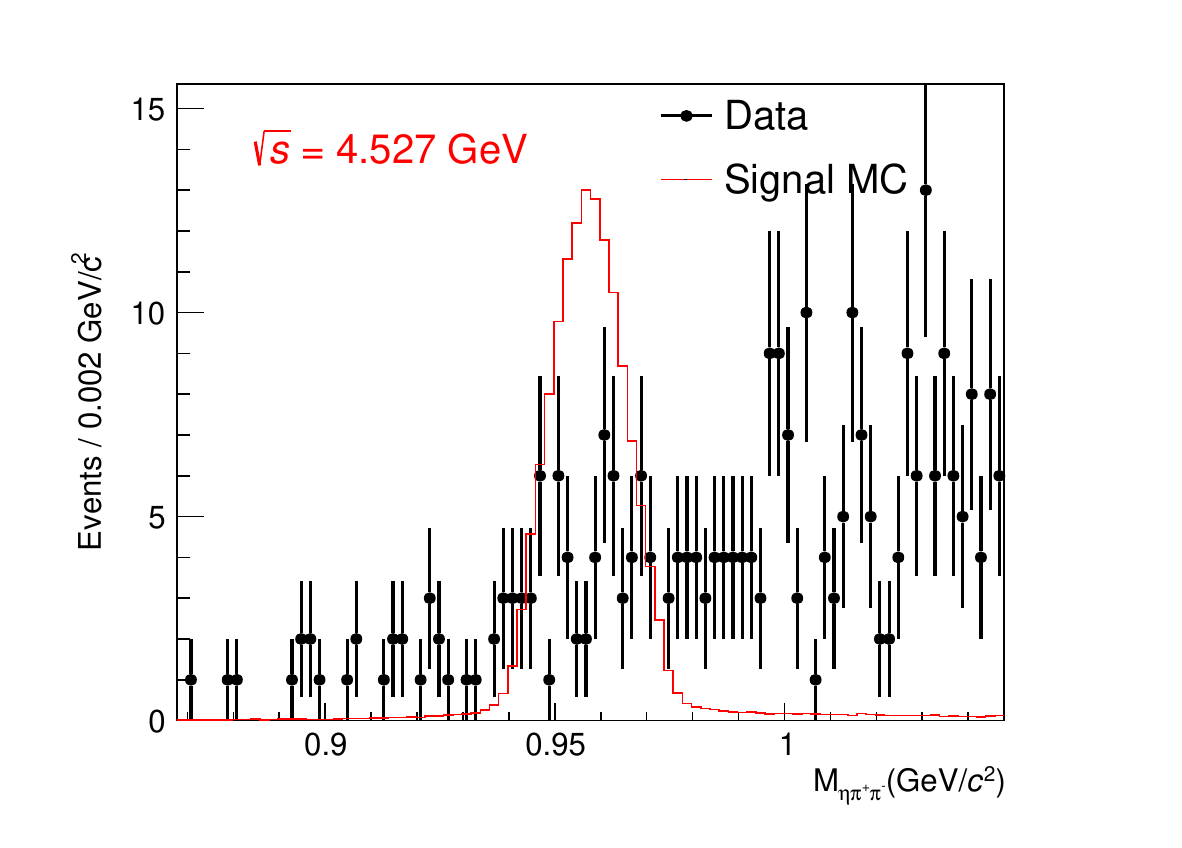}
        \captionsetup{skip=-7pt,font=normalsize}
    \end{subfigure}
    \begin{subfigure}{0.32\textwidth}
        \includegraphics[width=\linewidth]{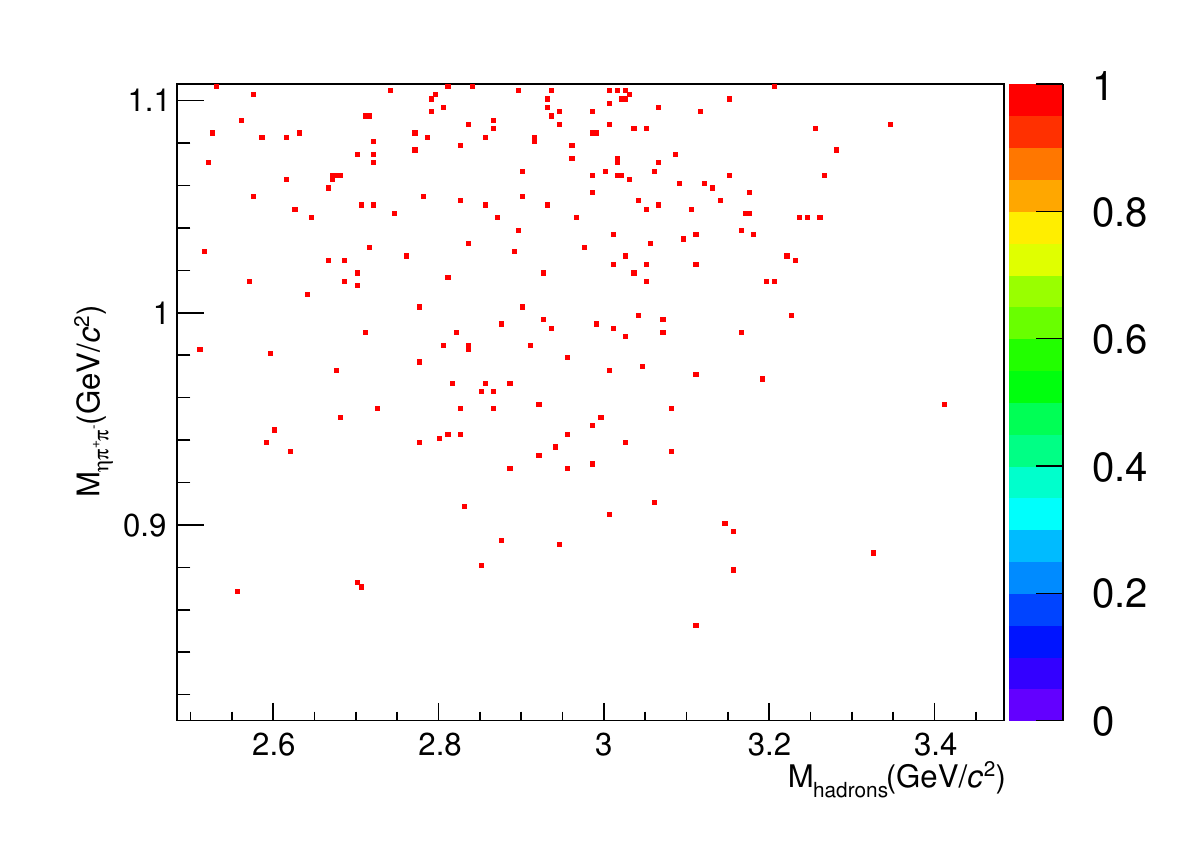}
        \captionsetup{skip=-7pt,font=normalsize}
    \end{subfigure}
    \begin{subfigure}{0.32\textwidth}
        \includegraphics[width=\linewidth]{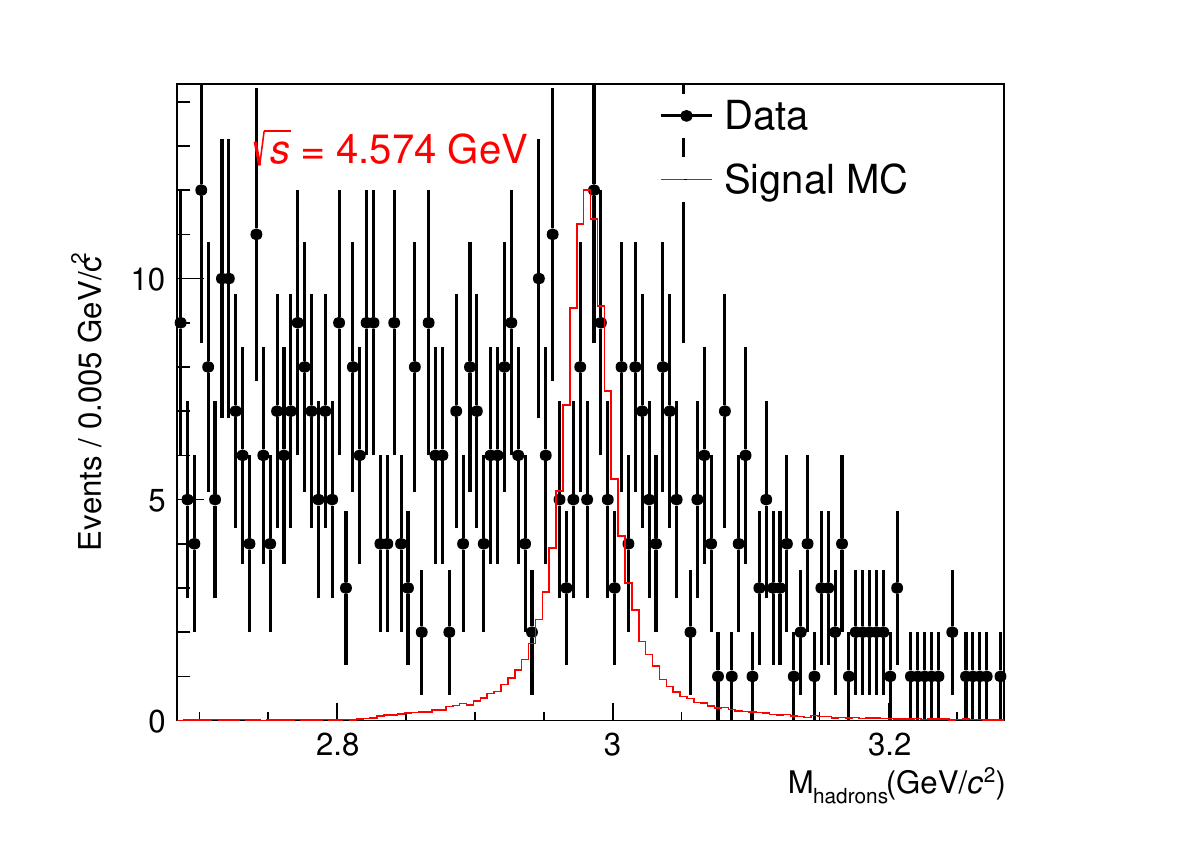}
        \captionsetup{skip=-7pt,font=normalsize}
    \end{subfigure}
    \begin{subfigure}{0.32\textwidth}
        \includegraphics[width=\linewidth]{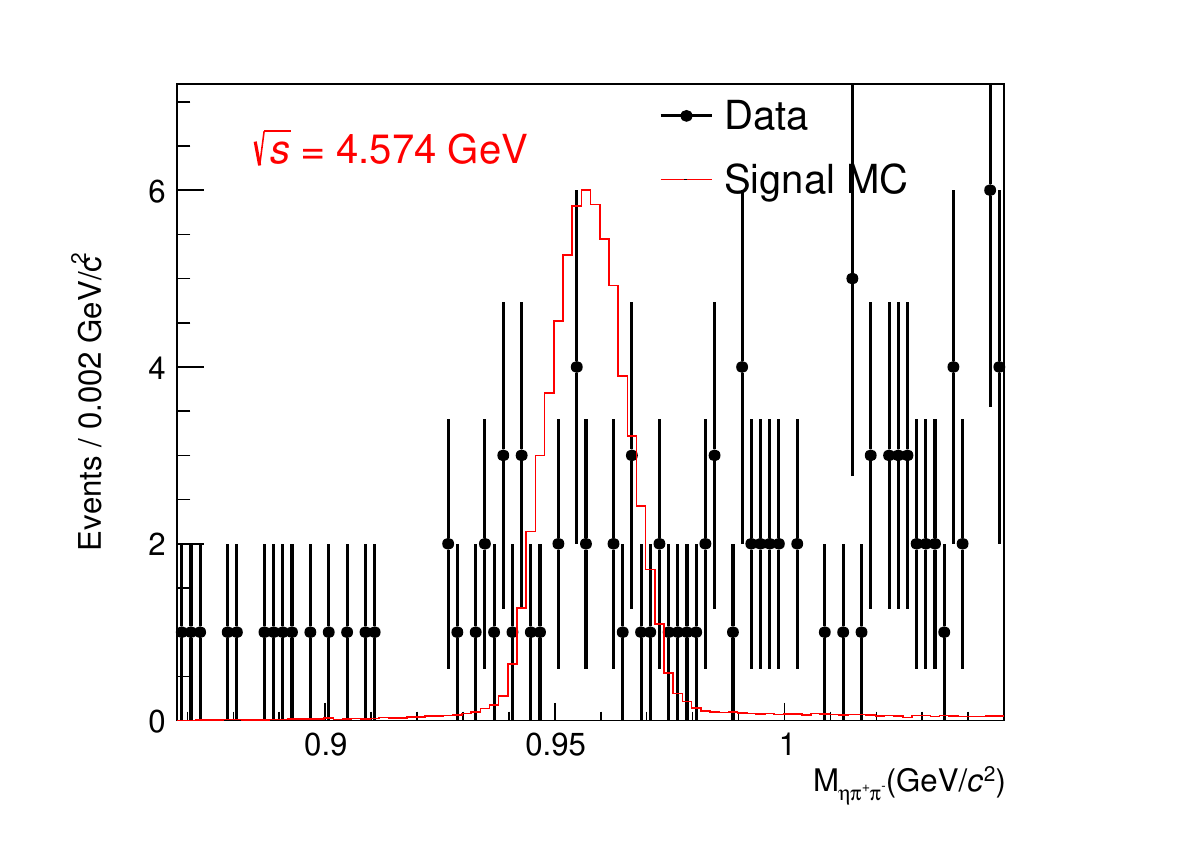}
        \captionsetup{skip=-7pt,font=normalsize}
    \end{subfigure}
\captionsetup{justification=raggedright}
\caption{The distributions of (Left) $M_{hadrons}$ versus $M_{\gamma\gamma}$, (Middle) $M_{hadrons}$, and (Right) $M_{\gamma\gamma}$ at $\sqrt s=4.436-4.574$~GeV.}
\label{fig:normal10}
\end{figure*}

\begin{figure*}[htbp]
    \begin{subfigure}{0.32\textwidth}
        \includegraphics[width=\linewidth]{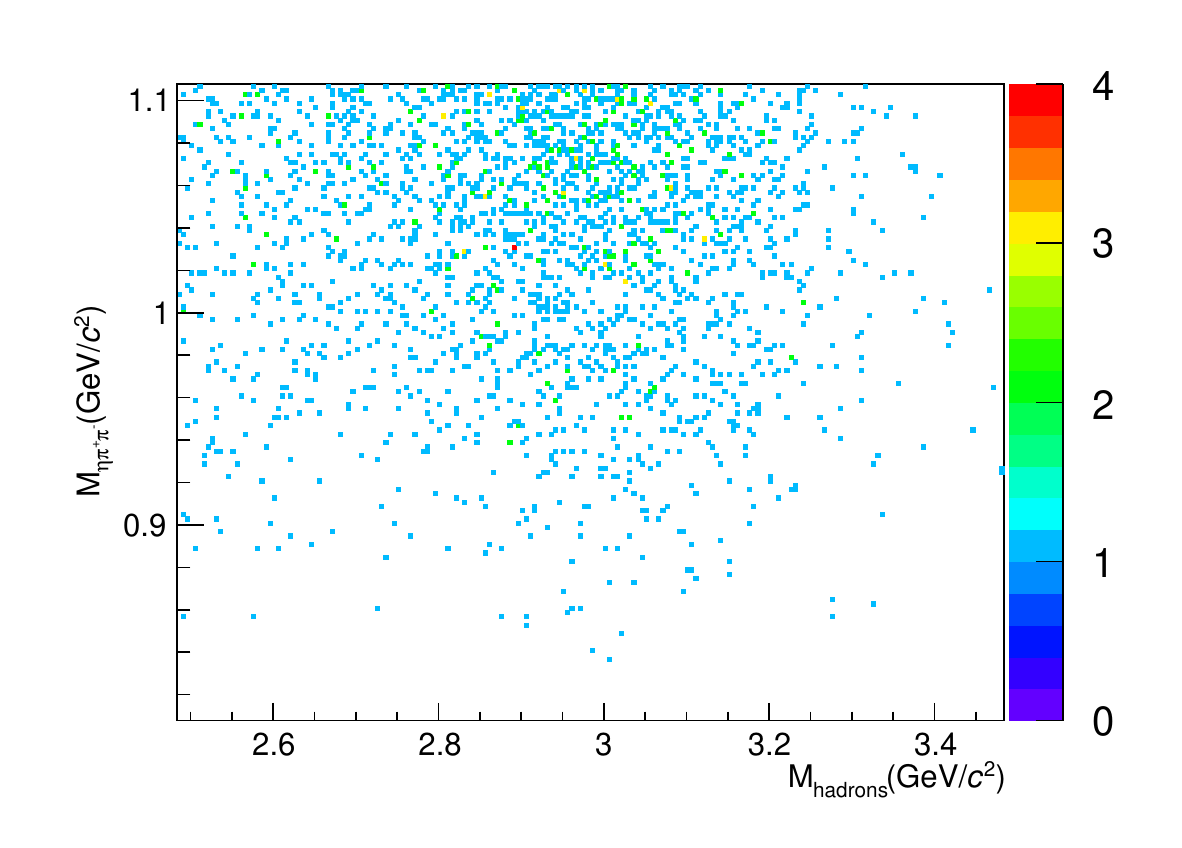}
        \captionsetup{skip=-7pt,font=normalsize}
    \end{subfigure}
    \begin{subfigure}{0.32\textwidth}
        \includegraphics[width=\linewidth]{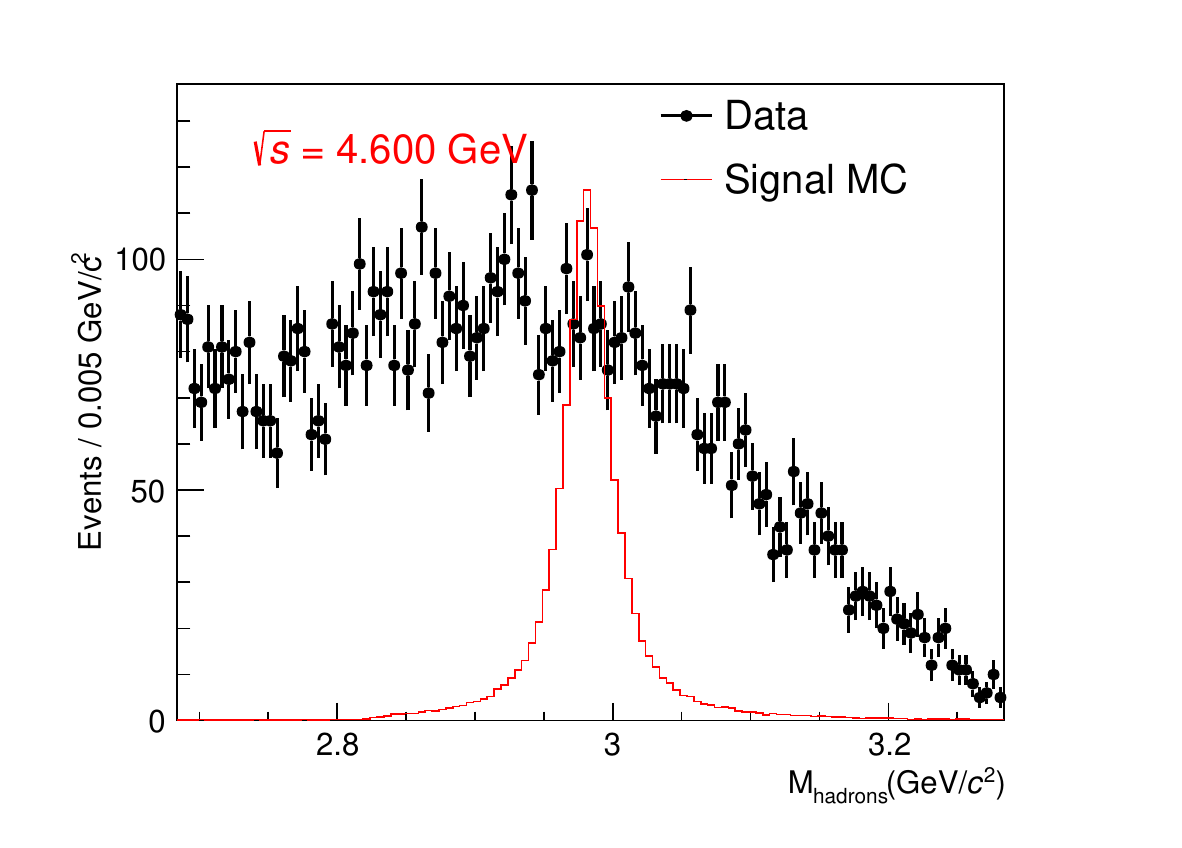}
        \captionsetup{skip=-7pt,font=normalsize}
    \end{subfigure}
    \begin{subfigure}{0.32\textwidth}
        \includegraphics[width=\linewidth]{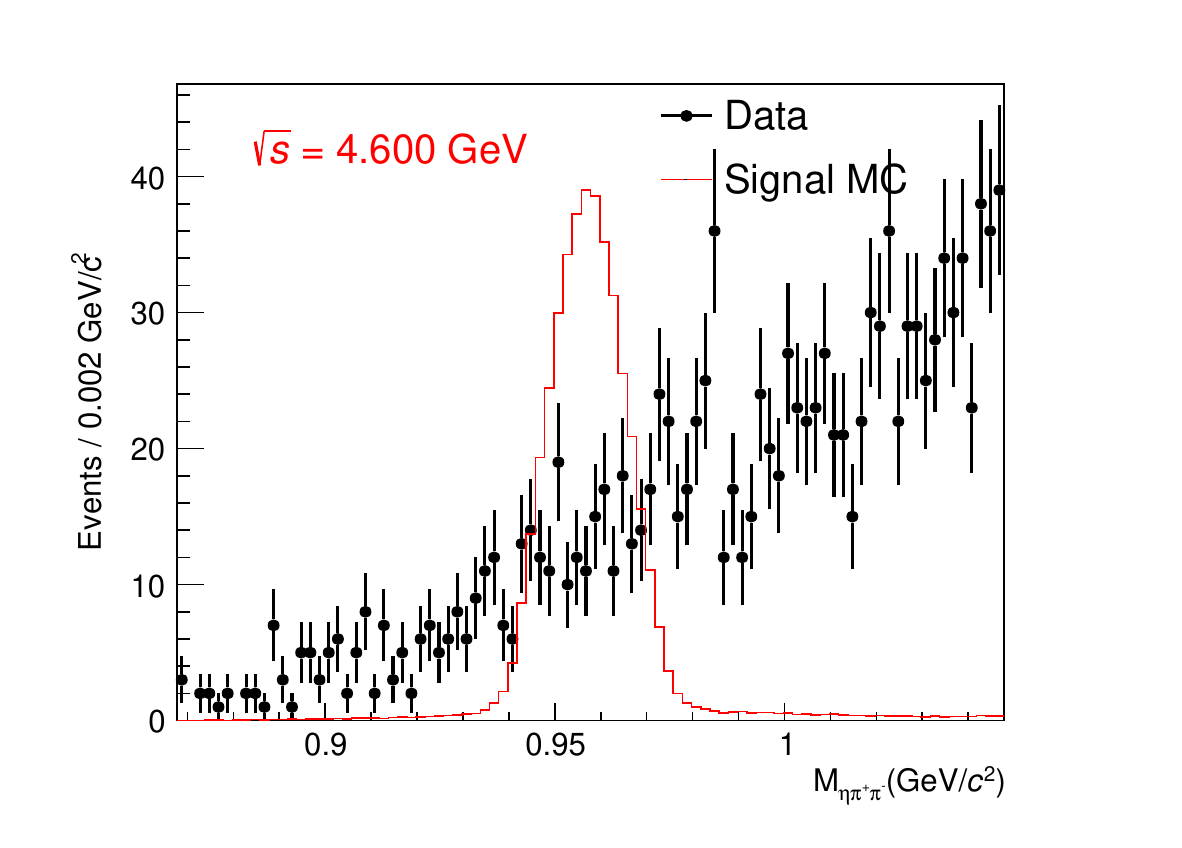}
        \captionsetup{skip=-7pt,font=normalsize}
    \end{subfigure}
    \begin{subfigure}{0.32\textwidth}
        \includegraphics[width=\linewidth]{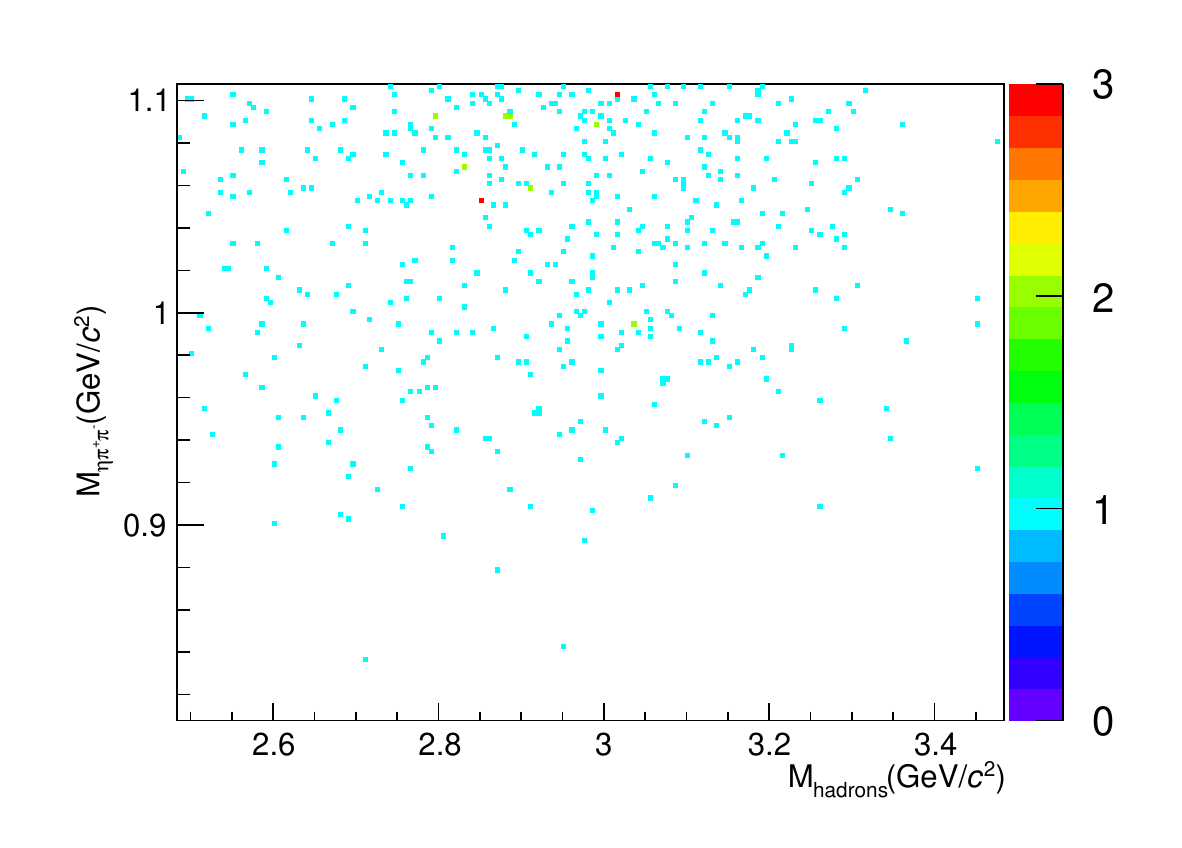}
        \captionsetup{skip=-7pt,font=normalsize}
    \end{subfigure}
    \begin{subfigure}{0.32\textwidth}
        \includegraphics[width=\linewidth]{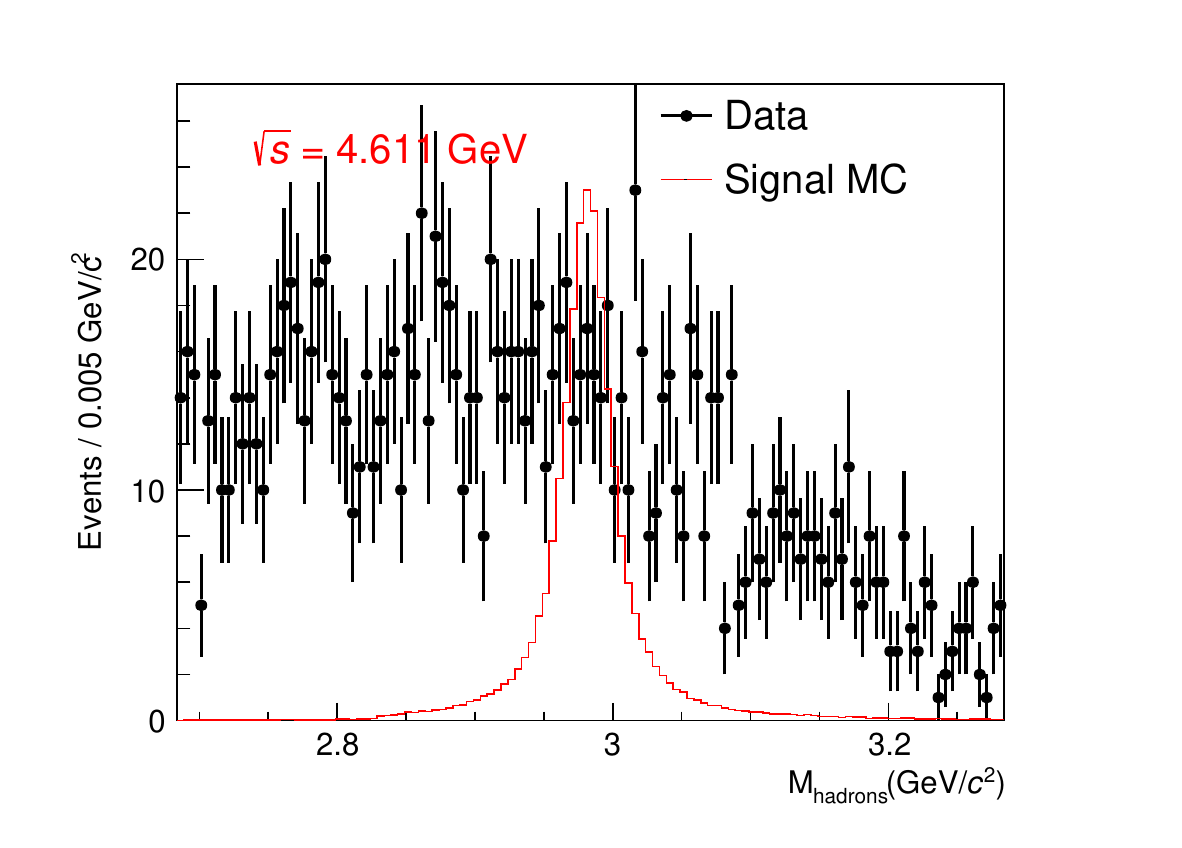}
        \captionsetup{skip=-7pt,font=normalsize}
    \end{subfigure}
    \begin{subfigure}{0.32\textwidth}
        \includegraphics[width=\linewidth]{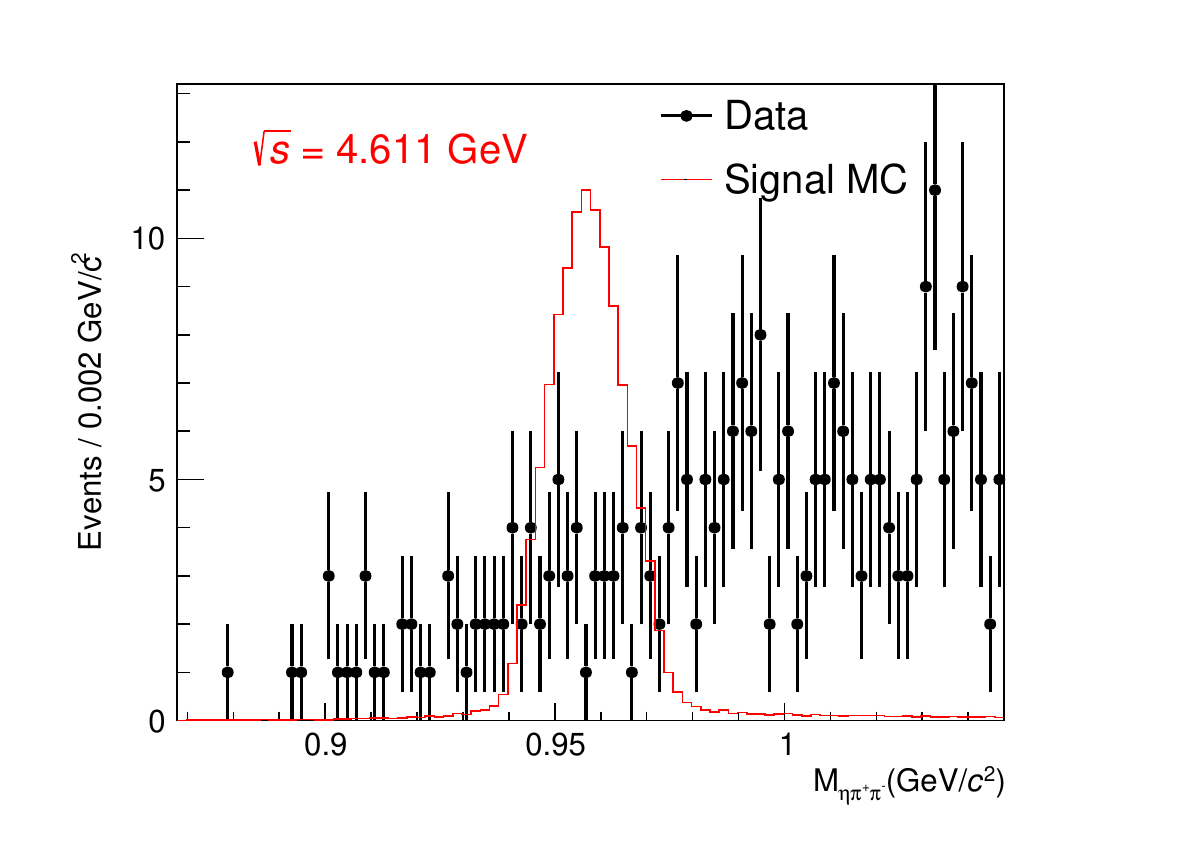}
        \captionsetup{skip=-7pt,font=normalsize}
    \end{subfigure}
    \begin{subfigure}{0.32\textwidth}
        \includegraphics[width=\linewidth]{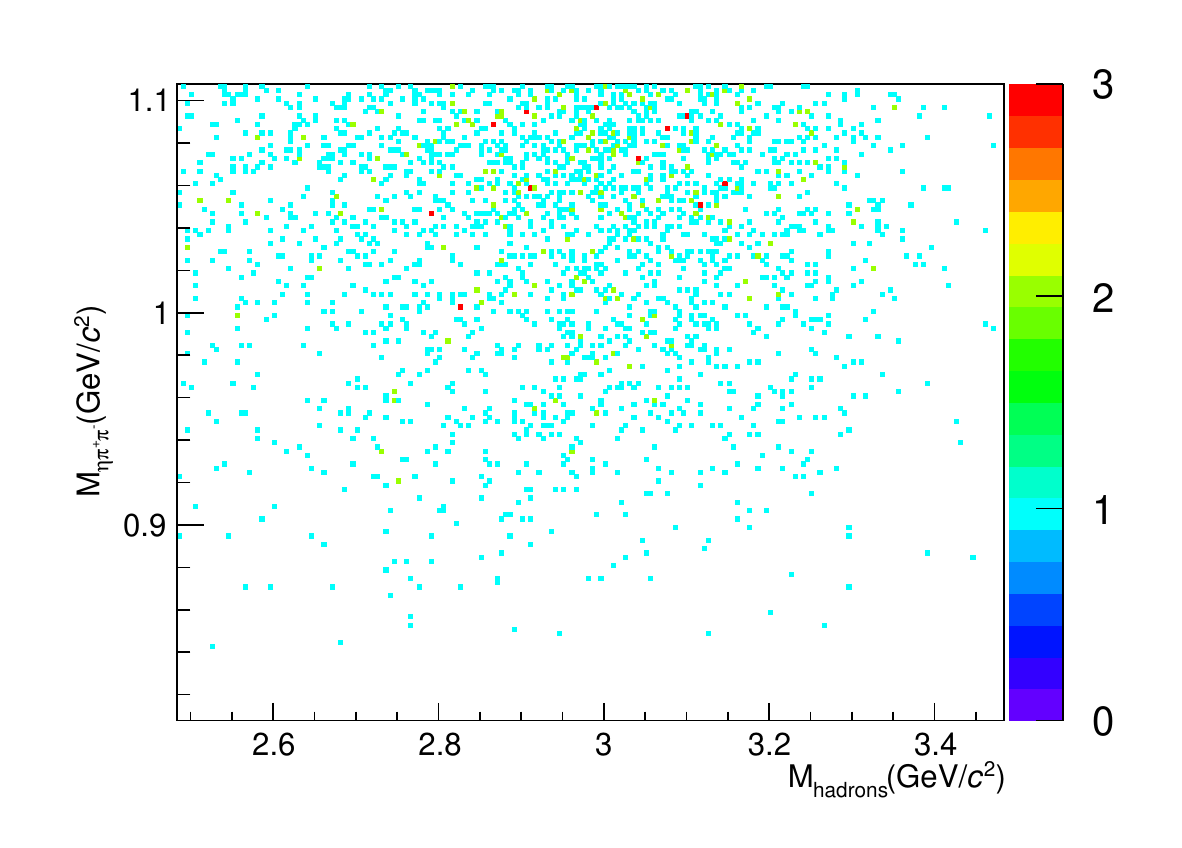}
        \captionsetup{skip=-7pt,font=normalsize}
    \end{subfigure}
    \begin{subfigure}{0.32\textwidth}
        \includegraphics[width=\linewidth]{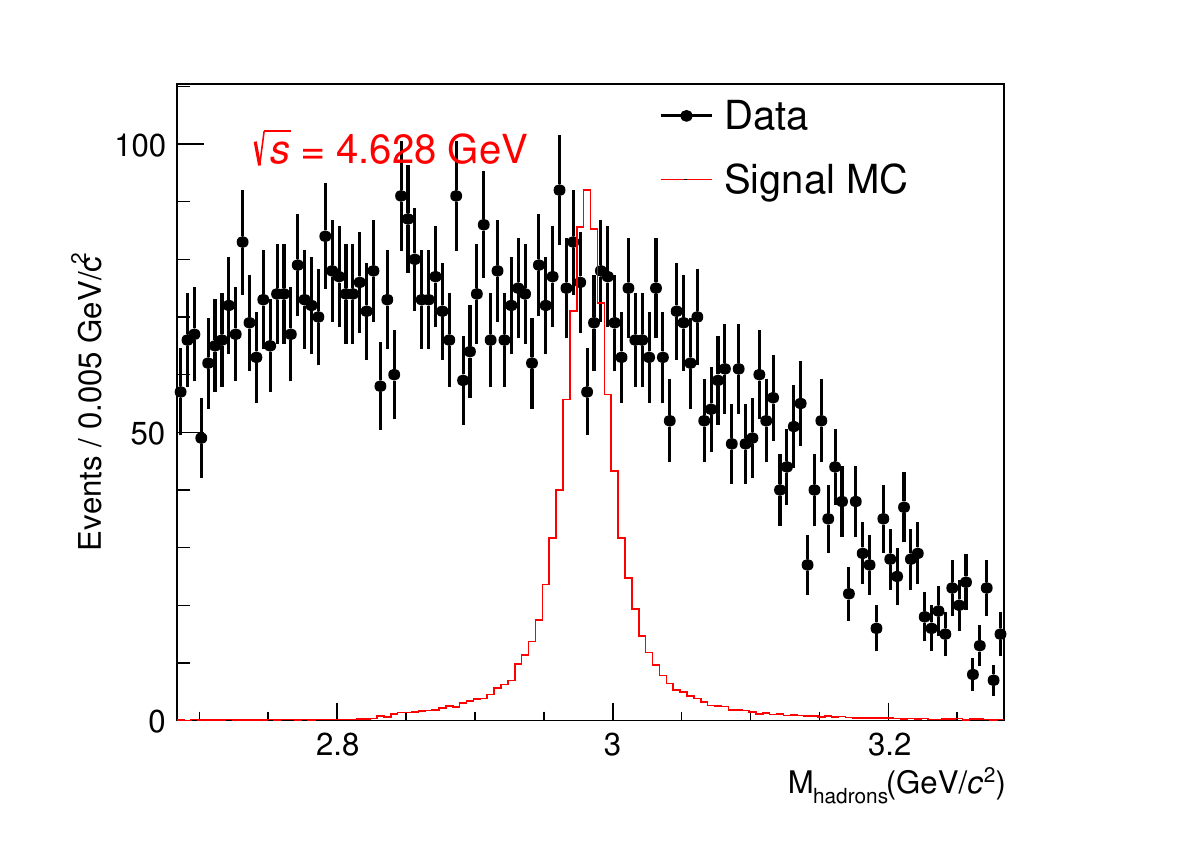}
        \captionsetup{skip=-7pt,font=normalsize}
    \end{subfigure}
    \begin{subfigure}{0.32\textwidth}
        \includegraphics[width=\linewidth]{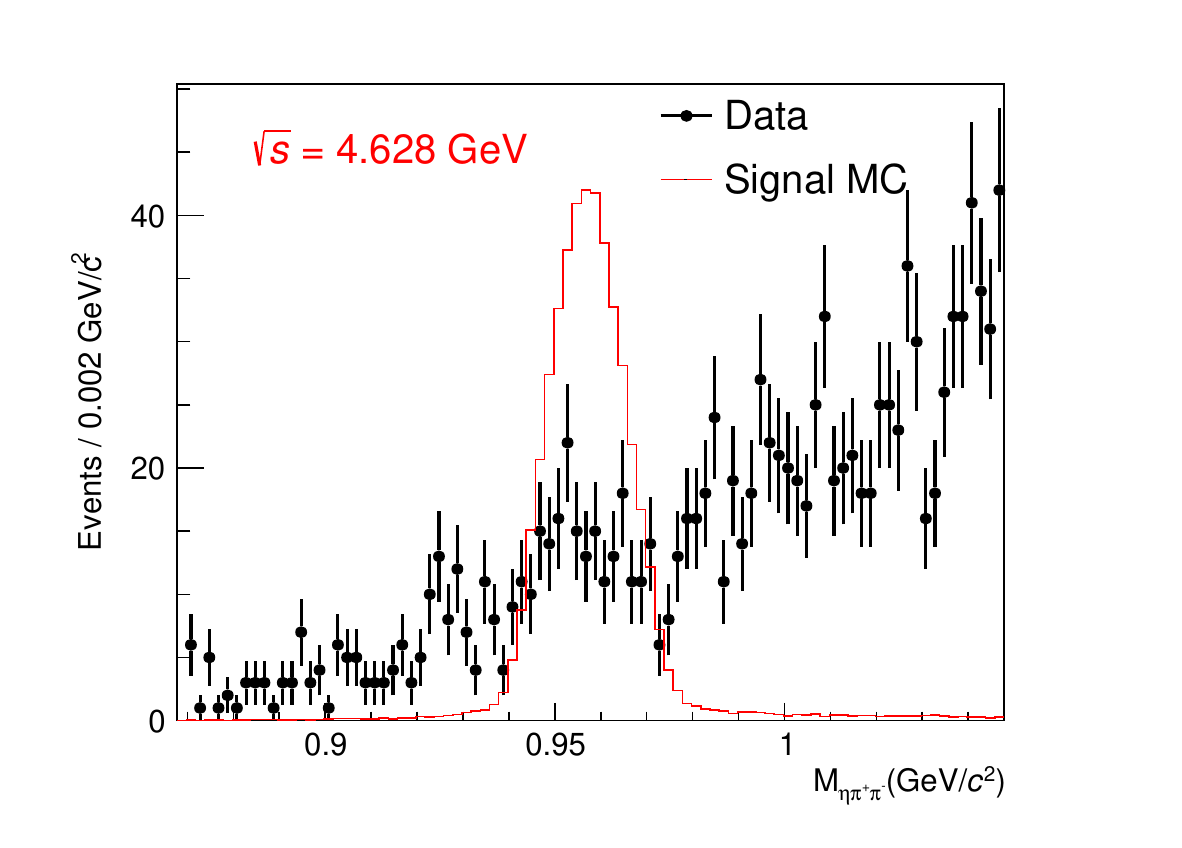}
        \captionsetup{skip=-7pt,font=normalsize}
    \end{subfigure}
    \begin{subfigure}{0.32\textwidth}
        \includegraphics[width=\linewidth]{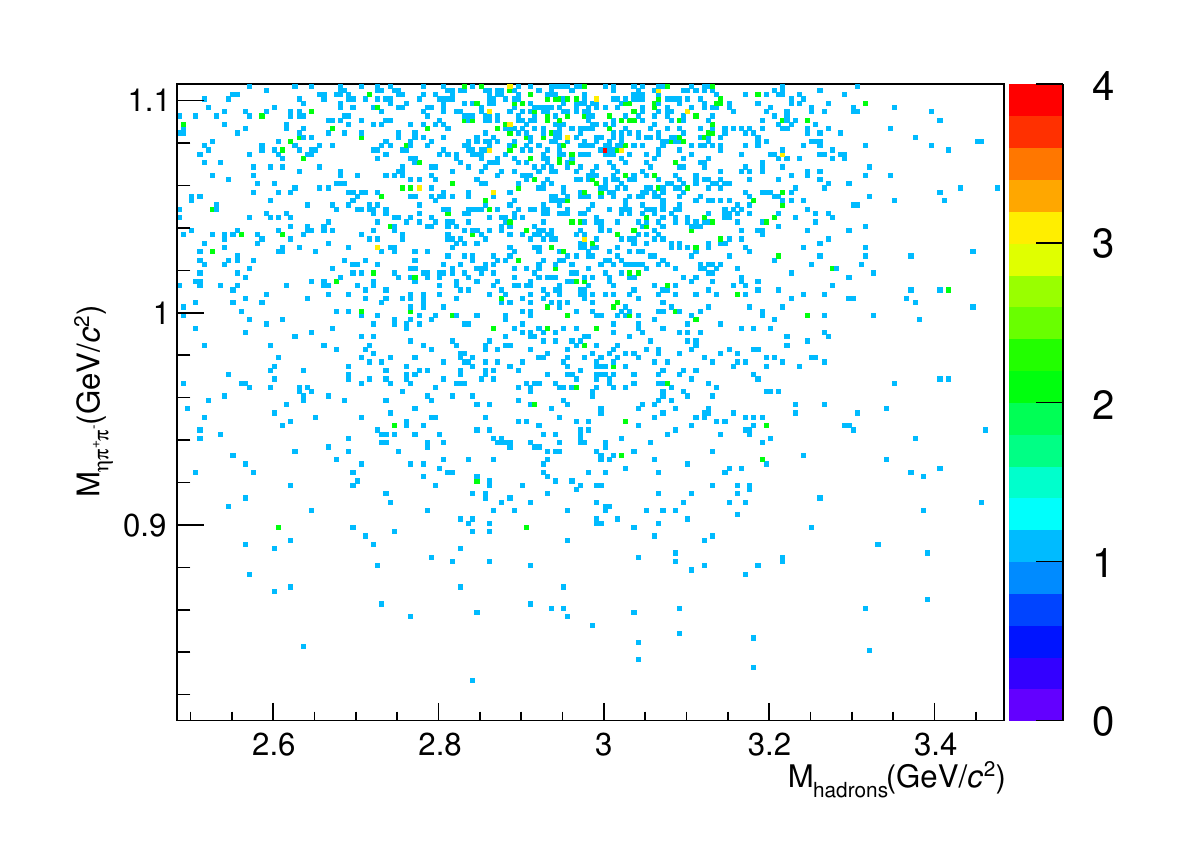}
        \captionsetup{skip=-7pt,font=normalsize}
    \end{subfigure}
    \begin{subfigure}{0.32\textwidth}
        \includegraphics[width=\linewidth]{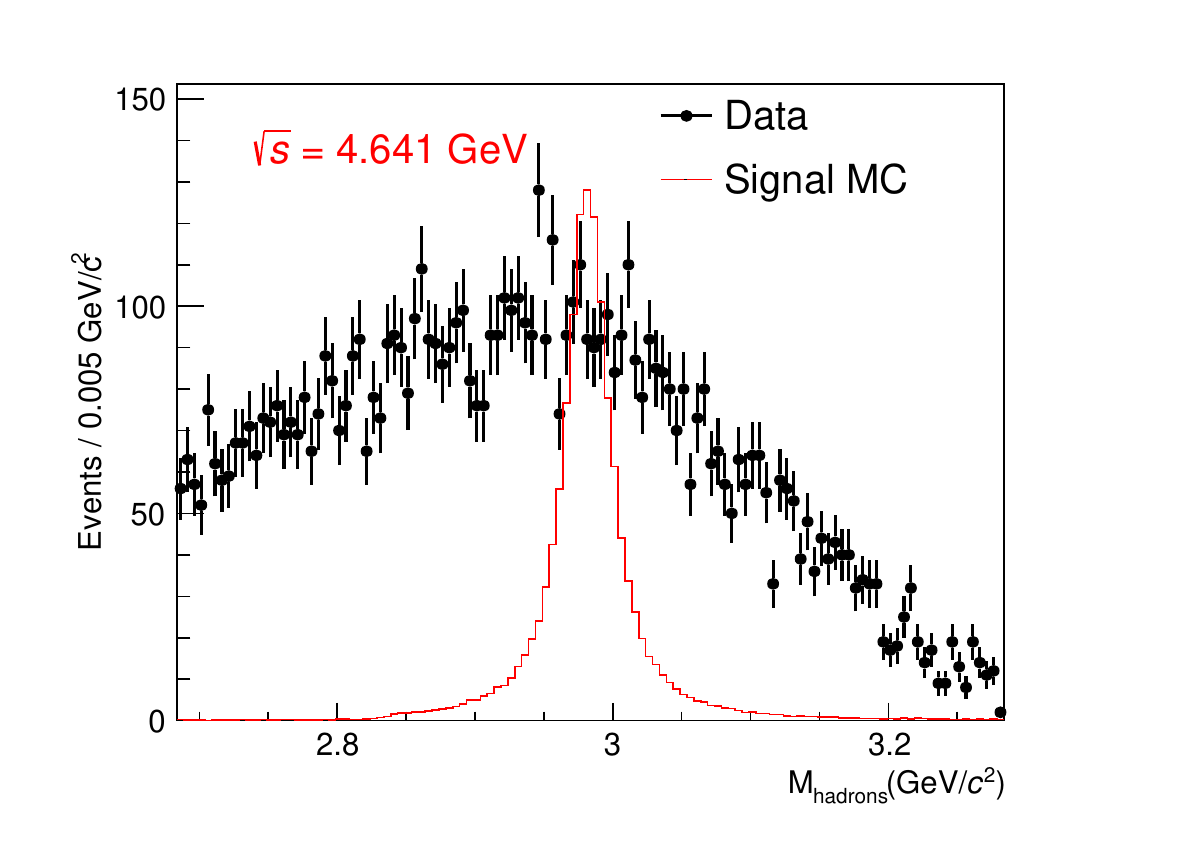}
        \captionsetup{skip=-7pt,font=normalsize}
    \end{subfigure}
    \begin{subfigure}{0.32\textwidth}
        \includegraphics[width=\linewidth]{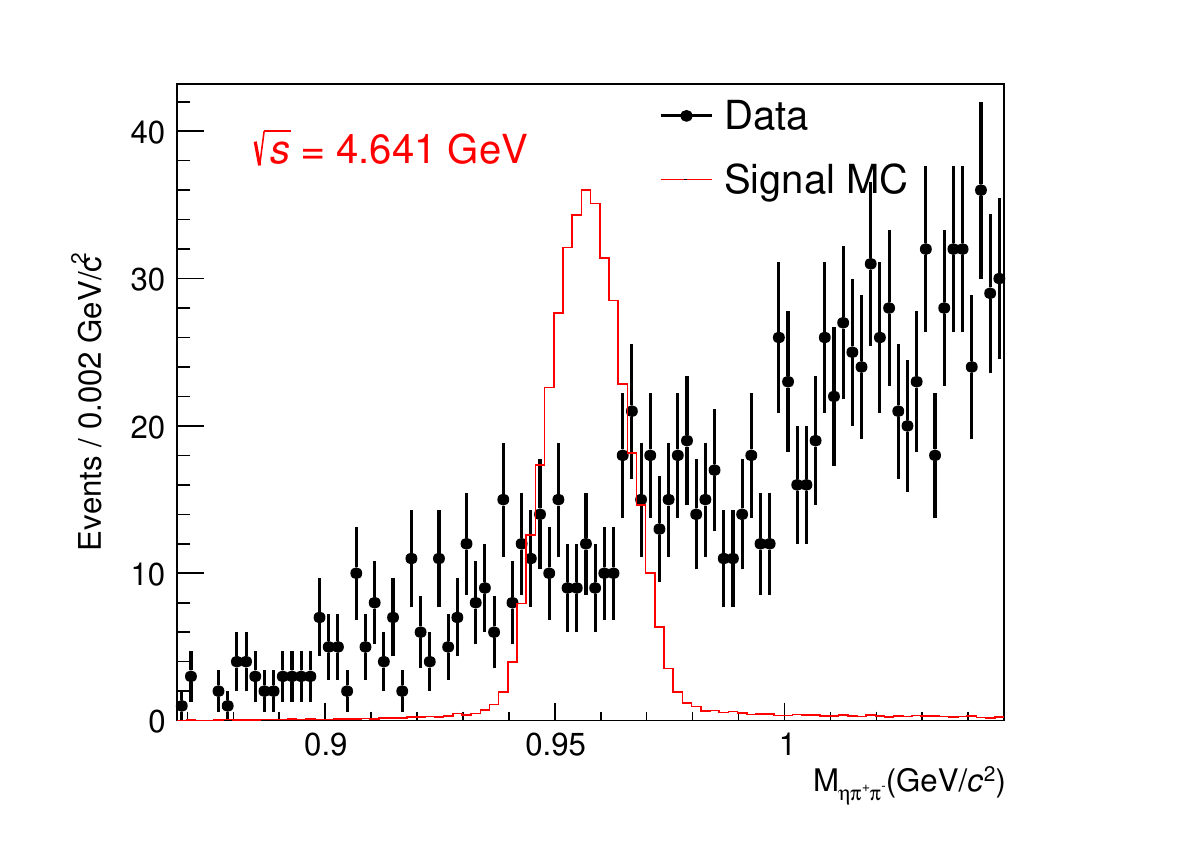}
        \captionsetup{skip=-7pt,font=normalsize}
    \end{subfigure}
    \begin{subfigure}{0.32\textwidth}
        \includegraphics[width=\linewidth]{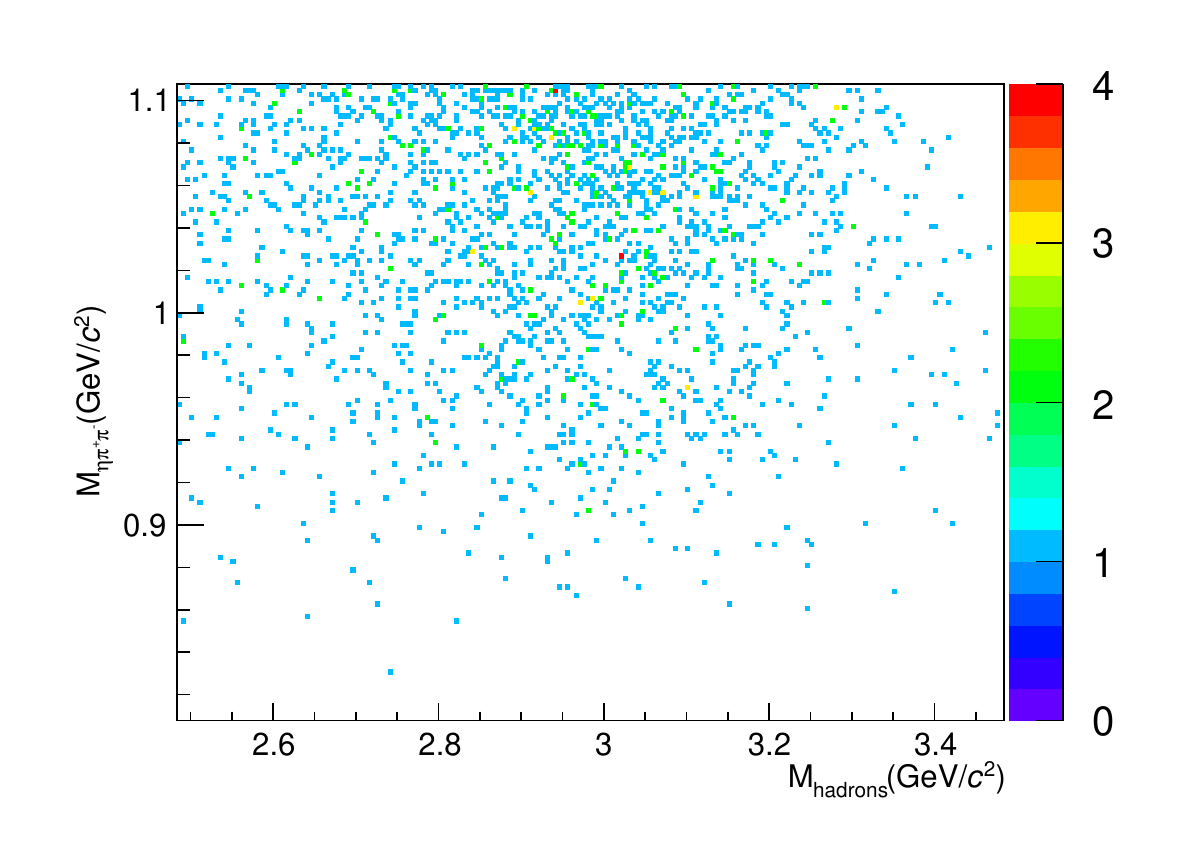}
        \captionsetup{skip=-7pt,font=normalsize}
    \end{subfigure}
    \begin{subfigure}{0.32\textwidth}
        \includegraphics[width=\linewidth]{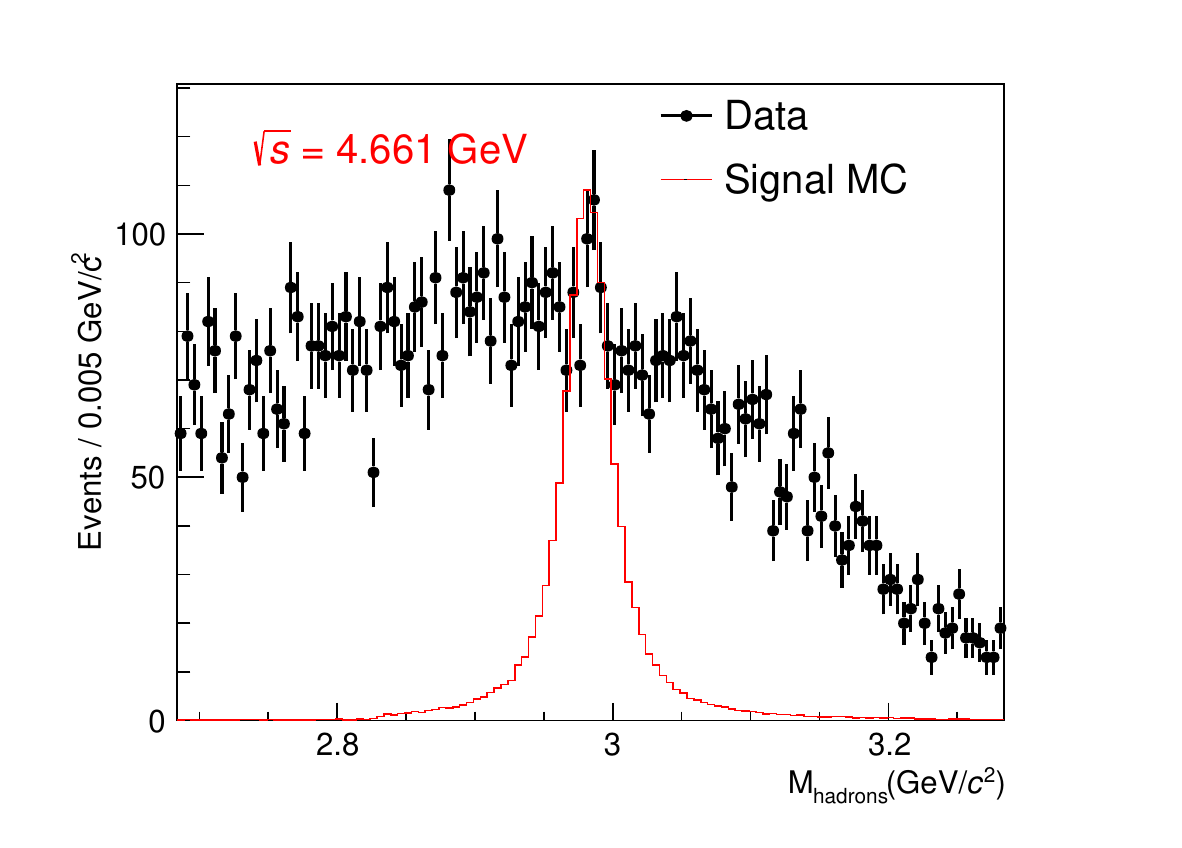}
        \captionsetup{skip=-7pt,font=normalsize}
    \end{subfigure}
    \begin{subfigure}{0.32\textwidth}
        \includegraphics[width=\linewidth]{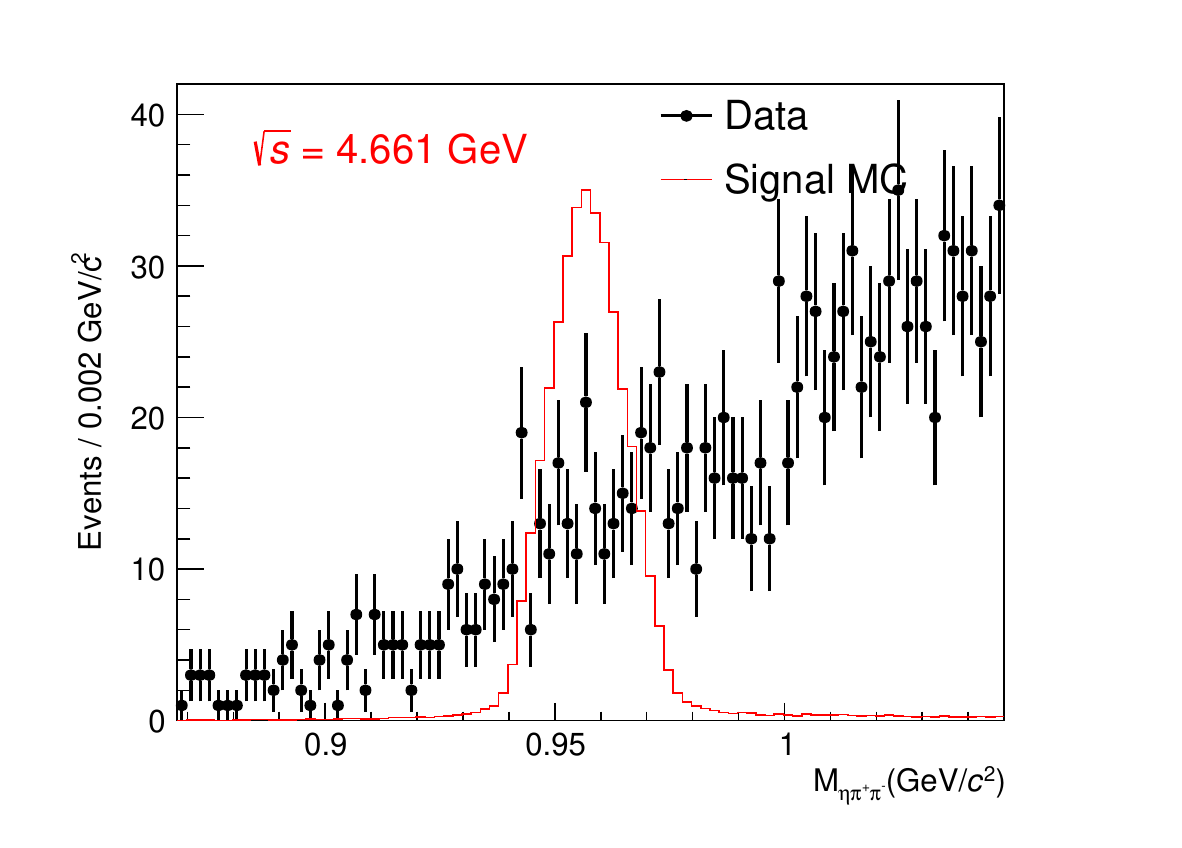}
        \captionsetup{skip=-7pt,font=normalsize}
    \end{subfigure}
\captionsetup{justification=raggedright}
\caption{The distributions of (Left) $M_{hadrons}$ versus $M_{\gamma\gamma}$, (Middle) $M_{hadrons}$, and (Right) $M_{\gamma\gamma}$ at $\sqrt s=4.600-4.661$~GeV.}
\label{fig:normal12}
\end{figure*}

\begin{figure*}[htbp]
\begin{subfigure}{0.24\textwidth}
  \centering
  \includegraphics[width=\textwidth]{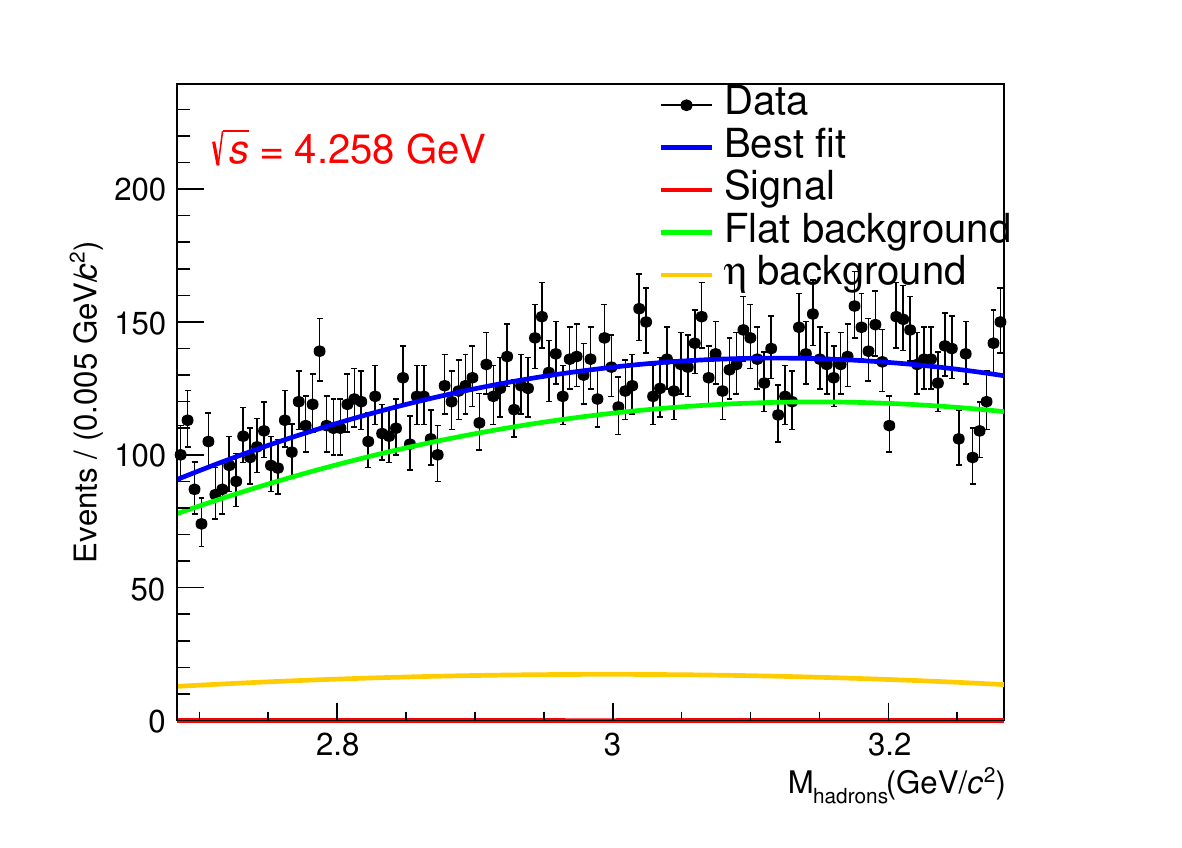}
  \captionsetup{skip=-10pt,font=large}
\end{subfigure}
\begin{subfigure}{0.24\textwidth}
  \centering
  \includegraphics[width=\textwidth]{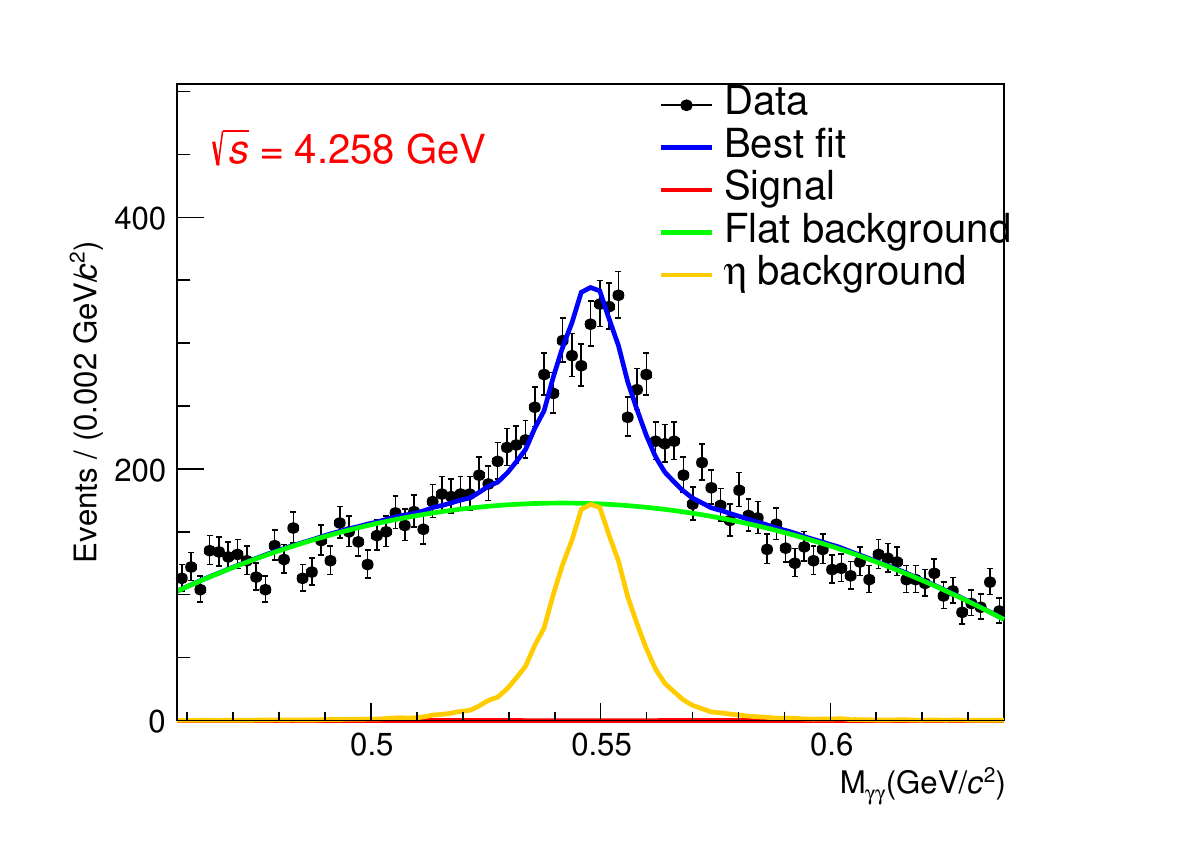}
  \captionsetup{skip=-10pt,font=large}
\end{subfigure}
 \begin{subfigure}{0.24\textwidth}
  \centering
  \includegraphics[width=\textwidth]{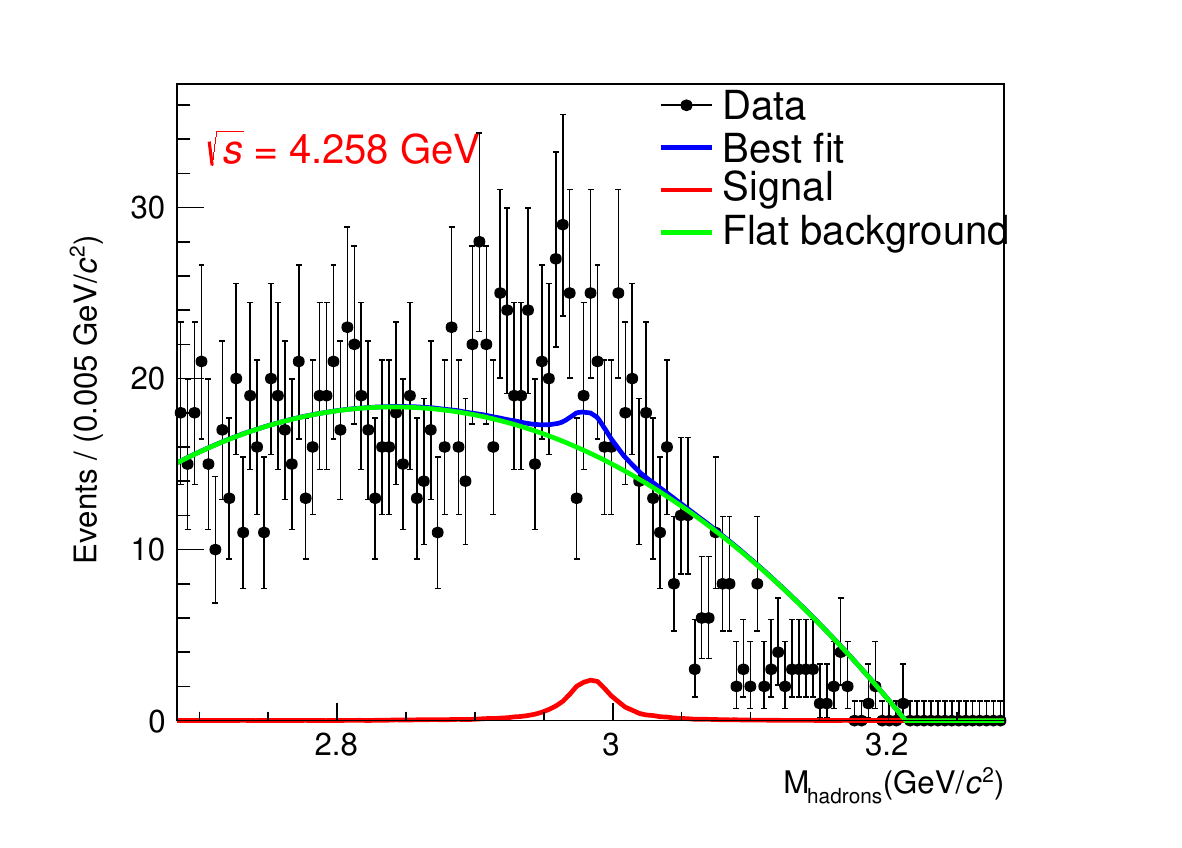}
  \captionsetup{skip=-10pt,font=large}
\end{subfigure}
  \begin{subfigure}{0.24\textwidth}
  \centering
  \includegraphics[width=\textwidth]{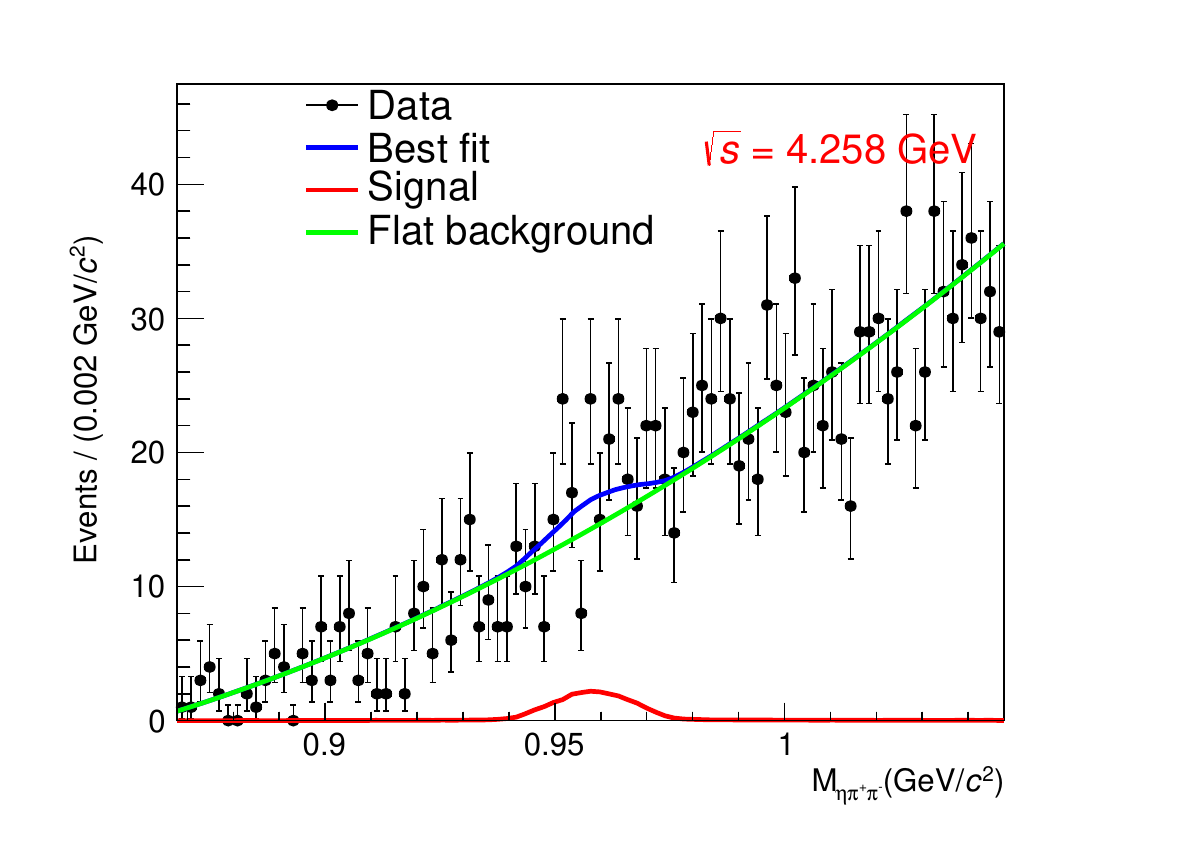}
  \captionsetup{skip=-10pt,font=large}
\end{subfigure}
\begin{subfigure}{0.24\textwidth}
  \centering
  \includegraphics[width=\textwidth]{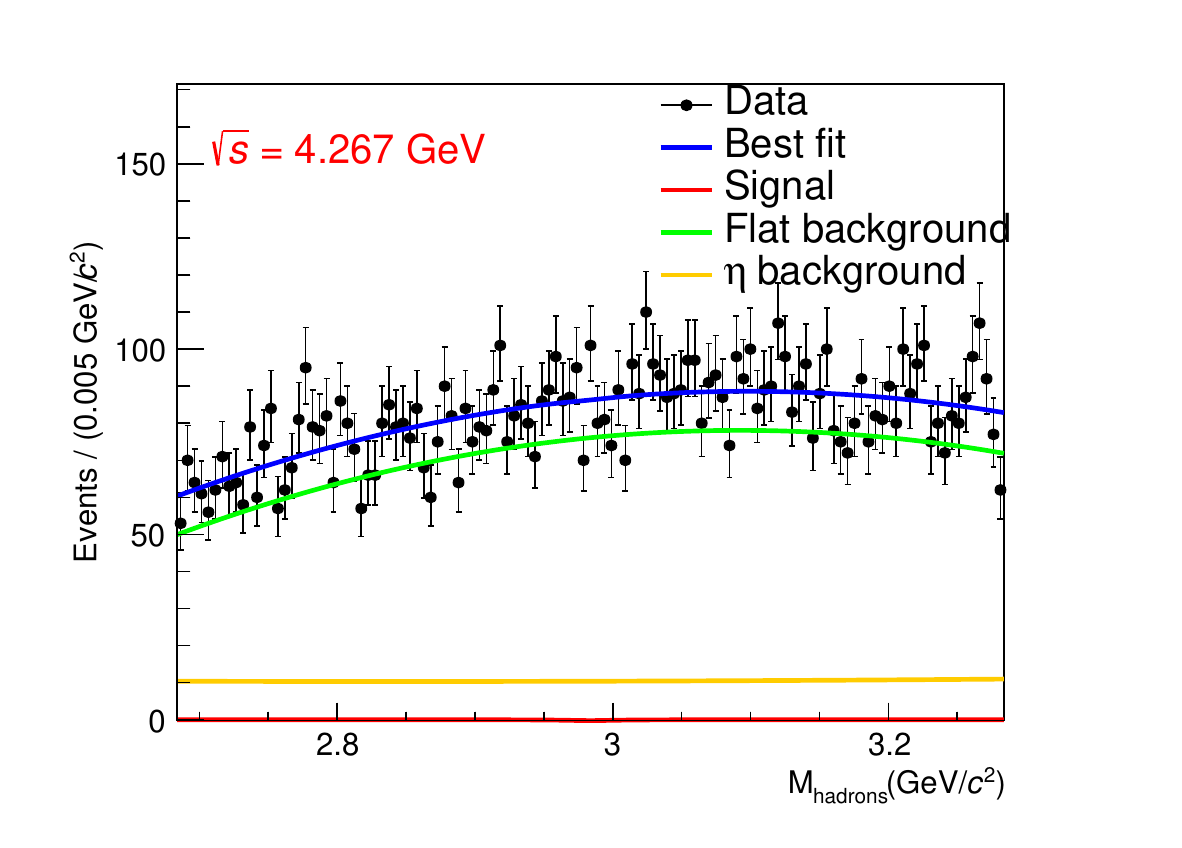}
  \captionsetup{skip=-10pt,font=large}
\end{subfigure}
\begin{subfigure}{0.24\textwidth}
  \centering
  \includegraphics[width=\textwidth]{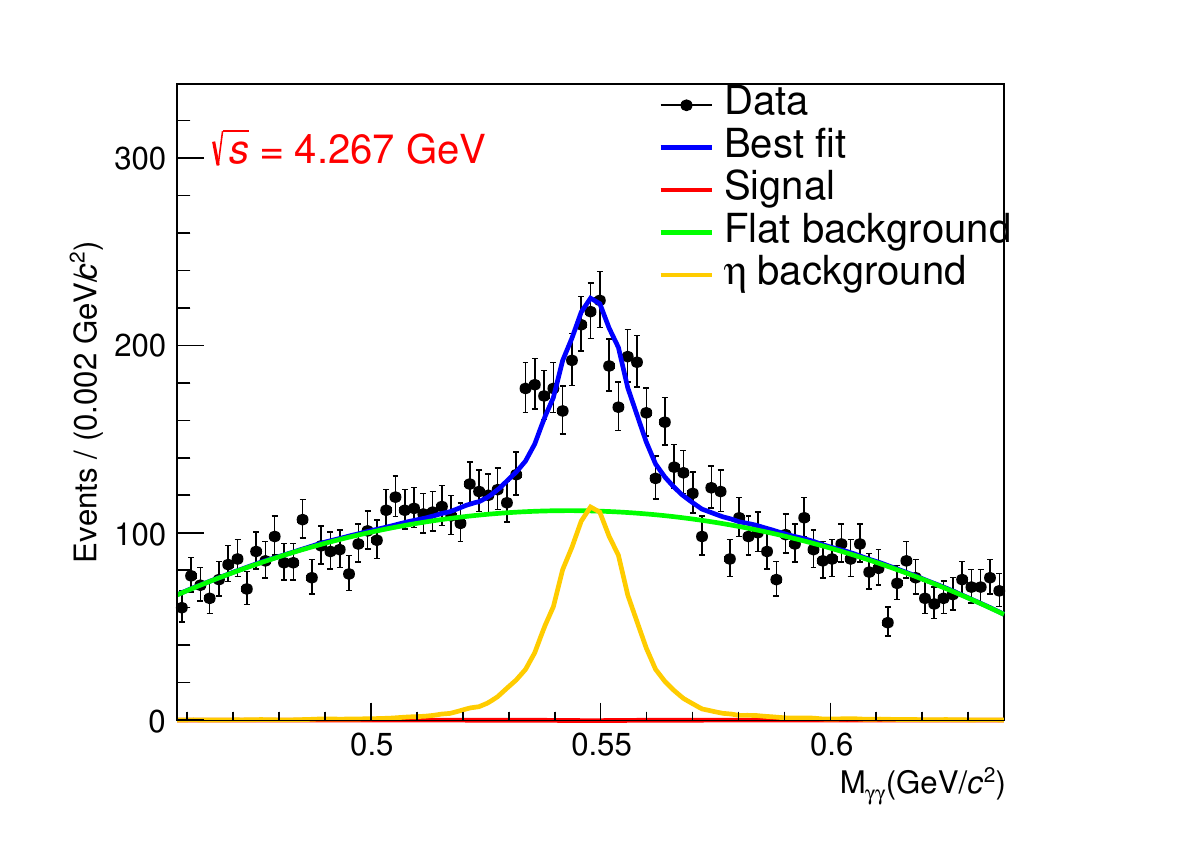}
  \captionsetup{skip=-10pt,font=large}
\end{subfigure}
 \begin{subfigure}{0.24\textwidth}
  \centering
  \includegraphics[width=\textwidth]{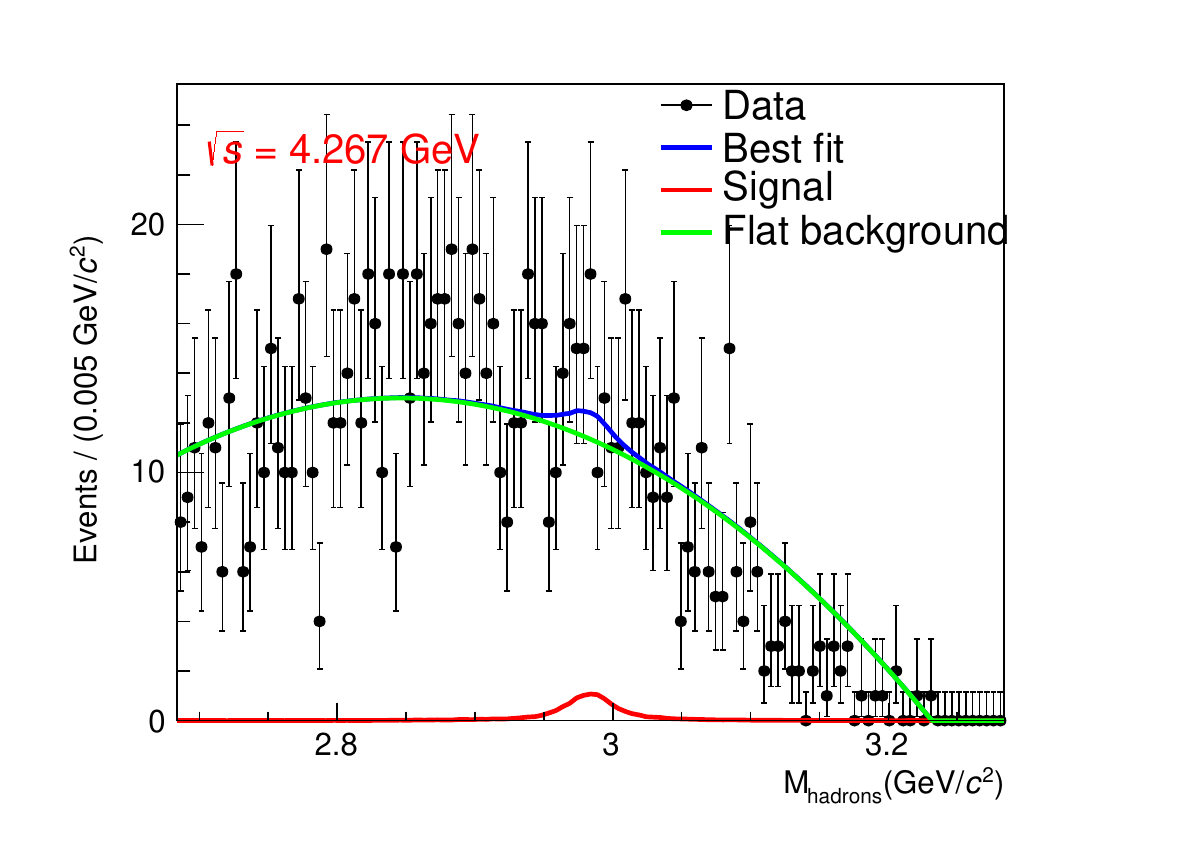}
  \captionsetup{skip=-10pt,font=large}
\end{subfigure}
  \begin{subfigure}{0.24\textwidth}
  \centering
  \includegraphics[width=\textwidth]{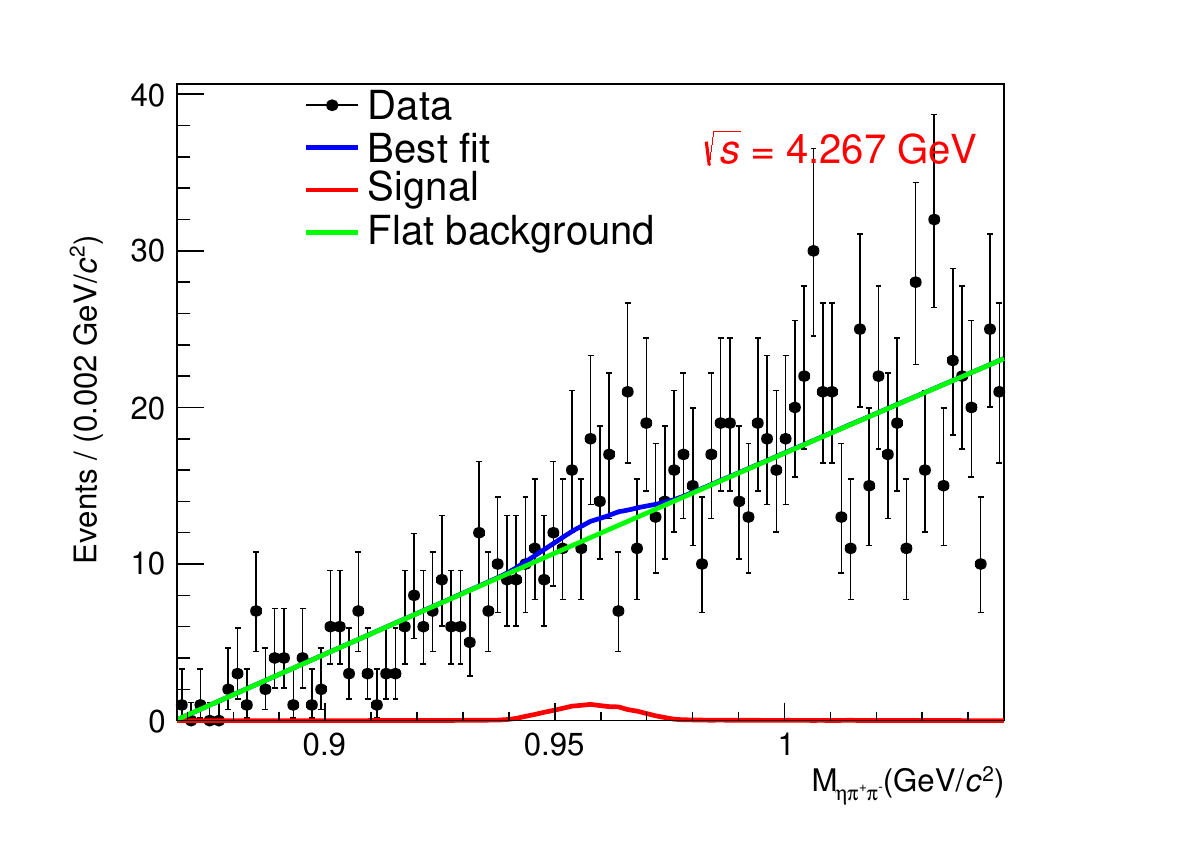}
  \captionsetup{skip=-10pt,font=large}
\end{subfigure}
\begin{subfigure}{0.24\textwidth}
  \centering
  \includegraphics[width=\textwidth]{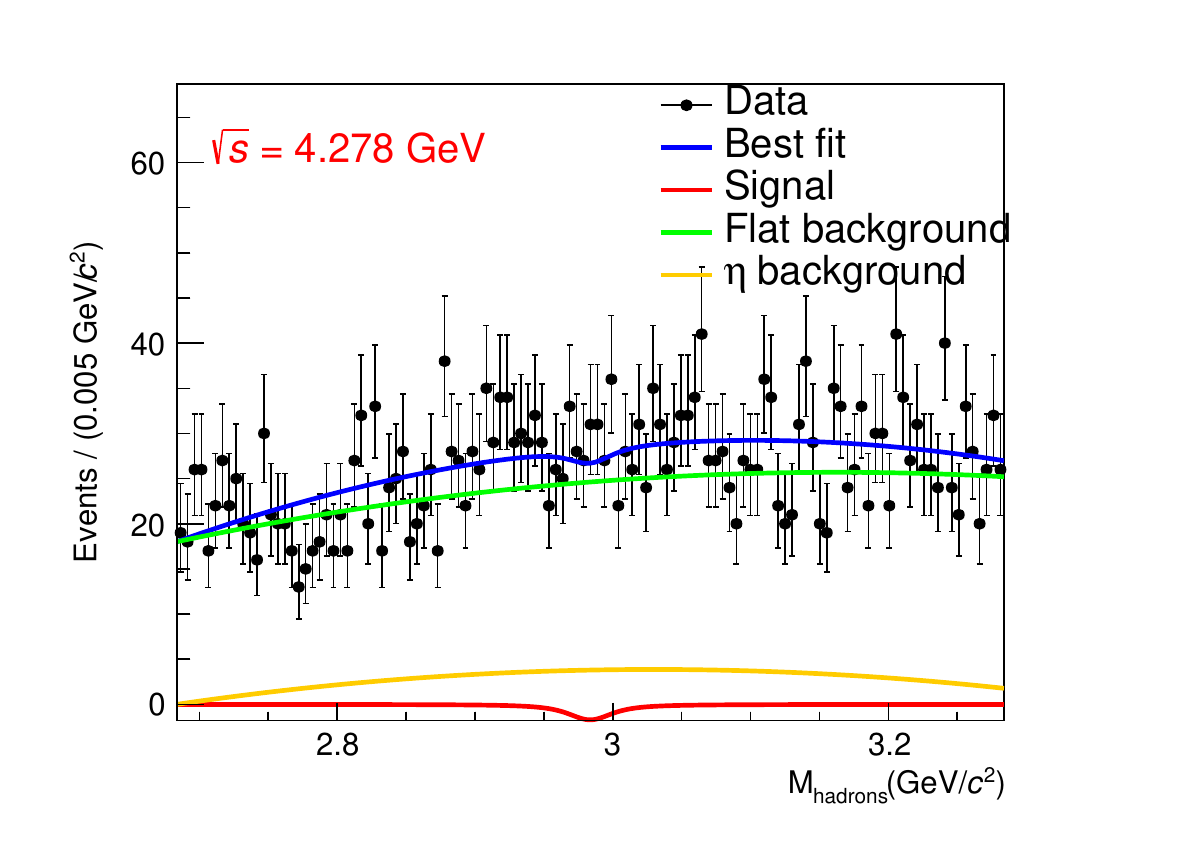}
  \captionsetup{skip=-10pt,font=large}
\end{subfigure}
\begin{subfigure}{0.24\textwidth}
  \centering
  \includegraphics[width=\textwidth]{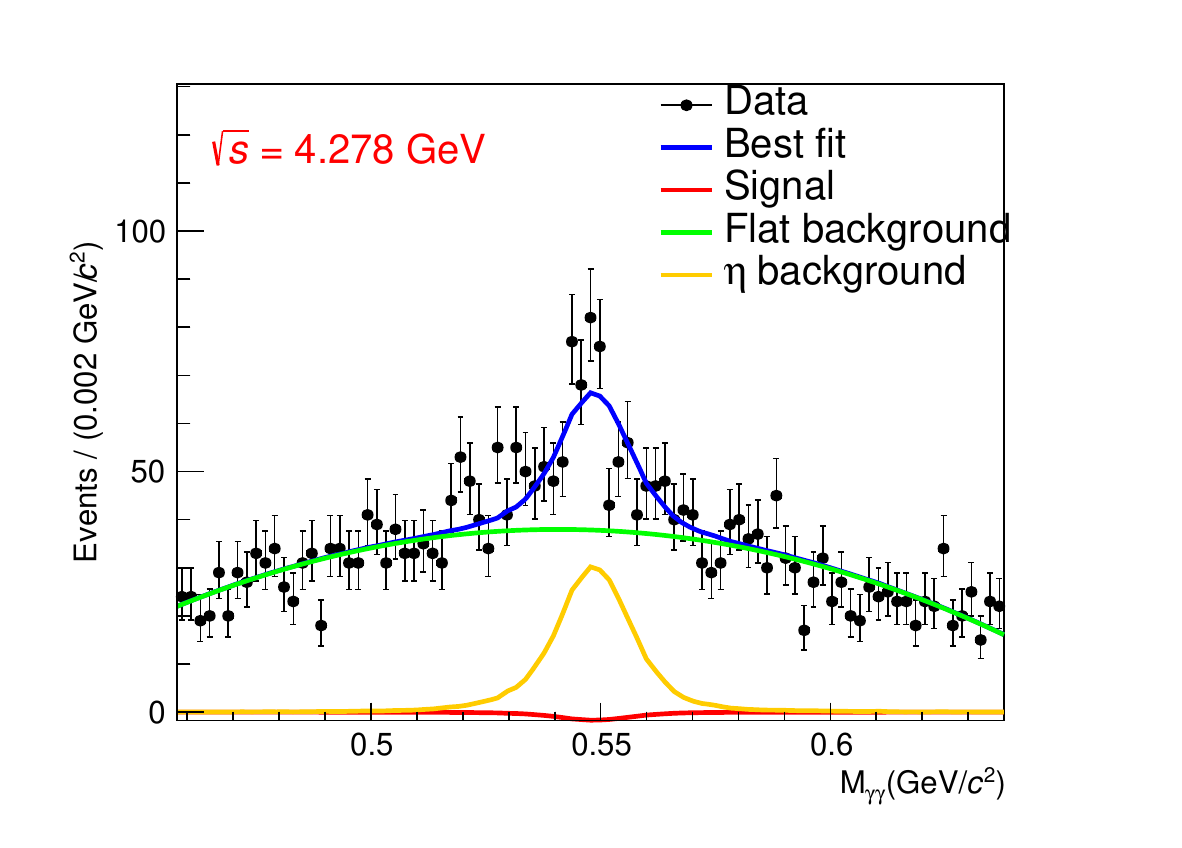}
  \captionsetup{skip=-10pt,font=large}
\end{subfigure}
 \begin{subfigure}{0.24\textwidth}
  \centering
  \includegraphics[width=\textwidth]{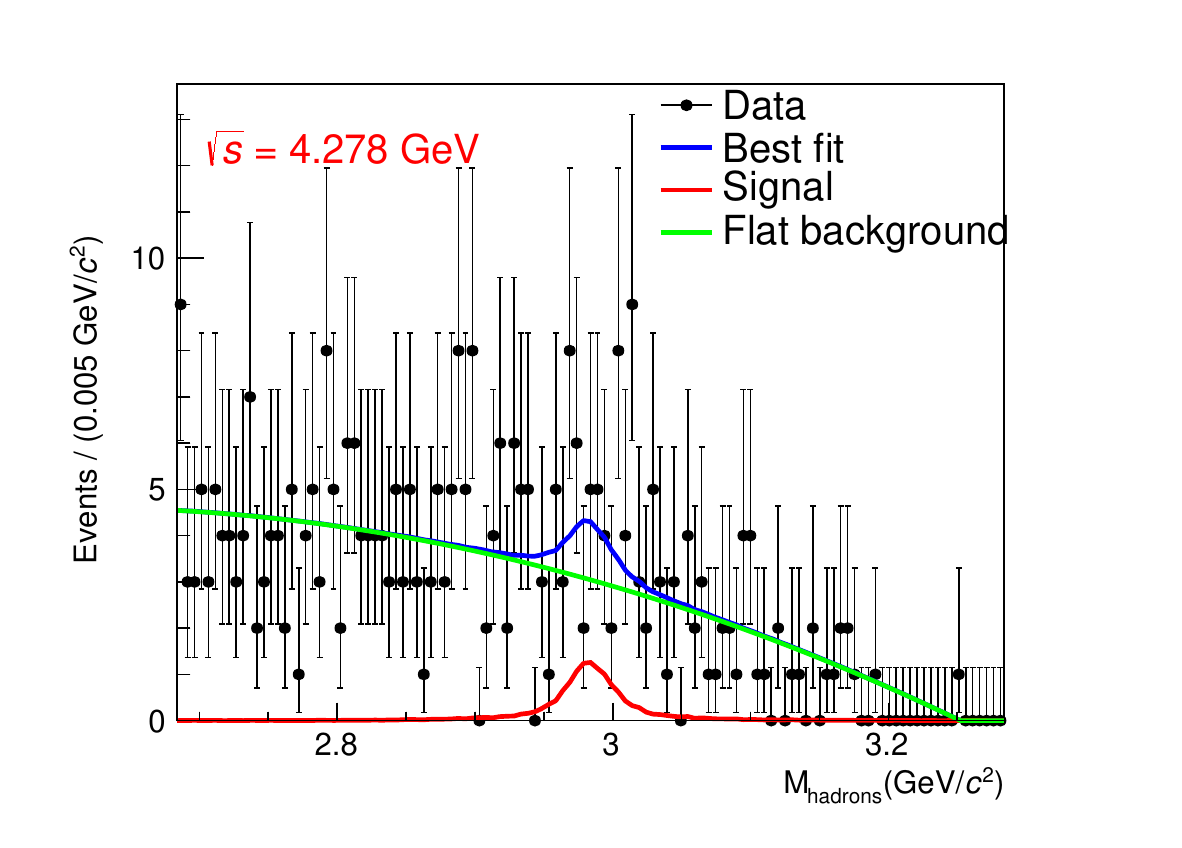}
  \captionsetup{skip=-10pt,font=large}
\end{subfigure}
  \begin{subfigure}{0.24\textwidth}
  \centering
  \includegraphics[width=\textwidth]{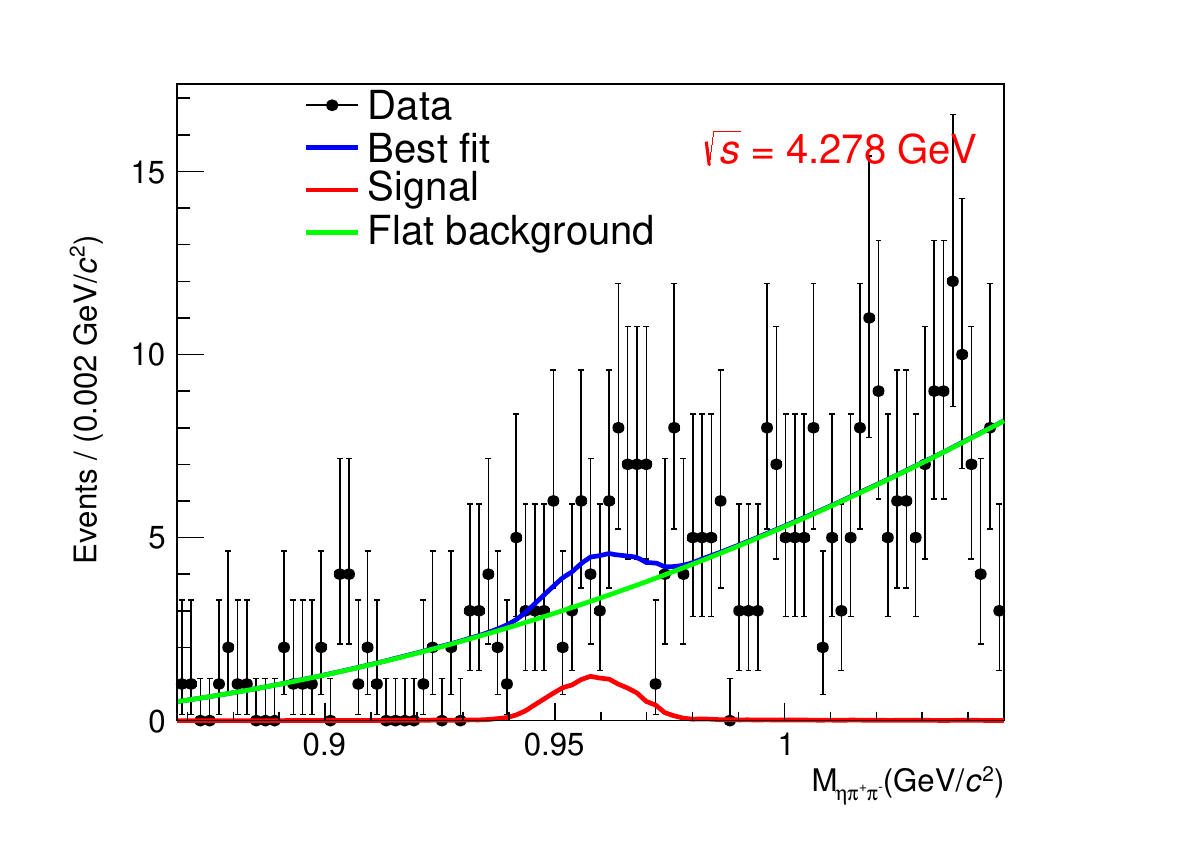}
  \captionsetup{skip=-10pt,font=large}
\end{subfigure}
\begin{subfigure}{0.24\textwidth}
  \centering
  \includegraphics[width=\textwidth]{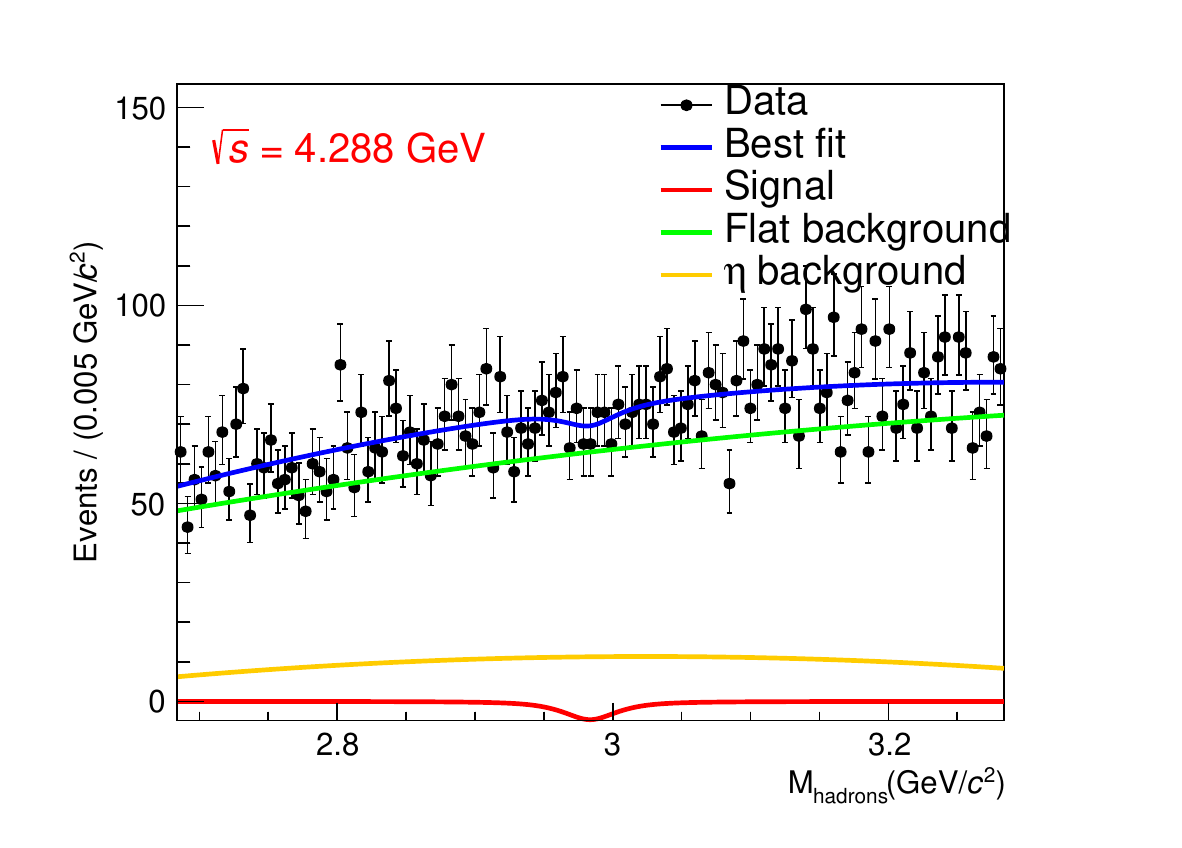}
  \captionsetup{skip=-10pt,font=large}
\end{subfigure}
\begin{subfigure}{0.24\textwidth}
  \centering
  \includegraphics[width=\textwidth]{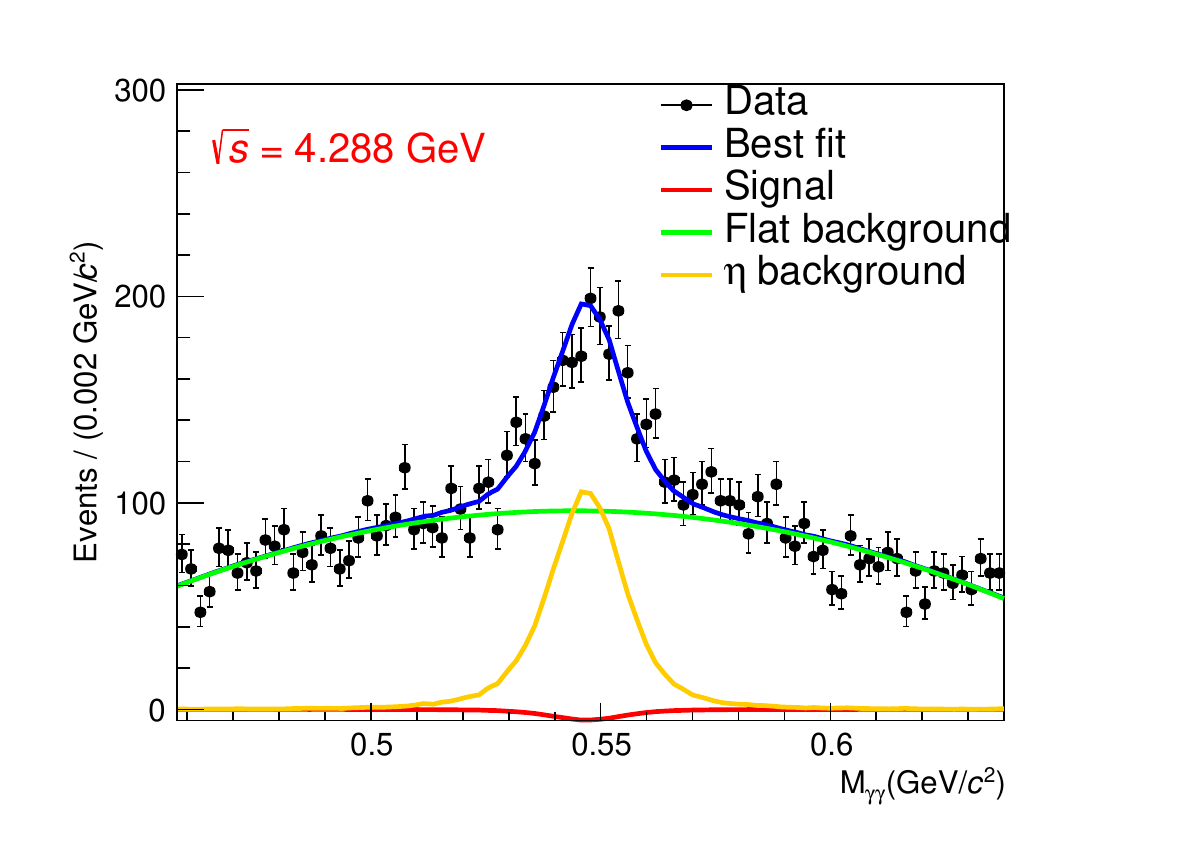}
  \captionsetup{skip=-10pt,font=large}
\end{subfigure}
 \begin{subfigure}{0.24\textwidth}
  \centering
  \includegraphics[width=\textwidth]{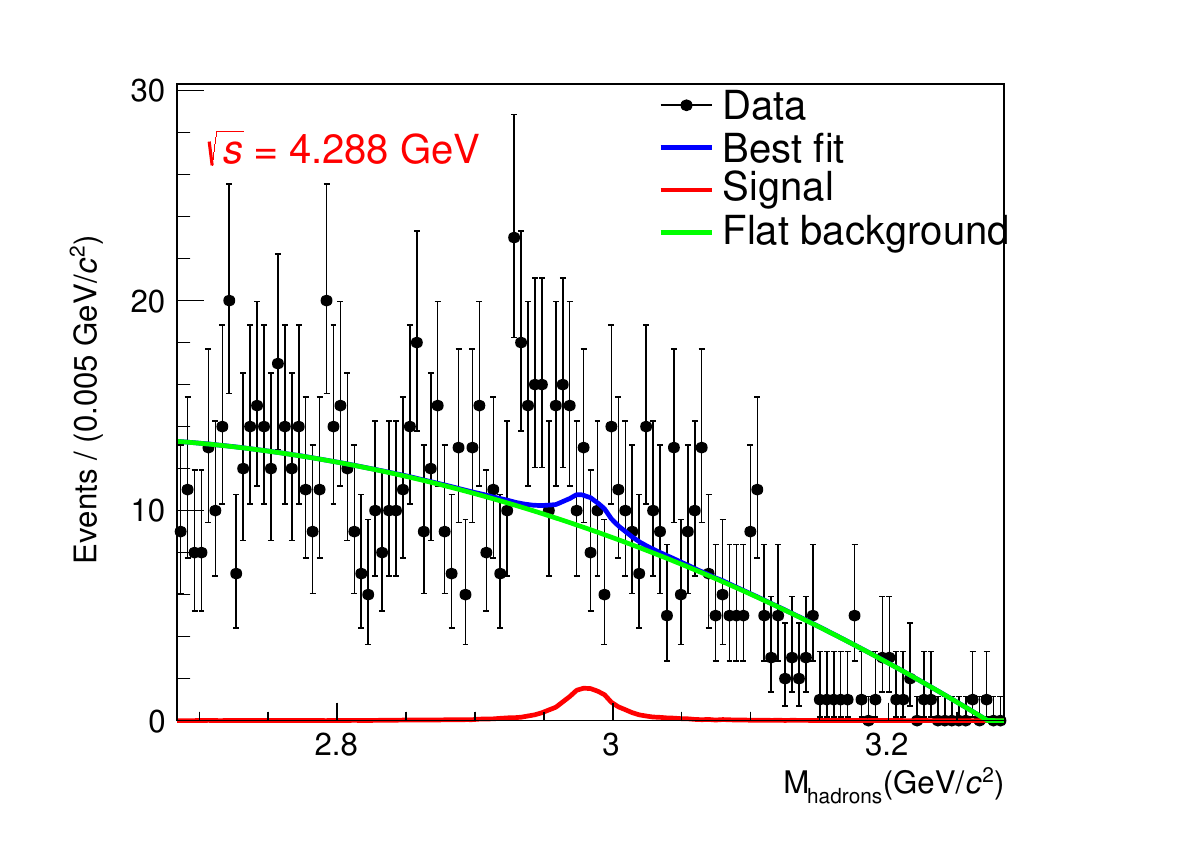}
  \captionsetup{skip=-10pt,font=large}
\end{subfigure}
  \begin{subfigure}{0.24\textwidth}
  \centering
  \includegraphics[width=\textwidth]{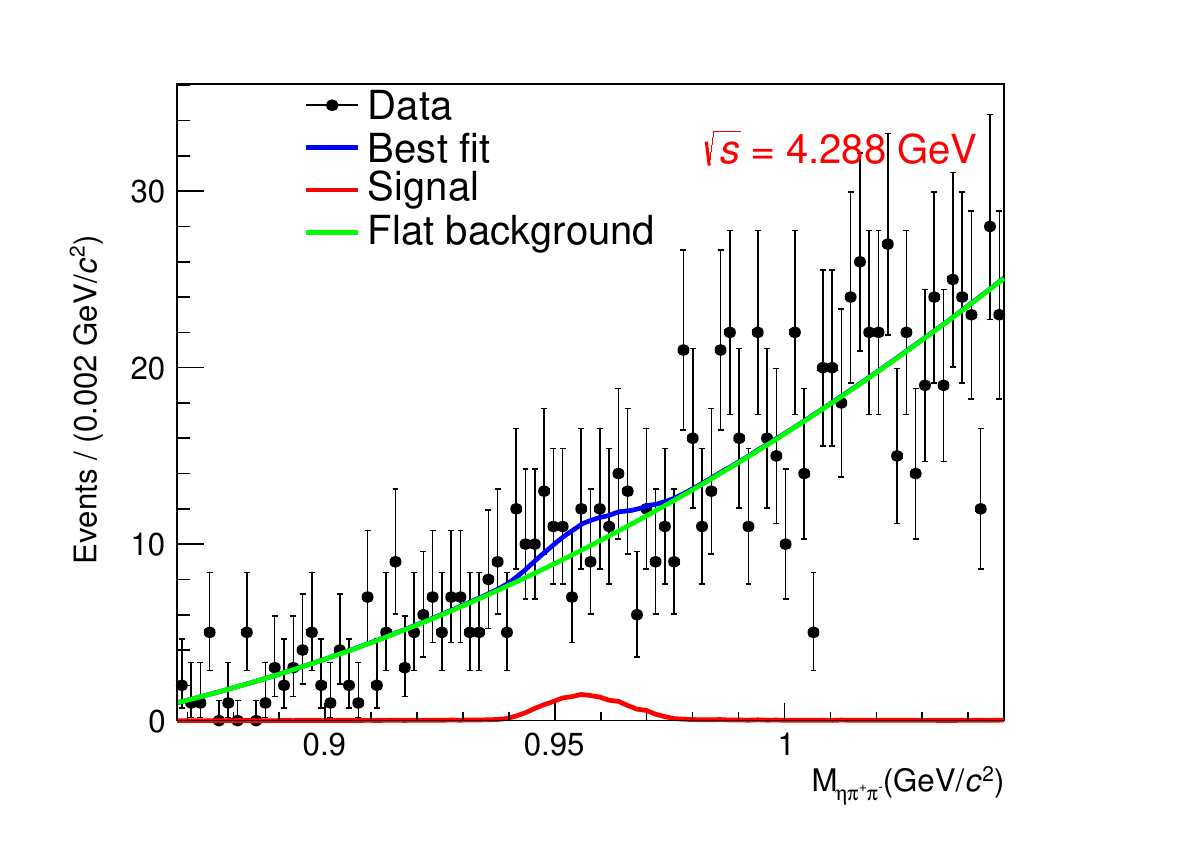}
  \captionsetup{skip=-10pt,font=large}
\end{subfigure}
\begin{subfigure}{0.24\textwidth}
  \centering
  \includegraphics[width=\textwidth]{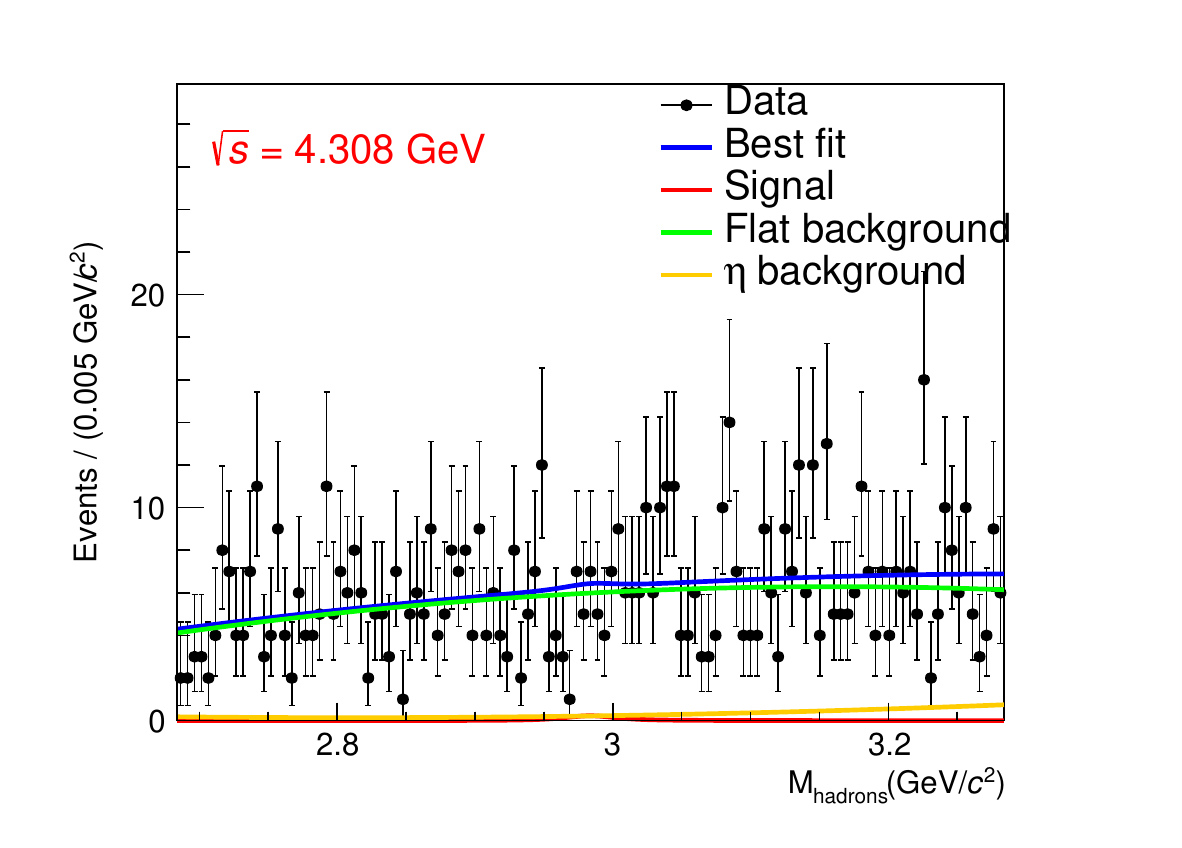}
  \captionsetup{skip=-10pt,font=large}
\end{subfigure}
\begin{subfigure}{0.24\textwidth}
  \centering
  \includegraphics[width=\textwidth]{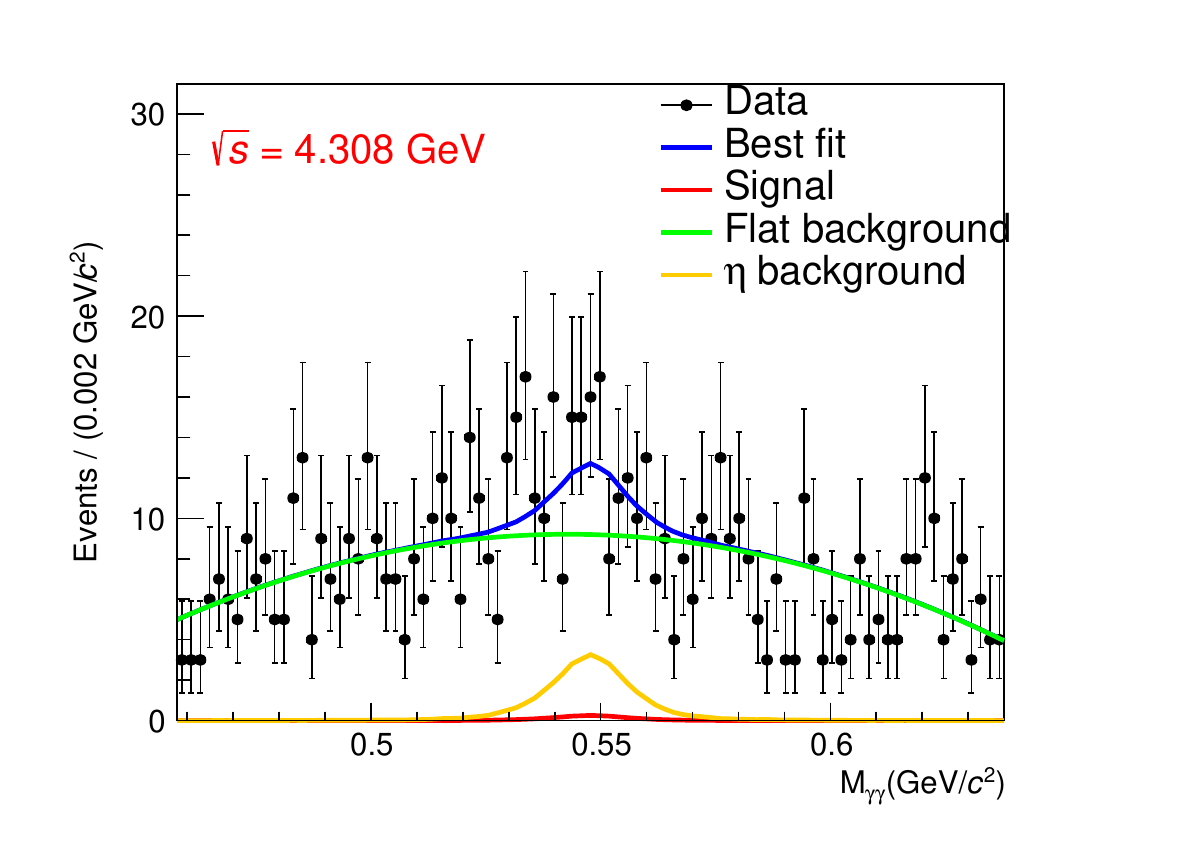}
  \captionsetup{skip=-10pt,font=large}
\end{subfigure}
 \begin{subfigure}{0.24\textwidth}
  \centering
  \includegraphics[width=\textwidth]{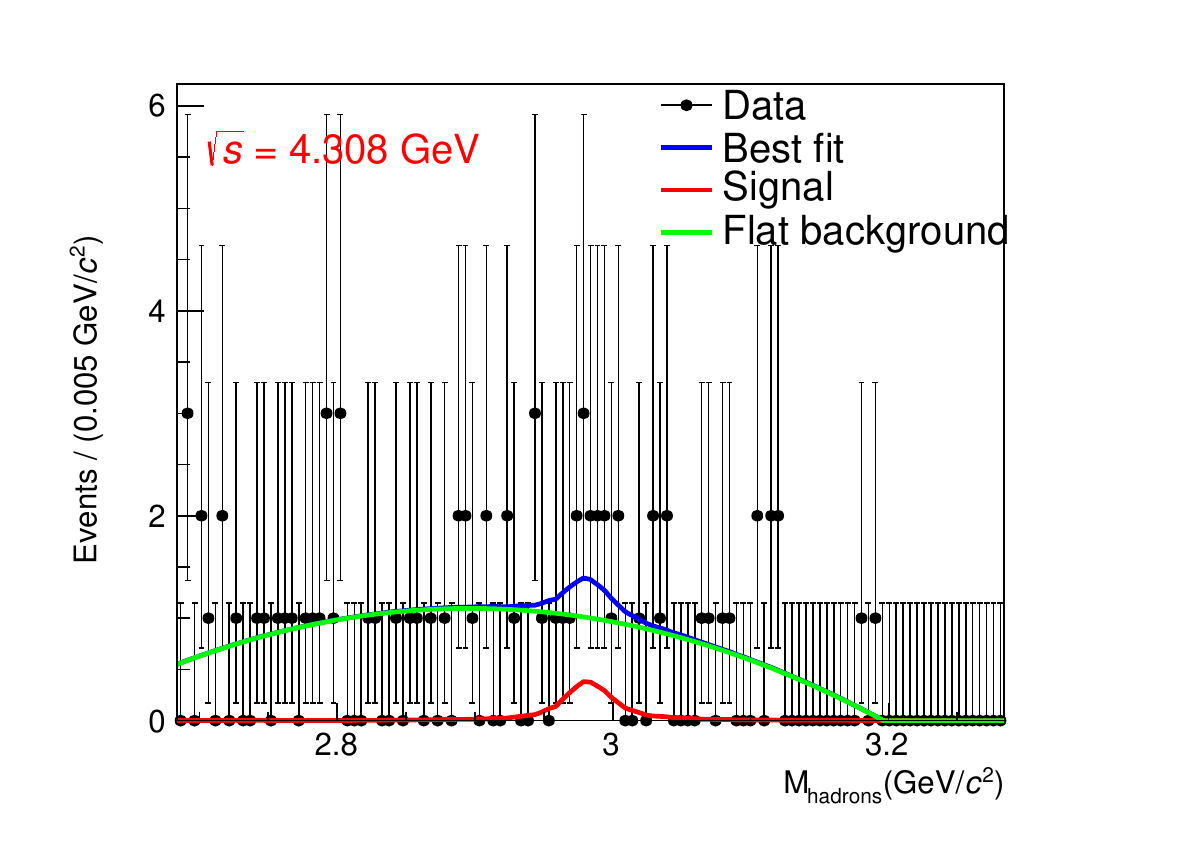}
  \captionsetup{skip=-10pt,font=large}
\end{subfigure}
  \begin{subfigure}{0.24\textwidth}
  \centering
  \includegraphics[width=\textwidth]{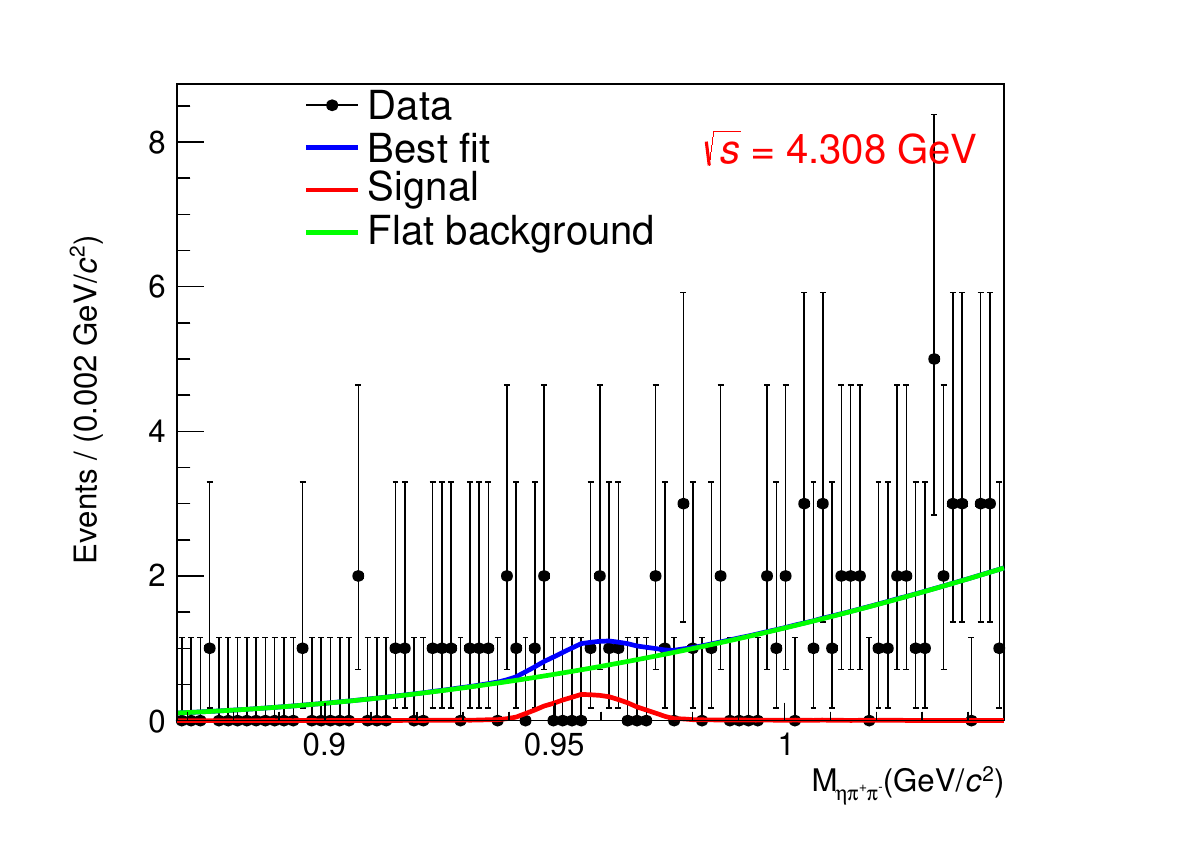}
  \captionsetup{skip=-10pt,font=large}
\end{subfigure}
\captionsetup{justification=raggedright}
\caption{Fits to the invariant mass distributions of (Left)(Middle-Right)$M(hadrons)$, (Middle-Left)$M(\gamma\gamma)$ and (Right)$M(\eta\pi^{+}\pi^{-})$ at $\sqrt(s)=4.258-4.308$~GeV.}
\label{fig:fit1}
\end{figure*}

\begin{figure*}[htbp]
\begin{subfigure}{0.24\textwidth}
  \centering
  \includegraphics[width=\textwidth]{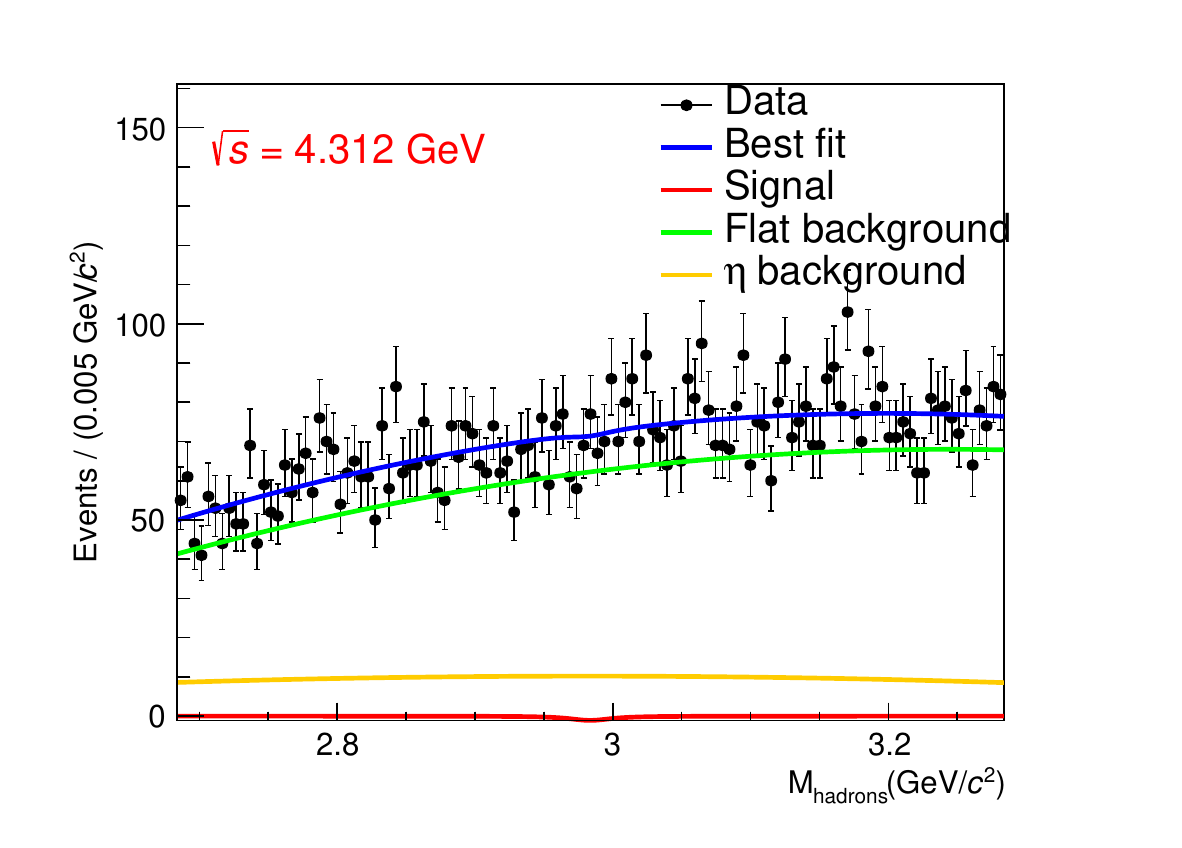}
  \captionsetup{skip=-10pt,font=large}
\end{subfigure}
\begin{subfigure}{0.24\textwidth}
  \centering
  \includegraphics[width=\textwidth]{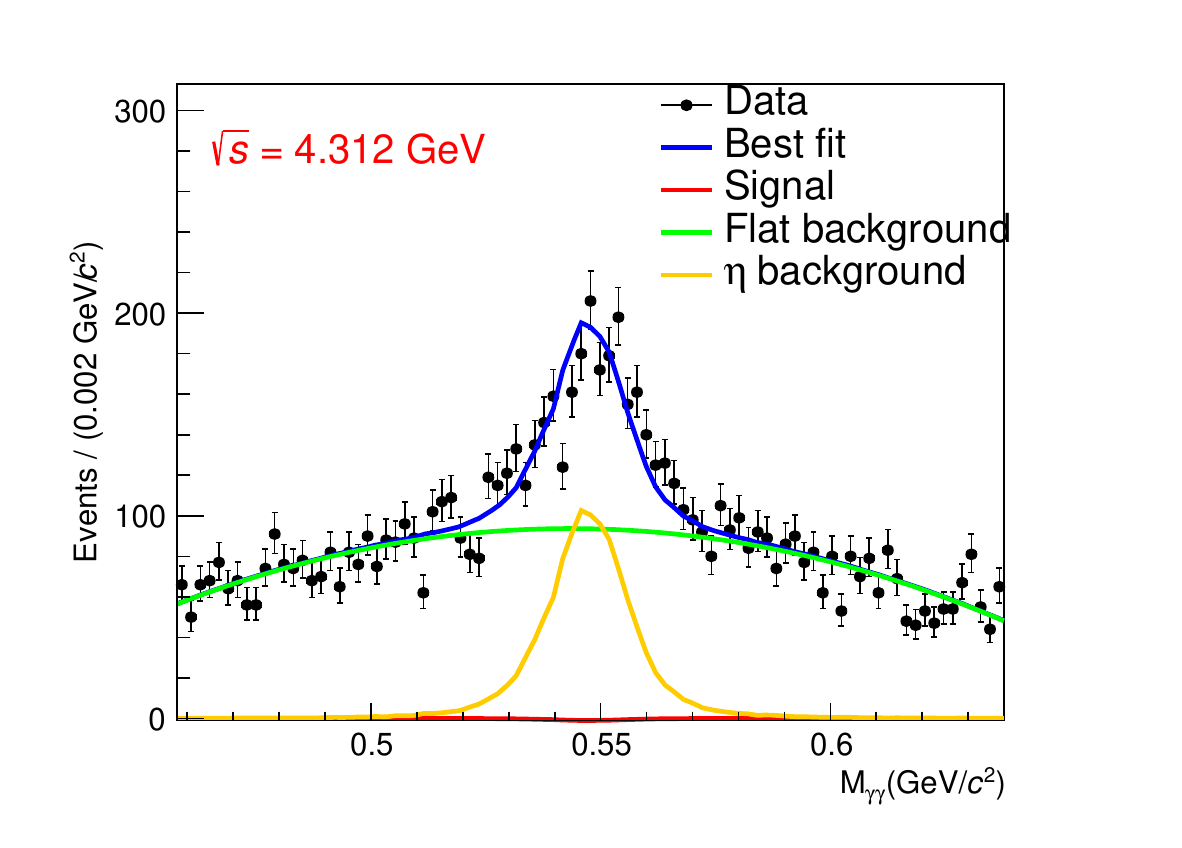}
  \captionsetup{skip=-10pt,font=large}
\end{subfigure}
 \begin{subfigure}{0.24\textwidth}
  \centering
  \includegraphics[width=\textwidth]{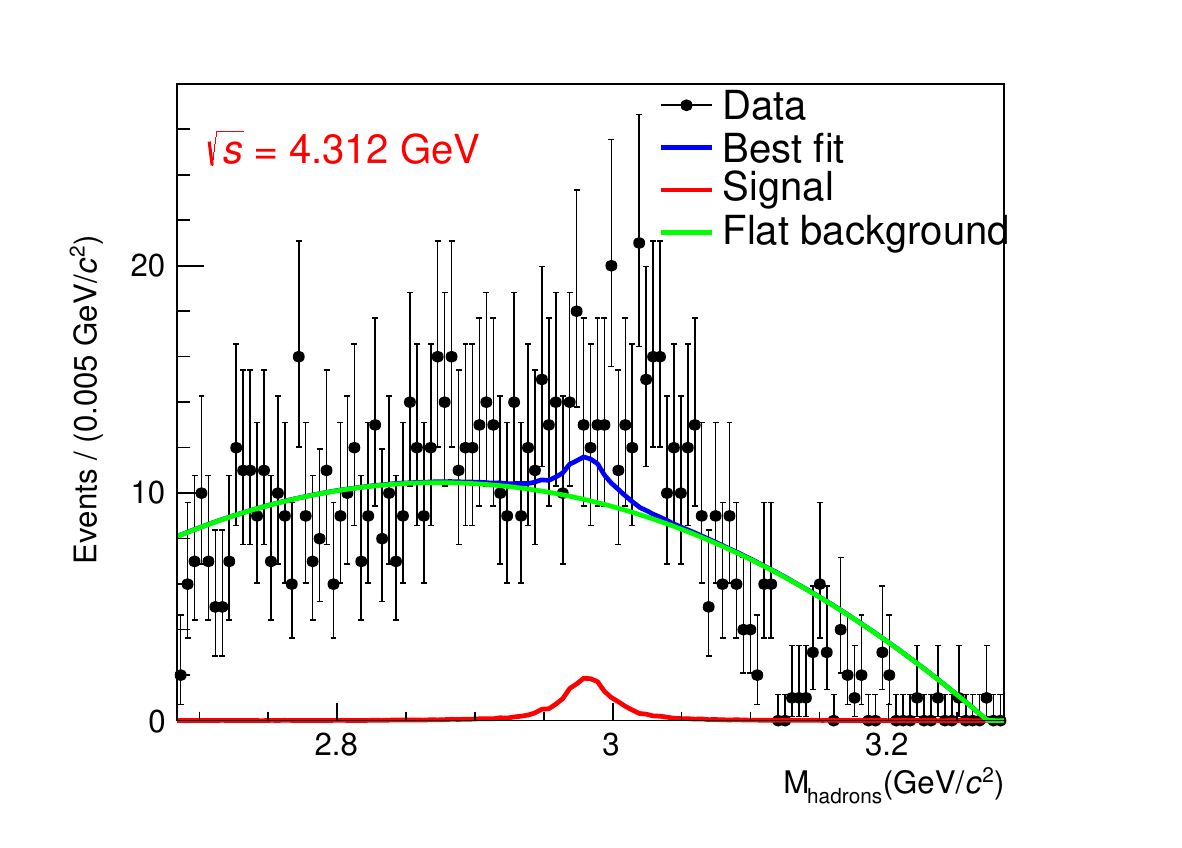}
  \captionsetup{skip=-10pt,font=large}
\end{subfigure}
  \begin{subfigure}{0.24\textwidth}
  \centering
  \includegraphics[width=\textwidth]{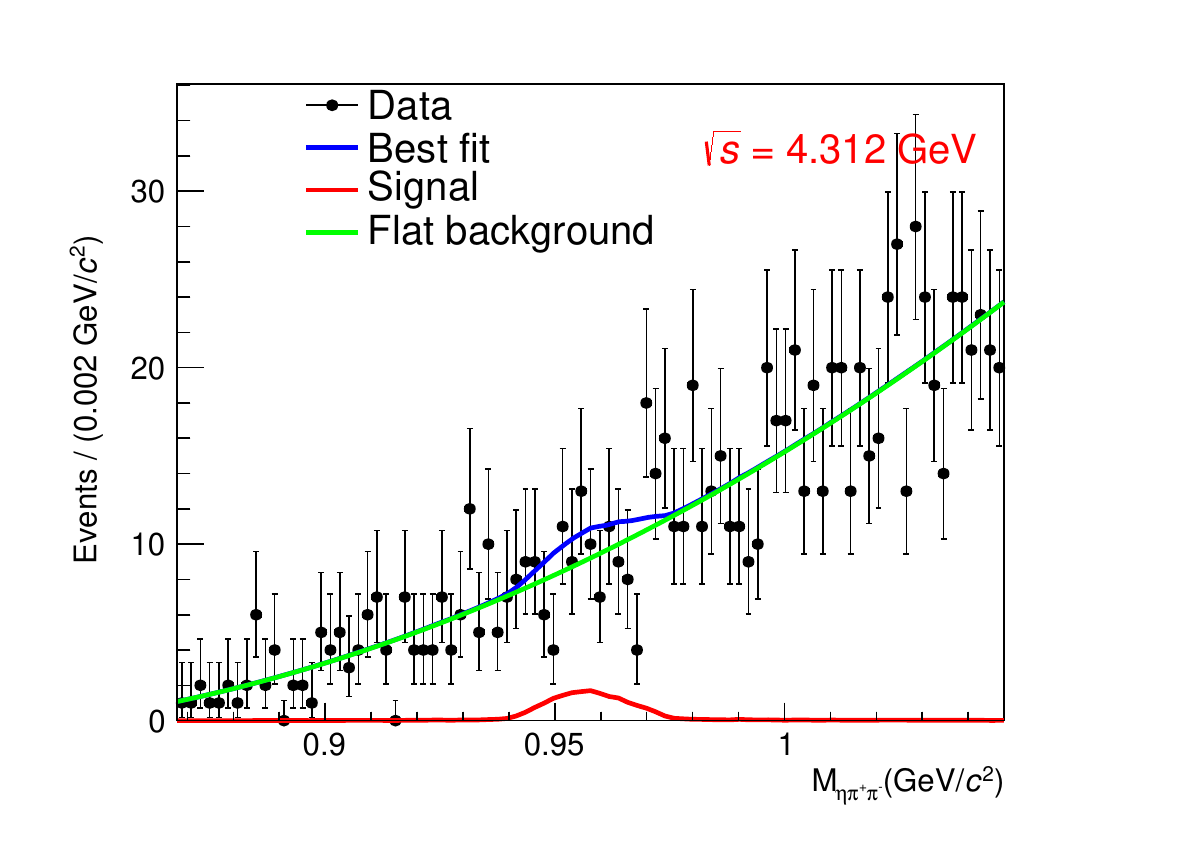}
  \captionsetup{skip=-10pt,font=large}
\end{subfigure}
\begin{subfigure}{0.24\textwidth}
  \centering
  \includegraphics[width=\textwidth]{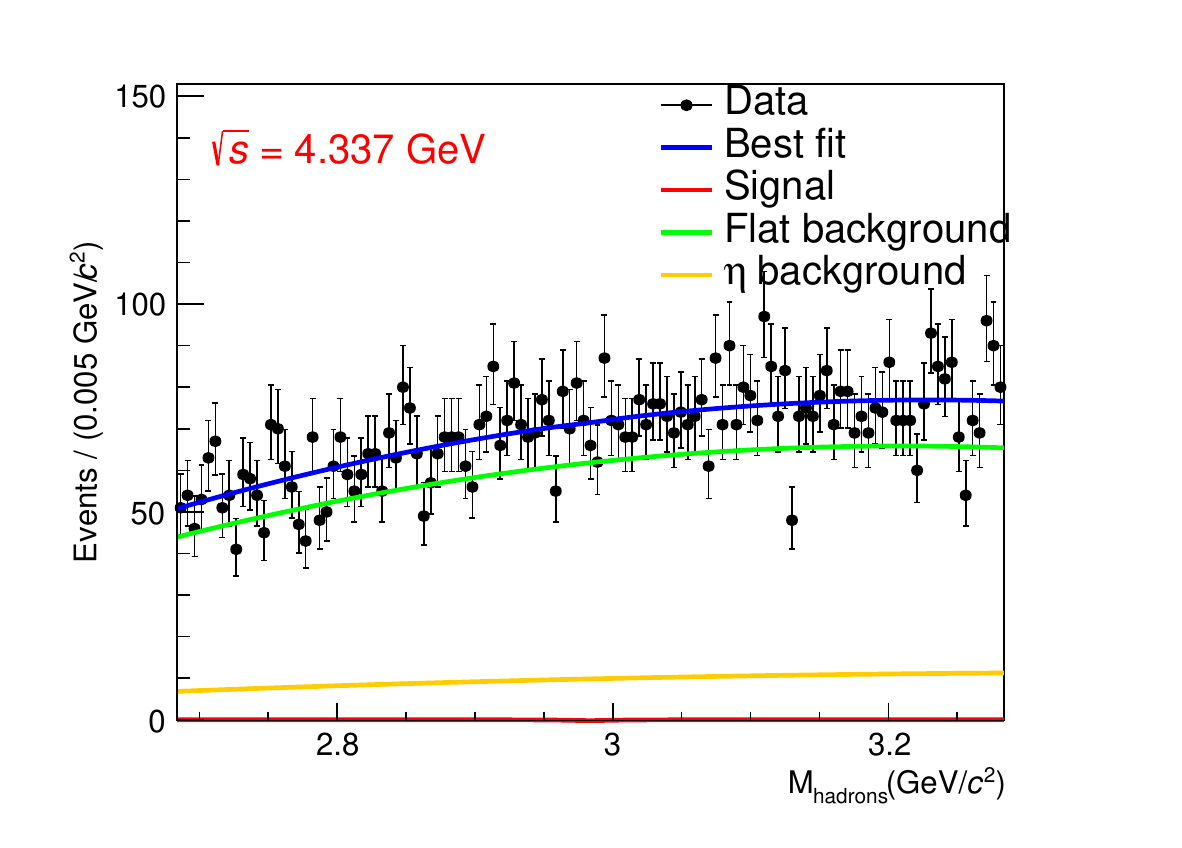}
  \captionsetup{skip=-10pt,font=large}
\end{subfigure}
\begin{subfigure}{0.24\textwidth}
  \centering
  \includegraphics[width=\textwidth]{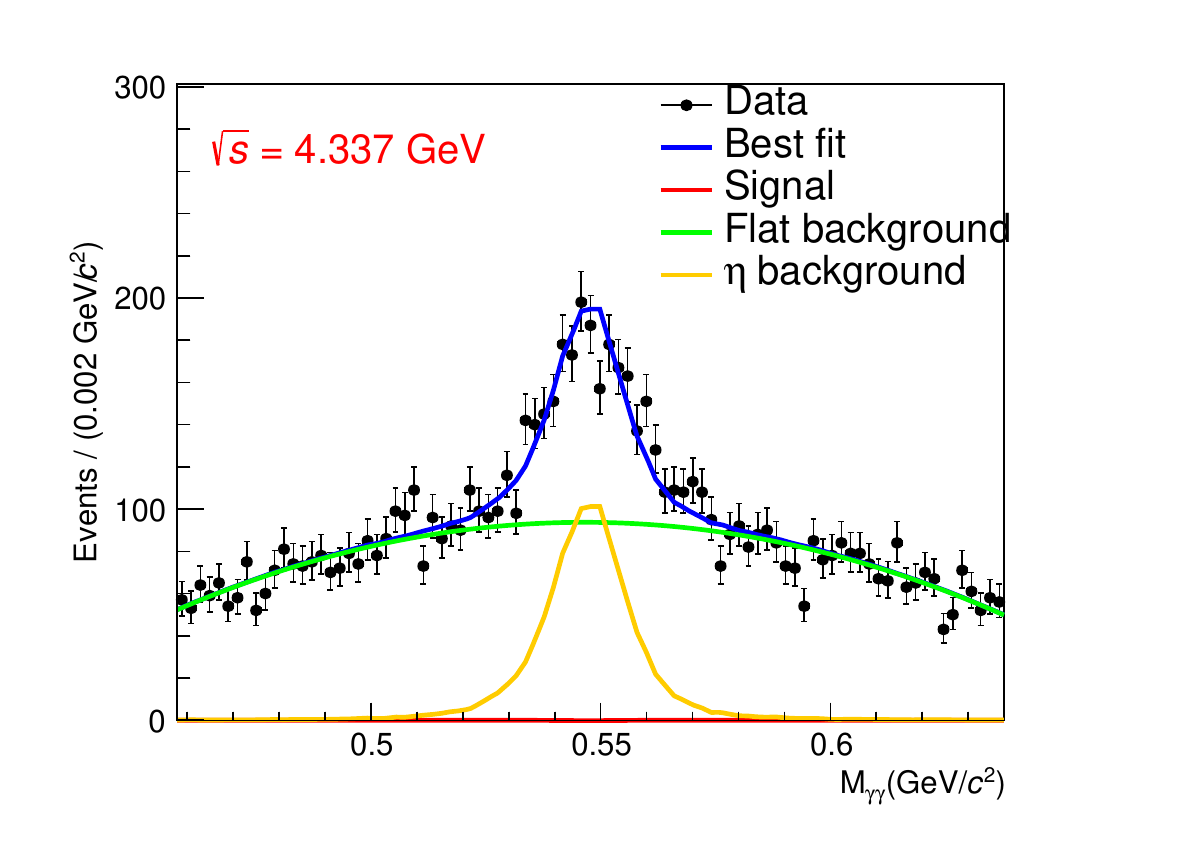}
  \captionsetup{skip=-10pt,font=large}
\end{subfigure}
 \begin{subfigure}{0.24\textwidth}
  \centering
  \includegraphics[width=\textwidth]{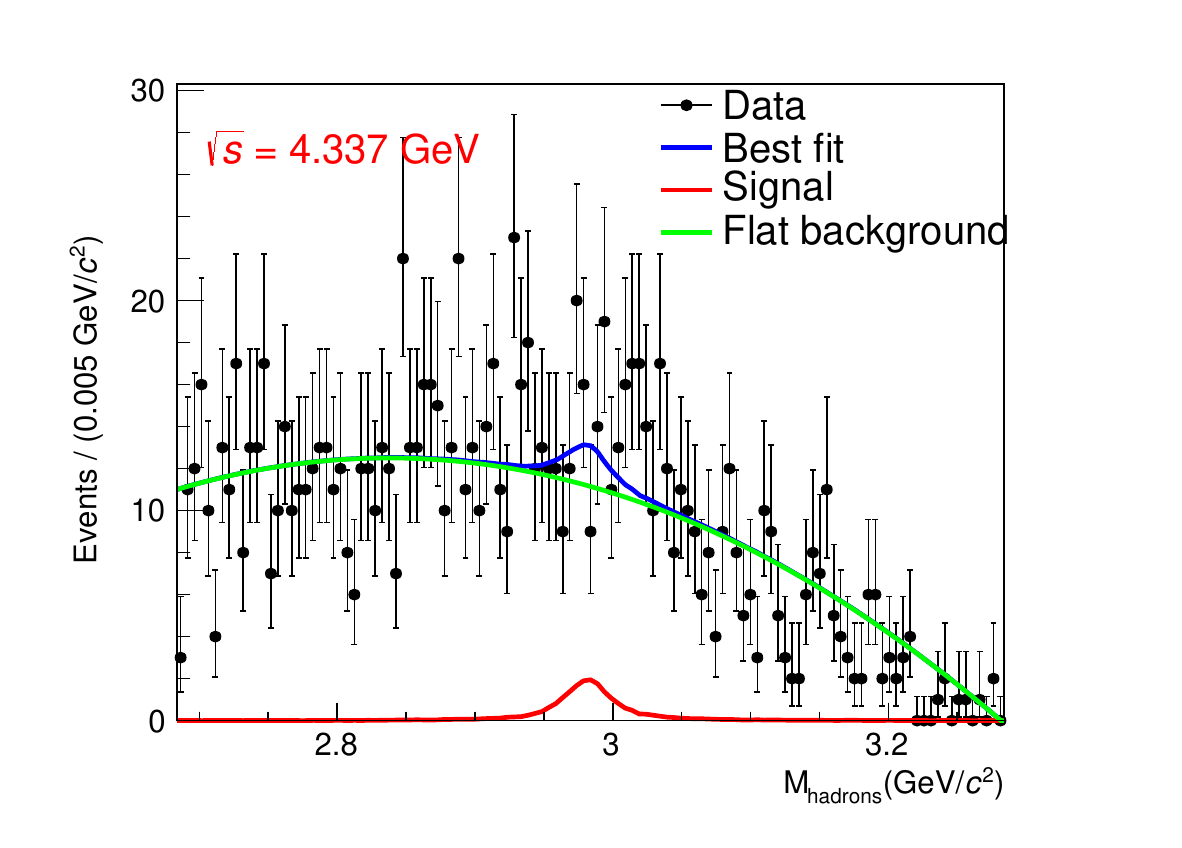}
  \captionsetup{skip=-10pt,font=large}
\end{subfigure}
  \begin{subfigure}{0.24\textwidth}
  \centering
  \includegraphics[width=\textwidth]{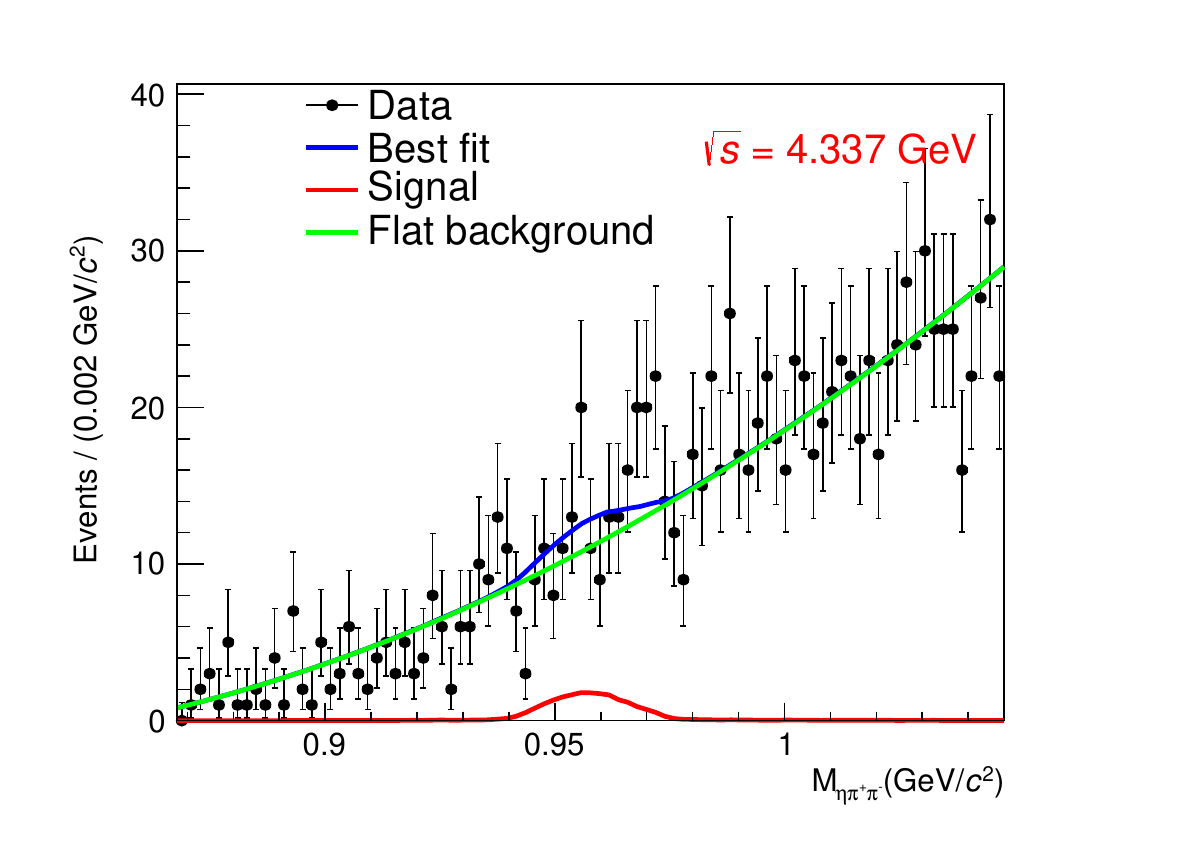}
  \captionsetup{skip=-10pt,font=large}
\end{subfigure}
\begin{subfigure}{0.24\textwidth}
  \centering
  \includegraphics[width=\textwidth]{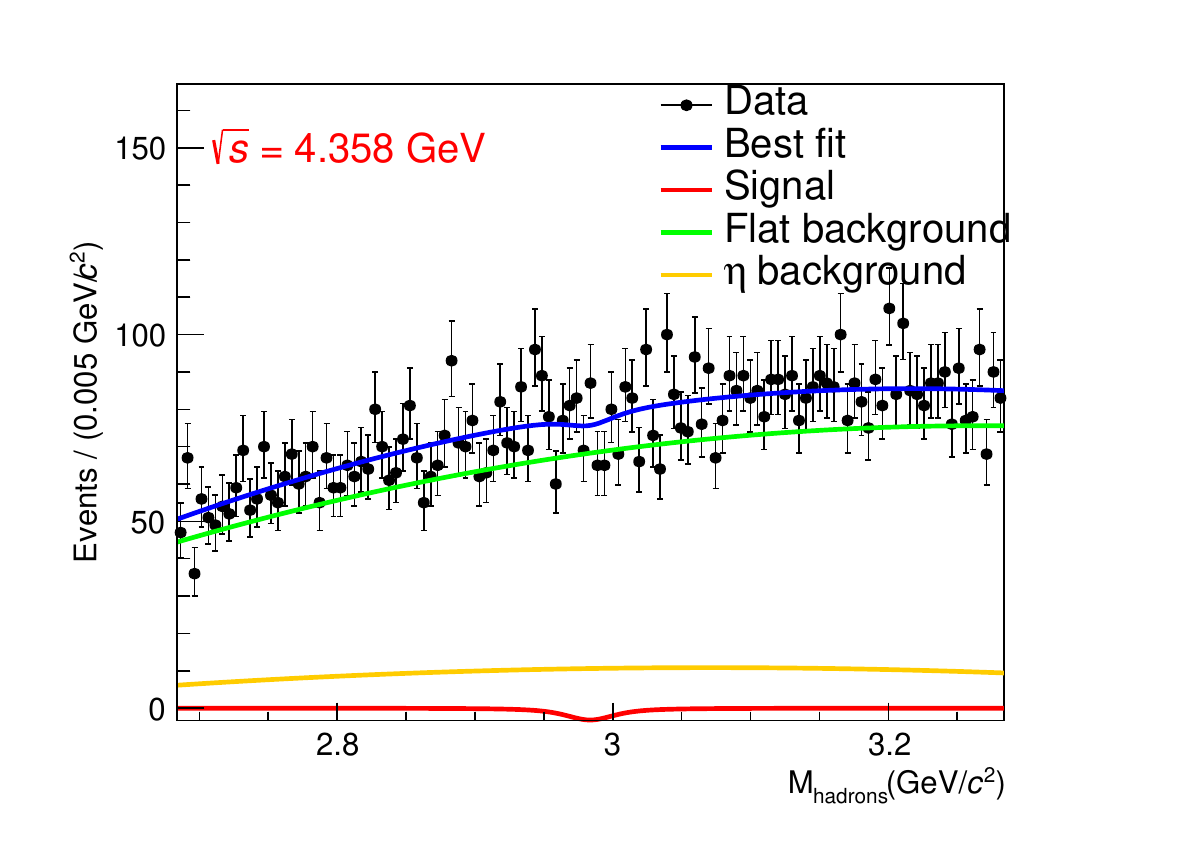}
  \captionsetup{skip=-10pt,font=large}
\end{subfigure}
\begin{subfigure}{0.24\textwidth}
  \centering
  \includegraphics[width=\textwidth]{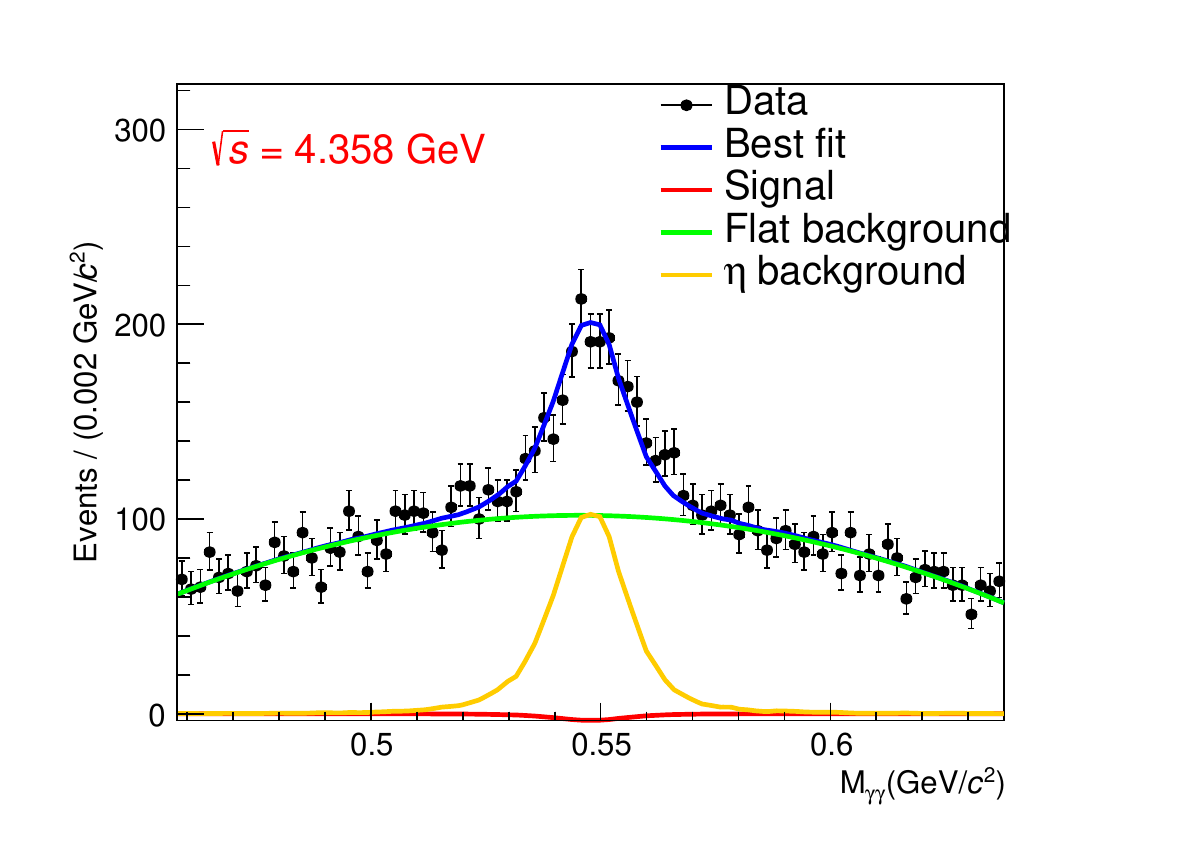}
  \captionsetup{skip=-10pt,font=large}
\end{subfigure}
 \begin{subfigure}{0.24\textwidth}
  \centering
  \includegraphics[width=\textwidth]{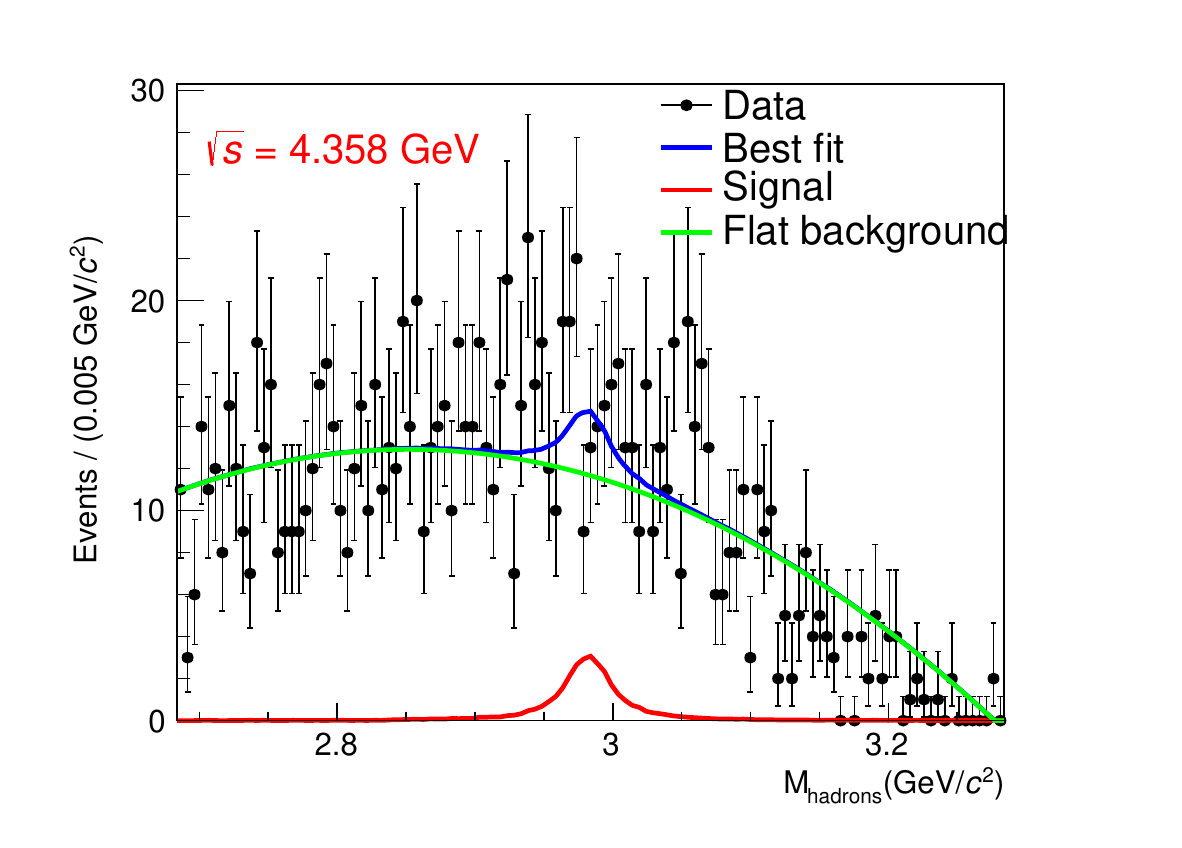}
  \captionsetup{skip=-10pt,font=large}
\end{subfigure}
  \begin{subfigure}{0.24\textwidth}
  \centering
  \includegraphics[width=\textwidth]{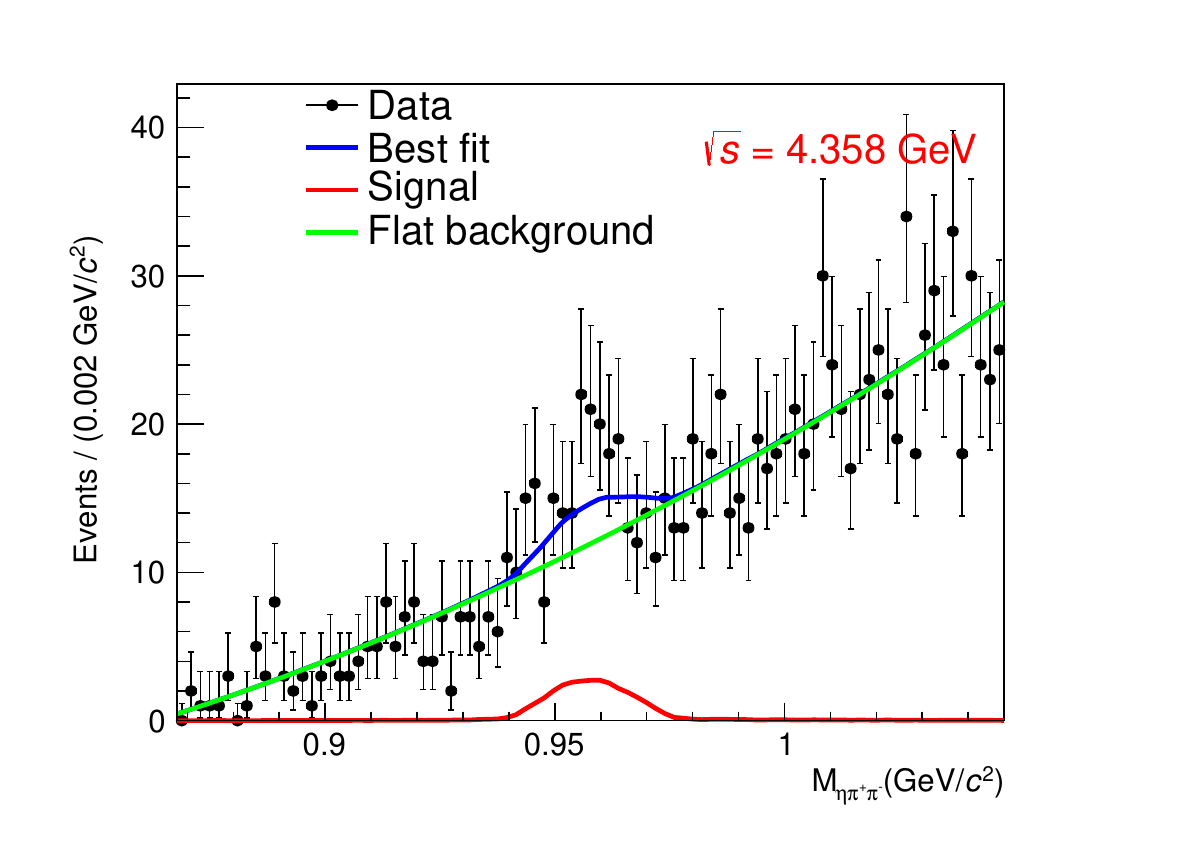}
  \captionsetup{skip=-10pt,font=large}
\end{subfigure}
\begin{subfigure}{0.24\textwidth}
  \centering
  \includegraphics[width=\textwidth]{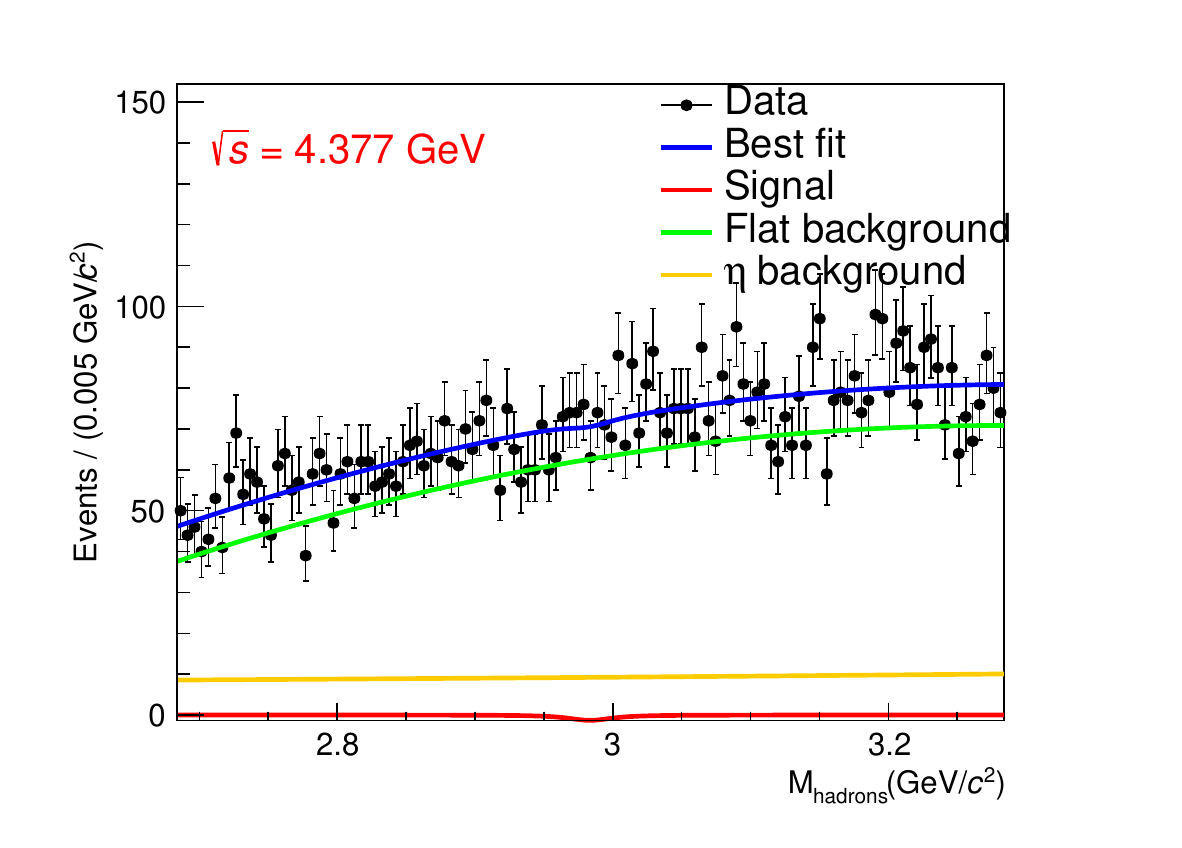}
  \captionsetup{skip=-10pt,font=large}
\end{subfigure}
\begin{subfigure}{0.24\textwidth}
  \centering
  \includegraphics[width=\textwidth]{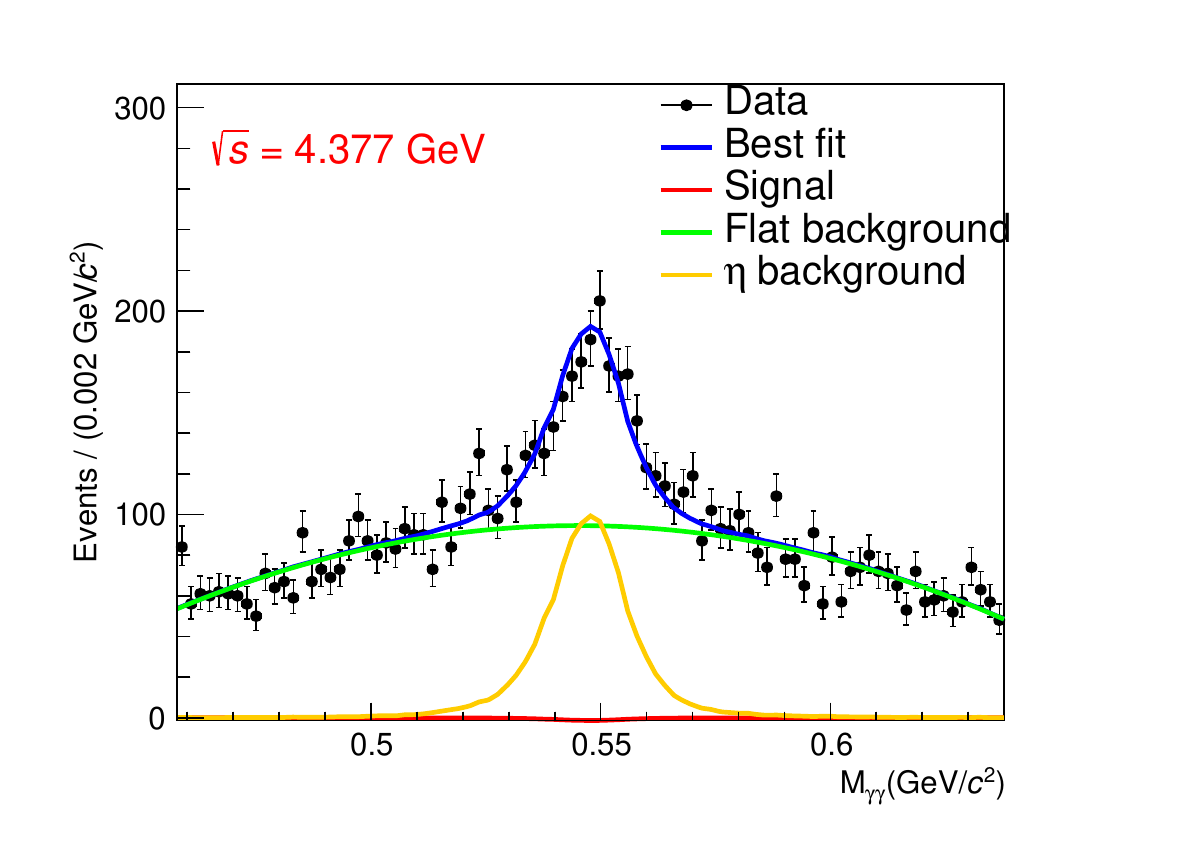}
  \captionsetup{skip=-10pt,font=large}
\end{subfigure}
 \begin{subfigure}{0.24\textwidth}
  \centering
  \includegraphics[width=\textwidth]{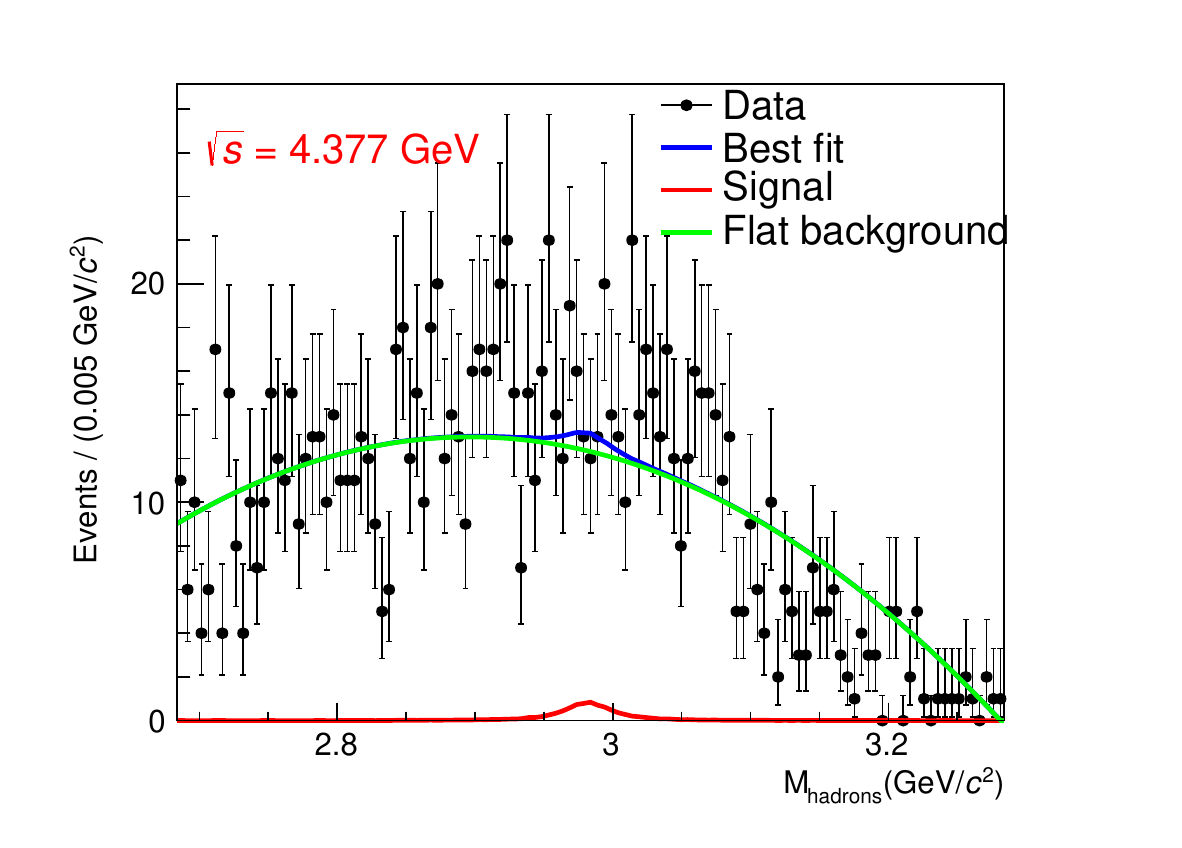}
  \captionsetup{skip=-10pt,font=large}
\end{subfigure}
  \begin{subfigure}{0.24\textwidth}
  \centering
  \includegraphics[width=\textwidth]{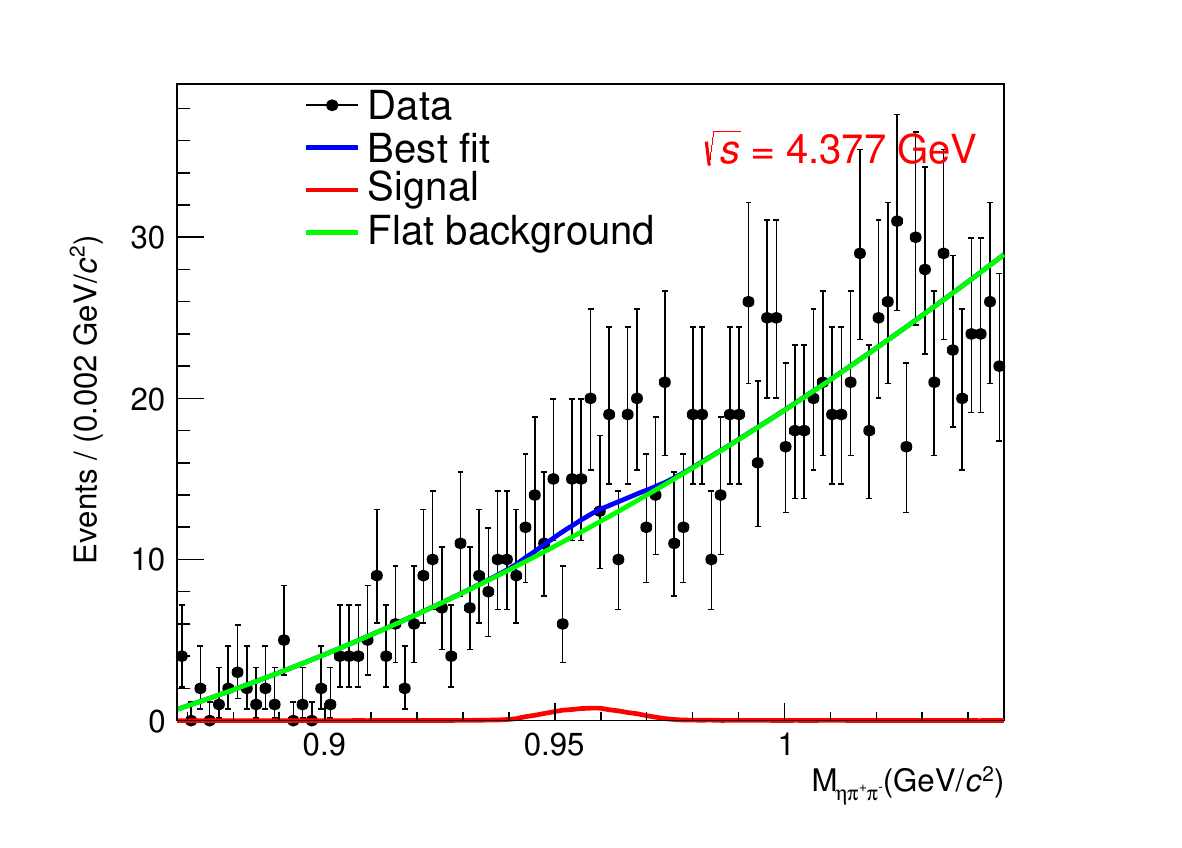}
  \captionsetup{skip=-10pt,font=large}
\end{subfigure}
\begin{subfigure}{0.24\textwidth}
  \centering
  \includegraphics[width=\textwidth]{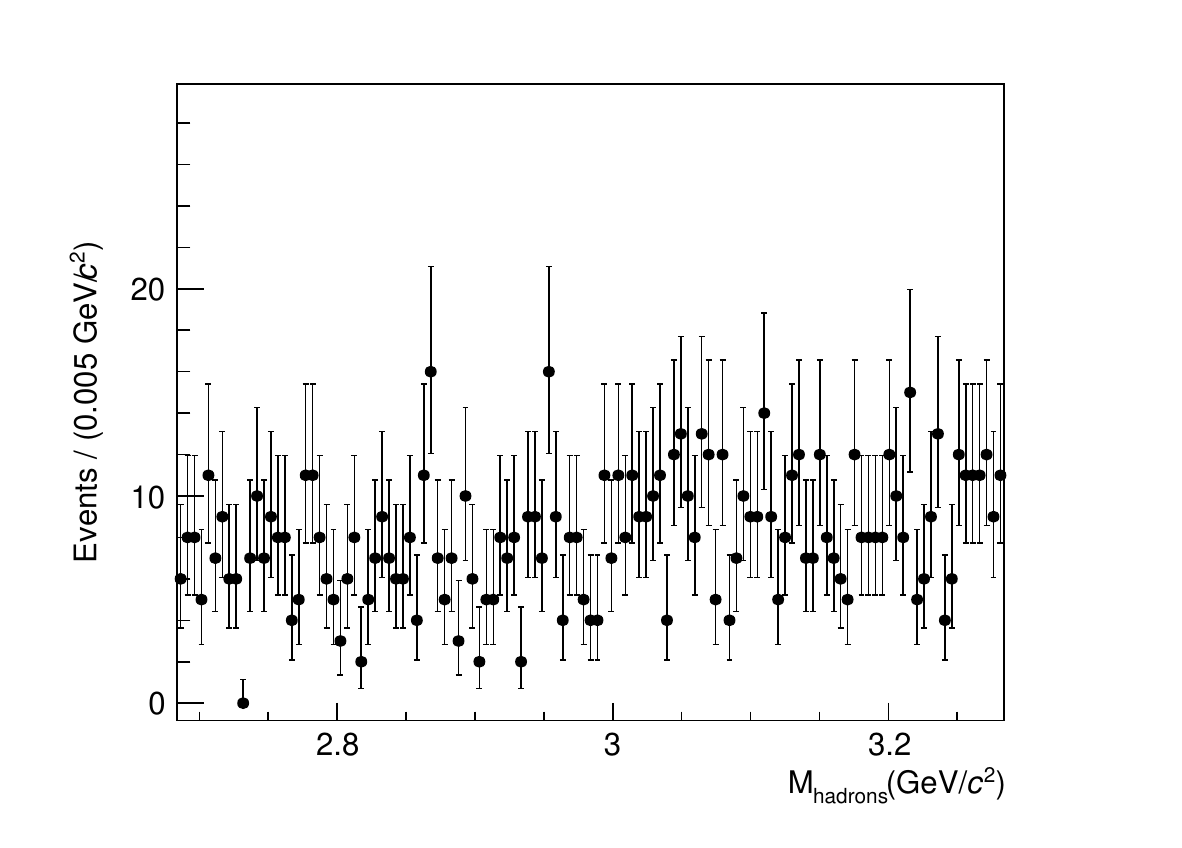}
  \captionsetup{skip=-10pt,font=large}
\end{subfigure}
\begin{subfigure}{0.24\textwidth}
  \centering
  \includegraphics[width=\textwidth]{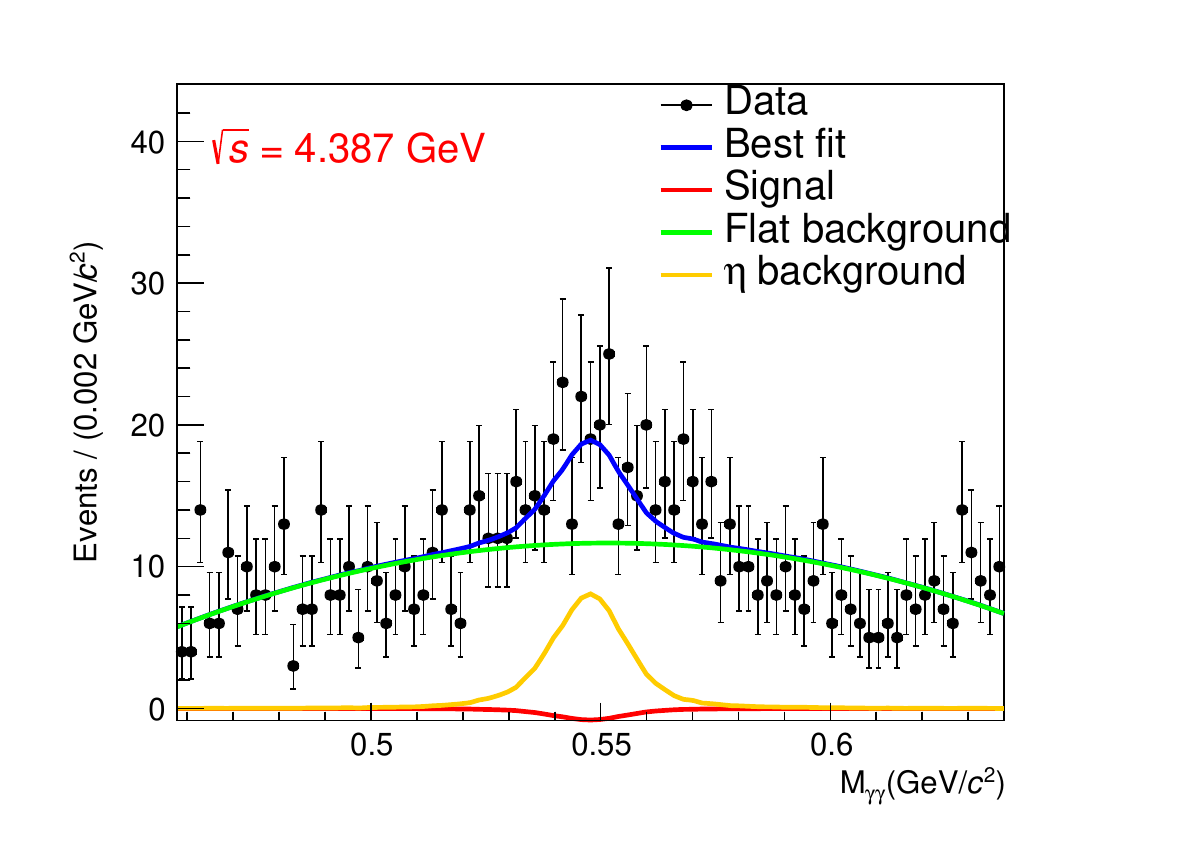}
  \captionsetup{skip=-10pt,font=large}
\end{subfigure}
 \begin{subfigure}{0.24\textwidth}
  \centering
  \includegraphics[width=\textwidth]{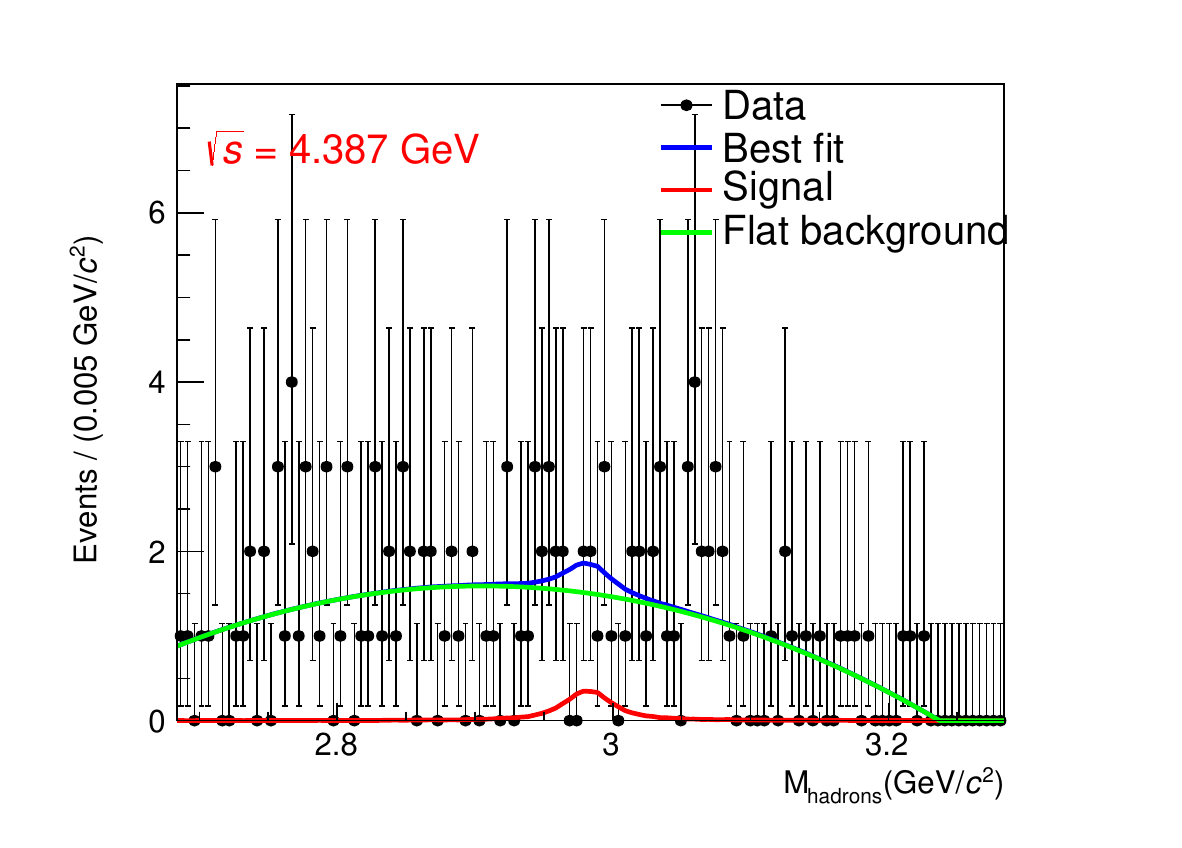}
  \captionsetup{skip=-10pt,font=large}
\end{subfigure}
  \begin{subfigure}{0.24\textwidth}
  \centering
  \includegraphics[width=\textwidth]{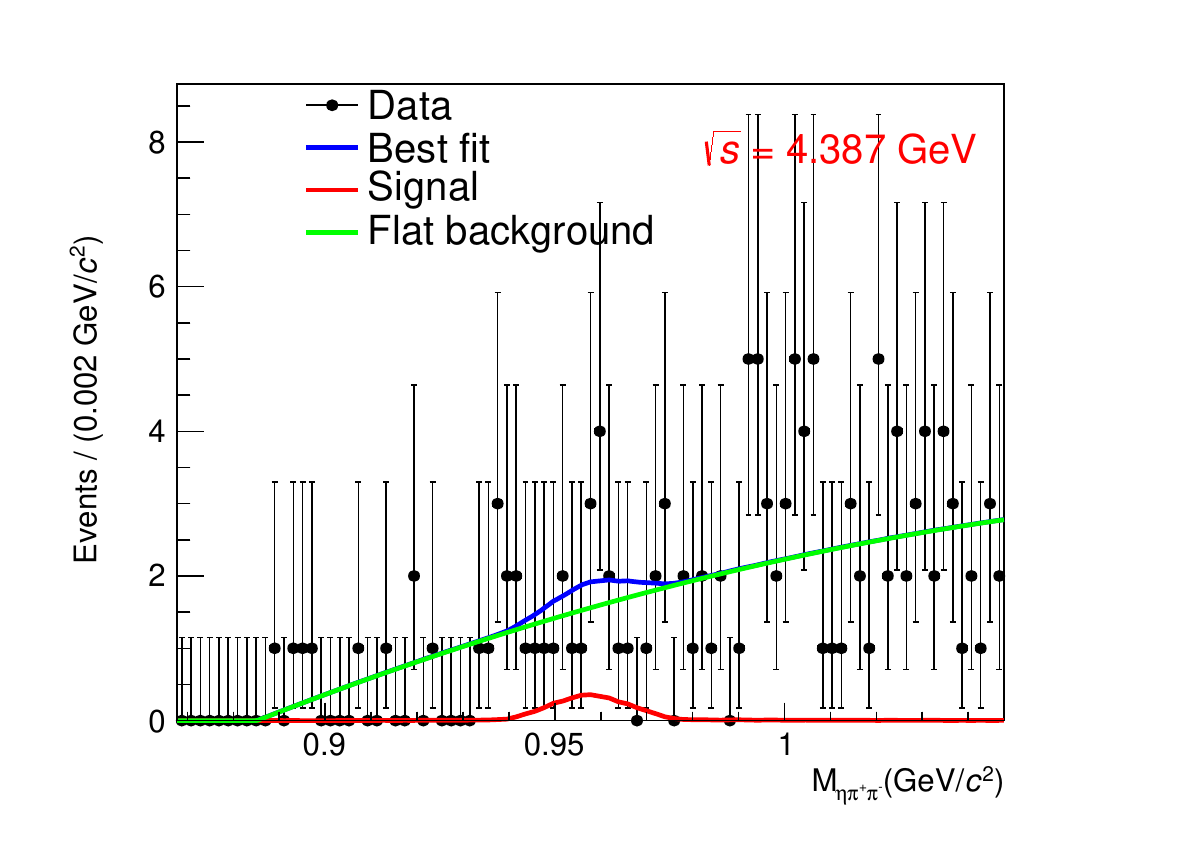}
  \captionsetup{skip=-10pt,font=large}
\end{subfigure}
\captionsetup{justification=raggedright}
\caption{Fits to the invariant mass distributions of (Left)(Middle-Right)$M(hadrons)$, (Middle-Left)$M(\gamma\gamma)$ and (Right)$M(\eta\pi^{+}\pi^{-})$ at $\sqrt(s)=4.312-4.387$~GeV.}
\label{fig:fit2}
\end{figure*}

\begin{figure*}[htbp]
\begin{subfigure}{0.24\textwidth}
  \centering
  \includegraphics[width=\textwidth]{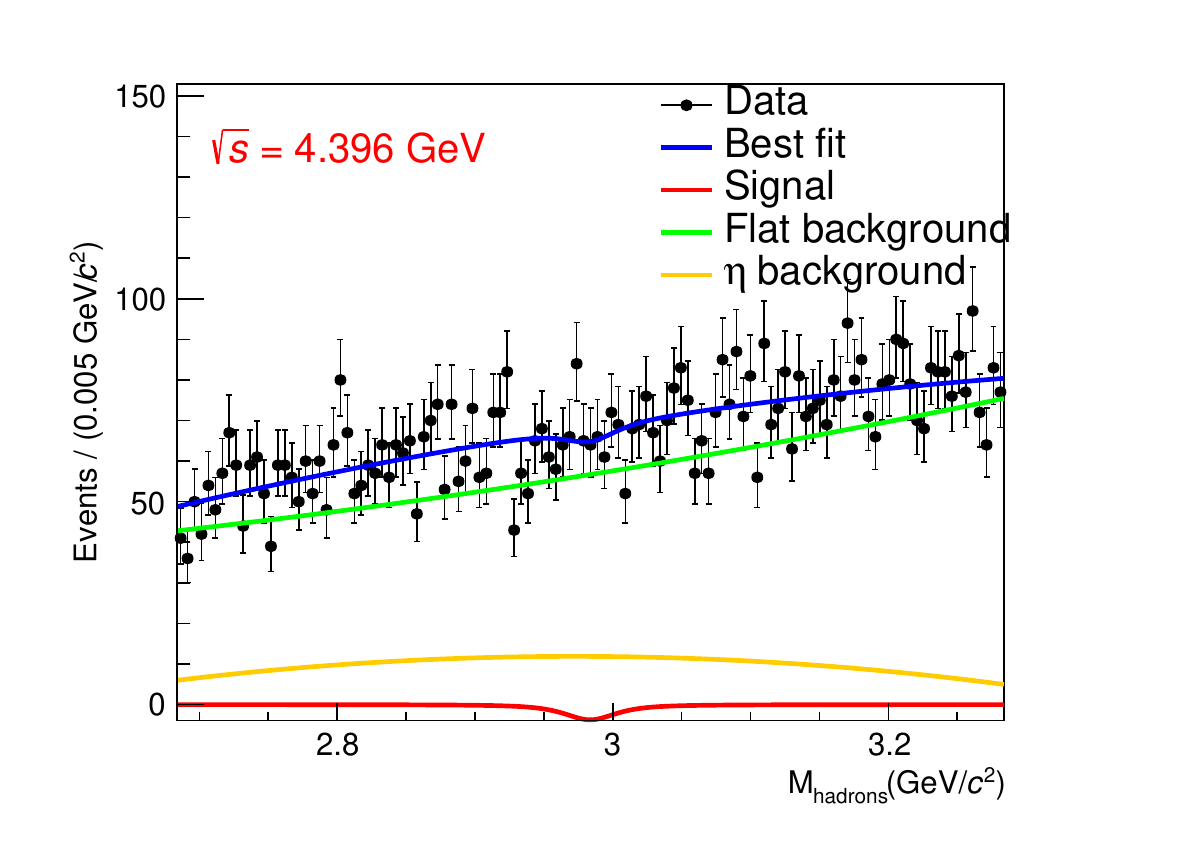}
  \captionsetup{skip=-10pt,font=large}
\end{subfigure}
\begin{subfigure}{0.24\textwidth}
  \centering
  \includegraphics[width=\textwidth]{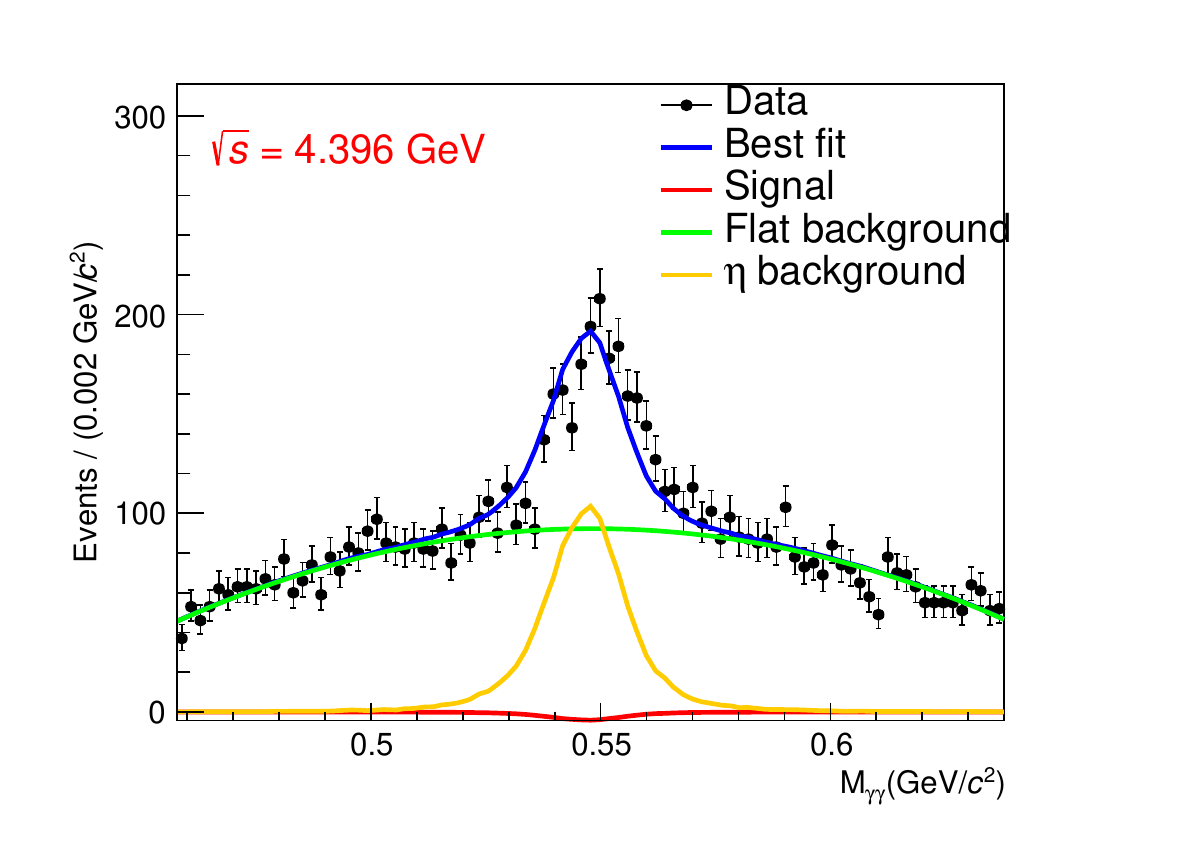}
  \captionsetup{skip=-10pt,font=large}
\end{subfigure}
 \begin{subfigure}{0.24\textwidth}
  \centering
  \includegraphics[width=\textwidth]{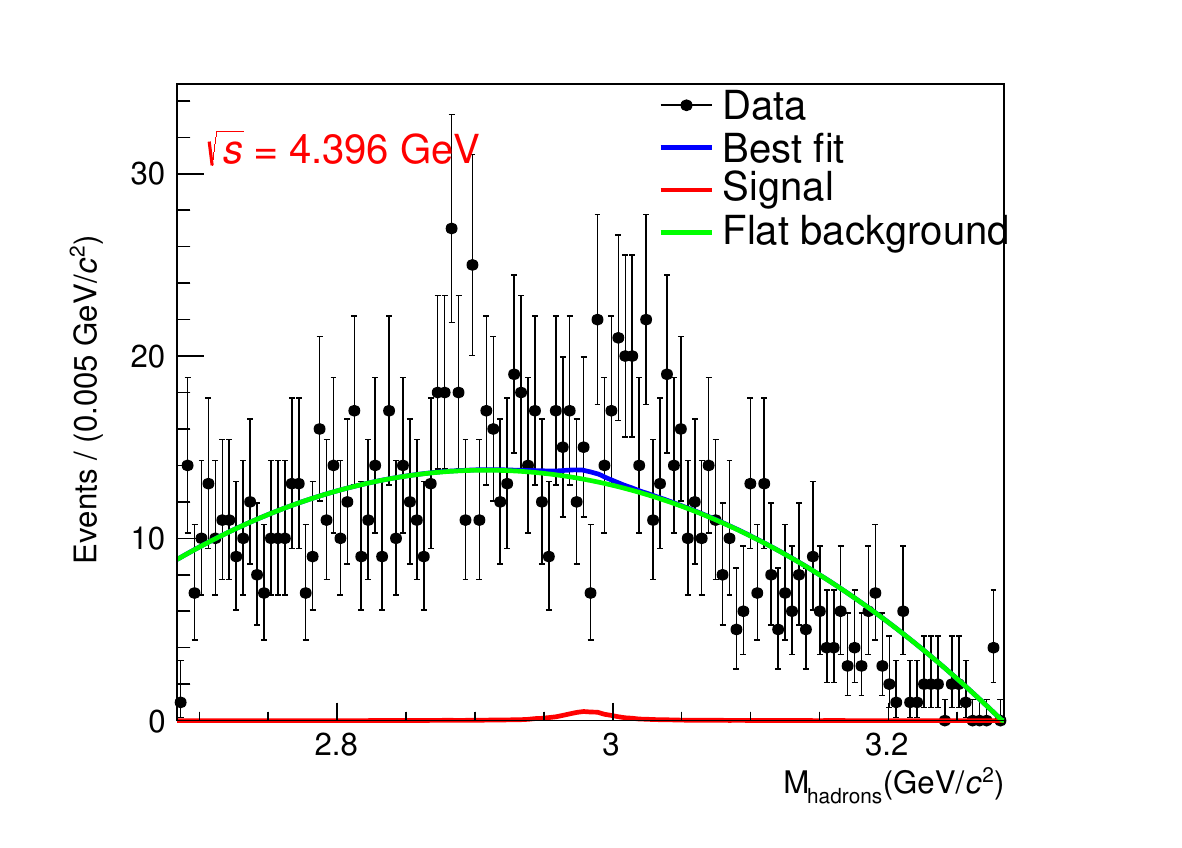}
  \captionsetup{skip=-10pt,font=large}
\end{subfigure}
  \begin{subfigure}{0.24\textwidth}
  \centering
  \includegraphics[width=\textwidth]{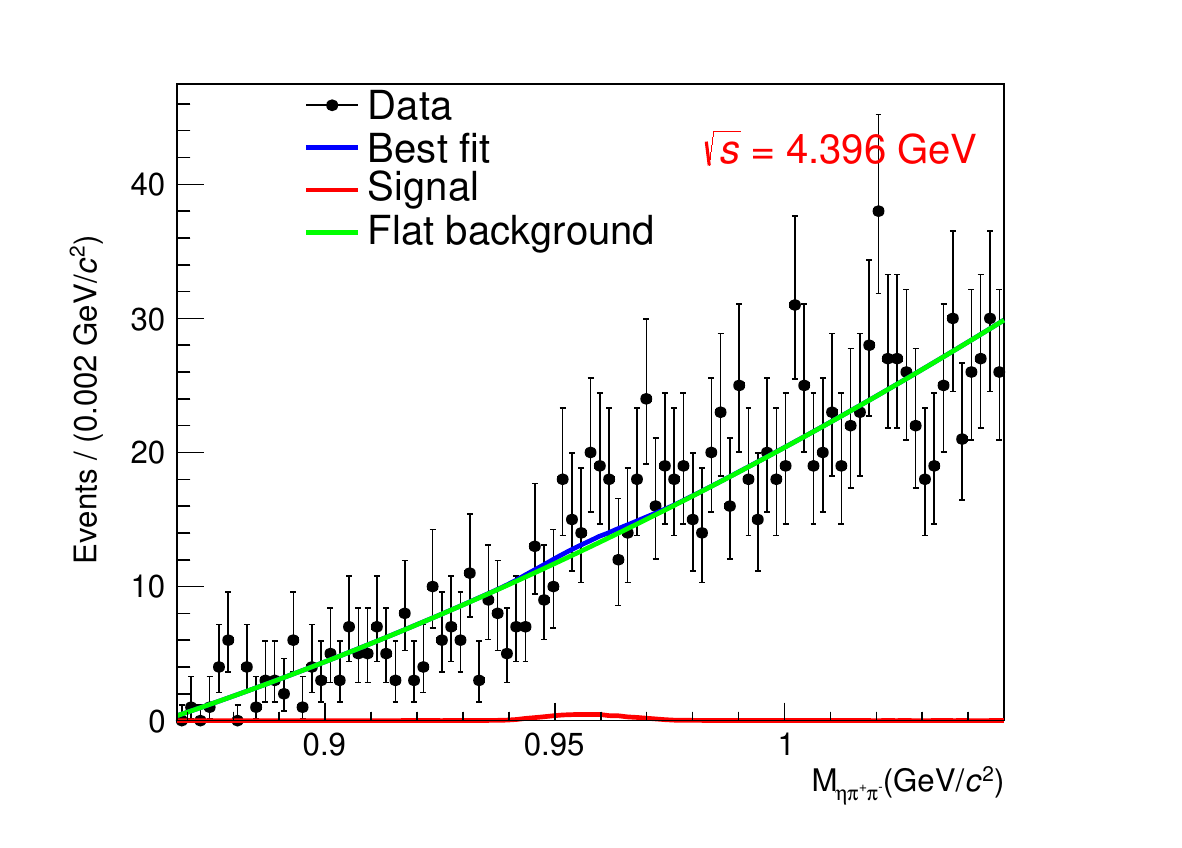}
  \captionsetup{skip=-10pt,font=large}
\end{subfigure}
\begin{subfigure}{0.24\textwidth}
  \centering
  \includegraphics[width=\textwidth]{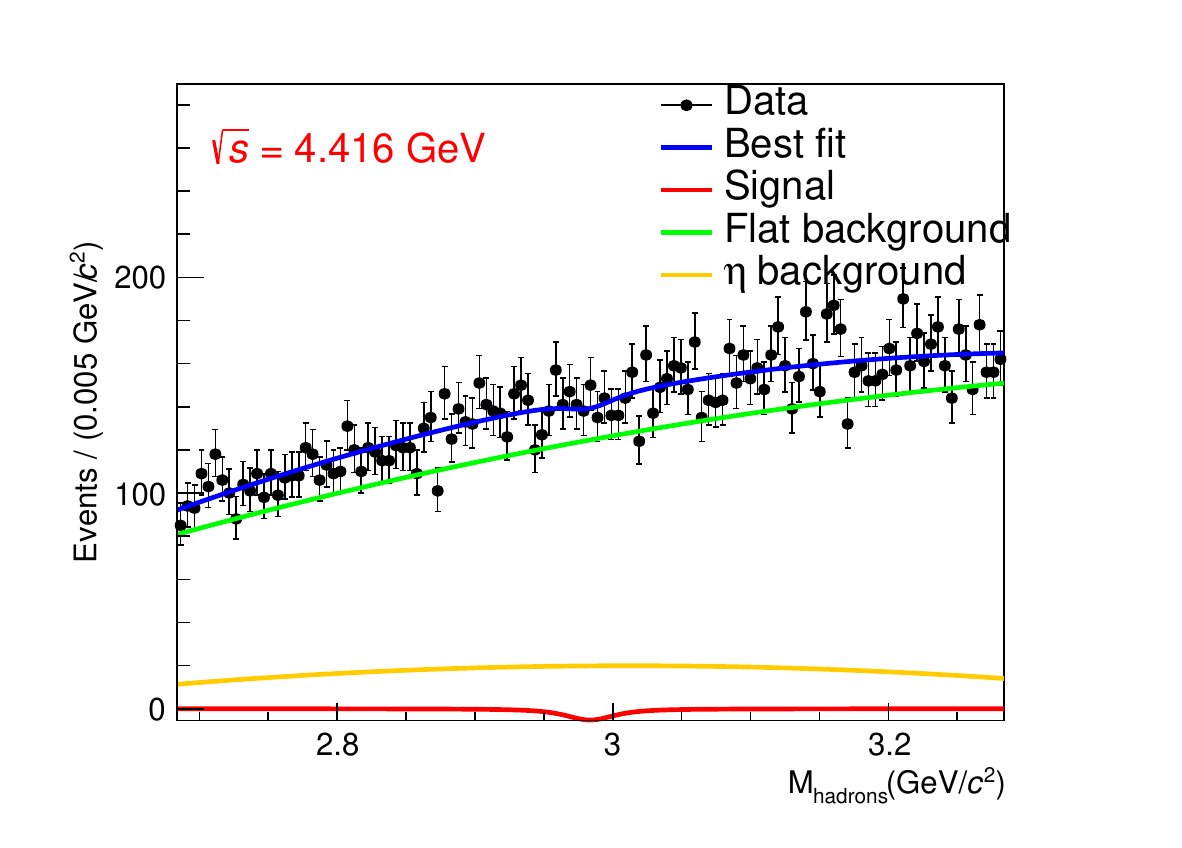}
  \captionsetup{skip=-10pt,font=large}
\end{subfigure}
\begin{subfigure}{0.24\textwidth}
  \centering
  \includegraphics[width=\textwidth]{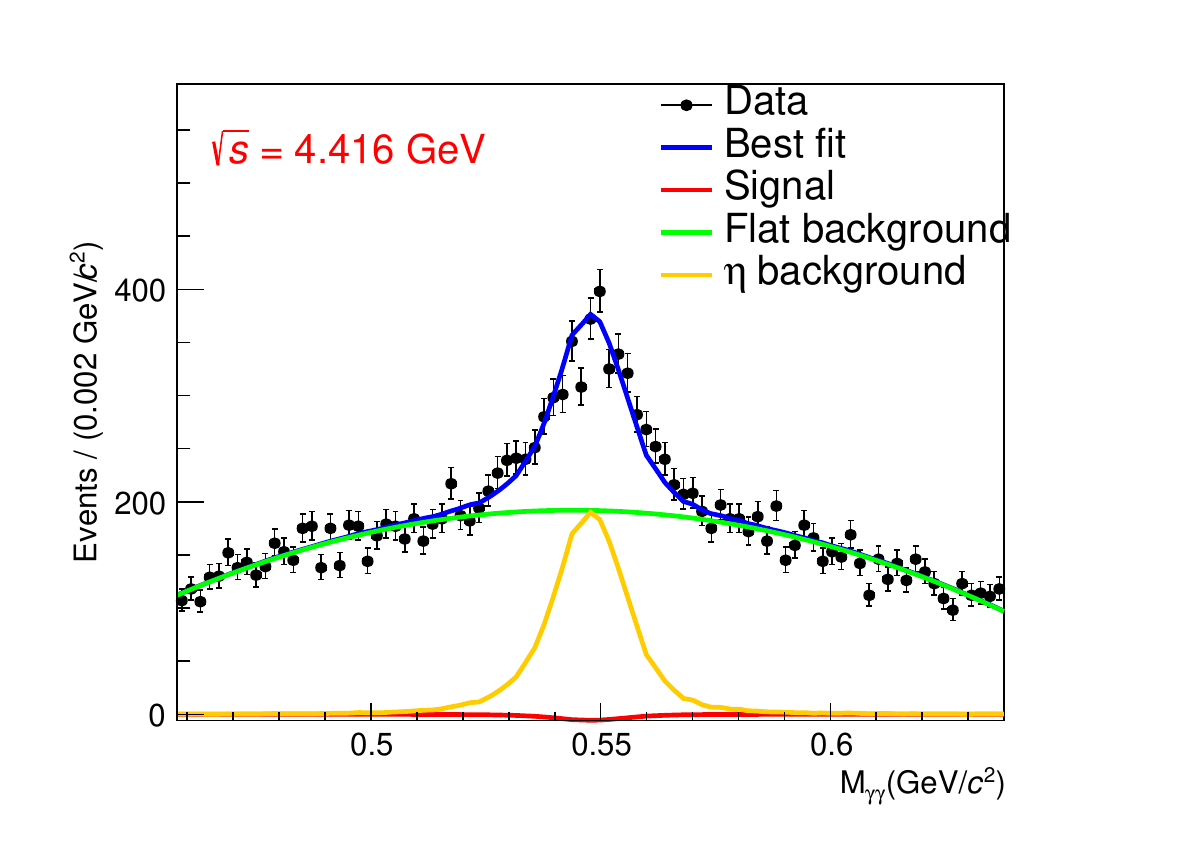}
  \captionsetup{skip=-10pt,font=large}
\end{subfigure}
 \begin{subfigure}{0.24\textwidth}
  \centering
  \includegraphics[width=\textwidth]{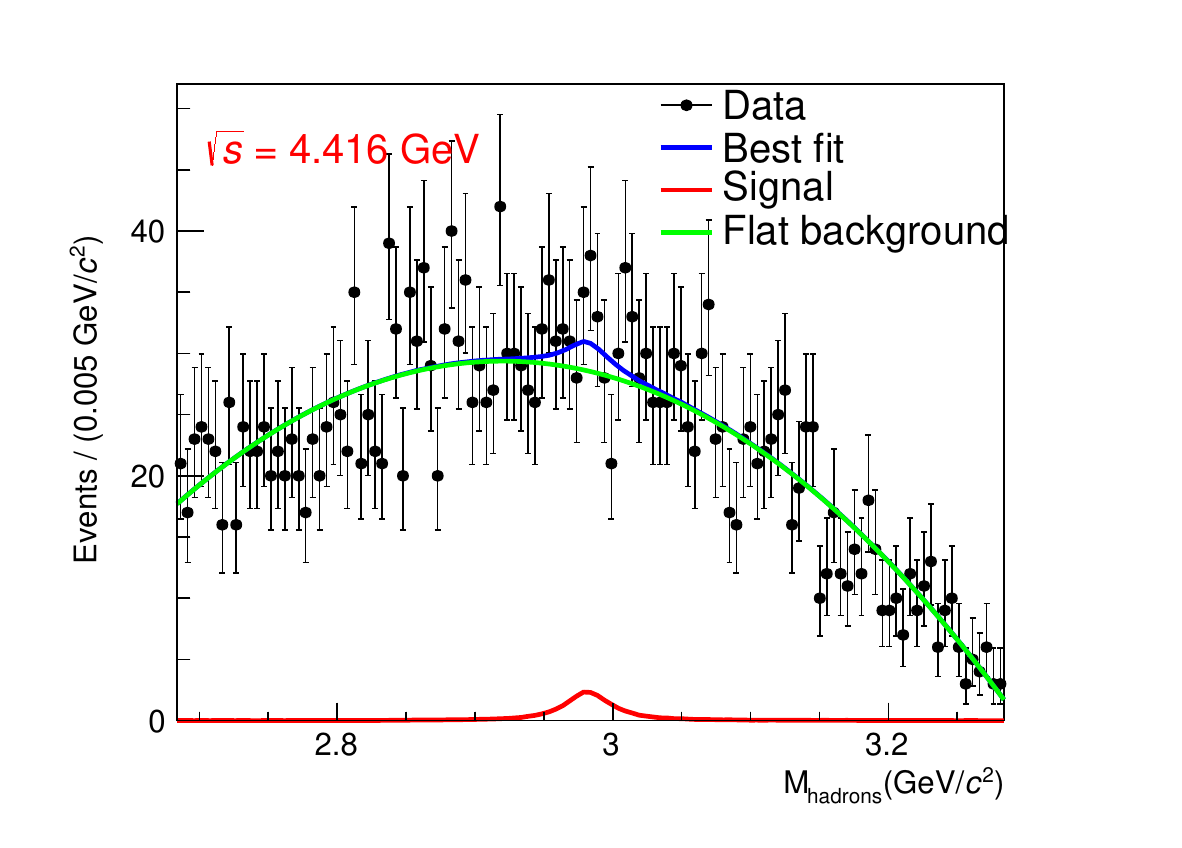}
  \captionsetup{skip=-10pt,font=large}
\end{subfigure}
  \begin{subfigure}{0.24\textwidth}
  \centering
  \includegraphics[width=\textwidth]{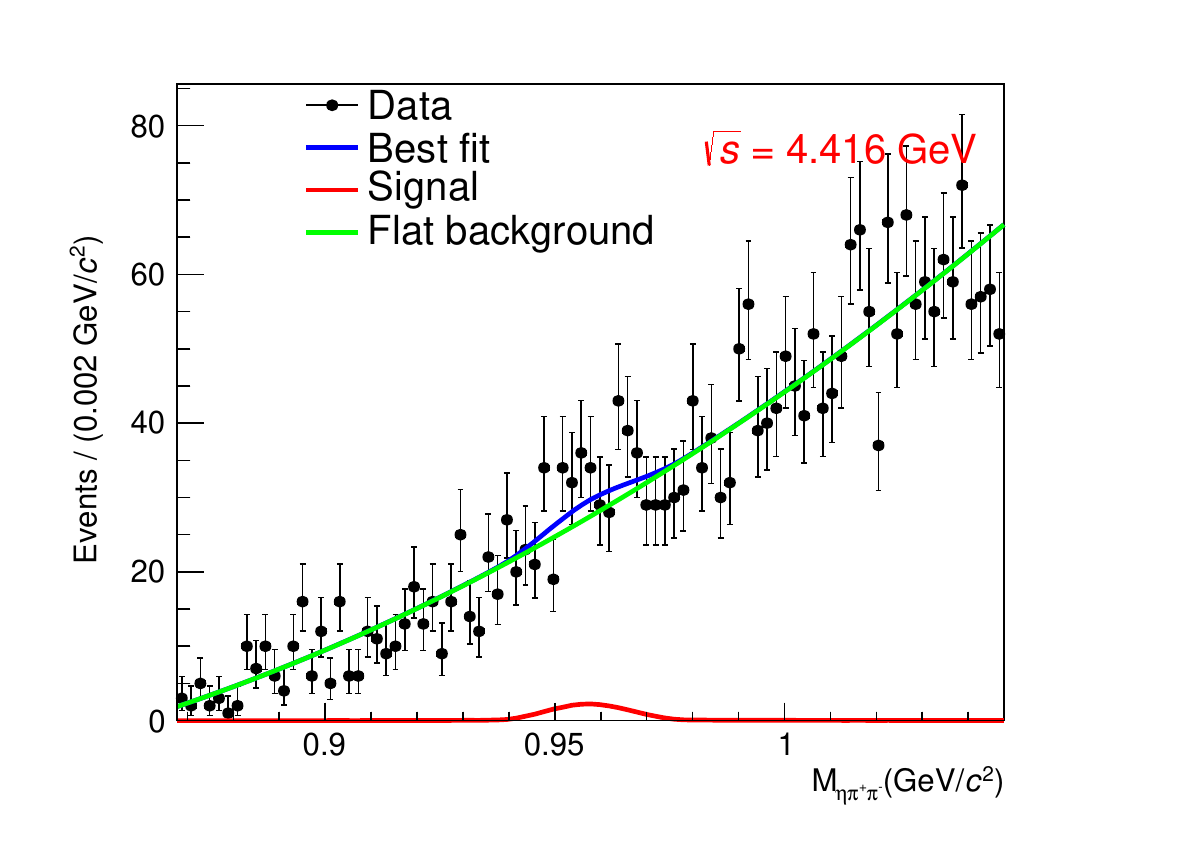}
  \captionsetup{skip=-10pt,font=large}
\end{subfigure}
\begin{subfigure}{0.24\textwidth}
  \centering
  \includegraphics[width=\textwidth]{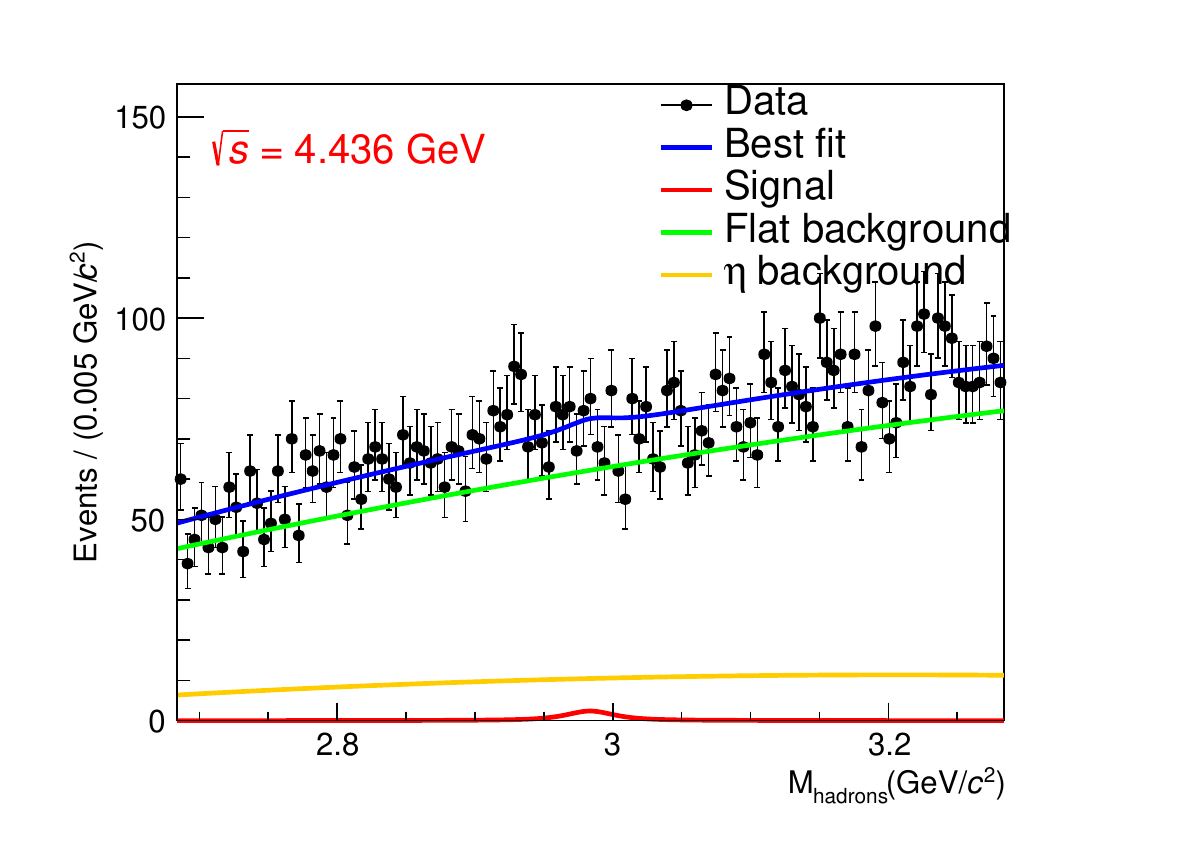}
  \captionsetup{skip=-10pt,font=large}
\end{subfigure}
\begin{subfigure}{0.24\textwidth}
  \centering
  \includegraphics[width=\textwidth]{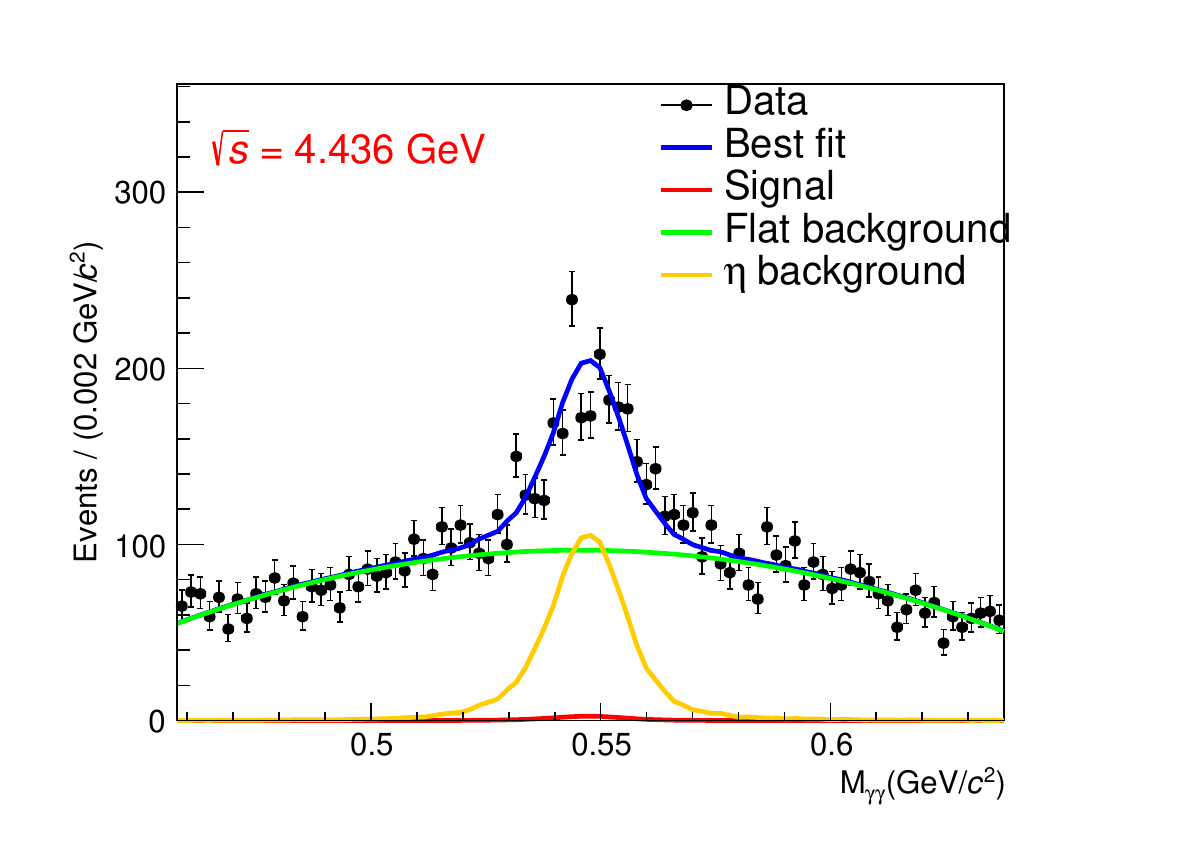}
  \captionsetup{skip=-10pt,font=large}
\end{subfigure}
 \begin{subfigure}{0.24\textwidth}
  \centering
  \includegraphics[width=\textwidth]{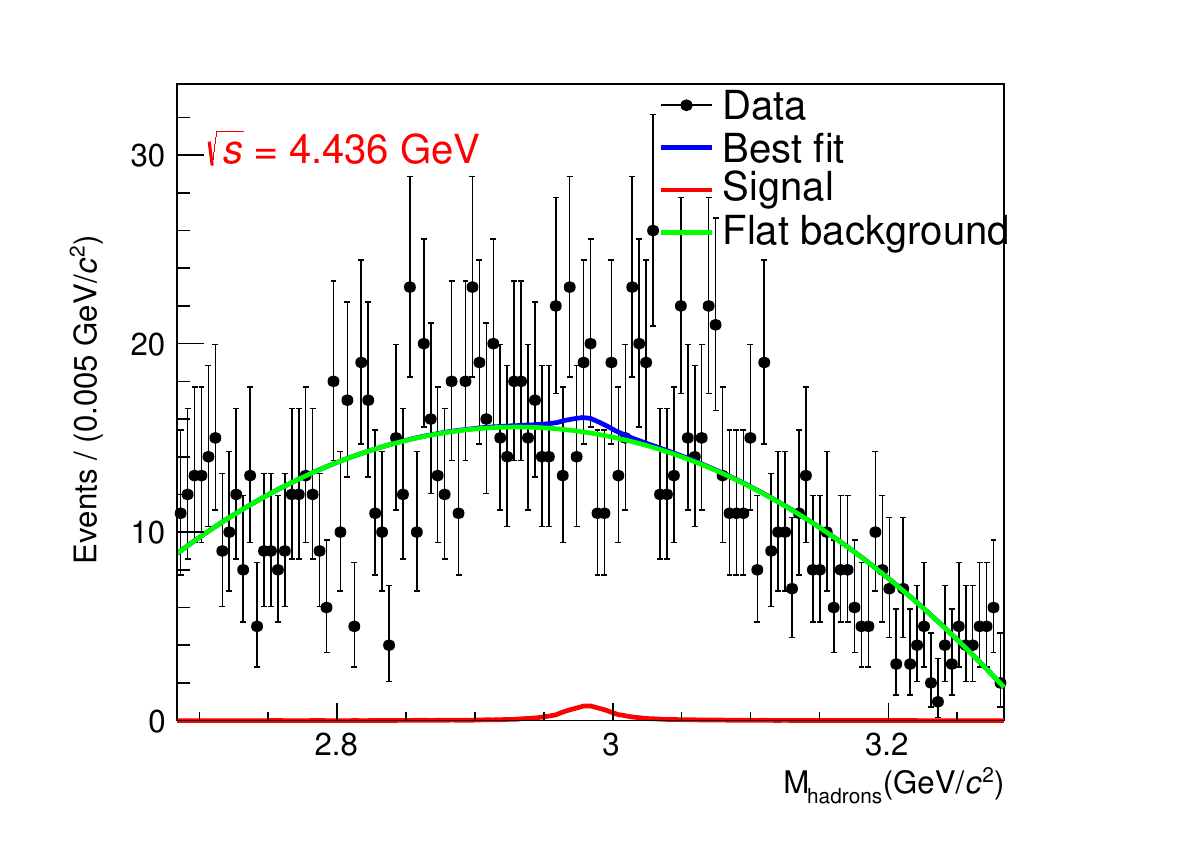}
  \captionsetup{skip=-10pt,font=large}
\end{subfigure}
  \begin{subfigure}{0.24\textwidth}
  \centering
  \includegraphics[width=\textwidth]{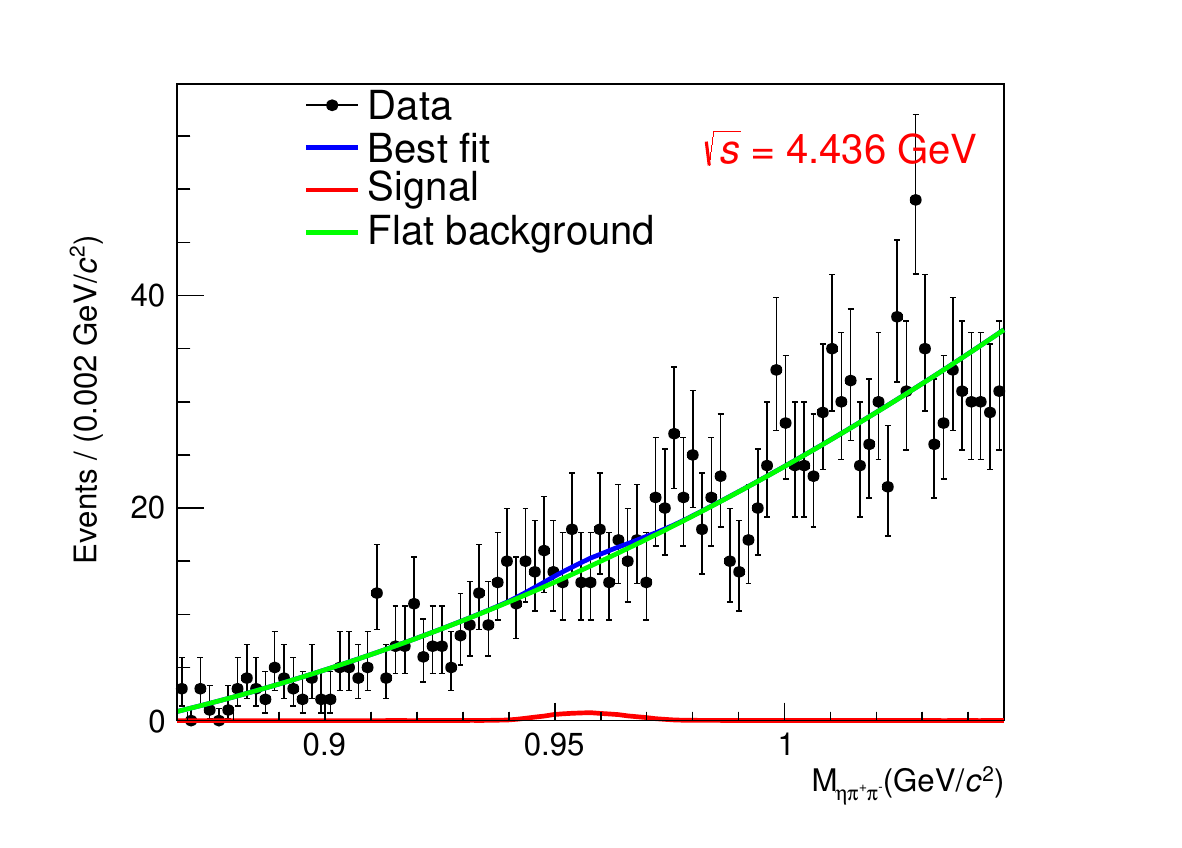}
  \captionsetup{skip=-10pt,font=large}
\end{subfigure}
\begin{subfigure}{0.24\textwidth}
  \centering
  \includegraphics[width=\textwidth]{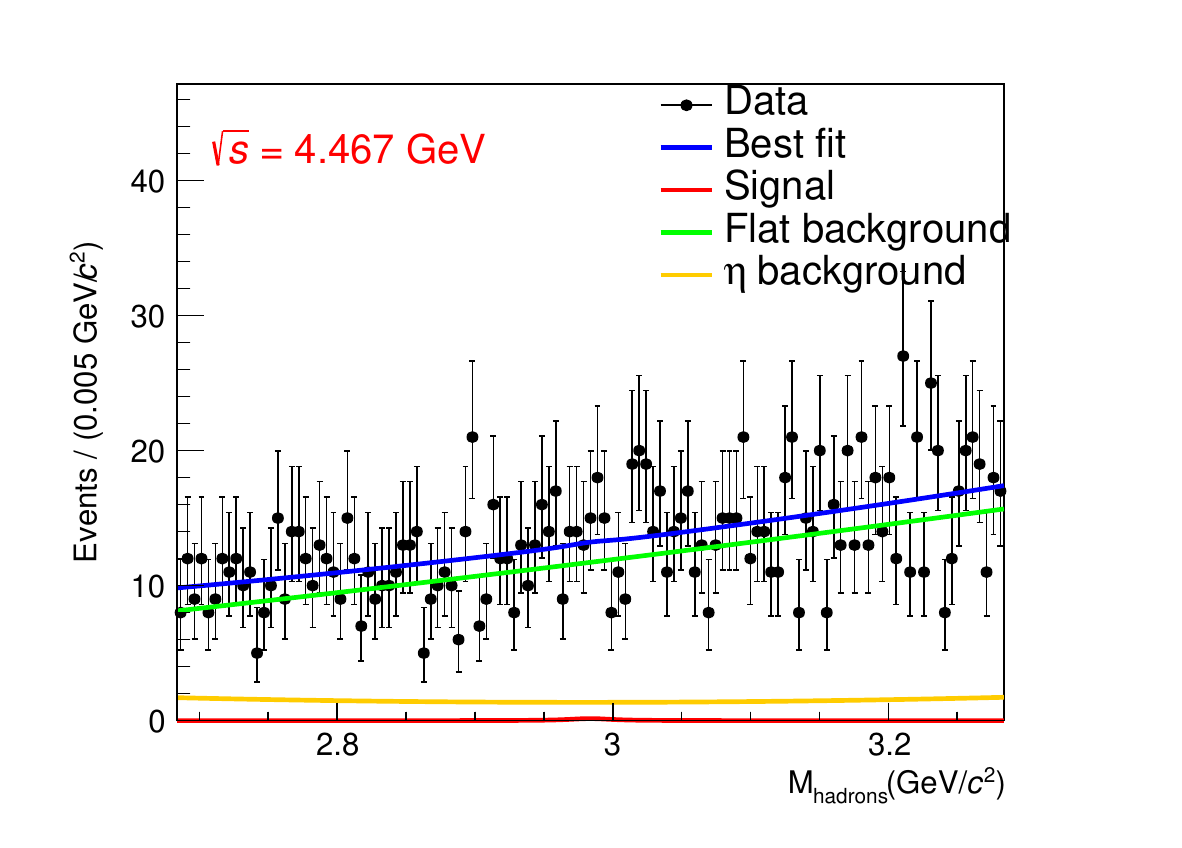}
  \captionsetup{skip=-10pt,font=large}
\end{subfigure}
\begin{subfigure}{0.24\textwidth}
  \centering
  \includegraphics[width=\textwidth]{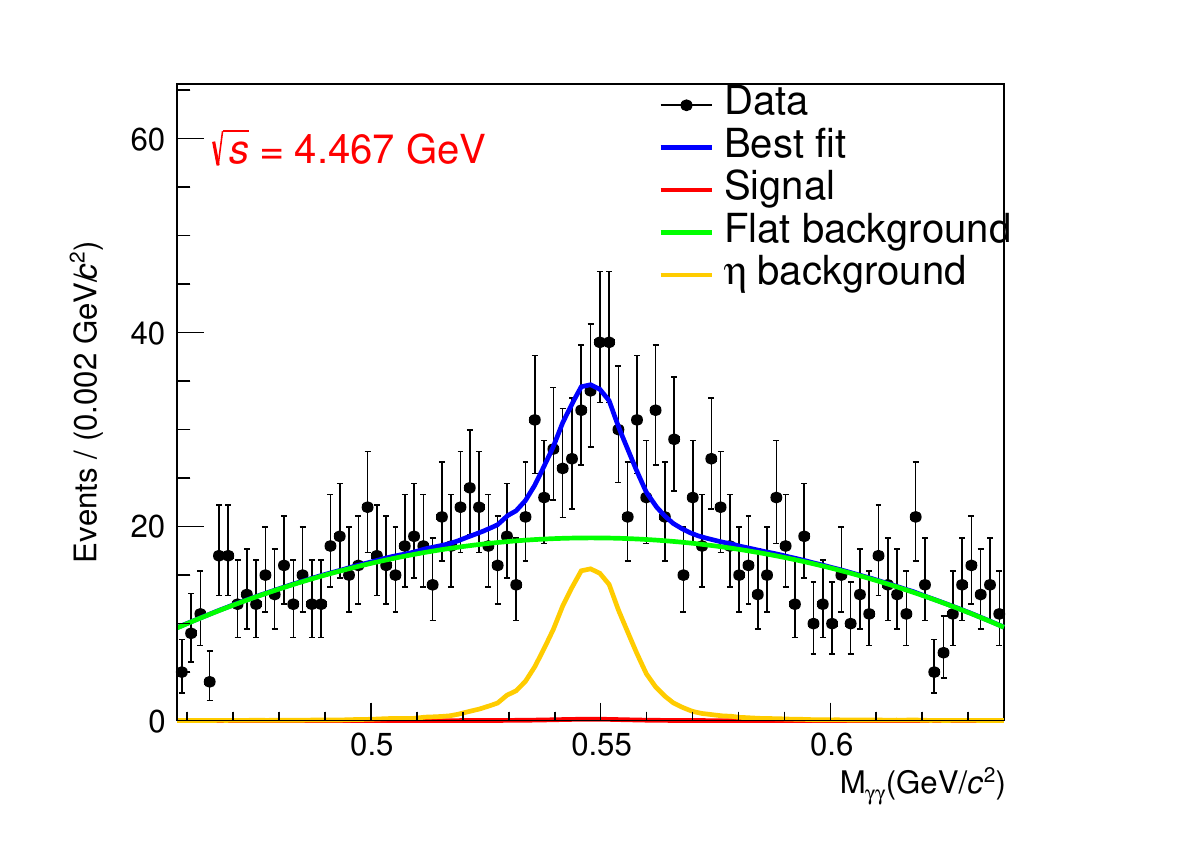}
  \captionsetup{skip=-10pt,font=large}
\end{subfigure}
 \begin{subfigure}{0.24\textwidth}
  \centering
  \includegraphics[width=\textwidth]{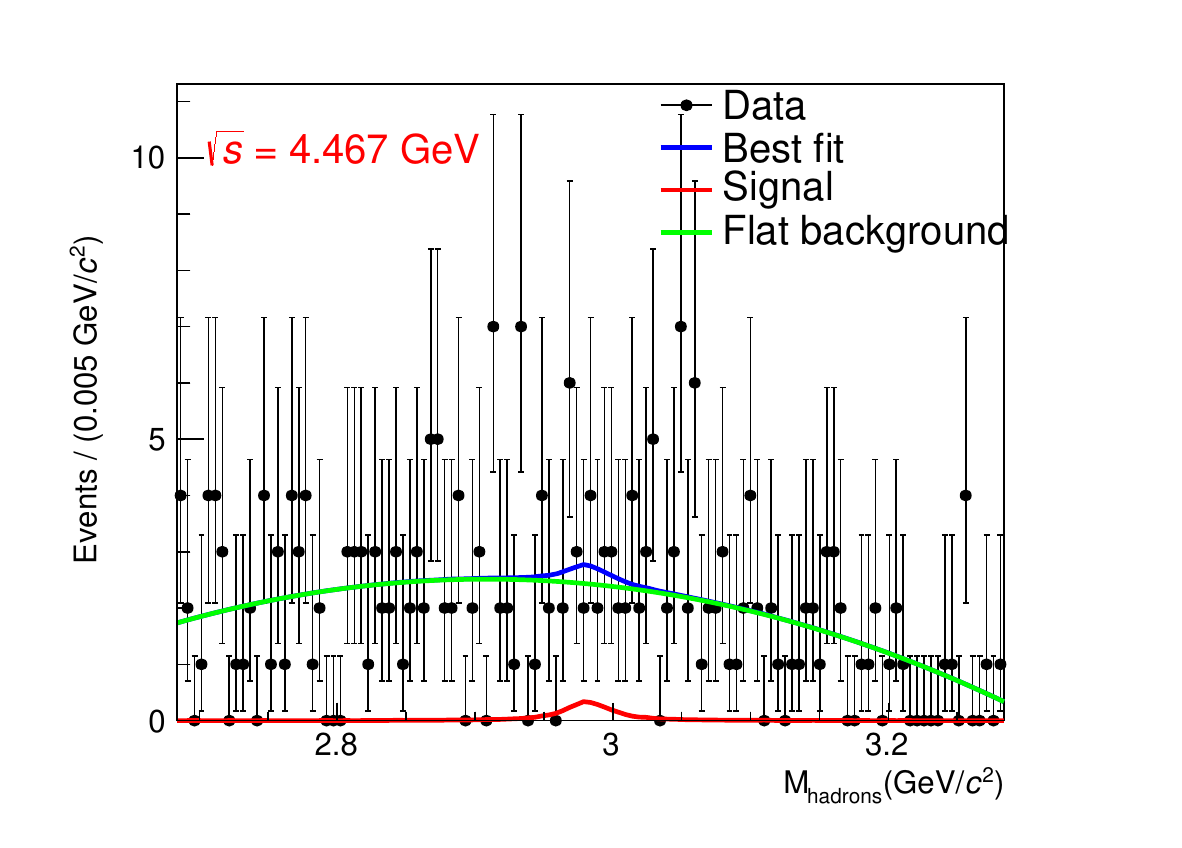}
  \captionsetup{skip=-10pt,font=large}
\end{subfigure}
  \begin{subfigure}{0.24\textwidth}
  \centering
  \includegraphics[width=\textwidth]{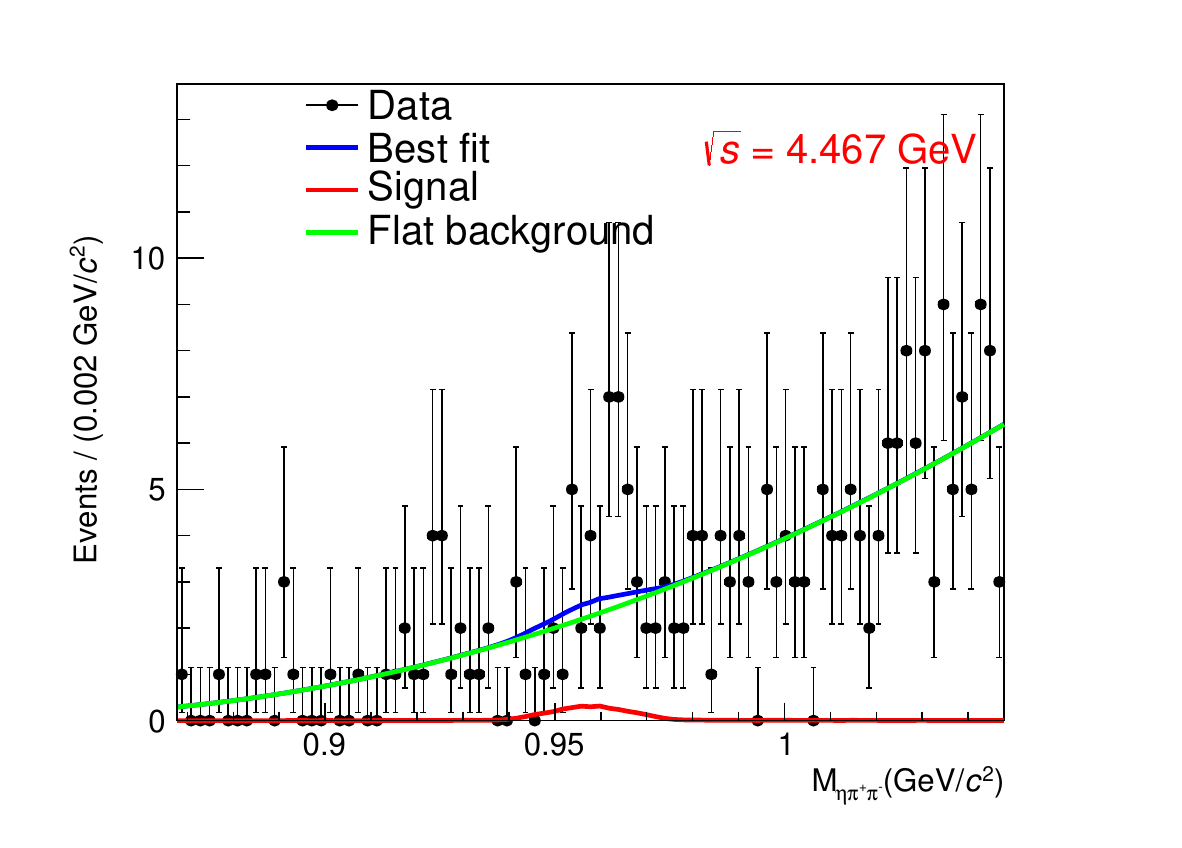}
  \captionsetup{skip=-10pt,font=large}
\end{subfigure}
\begin{subfigure}{0.24\textwidth}
  \centering
  \includegraphics[width=\textwidth]{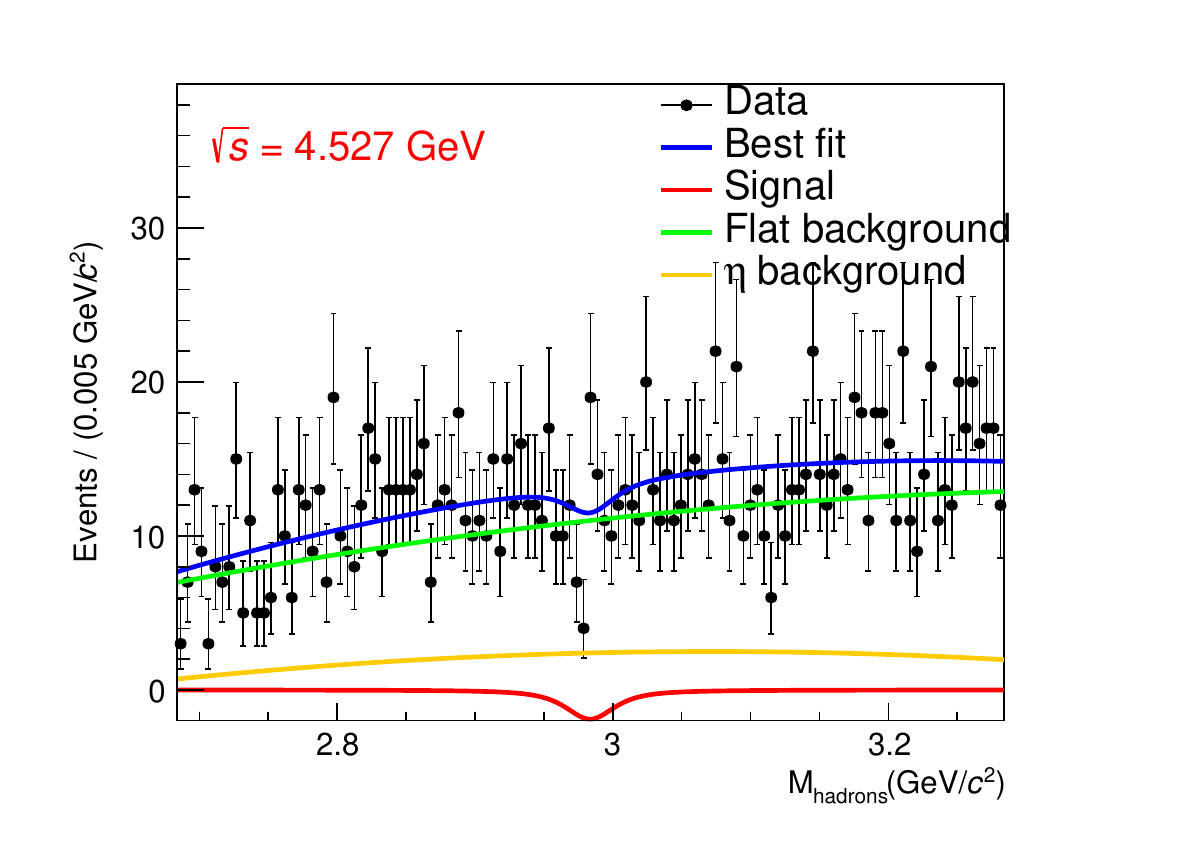}
  \captionsetup{skip=-10pt,font=large}
\end{subfigure}
\begin{subfigure}{0.24\textwidth}
  \centering
  \includegraphics[width=\textwidth]{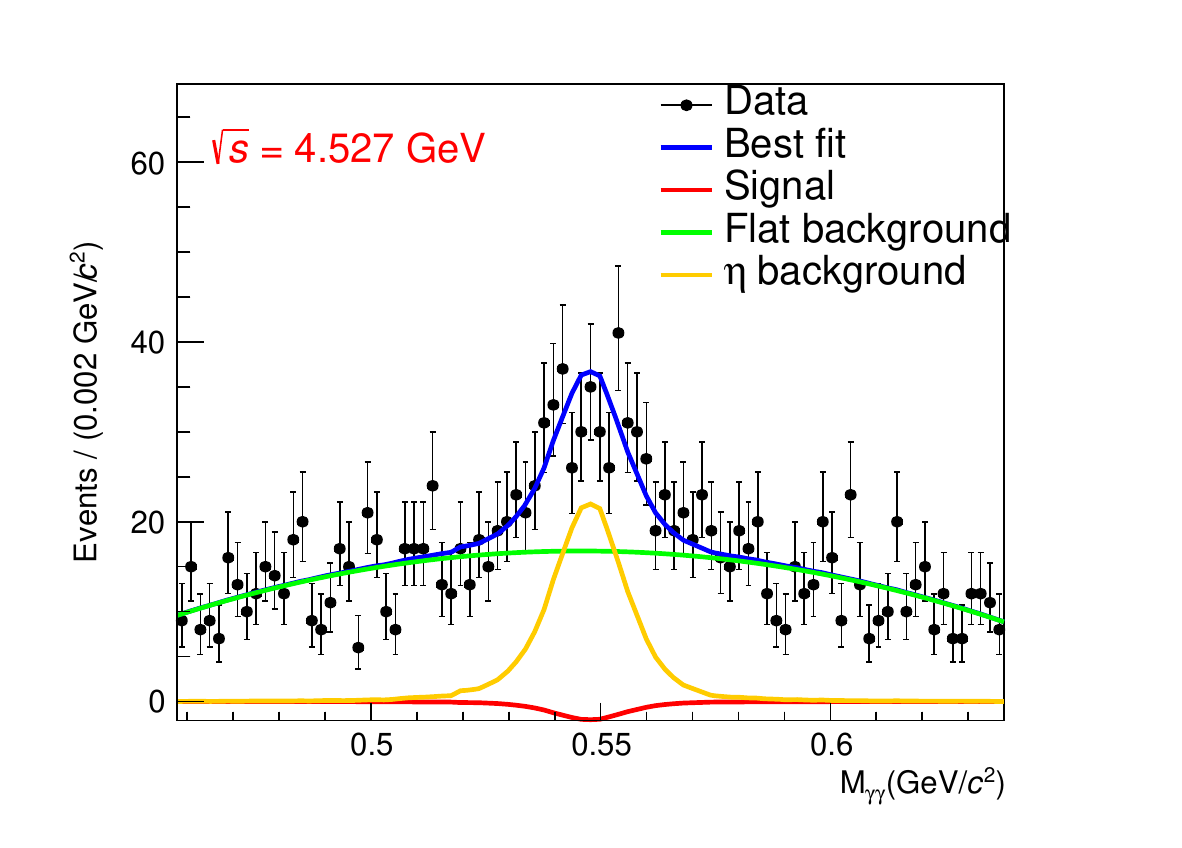}
  \captionsetup{skip=-10pt,font=large}
\end{subfigure}
 \begin{subfigure}{0.24\textwidth}
  \centering
  \includegraphics[width=\textwidth]{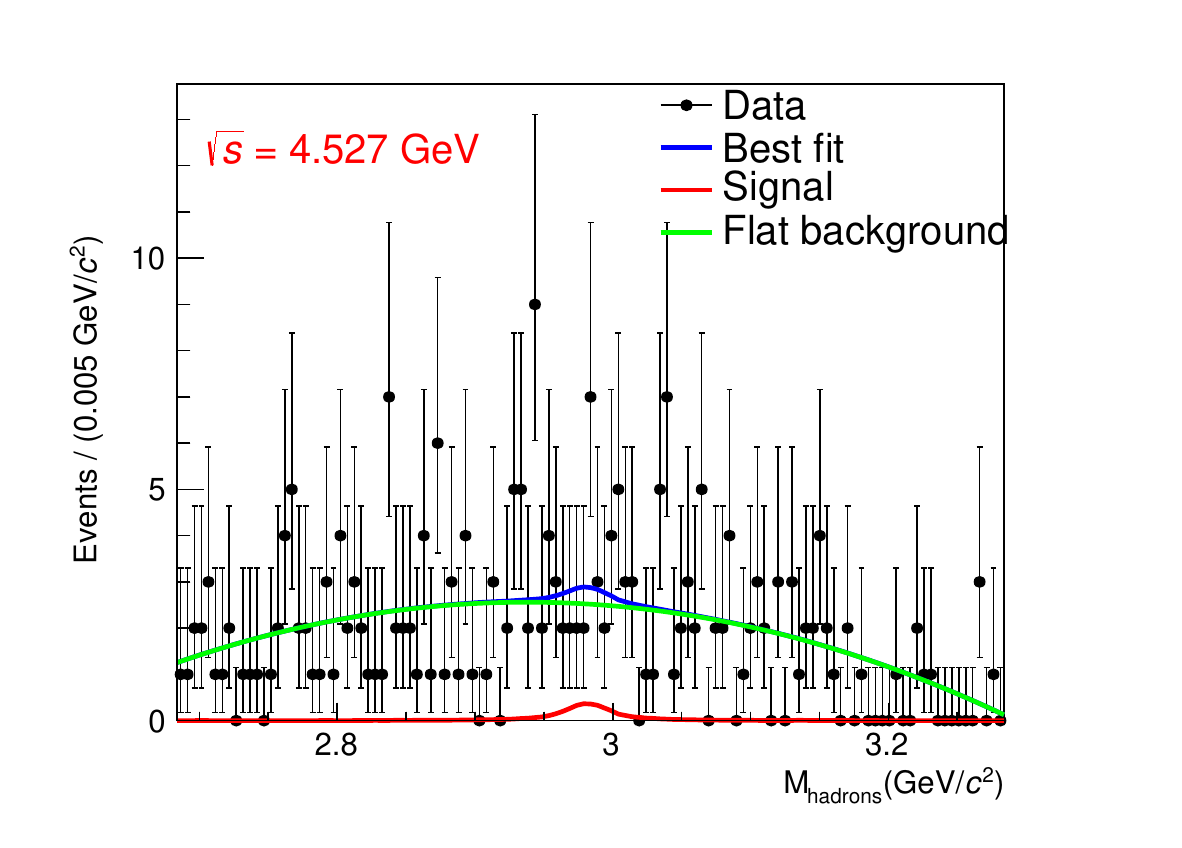}
  \captionsetup{skip=-10pt,font=large}
\end{subfigure}
  \begin{subfigure}{0.24\textwidth}
  \centering
  \includegraphics[width=\textwidth]{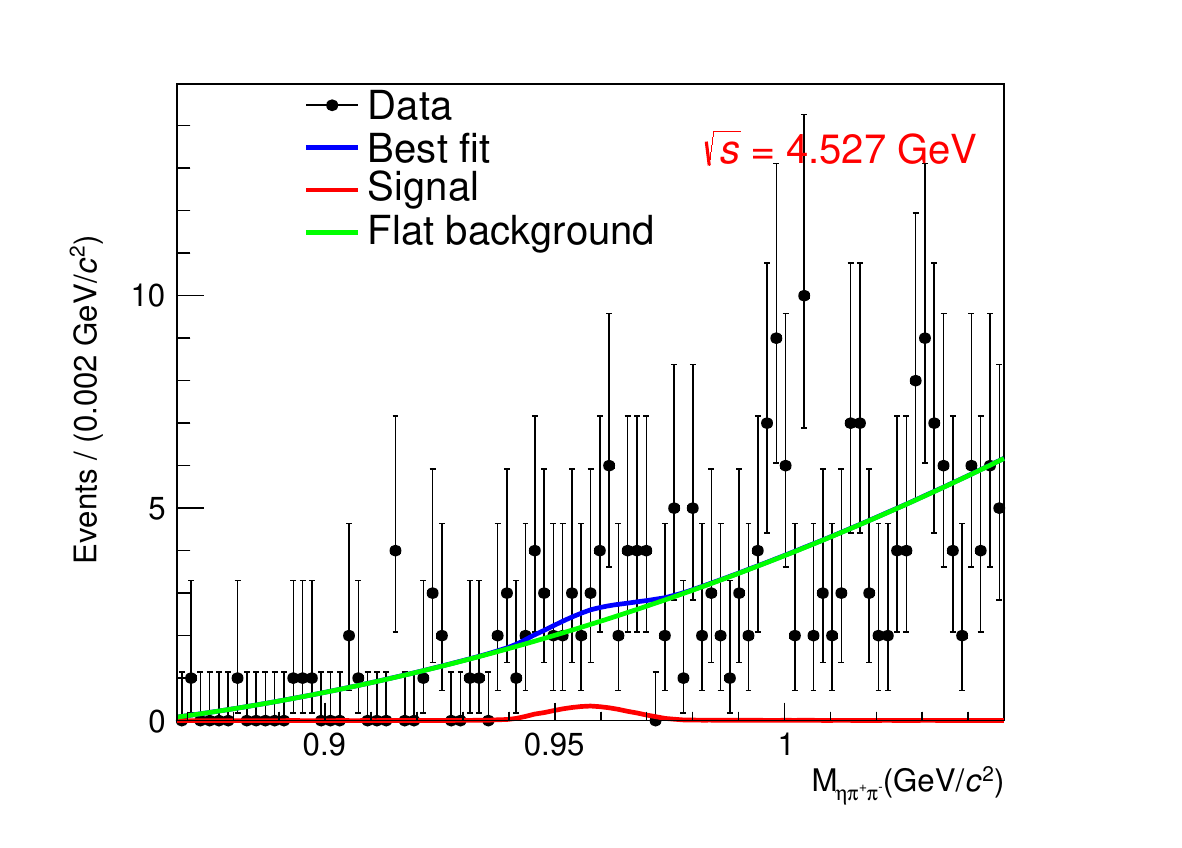}
  \captionsetup{skip=-10pt,font=large}
\end{subfigure}
\captionsetup{justification=raggedright}
\caption{Fits to the invariant mass distributions of (Left)(Middle-Right)$M(hadrons)$, (Middle-Left)$M(\gamma\gamma)$ and (Right)$M(\eta\pi^{+}\pi^{-})$ at $\sqrt(s)=4.396-4.527$~GeV.}
\label{fig:fit3}
\end{figure*}

Figs~\ref{fig:normal1}-\ref{fig:normal6} display the distributions of $M_{\gamma\gamma}$ versus $M_{hadrons}$, $M_{hadrons}$, and $M_{\gamma\gamma}$ for each data sample. Figs~\ref{fig:normal7}-\ref{fig:normal12} display the distributions of $M_{\eta\pippim}$ versus $M_{hadrons}$, $M_{hadrons}$, and $M_{\eta\pippim}$ for each data sample. And Figs~\ref{fig:fit1}-\ref{fig:fit3} are the fit results to the invariant mass distributions for each data sample.



\begin{figure*}[htbp]
\begin{subfigure}{0.24\textwidth}
  \centering
  \includegraphics[width=\textwidth]{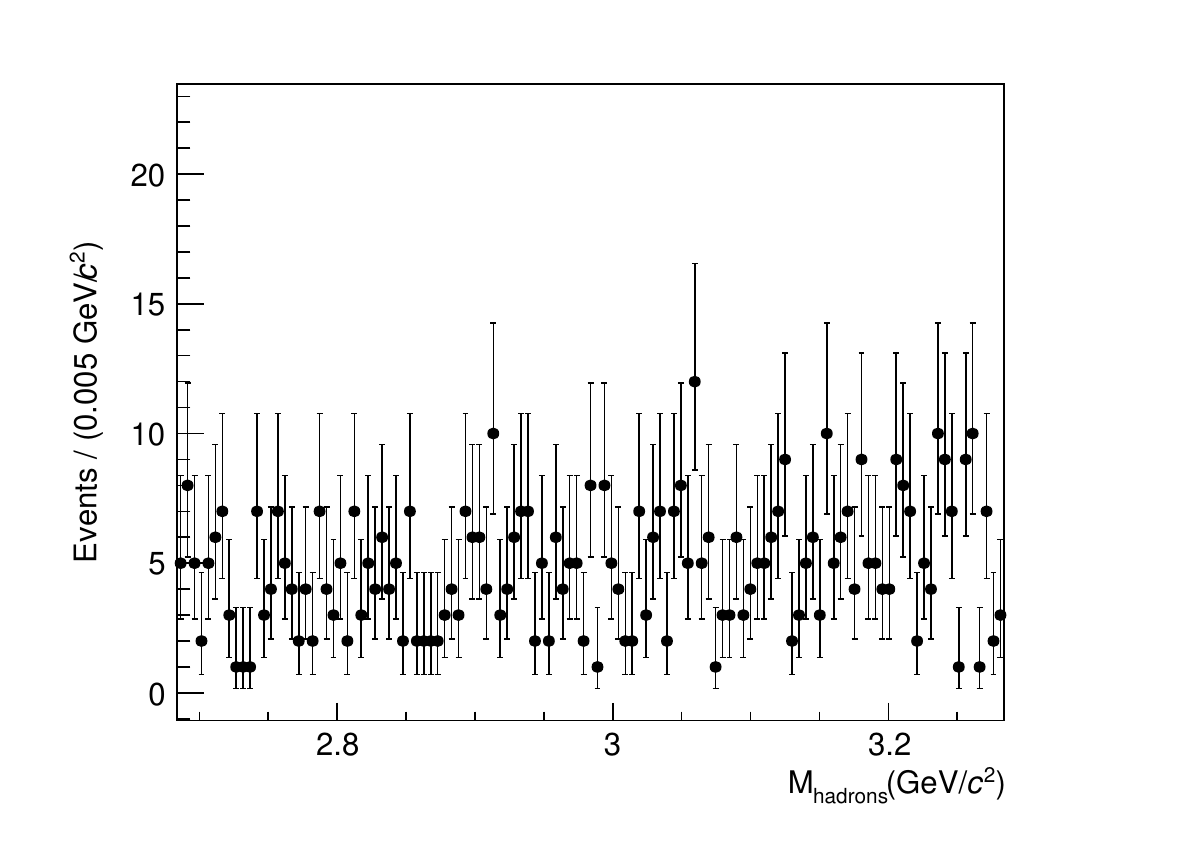}
  \captionsetup{skip=-10pt,font=large}
\end{subfigure}
\begin{subfigure}{0.24\textwidth}
  \centering
  \includegraphics[width=\textwidth]{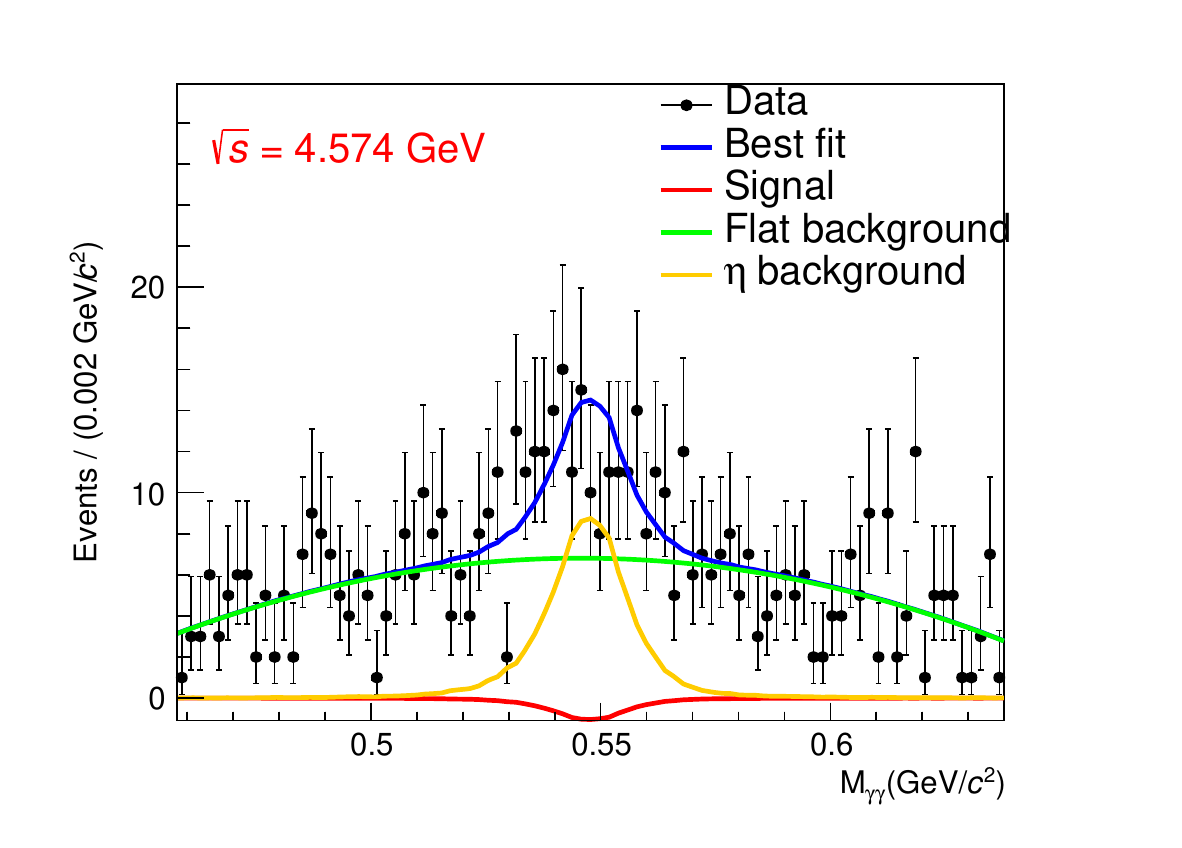}
  \captionsetup{skip=-10pt,font=large}
\end{subfigure}
 \begin{subfigure}{0.24\textwidth}
  \centering
  \includegraphics[width=\textwidth]{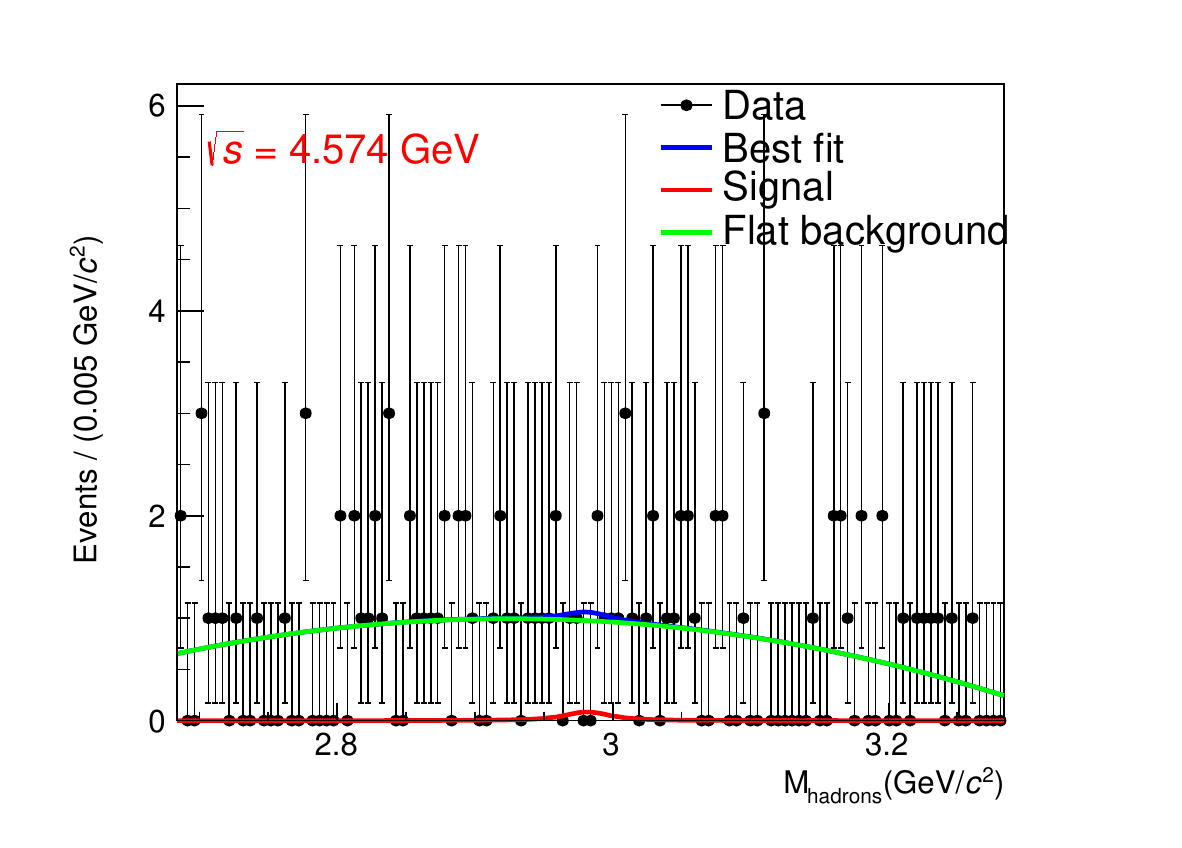}
  \captionsetup{skip=-10pt,font=large}
\end{subfigure}
  \begin{subfigure}{0.24\textwidth}
  \centering
  \includegraphics[width=\textwidth]{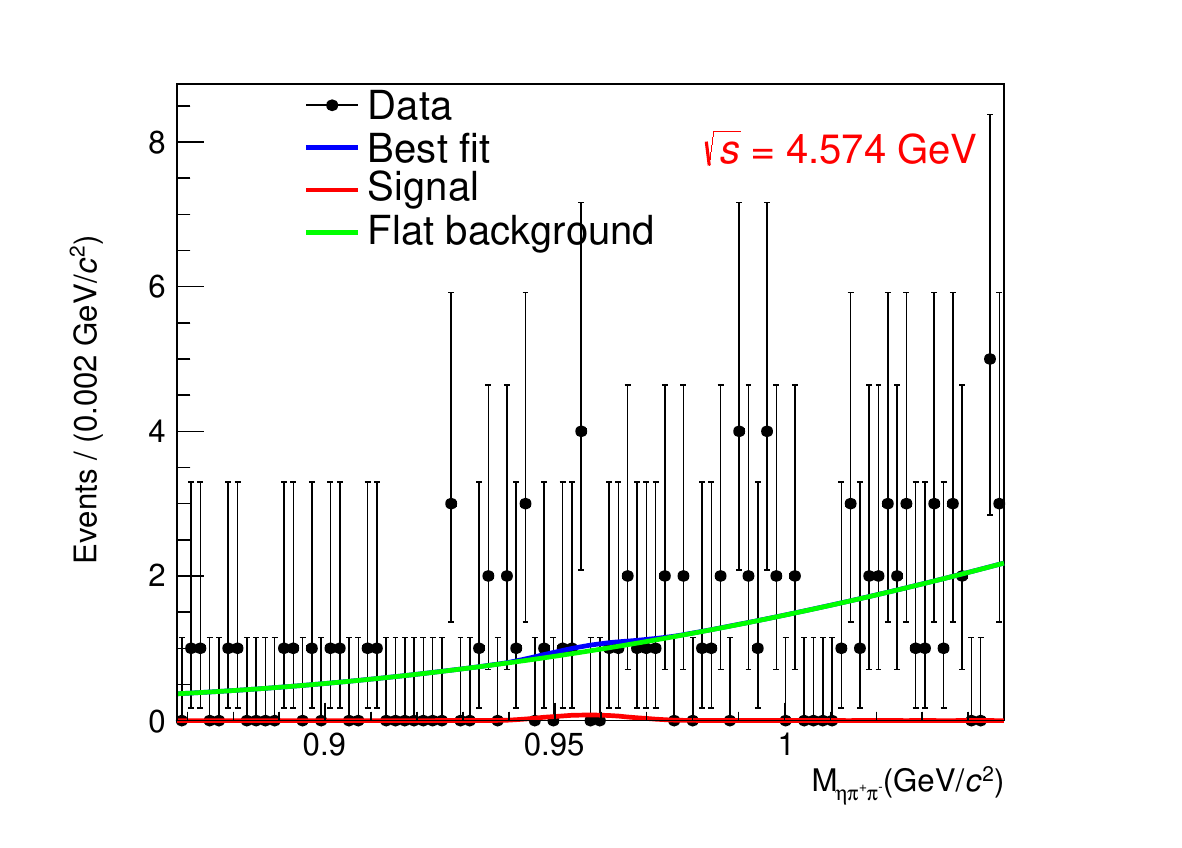}
  \captionsetup{skip=-10pt,font=large}
\end{subfigure}
\begin{subfigure}{0.24\textwidth}
  \centering
  \includegraphics[width=\textwidth]{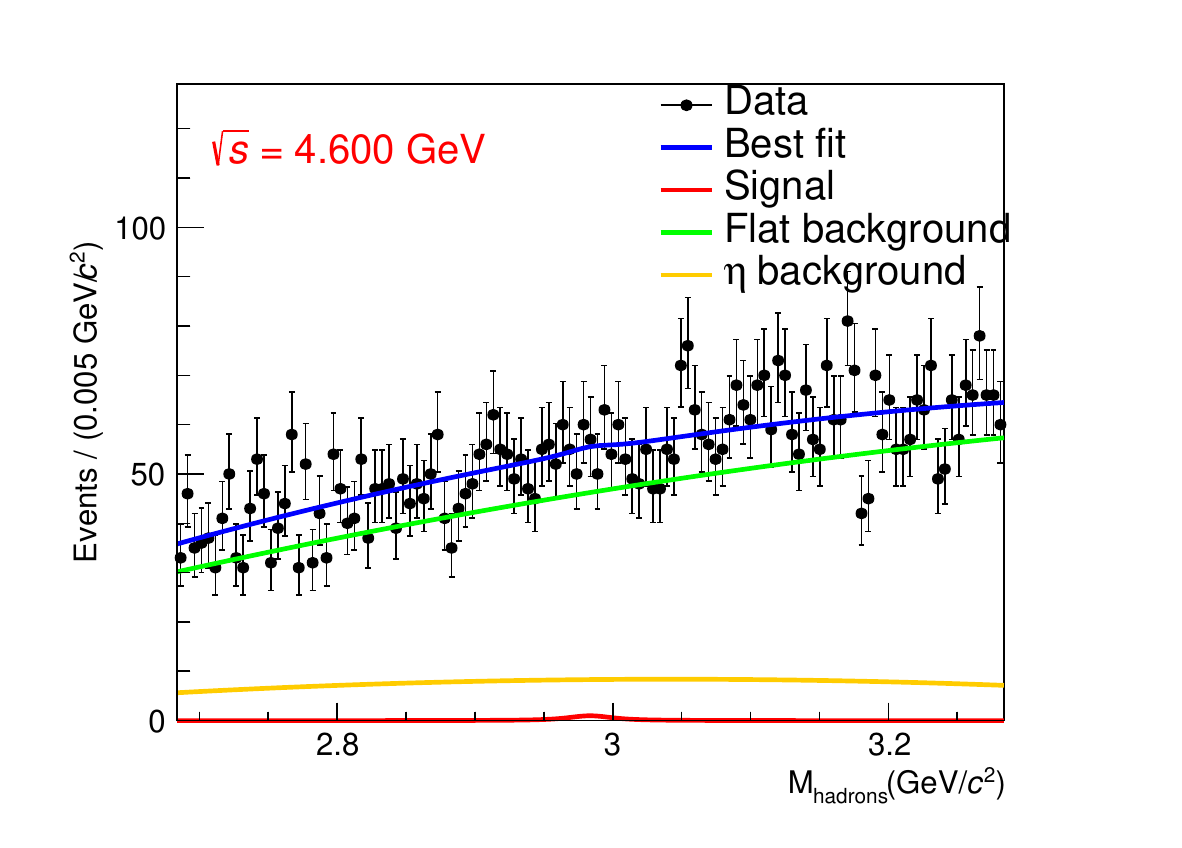}
  \captionsetup{skip=-10pt,font=large}
\end{subfigure}
\begin{subfigure}{0.24\textwidth}
  \centering
  \includegraphics[width=\textwidth]{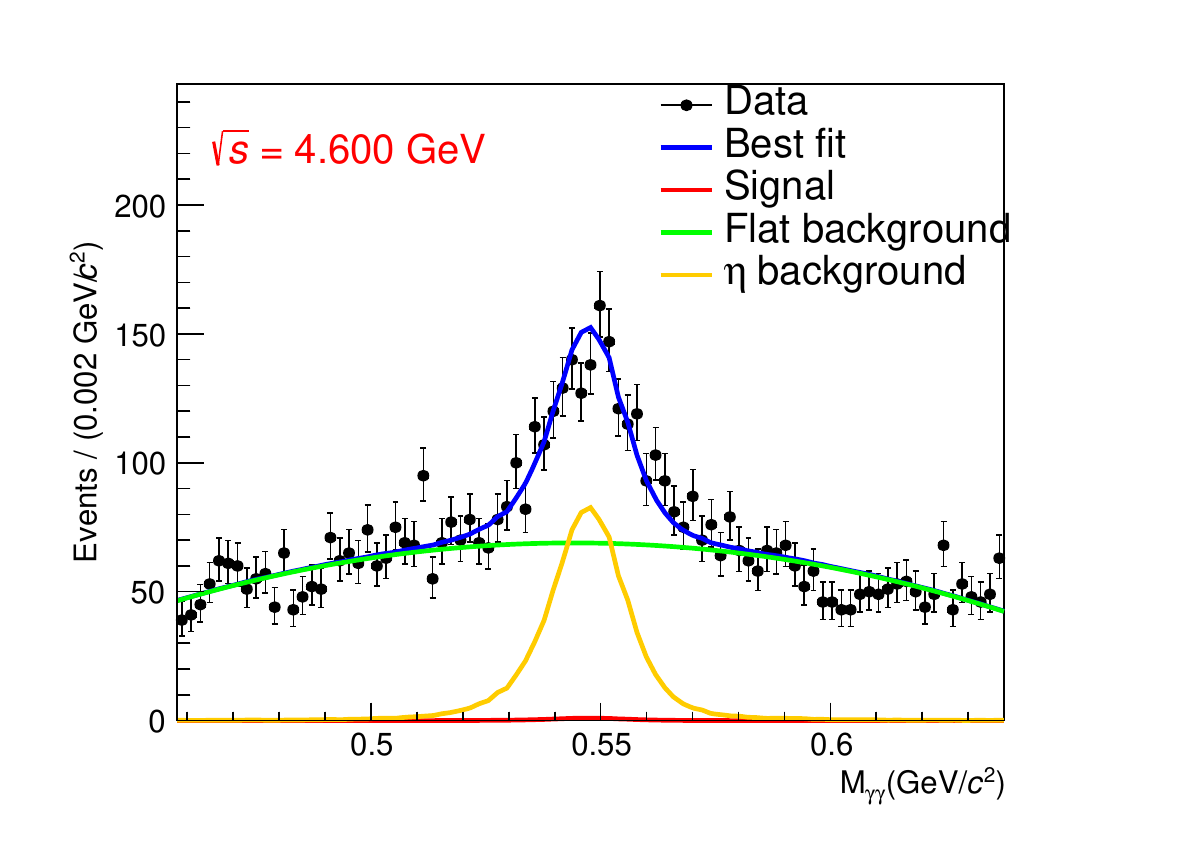}
  \captionsetup{skip=-10pt,font=large}
\end{subfigure}
 \begin{subfigure}{0.24\textwidth}
  \centering
  \includegraphics[width=\textwidth]{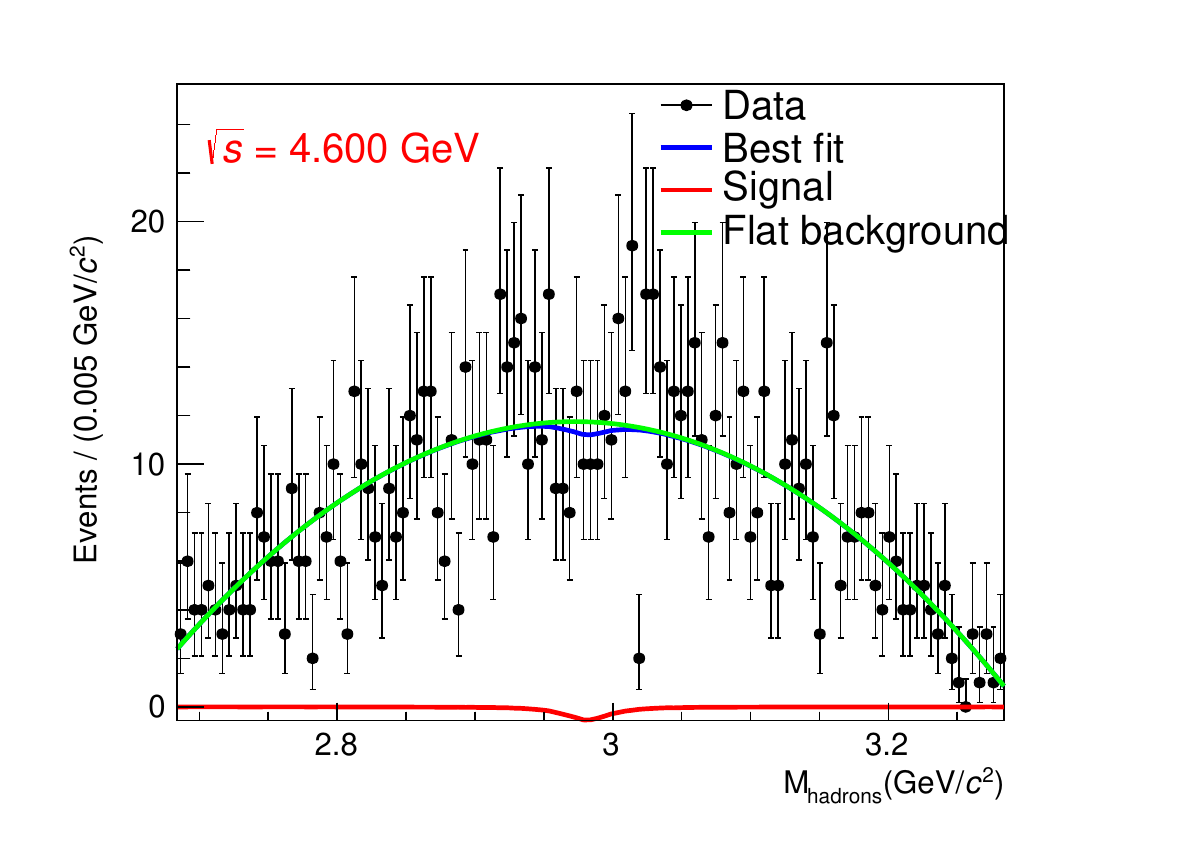}
  \captionsetup{skip=-10pt,font=large}
\end{subfigure}
  \begin{subfigure}{0.24\textwidth}
  \centering
  \includegraphics[width=\textwidth]{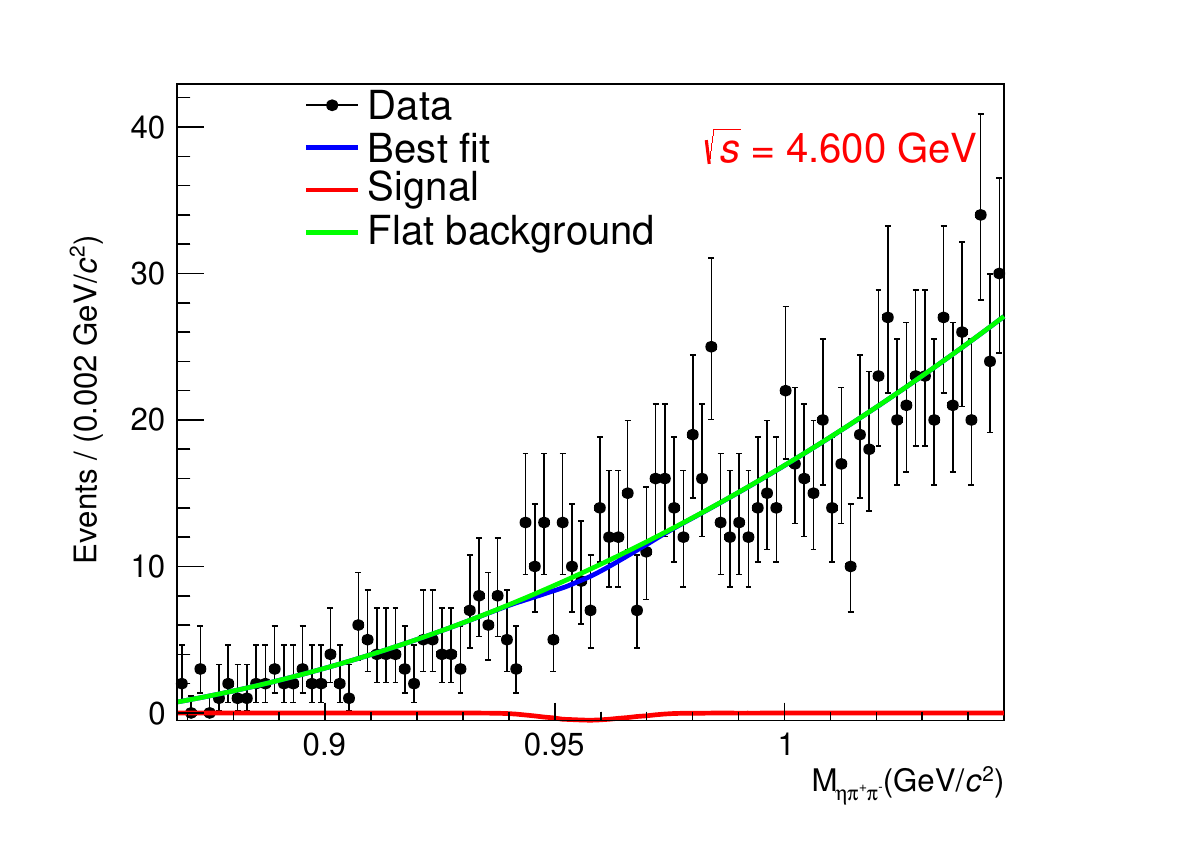}
  \captionsetup{skip=-10pt,font=large}
\end{subfigure}
\begin{subfigure}{0.24\textwidth}
  \centering
  \includegraphics[width=\textwidth]{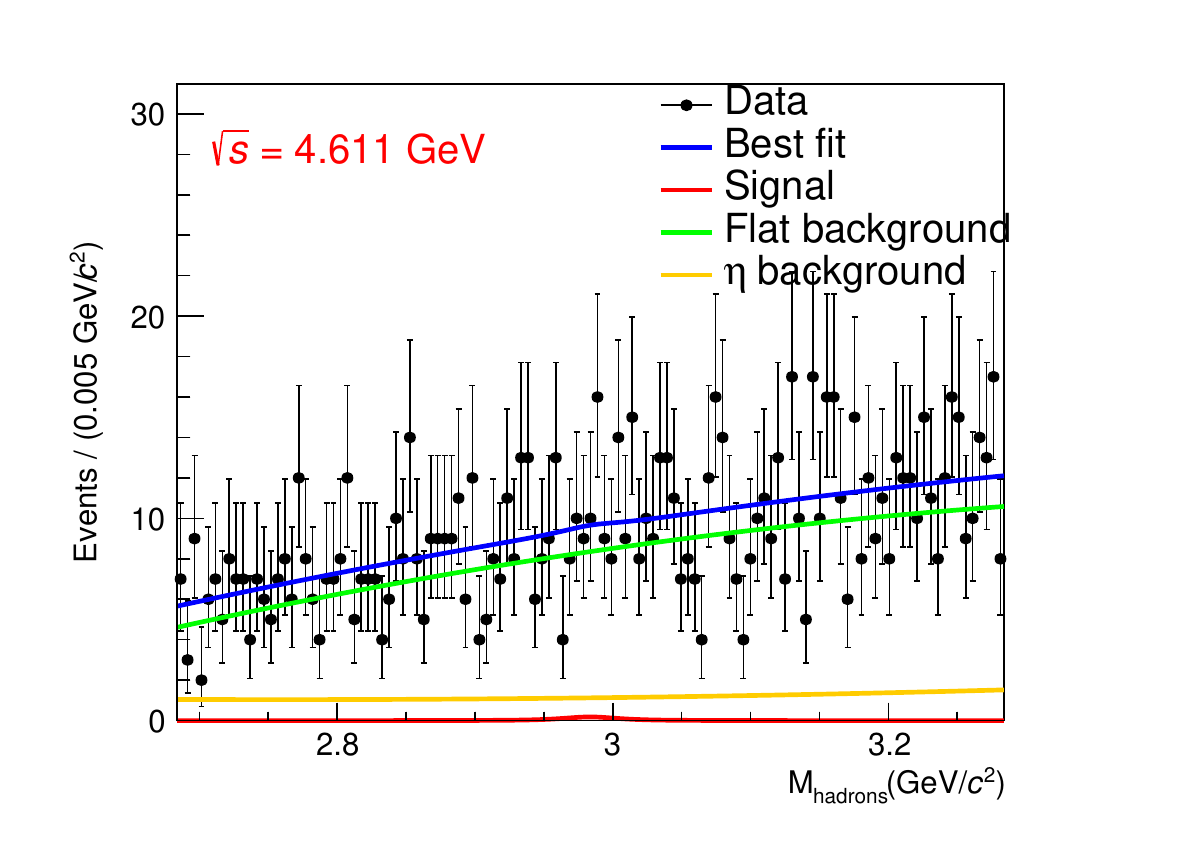}
  \captionsetup{skip=-10pt,font=large}
\end{subfigure}
\begin{subfigure}{0.24\textwidth}
  \centering
  \includegraphics[width=\textwidth]{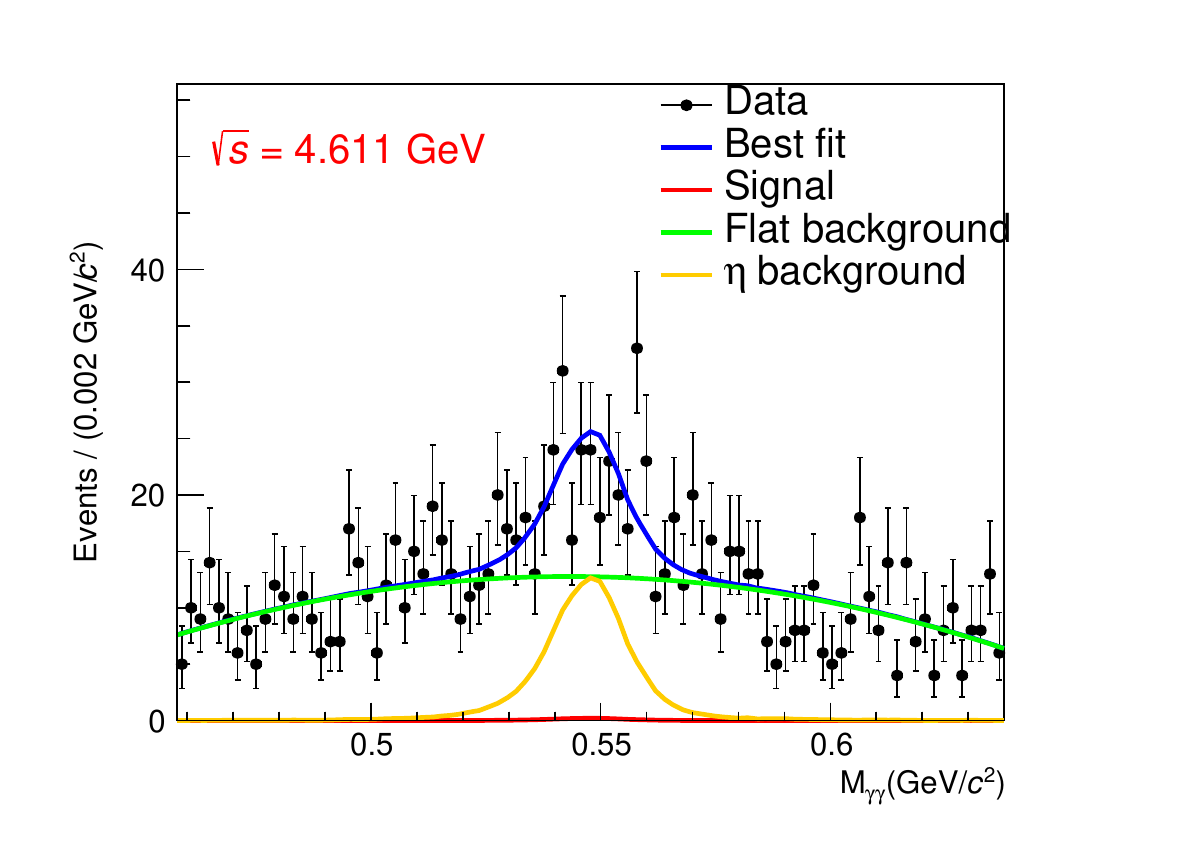}
  \captionsetup{skip=-10pt,font=large}
\end{subfigure}
 \begin{subfigure}{0.24\textwidth}
  \centering
  \includegraphics[width=\textwidth]{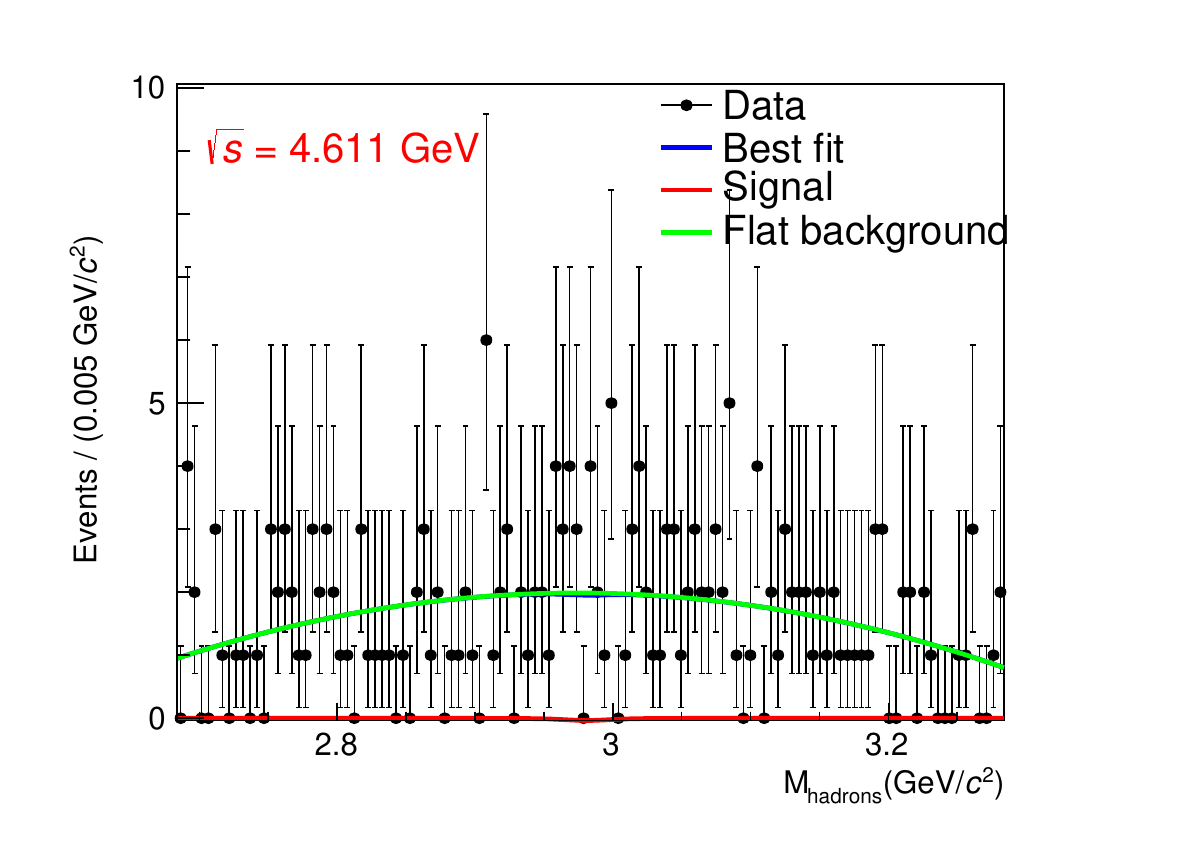}
  \captionsetup{skip=-10pt,font=large}
\end{subfigure}
  \begin{subfigure}{0.24\textwidth}
  \centering
  \includegraphics[width=\textwidth]{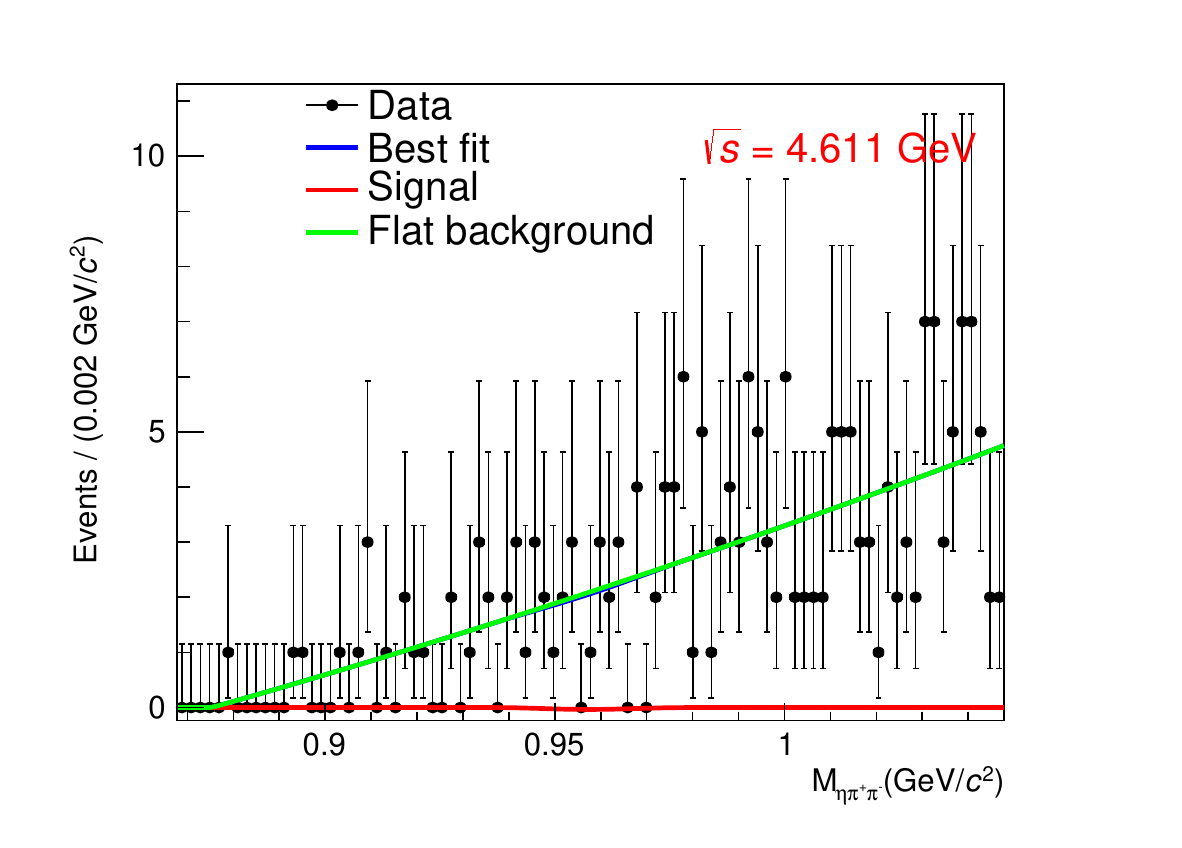}
  \captionsetup{skip=-10pt,font=large}
\end{subfigure}
\begin{subfigure}{0.24\textwidth}
  \centering
  \includegraphics[width=\textwidth]{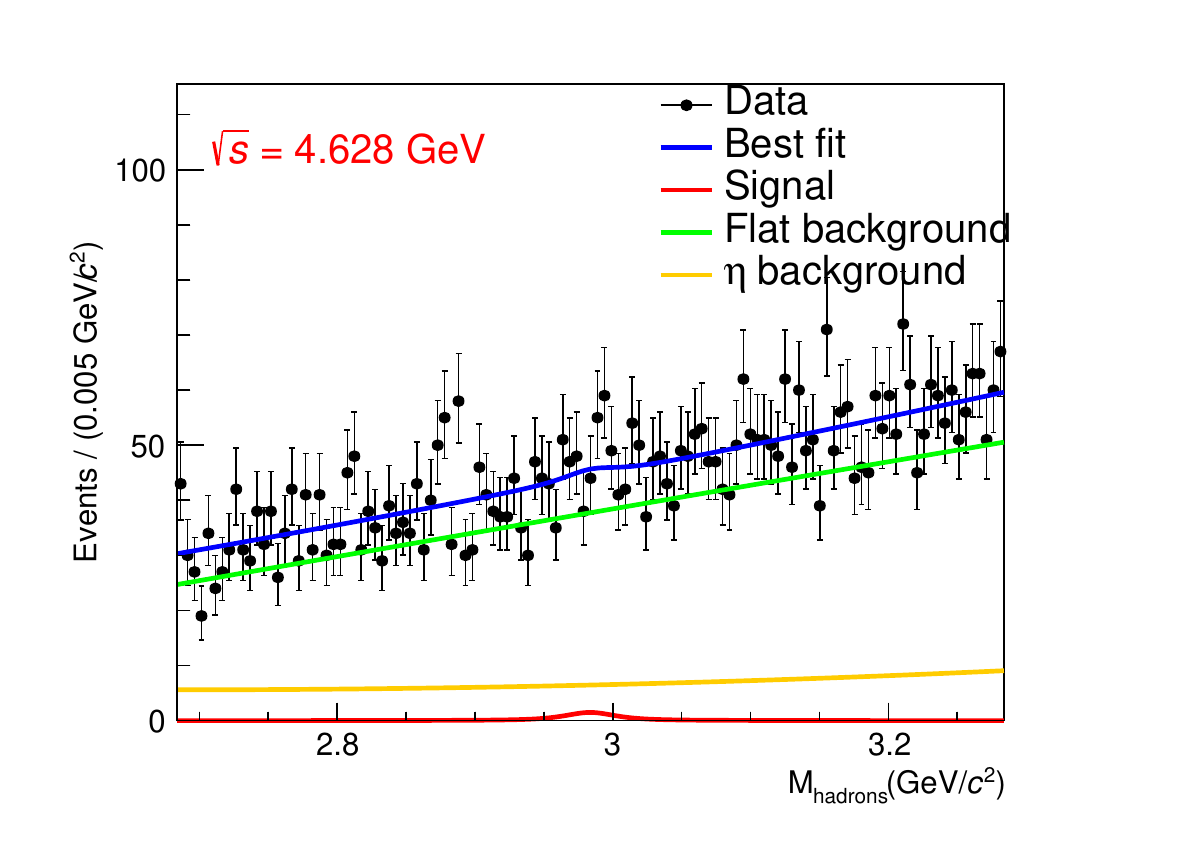}
  \captionsetup{skip=-10pt,font=large}
\end{subfigure}
\begin{subfigure}{0.24\textwidth}
  \centering
  \includegraphics[width=\textwidth]{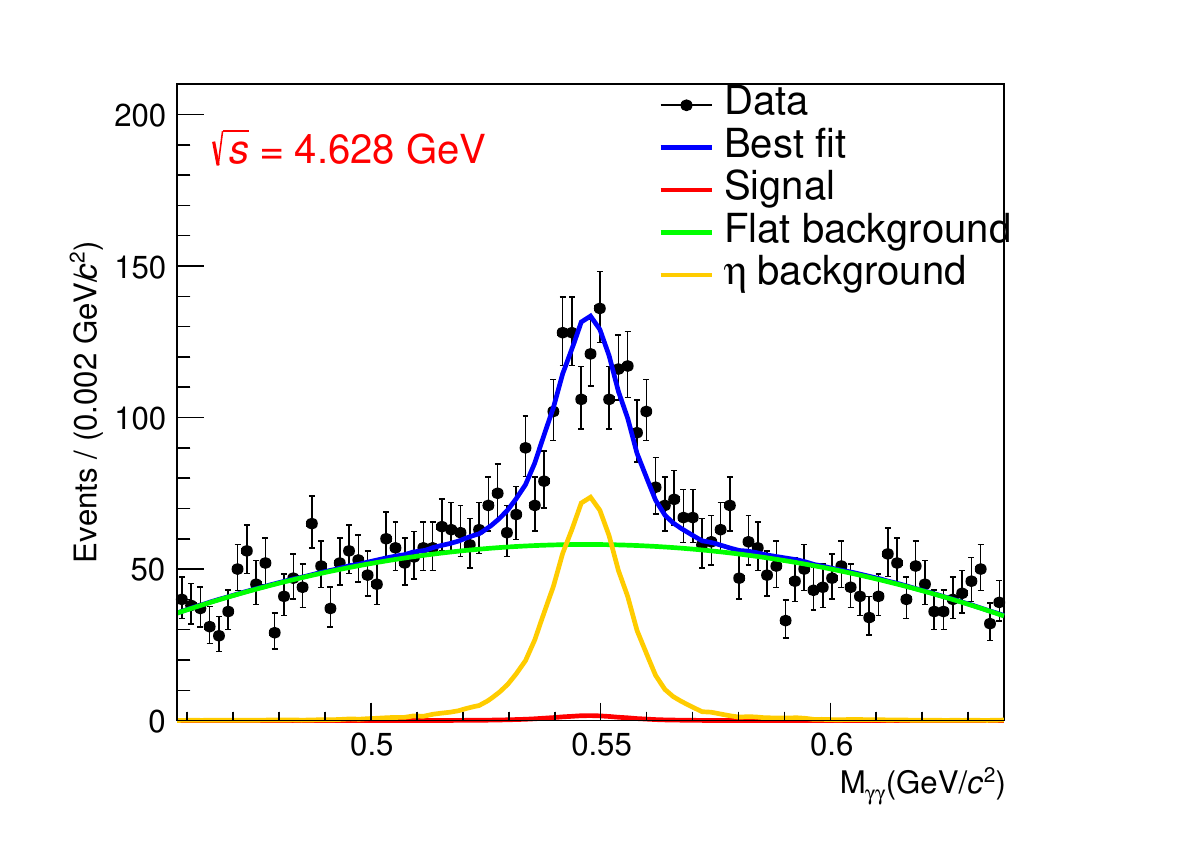}
  \captionsetup{skip=-10pt,font=large}
\end{subfigure}
 \begin{subfigure}{0.24\textwidth}
  \centering
  \includegraphics[width=\textwidth]{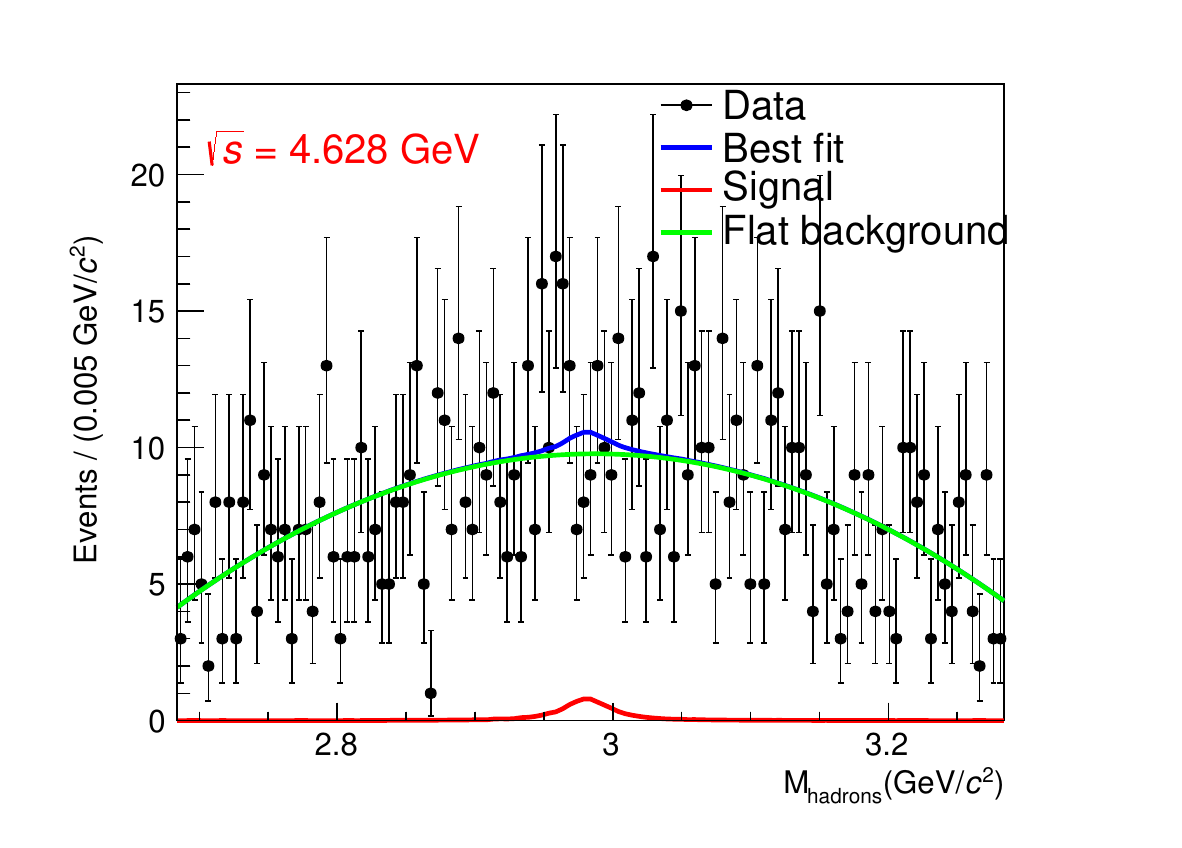}
  \captionsetup{skip=-10pt,font=large}
\end{subfigure}
  \begin{subfigure}{0.24\textwidth}
  \centering
  \includegraphics[width=\textwidth]{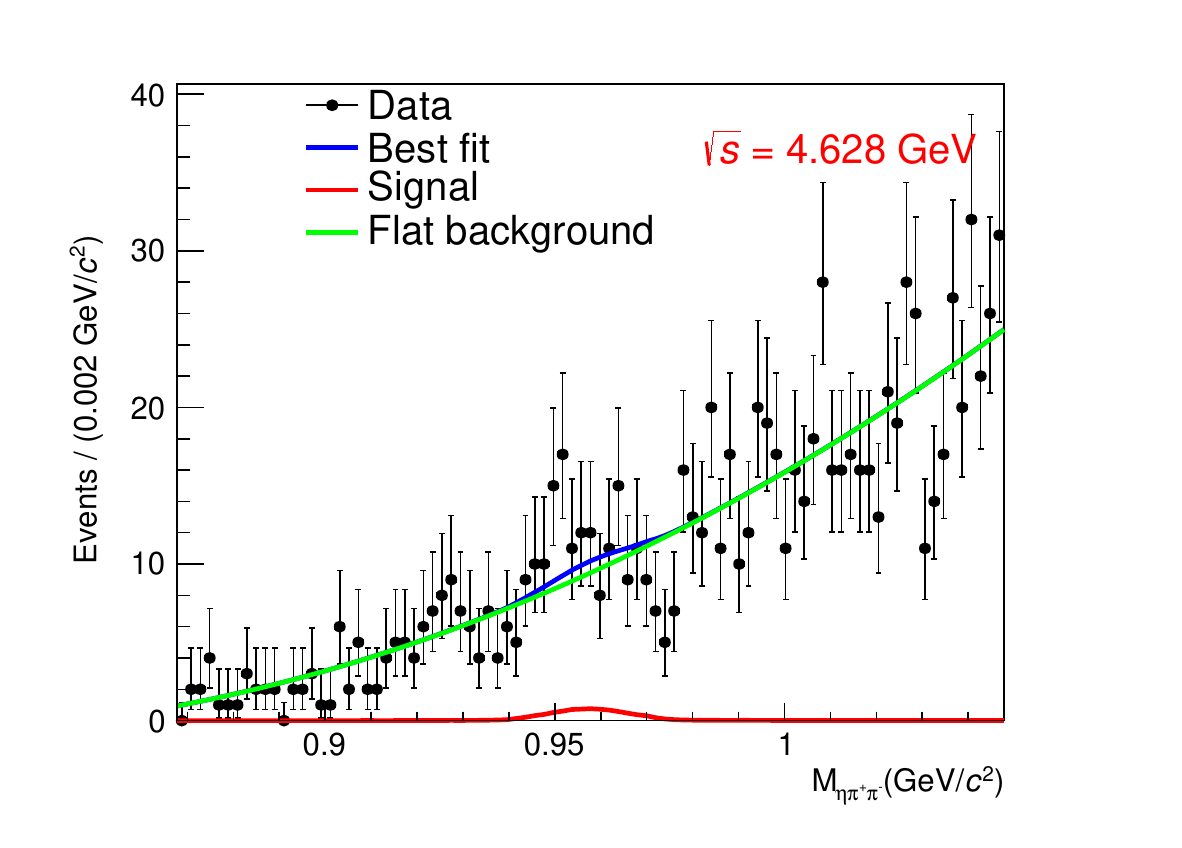}
  \captionsetup{skip=-10pt,font=large}
\end{subfigure}
\begin{subfigure}{0.24\textwidth}
  \centering
  \includegraphics[width=\textwidth]{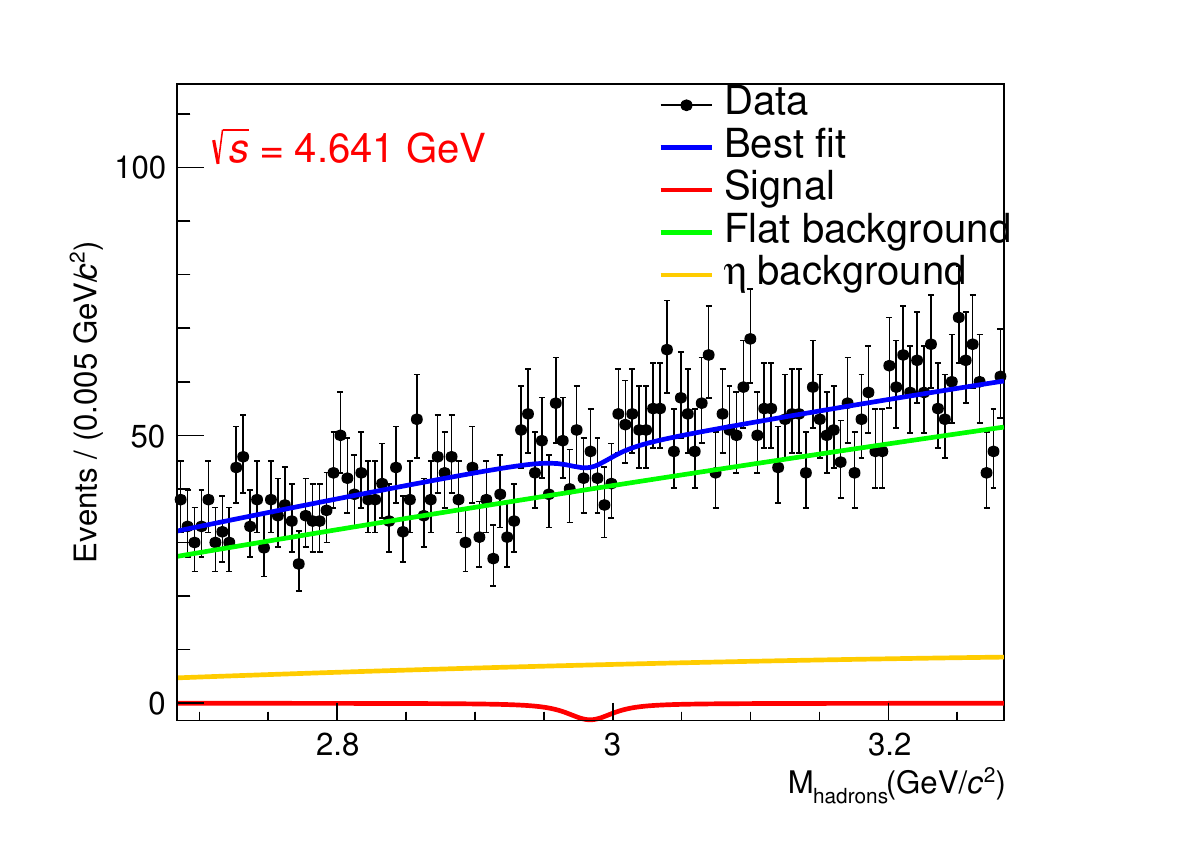}
  \captionsetup{skip=-10pt,font=large}
\end{subfigure}
\begin{subfigure}{0.24\textwidth}
  \centering
  \includegraphics[width=\textwidth]{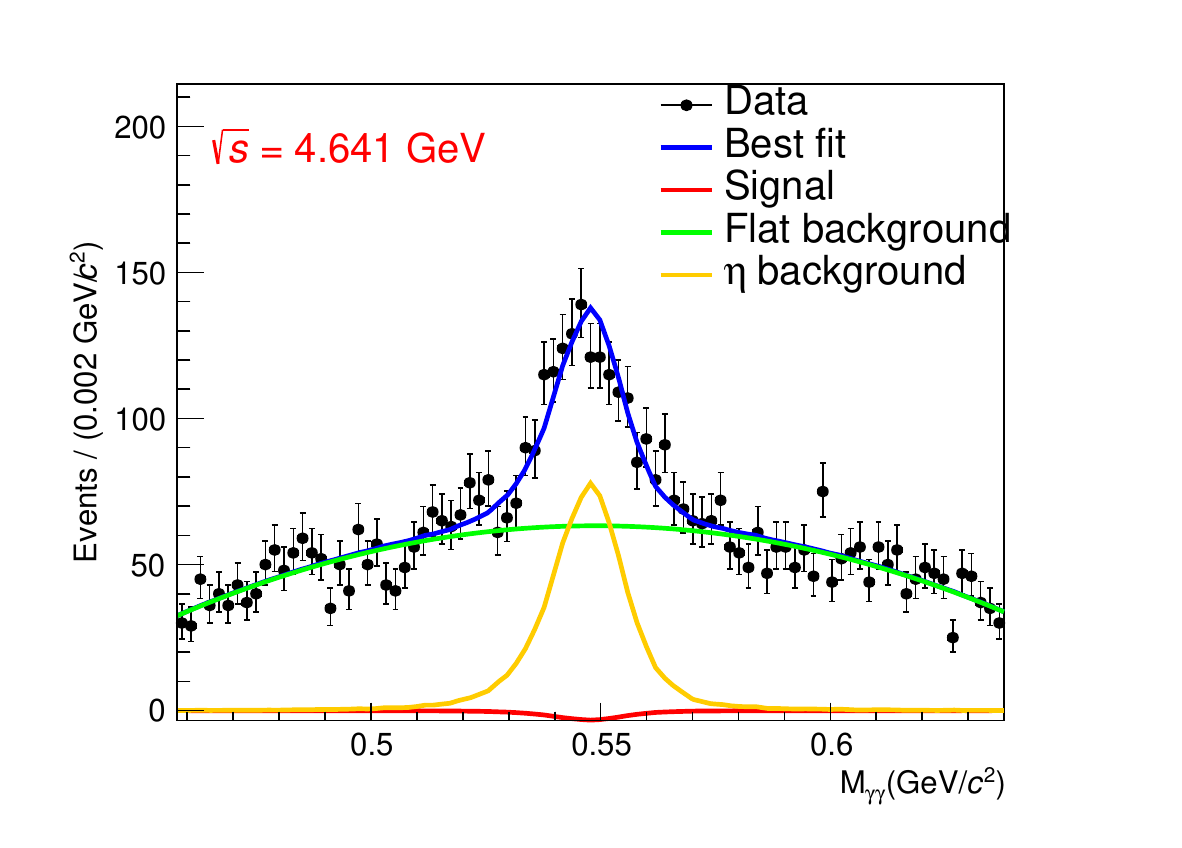}
  \captionsetup{skip=-10pt,font=large}
\end{subfigure}
 \begin{subfigure}{0.24\textwidth}
  \centering
  \includegraphics[width=\textwidth]{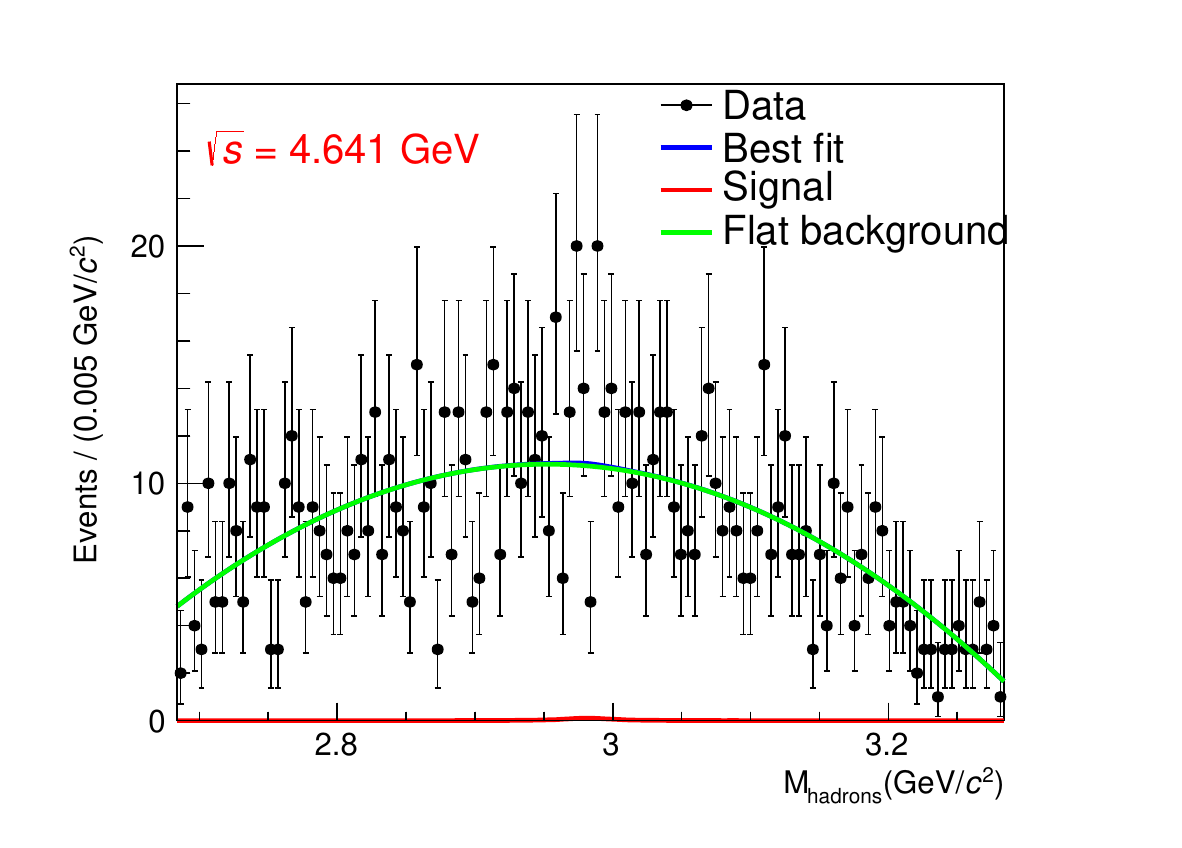}
  \captionsetup{skip=-10pt,font=large}
\end{subfigure}
  \begin{subfigure}{0.24\textwidth}
  \centering
  \includegraphics[width=\textwidth]{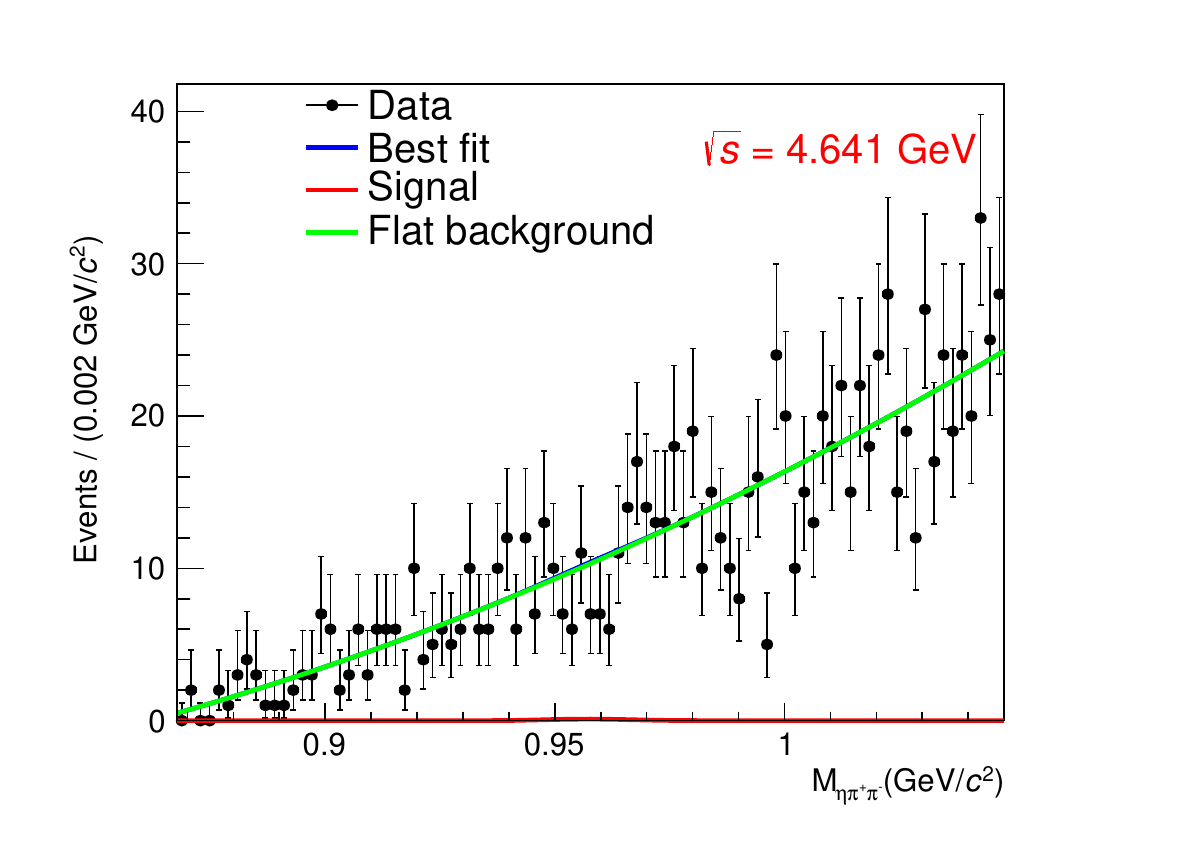}
  \captionsetup{skip=-10pt,font=large}
\end{subfigure}
\begin{subfigure}{0.24\textwidth}
  \centering
  \includegraphics[width=\textwidth]{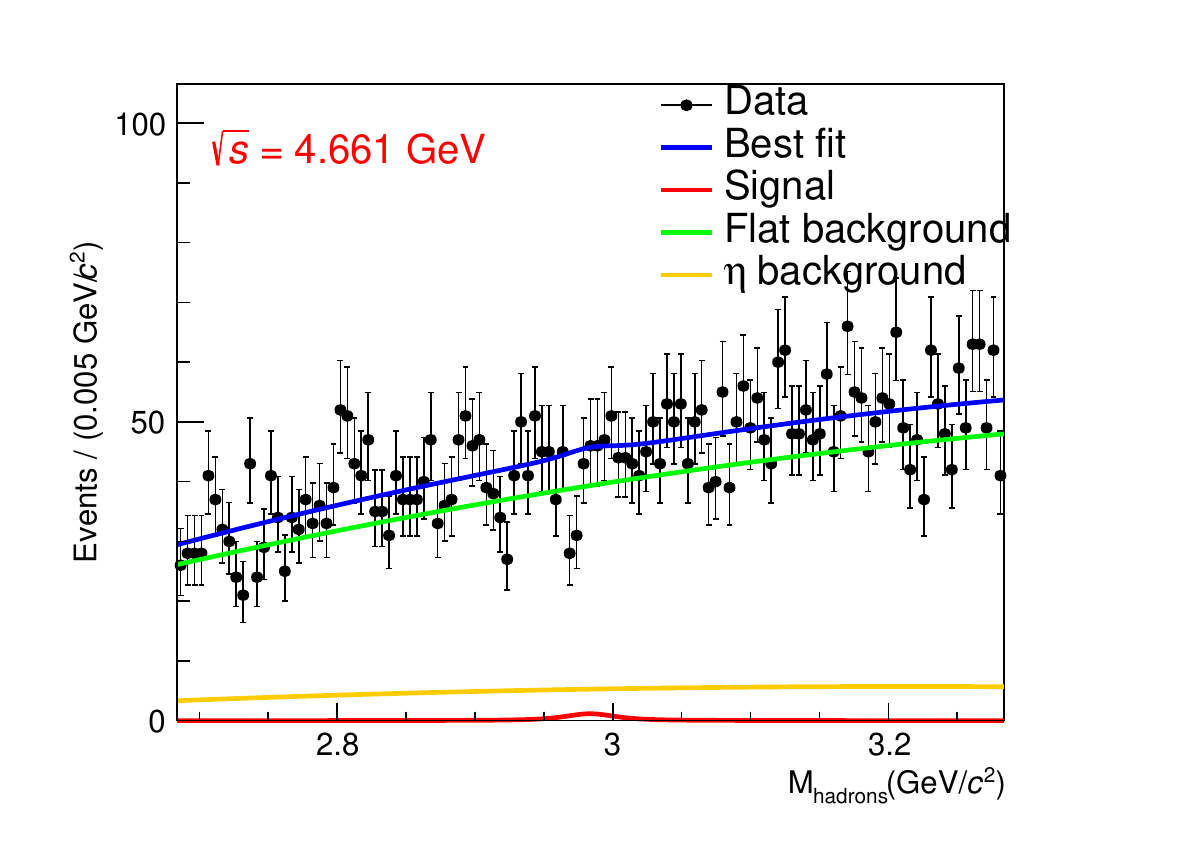}
  \captionsetup{skip=-10pt,font=large}
\end{subfigure}
\begin{subfigure}{0.24\textwidth}
  \centering
  \includegraphics[width=\textwidth]{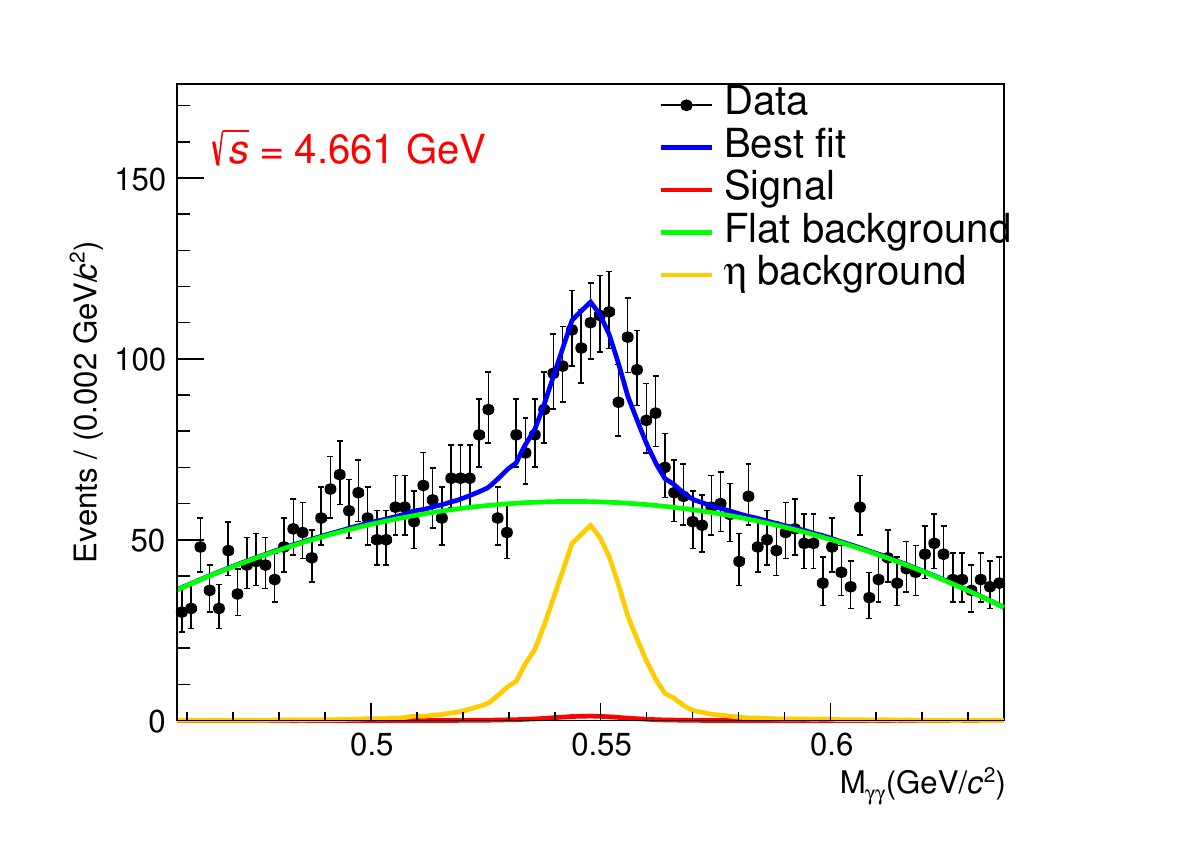}
  \captionsetup{skip=-10pt,font=large}
\end{subfigure}
 \begin{subfigure}{0.24\textwidth}
  \centering
  \includegraphics[width=\textwidth]{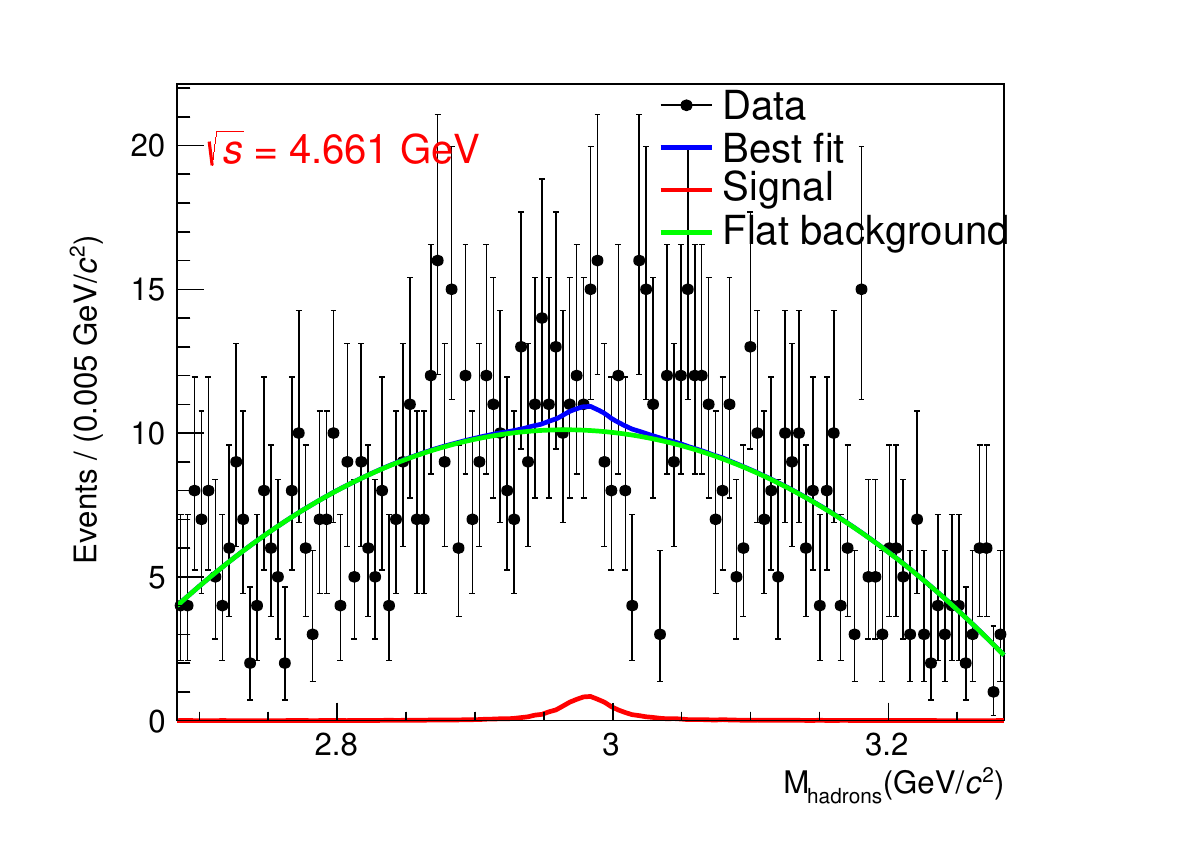}
  \captionsetup{skip=-10pt,font=large}
\end{subfigure}
  \begin{subfigure}{0.24\textwidth}
  \centering
  \includegraphics[width=\textwidth]{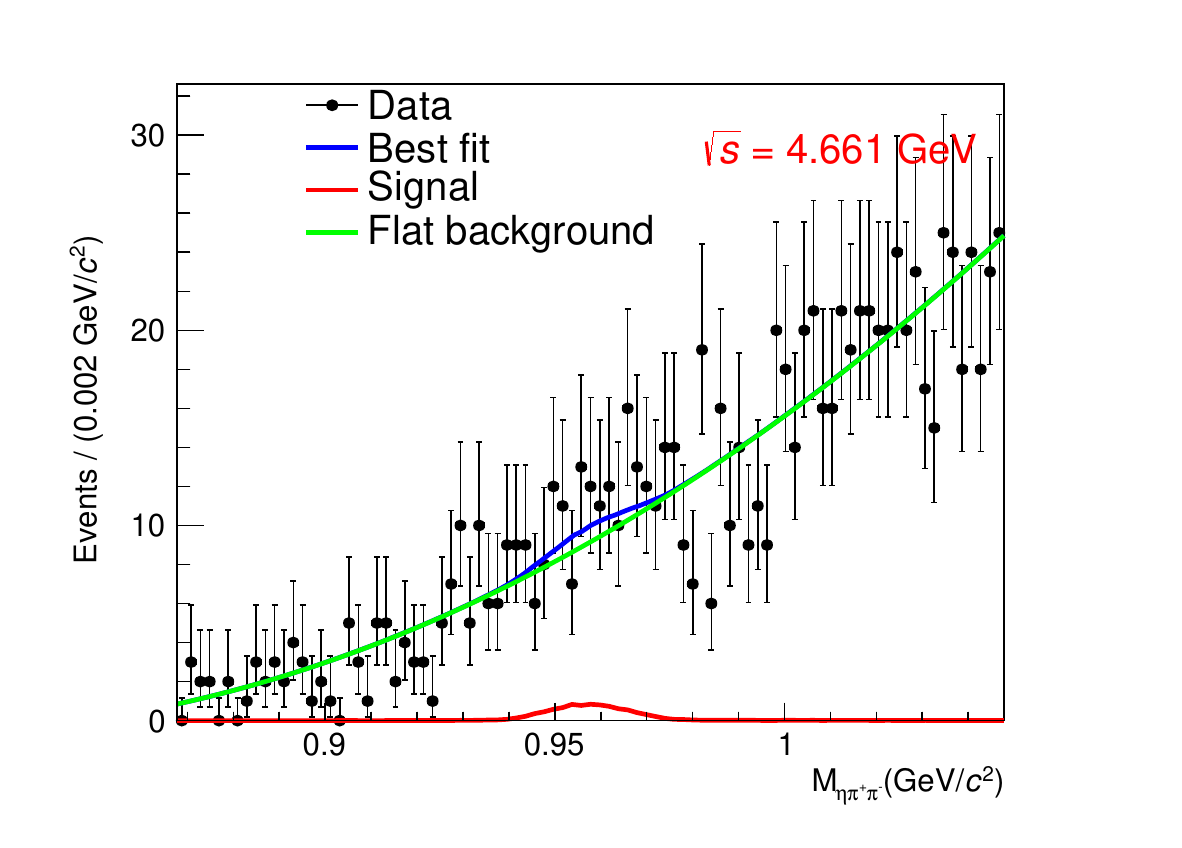}
  \captionsetup{skip=-10pt,font=large}
\end{subfigure}
\captionsetup{justification=raggedright}
\caption{Fits to the invariant mass distributions of (Left)(Middle-Right)$M(hadrons)$, (Middle-Left)$M(\gamma\gamma)$ and (Right)$M(\eta\pi^{+}\pi^{-})$ at $\sqrt(s)=4.574-4.661$~GeV.}
\label{fig:fit4}
\end{figure*}

\section{\label{sec:ACKNOWLEDGMENT}ACKNOWLEDGMENT}

The BESIII Collaboration thanks the staff of BEPCII (https://cstr.cn/31109.02.BEPC) and the IHEP computing center for their strong support. This work is supported in part by National Key R\&D Program of China under Contracts Nos. 2020YFA0406300, 2020YFA0406400, 2023YFA1606000, 2023YFA1606704; National Natural Science Foundation of China (NSFC) under Contracts Nos. 11635010, 11935015, 11935016, 11935018, 12025502, 12035009, 12035013, 12061131003, 12192260, 12192261, 12192262, 12192263, 12192264, 12192265, 12221005, 12225509, 12235017, 12361141819; the Chinese Academy of Sciences (CAS) Large-Scale Scientific Facility Program; CAS under Contract No. YSBR-101; 100 Talents Program of CAS; The Institute of Nuclear and Particle Physics (INPAC) and Shanghai Key Laboratory for Particle Physics and Cosmology; Agencia Nacional de Investigación y Desarrollo de Chile (ANID), Chile under Contract No. ANID PIA/APOYO AFB230003; German Research Foundation DFG under Contract No. FOR5327; Istituto Nazionale di Fisica Nucleare, Italy; Knut and Alice Wallenberg Foundation under Contracts Nos. 2021.0174, 2021.0299; Ministry of Development of Turkey under Contract No. DPT2006K-120470; National Research Foundation of Korea under Contract No. NRF-2022R1A2C1092335; National Science and Technology fund of Mongolia; National Science Research and Innovation Fund (NSRF) via the Program Management Unit for Human Resources \& Institutional Development, Research and Innovation of Thailand under Contract No. B50G670107; Swedish Research Council under Contract No. 2019.04595; U. S. Department of Energy under Contract No. DE-FG02-05ER41374



\clearpage
\bibliographystyle{apsrev4-2}
\bibliography{reference}

\end{document}